\documentclass{article}

\usepackage{arxiv}
\usepackage[utf8]{inputenc} 
\usepackage[T1]{fontenc}    
\usepackage{url}            
\usepackage{booktabs}       
\usepackage{amsfonts}       
\usepackage{nicefrac}       
\usepackage{microtype}      

\usepackage{amsmath,amssymb}
\usepackage[colorlinks,linkcolor=red,citecolor=blue,urlcolor=blue]{hyperref}

\usepackage{anyfontsize}

\usepackage{cite}
\usepackage{booktabs} 
\usepackage{array} 
\usepackage{tikz} 
\usepackage{color}
\usepackage{graphicx}
\usepackage{bm}
\usepackage[colorlinks,linkcolor=red,citecolor=blue,urlcolor=blue]{hyperref}
\usepackage{pifont} 
\usepackage{tabularx}
\newcolumntype{C}[1]{>{\centering\arraybackslash}p{#1}}
\usepackage{subcaption}
\usepackage{caption} 
\usepackage{tabularx}
\usepackage{bm}

\usepackage{enumitem}

\usepackage{subcaption}

\usepackage{lastpage,fancyhdr,graphicx}
\usepackage{epstopdf}
\fancyheadoffset[L]{2.25in}
\fancyfootoffset[L]{2.25in}
\allowdisplaybreaks  

\usepackage{listings}
\usepackage{xcolor}
\definecolor{whitesmoke}{rgb}{0.96, 0.96, 0.96}
\newlength\savedwidth

\usepackage[aboveskip=5pt,labelfont=bf,labelsep=period]{caption}

\makeatletter
\renewcommand{\@biblabel}[1]{\quad#1.}
\makeatother

\makeatletter
\newcommand*{\textoverline}[1]{$\overline{\hbox{#1}}\m@th$}
\makeatother

\newcommand{\CC}[1]{\textcolor{black}{#1}}
\newcommand{\REV}[1]{\textcolor{black}{#1}}

\usepackage{authblk}  

\title{\large{GPU-based compressible lattice Boltzmann simulations on non-uniform grids using standard C++ parallelism: From best practices to aerodynamics, aeroacoustics and supersonic flow simulations}}

\author{Christophe Coreixas\textsuperscript{1,2,}\thanks{Corresponding author: christophecoreixas@uic.edu.cn, christophe.coreixas@unige.ch}~}
\author{Jonas Latt\textsuperscript{2,}\thanks{Email: jonas.latt@unige.ch}}
\affil{\textsuperscript{1}Institute for Advanced Study, Beijing Normal - Hong Kong Baptist University, Zhuhai, China\\
\textsuperscript{2}Computer Science Department, University of Geneva, Carouge, Switzerland}

\date{}

\begin{document}
\maketitle

\begin{abstract}

Despite decades of research, creating accurate, robust, and efficient lattice Boltzmann methods (LBM) on non-uniform grids with seamless GPU acceleration remains challenging. This work introduces a novel strategy to address this challenge by integrating simple yet effective components: (1) parallel algorithms in modern C++, (2) conservative cell-centered grid refinement, (3) local boundary conditions, and (4) robust collision models. Our framework supports multiple lattices (D2Q9, D2Q13, D2Q21, D2Q37\CC{, D3Q27, etc}) tailored to various flow conditions. It includes collision models with polynomial and numerical equilibria, a second distribution for polyatomic behavior, a Jameson-like shock sensor, and generalizes Rohde's refinement strategy.

The framework's accuracy and robustness is validated across diverse benchmarks, including lid-driven cavity flows, Aeolian noise, 30P30N airfoil aerodynamics, inviscid Riemann problems, and viscous flows past a NACA airfoil in transonic and supersonic regimes. Modern C++ further enables our framework to reach GPU-native performance, while ensuring high portability, modularity, and ease of implementation. Notably, weakly compressible LBMs achieve state-of-the-art GPU efficiency on non-uniform grids, while fully compressible LBMs benefit from acceleration equivalent to thousands of CPU cores in the most compute-intensive cases. Our advanced performance models incorporate neighbor-list and asynchronous time-stepping effects, providing new insights into the performance decomposition of LB simulations on non-uniform grids.

Overall, this study sets a new standard for portable, tree-based LBMs, demonstrating that a combination of well-chosen components can achieve high performance, accuracy, and robustness across various flow conditions. As a final proof-of-concept, adaptive mesh refinement is proposed for subsonic and supersonic applications.
\end{abstract}

\keywords{Compressible LBM \and C++ based GPU Acceleration \and Conservative Grid Refinement \and Aerodynamics \and Aeroacoustics \and Supersonic}

\section{Introduction~\label{sec:intro}}

Over the past thirty years, the lattice Boltzmann method (LBM) has proven highly effective in computational fluid dynamics (CFD), becoming a strong competitor to traditional Navier-Stokes solvers for simulating subsonic unsteady flows around realistic geometries~\cite{MANOHA_AIAA_2846_2015,ASTOUL_PhD_2021,DEGRIGNY_PhD_2021,ANIELLO_CF_241_2022,KIRIS_AIAA_WMLES_LBM_2022}. This is explained by the following three reasons. 
First, LBM usually relies on Cartesian Octree grids, which drastically reduces the complexity of meshing realistic geometries. Second, LBMs dedicated to low-speed simulations have excellent spectral properties, particularly reduced numerical dissipation for acoustic wave propagation~\cite{MARIE_JCP_228_2009,WISSOCQ_PRE_102_2020,SUSS_JCP_485_2023}. Third, the number of floating point operations required to advance the LB scheme in time is very low. This makes LBM particularly suitable for high-performance computing on both CPU and GPU architectures. Especially on the latter platforms, the computational expense becomes so low that memory bandwidth becomes the only limitation to its performance~\cite{HOLZER_IJHPCA_2021,HOLZER_PhD_2025,LATT_PLOSONE_16_2021,LATT_ARXIV_09242_2025}.

Nonetheless, developing a GPU-accelerated Octree-based LB solver for industrial CFD is far from trivial. Assuming the LB solver already meets the industry's criteria for accuracy, robustness and efficiency, two additional aspects remain to be addressed: (1) implementing the tree structure and its traversal in a parallel and thread-safe manner, and (2) finding an accurate and robust refinement strategy for the LB solver that also minimizes memory accesses.
For the first point, it is crucial to ensure proper synchronization of all threads during the tree construction, which is complex due to the recursive nature of its structure. Additionally, maintaining spatial data locality is essential for ensuring efficient memory access patterns, which in turn lead to near-peak performance. As many grid refinement strategies rely on non-local space or time interpolations, this significantly restricts viable options for GPU-accelerated LB solvers.

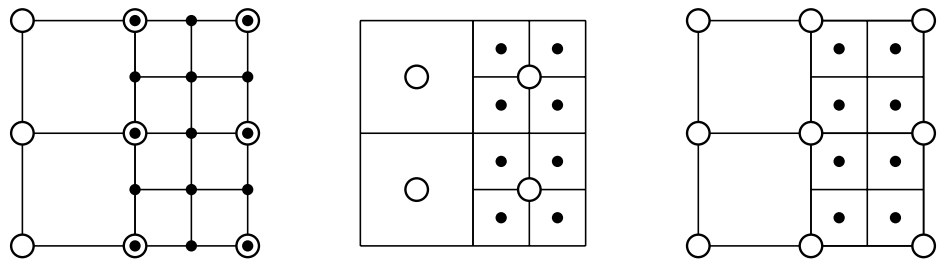
\begin{figure}[b]
  \centering
  \resizebox{0.75\textwidth}{!}{
    \begin{tikzpicture}[scale=1]
        \draw [thick, black] (0,0) grid[step=2.] (2,4);
        \draw [thick, black] (2,0) grid[step=1.] (4,4);
        \foreach \x in {0,2,4}
            \foreach \y in {0,2,4}{
                \draw [black, very thick, fill=white] (\x,\y) circle (0.2);
            }
        \foreach \x in {2,3,4}
            \foreach \y in {0,1,2,3,4}{
                \fill [fill=black] (\x,\y) circle (0.1);
            }
        \begin{scope}[shift={(6,0)}]
            \draw [thick, black] (0,0) grid[step=2.] (2,4);
            \draw [thick, black] (2,0) grid[step=1.] (4,4);
            \foreach \x in {1, 3}
                \foreach \y in {1,3}{
                    \draw [black, very thick, fill=white] (\x,\y) circle (0.2);
                }
            \foreach \x in {2.5,3.5}
                \foreach \y in {0.5,1.5,2.5,3.5}{
                    \fill (\x,\y) circle (0.1);
                }
        \end{scope}
        \begin{scope}[shift={(12,0)}]
            \draw [thick, black] (2,0) grid[step=1.] (4,4);
            \foreach \x in {2.5,3.5}
                \foreach \y in {0.5,1.5,2.5,3.5}{
                    \fill (\x,\y) circle (0.1);
                }
            \draw [thick, black] (0,0) grid[step=2.] (4,4);
            \foreach \x in {0,2,4}
                \foreach \y in {0,2,4}{
                    \draw [black, very thick, fill=white] (\x,\y) circle (0.2);
                }
        \end{scope}
    \end{tikzpicture}
    }
\caption{2D illustration of the three families of grid refinement strategies available in the LB context~\cite{SCHUKMANN_FLUIDS_8_2023,SCHUKMANN_FLUIDS_10_2025}. From left to right: vertex-centered (VC), cell-centered (CC), and combined (CM) approaches. Filled black circles and unfilled circles indicate locations where fine and coarse LB data is stored, respectively. When circles overlap, as in the vertex-centered approach, LB data is available at both coarse and fine levels.}
\label{fig:vc_cc_co}
\end{figure}

In more detail, the LB literature typically considers three main grid refinement strategies~\cite{SCHUKMANN_FLUIDS_8_2023,SCHUKMANN_FLUIDS_10_2025}. They are respectively known as \emph{vertex-centered}~\cite{FILIPPOVA_JCP_147_1998,YU_IJNMF_39_2002,DUPUIS_PRE_67_2003,LAGRAVA_JCP_231_2012,GENDRE_PRE_96_2017, ASTOUL_JCP_418_2020,ASTOUL_JCP_447_2021,FENG_PRE_101_2020,LYU_PoF_35_2023} (also called \emph{cell-vertex}), \emph{cell-centered}~\cite{ROHDE_IJNMF_51_2006,CHENb_PA_362_2006,LI_EPJST_171_2009,DEITERDING_NRMEFMX_2016,SCHORNBAUM_SIAM_40_2018,
LI_PhD_2011,FARES_AIAA_0952_2014,RIBEIRO_AIAA_1438_2017,GONZALEZMARTINO_AIAA_3919_2018,
KOPRIVA_AIAA_3929_2019,LI_AIAAJ_J063199_2023,YU_JCP_228_2009,WERNER_IJNMF_93_2021,WATANABE_CPC_264_2021}, and \emph{combined} approaches~\cite{GEIER_EPJST_171_2009b,FAR_CMWA_79_2020,FEUCHTER_CF_224_2021}. 
Hereafter, they are referred to as VC, CC and CM grid refinement strategies, and illustrated on Figure~\ref{fig:vc_cc_co}.
Without delving too deeply into technical details, the reconstruction of missing information at the interface between two refinement levels requires tackling two primary challenges: (i) violation of exact conservation rules and (ii) improper rescaling of LB data. Most grid refinement algorithms based on VC and CM approaches pay much attention to the second issue, often overlooking accuracy problems caused by improper mass, momentum, and energy conservation at the interface. Conversely, CC strategies, such as the one proposed by \CC{Rohde} et al.~\cite{ROHDE_IJNMF_51_2006}, automatically enforce conservation rules.
The conservation issue was only recently investigated by Astoul et al.~\cite{ASTOUL_JCP_447_2021,ASTOUL_PhD_2021} for VC algorithms, and more broadly by Schukmann et al.~\cite{SCHUKMANN_FLUIDS_8_2023,SCHUKMANN_FLUIDS_10_2025} for CC, VC, and CM algorithms. Aggregating the results of these studies, it appears that ensuring exact conservation rules at the interface significantly improves accuracy. This property brings back \CC{Rohde}'s approach~\cite{ROHDE_IJNMF_51_2006} to the forefront of discussions, as, additionally to its natural conservation properties, the implementation of the model is particularly straightforward and does not require any explicit interpolation scheme, or handling of special cases near boundaries. Furthermore, it can easily be extended to any velocity discretization, which is essential for simulating supersonic flows with high-order LBMs.. All of this leads to the presumption, further substantiated by the results presented in this article, that \CC{Rohde}'s approach and its high-order extension are very good candidates for the efficient implementation of general LB solvers on non-uniform grids.  

Once the grid refinement strategy has been chosen, there exist several ways to accelerate the resulting LB solver on dedicated hardware. These include hardware-specific and low-level coding languages (CUDA~\cite{NICKOLLS_QUEUE_2008}, HIP~\cite{HIP_WEBSITE}), ``pragmas-based'' local acceleration strategies (OpenACC~\cite{WIENKE_EUROPAR_2012}, OpenMP~\cite{CHANDRA_Book_2001}), dedicated libraries and frameworks (OpenCL~\cite{MUNSHI_IEEE_2009}, SYCL~\cite{ALPAY_PIWOCL_2020}, Thrust~\cite{HWU_Chapter_2012}, Kokkos~\cite{CARTEREDWRADS_JPDC_74_2014}), and ISO languages like C++, Fortran, and Python~\cite{LARKIN_GTC_2022,LARKIN_GTC_2024}. Relying on ISO languages offers the advantage of simplicity, modularity and portability, hence providing a uniform way to develop code for both CPUs and accelerated hardware such as GPU cards. In that context, switching to different target hardware is typically achieved by a simple change of compiler or compilation flag~\cite{LARKIN_GTC_2022,LARKIN_GTC_2024}. This strategy has recently been used to accelerate lattice Boltzmann~\cite{LATT_PLOSONE_16_2021,THYAGARAJAN_PoF_35_2023,LATT_ARXIV_09242_2025}, finite difference~\cite{COREIXAS_HIFILED_LBFDGPU_2022,CAPLAN_SC_2025,CAPLAN_ARXIV_2501_2025}, and discrete element methods on NVIDIA GPU cards~\cite{MAGGIOAPRILE_Master_2023}. Interestingly, the same strategy was applied to accelerate mini-apps~\cite{LIN_IEEE_2022,LIN_ARXIV_02680_2024} and fluid solvers~\cite{CAPLAN_SC_2025,CAPLAN_ARXIV_2501_2025} using GPU cards from different vendors such as AMD and Intel. This confirms that relying on ISO languages, such as C++, is a viable option for accelerating CFD tools while ensuring their portability over a wide range of hardware.

Building on these considerations, we propose a general framework for GPU-accelerated LBMs on non-uniform grids. Our grid refinement employs the CC approach by Rohde et al.~\cite{ROHDE_IJNMF_51_2006}, which offers efficient local memory access, ensures conservation rules, and seamlessly extends to high-order velocity discretizations in any number of physical dimensions. GPU acceleration is achieved thanks to the parallel algorithms available in the standard library (STL) of C++, similarly to the LB framework STLBM dedicated to fluid simulations on uniform grids~\cite{LATT_PLOSONE_16_2021} and the open-source solver Palabos~\cite{LATT_ARXIV_09242_2025}. 
As a proof of concept, we also present preliminary results on CPU-based adaptive mesh refinement (AMR) for both subosnic and supersonic flow simulations. In that context, AMR is used to dynamically adjusts mesh resolution based on either prescribed constraints or local flow features, thus optimizing computational resources by refining only where necessary. 

The paper is organized as follows. Sections~\ref{sec:LBM} and~\ref{sec:grid_refinement} provide a brief overview of LBMs and a detailed explanation of our grid-refinement strategy. Section~\ref{sec:implementation_and_c++} discusses implementation details, focusing on the use of parallel algorithms for the portable execution of tree-based LBMs on both CPUs and GPUs. Validation of the framework is presented in Sections~\ref{sec:validation_weakly_compressible} and~\ref{sec:validation_fully_compressible}, using well-established benchmarks for low-speed and high-speed flows in 2D, respectively. Performance analyses are conducted in Section~\ref{sec:perfo} for both weakly and fully compressible LBMs. 
Section~\ref{sec:amr} presents preliminary results on adaptive mesh refinement for subsonic and supersonic flow simulations using high-order LBMs.
\CC{A summary of our findings is proposed in Section~\ref{sec:findings}. Future research directions are addressed in Section~\ref{sec:perspectives} and illustrated with the 3D simulation of the turbulent flow past a sphere.}

Before diving into the core of our work, we emphasize that one of our goal is to establish best practices for developing portable LBMs for subsonic and supersonic flows on non-uniform grids. With this idea in mind, we provide detailed insights into key choices regarding numerical models, grid refinement, coding strategies, and validation cases. While our focus is on the latter three aspects, the numerical methods used in the bulk of the simulation domain remain relatively standard, with technical details deferred to~\ref{app:lbm}. Additional benchmarks and performance analyses supporting our findings are available in~\ref{app:complementary_data}.
Another goal of this work is to provide detailed answers to the following key questions:
\begin{enumerate}
\item Can standard LBMs based on appropriate collision models, CC grid refinement strategy, and simple boundary conditions, accurately simulate fluid flows across Reynolds numbers from 100 to over $10^6$, and for local Mach numbers up to 0.5-0.6?
\item Can our extension of the grid refinement strategy to high-order LBMs achieve similar levels of accuracy and robustness for transonic and supersonic regimes in both inviscid and viscous conditions?
\item How can performance models be derived to account for the impact of the grid refinement strategy (e.g., neighbor lists and time asynchronicity)? What is the corresponding theoretical performance loss compared to LBMs on matrix-like grids?
\item How does the performance of GPU-accelerated LBMs implemented with modern C++ compare to GPU-native CUDA implementations? How close is our framework to achieving peak GPU performance? What speedups can be achieved relative to CPU-based implementations?
\end{enumerate}

\section{General background on lattice Boltzmann methods~\label{sec:LBM}}

In practice, LBM solves a simplified version of the Boltzmann equation, namely, the discrete velocity Boltzmann equation (DVBE)
\begin{equation}
\partial_t f_i + \bm{\xi}_i\cdot\bm\nabla f_i = \Omega_{f_i}. 
\label{eq:LBE}
 \end{equation}
The latter describes the space and time evolution of $V$ groups of \emph{fictive} particles through their distribution function $f_i$ ($i\in \{0,...,V-1\}$), which can be interpreted as the \CC{mass density} of \emph{fictive} particles located at $(\bm{x},t)$ and propagating at a mesoscopic velocity $\bm{\xi}_i$. These particles are assumed to either collide or stream on a Cartesian grid, leading to the efficient collide-and-stream algorithm~\cite{DELLAR_CMA_65_2013}
\begin{equation}
f_i(\bm{x}+\bm{\xi}_i\Delta t, t+\Delta t) = f_i^{eq}(\bm{x}, t) + \Delta t (1-1/\tau) f_i^{neq}(\bm{x}, t), 
\label{eq:CollAndStream}
 \end{equation}     
where the right-hand side (collision) is computed locally whereas the left-hand side (streaming) is non-local by nature. For the sake of illustration, the collision term is modelled using a relaxation approach, known as the BGK approximation, and named after its authors Bhatnagar, Gross, and Krook~\cite{BHATNAGAR_PR_94_1954}. $f_i^{eq}$ is the discrete version of the equilibrium distribution function of Maxwell-Boltzmann~\cite{SHAN_PRL_80_1998,SHAN_JFM_550_2006}, used for the relaxation of populations during collisions, and $f_i^{neq}$ is defined as the non-equilibrium population $f_i-f_i^{eq}$. $\tau$ is the relaxation time related to diffusivity coefficients such as the shear viscosity $\nu$, the bulk viscosity $\nu_b$ and the thermal diffusivity $\nu_T = \nu / \mathrm{Pr}$ with $\mathrm{Pr}$ being the Prandtl number. For the BGK collision, the Prandtl number is fixed to unity but this defect can easily be corrected by introducing additional relaxation parameters~\cite{GUO_Book_2013}. $\Delta t$ is the time step, closely related to the space step $\Delta x$, through the coupling between the geometric and the velocity spaces using an on-grid condition for the velocity discretization of the Boltzmann equation~\cite{PHILIPPI_PRE_73_2006}.

The BGK-LBM is known to reach its stability limit when diffusivity coefficients are too low. This notably happens when simulating high-Reynolds number flows in under-resolved conditions. To solve these stability issues, a number of collision models have been proposed over the years~\cite{COREIXAS_PRE_100_2019}. While adopting a more robust collision model is sufficient for the simulation of low-speed flows, things are more complicated in the transonic and supersonic regimes. In that case, relying on coupled LB-LB methods~\cite{LIa_PRE_76_2007,JAMMALAMADAKA_AIAA_3055_2020,LATT_RSTA_378_2020,COREIXAS_PoF_32_2020}, hybrid LB-FD formulations~\cite{NIE_AIAA_139_2009,FARES_AIAA_0952_2014,RENARD_PhD_2021} or more robust numerical discretizations of the DVBE~\cite{GUO_Book_2013,GUO_AA_3_2021} becomes the new standard for the simulation of fully compressible flows in realistic (low viscosity) conditions. 

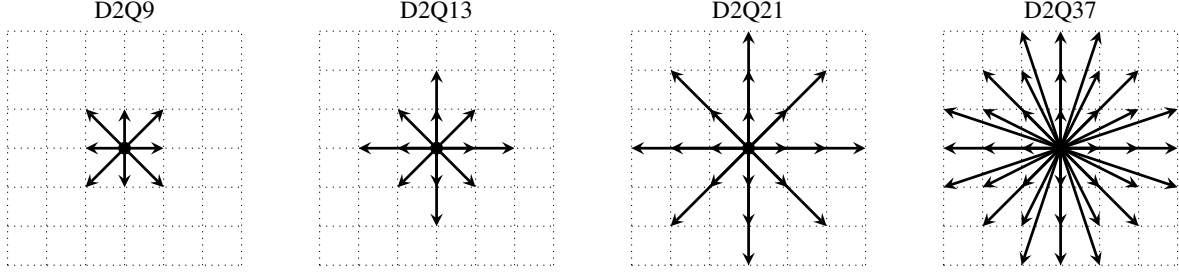
\begin{figure}
  \centering
  \resizebox{0.95\textwidth}{!}{
    \begin{tikzpicture}  
	\begin{scope}[xshift=0cm,yshift=0cm,scale=0.5]
	\node at (0,3.5) {\small D2Q9};
	\draw [black,dotted] (-3,-3) grid (3,3);
	\draw [fill=black] (0.,0.) circle[radius=1.5mm];
	\draw [black, line width=0.35mm,>=stealth, ->] (0,0) -- (1,1);
	\draw [black, line width=0.35mm,>=stealth, ->] (0,0) -- (1,0);
	\draw [black, line width=0.35mm,>=stealth, ->] (0,0) -- (1,-1);
	\draw [black, line width=0.35mm,>=stealth, ->] (0,0) -- (0,1);
	\draw [black, line width=0.35mm,>=stealth, ->] (0,0) -- (0,-1);
	\draw [black, line width=0.35mm,>=stealth, ->] (0,0) -- (-1,1);
	\draw [black, line width=0.35mm,>=stealth, ->] (0,0) -- (-1,0);
	\draw [black, line width=0.35mm,>=stealth, ->] (0,0) -- (-1,-1);
	\end{scope}

	\begin{scope}[xshift=4cm,yshift=0cm,scale=0.5]
	\node at (0,3.5) {\small D2Q13};
	\draw [black,dotted] (-3,-3) grid (3,3);
	\draw [fill=black] (0.,0.) circle[radius=1.5mm];
	\draw [black, line width=0.35mm,>=stealth, ->] (0,0) -- (2,0);
	\draw [black, line width=0.35mm,>=stealth, ->] (0,0) -- (1,1);
	\draw [black, line width=0.35mm,>=stealth, ->] (0,0) -- (1,0);
	\draw [black, line width=0.35mm,>=stealth, ->] (0,0) -- (1,-1);
	\draw [black, line width=0.35mm,>=stealth, ->] (0,0) -- (0,2);
	\draw [black, line width=0.35mm,>=stealth, ->] (0,0) -- (0,1);
	\draw [black, line width=0.35mm,>=stealth, ->] (0,0) -- (0,-1);
	\draw [black, line width=0.35mm,>=stealth, ->] (0,0) -- (0,-2);
	\draw [black, line width=0.35mm,>=stealth, ->] (0,0) -- (-2,0);
	\draw [black, line width=0.35mm,>=stealth, ->] (0,0) -- (-1,1);
	\draw [black, line width=0.35mm,>=stealth, ->] (0,0) -- (-1,0);
	\draw [black, line width=0.35mm,>=stealth, ->] (0,0) -- (-1,-1);
	\end{scope}

	\begin{scope}[xshift=8cm,yshift=0cm,scale=0.5]
	\node at (0,3.5) {\small D2Q21};
	\draw [black,dotted] (-3,-3) grid (3,3);
	\draw [fill=black] (0.,0.) circle[radius=1.5mm];
	\draw [black, line width=0.35mm,>=stealth, ->] (0,0) -- (3,0);
	\draw [black, line width=0.35mm,>=stealth, ->] (0,0) -- (2,2);
	\draw [black, line width=0.35mm,>=stealth, ->] (0,0) -- (2,0);
	\draw [black, line width=0.35mm,>=stealth, ->] (0,0) -- (2,-2);
	\draw [black, line width=0.35mm,>=stealth, ->] (0,0) -- (1,1);
	\draw [black, line width=0.35mm,>=stealth, ->] (0,0) -- (1,0);
	\draw [black, line width=0.35mm,>=stealth, ->] (0,0) -- (1,-1);
	\draw [black, line width=0.35mm,>=stealth, ->] (0,0) -- (0,3);
	\draw [black, line width=0.35mm,>=stealth, ->] (0,0) -- (0,2);
	\draw [black, line width=0.35mm,>=stealth, ->] (0,0) -- (0,1);
	\draw [black, line width=0.35mm,>=stealth, ->] (0,0) -- (0,-1);
	\draw [black, line width=0.35mm,>=stealth, ->] (0,0) -- (0,-2);
	\draw [black, line width=0.35mm,>=stealth, ->] (0,0) -- (0,-3);
	\draw [black, line width=0.35mm,>=stealth, ->] (0,0) -- (-3,0);
	\draw [black, line width=0.35mm,>=stealth, ->] (0,0) -- (-2,2);
	\draw [black, line width=0.35mm,>=stealth, ->] (0,0) -- (-2,0);
	\draw [black, line width=0.35mm,>=stealth, ->] (0,0) -- (-2,-2);
	\draw [black, line width=0.35mm,>=stealth, ->] (0,0) -- (-1,1);
	\draw [black, line width=0.35mm,>=stealth, ->] (0,0) -- (-1,0);
	\draw [black, line width=0.35mm,>=stealth, ->] (0,0) -- (-1,-1);
	\end{scope}

	\begin{scope}[xshift=12cm,yshift=0cm,scale=0.5]
	\node at (0,3.5) {\small D2Q37};
	\draw [black,dotted] (-3,-3) grid (3,3);
	\draw [fill=black] (0.,0.) circle[radius=1.5mm];
	\draw [black, line width=0.35mm,>=stealth, ->] (0,0) -- (3,1);
	\draw [black, line width=0.35mm,>=stealth, ->] (0,0) -- (3,0);
	\draw [black, line width=0.35mm,>=stealth, ->] (0,0) -- (3,-1);
	\draw [black, line width=0.35mm,>=stealth, ->] (0,0) -- (2,2);
	\draw [black, line width=0.35mm,>=stealth, ->] (0,0) -- (2,1);
	\draw [black, line width=0.35mm,>=stealth, ->] (0,0) -- (2,0);
	\draw [black, line width=0.35mm,>=stealth, ->] (0,0) -- (2,-1);
	\draw [black, line width=0.35mm,>=stealth, ->] (0,0) -- (2,-2);
	\draw [black, line width=0.35mm,>=stealth, ->] (0,0) -- (1,3);
	\draw [black, line width=0.35mm,>=stealth, ->] (0,0) -- (1,2);
	\draw [black, line width=0.35mm,>=stealth, ->] (0,0) -- (1,1);
	\draw [black, line width=0.35mm,>=stealth, ->] (0,0) -- (1,0);
	\draw [black, line width=0.35mm,>=stealth, ->] (0,0) -- (1,-1);
	\draw [black, line width=0.35mm,>=stealth, ->] (0,0) -- (1,-2);
	\draw [black, line width=0.35mm,>=stealth, ->] (0,0) -- (1,-3);
	\draw [black, line width=0.35mm,>=stealth, ->] (0,0) -- (0,3);
	\draw [black, line width=0.35mm,>=stealth, ->] (0,0) -- (0,2);
	\draw [black, line width=0.35mm,>=stealth, ->] (0,0) -- (0,1);
	\draw [black, line width=0.35mm,>=stealth, ->] (0,0) -- (0,-1);
	\draw [black, line width=0.35mm,>=stealth, ->] (0,0) -- (0,-2);
	\draw [black, line width=0.35mm,>=stealth, ->] (0,0) -- (0,-3);
	\draw [black, line width=0.35mm,>=stealth, ->] (0,0) -- (-3,1);
	\draw [black, line width=0.35mm,>=stealth, ->] (0,0) -- (-3,0);
	\draw [black, line width=0.35mm,>=stealth, ->] (0,0) -- (-3,-1);
	\draw [black, line width=0.35mm,>=stealth, ->] (0,0) -- (-2,2);
	\draw [black, line width=0.35mm,>=stealth, ->] (0,0) -- (-2,1);
	\draw [black, line width=0.35mm,>=stealth, ->] (0,0) -- (-2,0);
	\draw [black, line width=0.35mm,>=stealth, ->] (0,0) -- (-2,-1);
	\draw [black, line width=0.35mm,>=stealth, ->] (0,0) -- (-2,-2);
	\draw [black, line width=0.35mm,>=stealth, ->] (0,0) -- (-1,3);
	\draw [black, line width=0.35mm,>=stealth, ->] (0,0) -- (-1,2);
	\draw [black, line width=0.35mm,>=stealth, ->] (0,0) -- (-1,1);
	\draw [black, line width=0.35mm,>=stealth, ->] (0,0) -- (-1,0);
	\draw [black, line width=0.35mm,>=stealth, ->] (0,0) -- (-1,-1);
	\draw [black, line width=0.35mm,>=stealth, ->] (0,0) -- (-1,-2);
	\draw [black, line width=0.35mm,>=stealth, ->] (0,0) -- (-1,-3);
	\end{scope}
  \end{tikzpicture}
}
\caption{Standard and high-order lattices considered in this work: D2Q9~\cite{KRUGER_Book_2017}, D2Q13~\cite{WEIMAR_PA_224_1996}, D2Q21~\cite{ZHANG_PRE_74_2006} and D2Q37~\cite{PHILIPPI_PRE_73_2006}.}
\label{fig:lattices}
\end{figure}

In this work, we are interested in evaluating the ability of purely LB solvers to handle a wide range of physical phenomena at various Reynolds numbers and compressibility regimes on non-uniform grids. This implies relying on both standard and high-order velocity discretizations (see Figure~\ref{fig:lattices}), as well as, different collision models. According to academic~\cite{MALASPINAS_ARXIV_2015,BROGI_JASA_142_2017,COREIXAS_PRE_96_2017,COREIXAS_RSTA_378_2020,WISSOCQ_PRE_102_2020,JACOB_JT_19_2018,CHEN_PS_95_2020} and more applied works~\cite{ASTOUL_PhD_2021,DEGRIGNY_PhD_2021,NGUYEN_PhD_2023,GIANOLI_PhD_2024,WERNER_PoF_36_2024,CHEN_PATENT_Collision_2015,CASALINO_AIAA_1834_2019,KHORRAMI_CEAS_AERO_2019,ROMANI_PhD_2022}, the recursive regularization (RR) approach and its adds-on provide a good tradeoff of robustness and accuracy to achieve our goal. For high-speed flow simulations, we will rely on the general LB-LB coupling strategy described in our previous works~\cite{LATT_RSTA_378_2020,COREIXAS_PoF_32_2020}, where the discrete equilibrium $f_i^{eq}$ is computed via a Newton-Raphson-like approach to recover any physics in a stable and accurate manner while greatly benefiting from GPU acceleration~\cite{THYAGARAJAN_PoF_35_2023}. More technical details are available in~\ref{app:lbm}.

\section{Conservative grid refinement strategy for standard and high-order LBMs~\label{sec:grid_refinement}}

\subsection{Overview~\label{subsec:grid_refinement_overview}}

In the LB literature, three families of grid refinement approaches are available: vertex-centered (VC), cell-centered (CC) and combined (CM). Their main features are discussed in more details below. 

Starting with VC approaches~\cite{FILIPPOVA_JCP_147_1998,YU_IJNMF_39_2002,DUPUIS_PRE_67_2003,LAGRAVA_JCP_231_2012,GENDRE_PRE_96_2017, ASTOUL_JCP_418_2020,ASTOUL_JCP_447_2021,FENG_PRE_101_2020,LYU_PoF_35_2023,DEGRIGNY_PhD_2021,NGUYEN_PhD_2023,GIANOLI_PhD_2024,WERNER_PoF_36_2024}, LB data is stored at vertices located at cell corners, resulting in partially co-located coarse and fine cells along grid transition interfaces. In these methods, space and time interpolations are required to compute missing data, leading to an increased number of memory accesses. Although VC approaches are not natively conservative, recent developments aim to address this issue~\cite{ASTOUL_JCP_418_2020,ASTOUL_JCP_447_2021}. 
A major limitation of VC approaches is their lack of generality. The asymmetry in the treatment of interface cells and the reliance on interpolation require an exhaustive enumeration of geometric corner cases, which grows with the number of dimensions.
For instance, there is no single best way to adjust interpolation stencils when a refinement interface intersects a boundary condition or when moving from 2D to 3D simulations. As a result, determining the optimal combination of interpolation schemes quickly becomes a significant challenge.

Continuing with CC strategies, they propose to store data at the cell center and, as such, do not have co-located data. Contrary to VC approaches, the reconstruction of missing information is done in a finite-volume fashion, ensuring an exact implementation of conservation laws by balancing opposing fluxes going through cell interfaces. Their implementation is general and does not require any adjustments, regardless of the number of dimensions or the shape of the grid refinement interface, even if it intersects a boundary condition. Like VC approaches, CC-based LBMs have been used for a wide variety of applications including low-speed~\cite{ROHDE_IJNMF_51_2006,CHENb_PA_362_2006,LI_EPJST_171_2009,DEITERDING_NRMEFMX_2016,SCHORNBAUM_SIAM_40_2018}, high-speed~\cite{LI_PhD_2011,FARES_AIAA_0952_2014,RIBEIRO_AIAA_1438_2017,GONZALEZMARTINO_AIAA_3919_2018,KOPRIVA_AIAA_3929_2019,LI_AIAAJ_J063199_2023}, and multiphysics flows~\cite{YU_JCP_228_2009,WERNER_IJNMF_93_2021,WATANABE_CPC_264_2021}.

Finally, CM approaches store data at both the center and corners of fluid cells~\cite{GEIER_EPJST_171_2009b,FAR_CMWA_79_2020,FEUCHTER_CF_224_2021}, resulting in a non-co-located data layout. While this design was originally introduced to simplify spatial interpolations by enabling local computations of missing information, it can also be implemented using a CC arrangement~\cite{EITELAMOR_CF_75_2013}. Regardless of the implementation, the reliance on interpolations renders the algorithm non-conservative, presenting significant challenges, particularly when a refinement interface intersects boundary conditions. These limitations likely contribute to the limited adoption of CM approaches within the LB community.

\subsection{Present grid refinement framework~\label{subsec:grid_refinement_framework}}

Based on the comparative study proposed by Schukmann et al.~\cite{SCHUKMANN_FLUIDS_8_2023,SCHUKMANN_FLUIDS_10_2025}, and discussed in~\ref{app:covo_schukmann}, the CC approach proposed by Rohde et al.~\cite{ROHDE_IJNMF_51_2006} gathers numerous advantages in terms of efficiency, conservation rules and generality for computational aeroacoustics. In this work, we embed and further extend this approach in a fully featured computational framework for complex flows, demonstrating that it is both accurate and robust enough to simulate a broad range of physical phenomena through the use of various velocity discretizations and collision models.

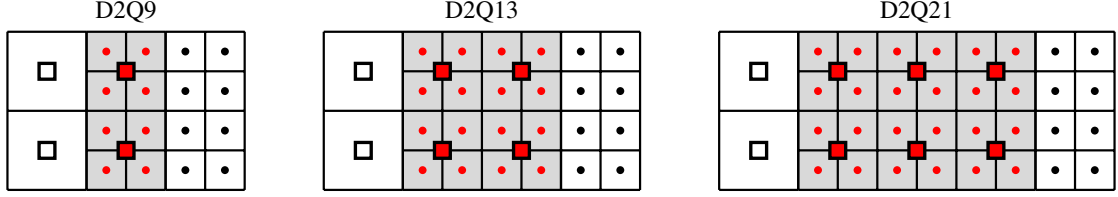
\begin{figure}
  \centering
  \resizebox{0.9\textwidth}{!}{
    \begin{tikzpicture}[scale=0.5]
        \begin{scope}[shift={(0,0)}]
            \node at (3,4.5) {\small D2Q9};
            \fill [gray!30] (2,0) rectangle (4,4);

            \draw [thick, black] (0,0) grid[step=2.] (2,4);
            \foreach \x in {1}
                \foreach \y in {1,3}{
                    \draw [black, very thick, fill=white] (\x-0.2,\y-0.2) rectangle (\x+0.2,\y+0.2) ;
                }
            \draw [thick, black] (2,0) grid[step=1.] (4,4);
            \foreach \x in {2.5,3.5}
                \foreach \y in {0.5,1.5,2.5,3.5}{
                    \fill [black, very thick,fill=red](\x,\y) circle (0.1);
                }
            \foreach \x in {3}
                \foreach \y in {1,3}{
                    \draw [black, very thick, fill=red] (\x-0.2,\y-0.2) rectangle (\x+0.2,\y+0.2) ;
                }
            \draw [thick, black] (4,0) grid[step=1.] (6,4);
            \foreach \x in {4.5,5.5}
                \foreach \y in {0.5,1.5,2.5,3.5}{
                    \fill (\x,\y) circle (0.1);
                }
        \end{scope}
 
        \begin{scope}[shift={(8,0)}]
            \node at (4,4.5) {\small D2Q13};
            \fill [gray!30] (2,0) rectangle (6,4);

            \draw [thick, black] (0,0) grid[step=2.] (2,4);
            \foreach \x in {1}
                \foreach \y in {1,3}{
                    \draw [black, very thick, fill=white] (\x-0.2,\y-0.2) rectangle (\x+0.2,\y+0.2) ;
                }
            \draw [thick, black] (2,0) grid[step=1.] (6,4);
            \foreach \x in {2.5,3.5,4.5,5.5}
                \foreach \y in {0.5,1.5,2.5,3.5}{
                    \fill [black, very thick,fill=red](\x,\y) circle (0.1);
                }
            \foreach \x in {3,5}
                \foreach \y in {1,3}{
                    \draw [black, very thick, fill=red] (\x-0.2,\y-0.2) rectangle (\x+0.2,\y+0.2) ;
                }
            \draw [thick, black] (6,0) grid[step=1.] (8,4);
            \foreach \x in {6.5,7.5}
                \foreach \y in {0.5,1.5,2.5,3.5}{
                    \fill (\x,\y) circle (0.1);
                }
        \end{scope}

        \begin{scope}[shift={(18,0)}]
            \node at (5,4.5) {\small D2Q21};
            \fill [gray!30] (2,0) rectangle (8,4);

            \draw [thick, black] (0,0) grid[step=2.] (2,4);
            \foreach \x in {1}
                \foreach \y in {1,3}{
                    \draw [black, very thick, fill=white] (\x-0.2,\y-0.2) rectangle (\x+0.2,\y+0.2) ;
                }
            \draw [thick, black] (2,0) grid[step=1.] (8,4);
            \foreach \x in {2.5,3.5,4.5,5.5,6.5,7.5}
                \foreach \y in {0.5,1.5,2.5,3.5}{
                    \fill [black, very thick,fill=red](\x,\y) circle (0.1);
                }
            \foreach \x in {3,5,7}
                \foreach \y in {1,3}{
                    \draw [black, very thick, fill=red] (\x-0.2,\y-0.2) rectangle (\x+0.2,\y+0.2) ;
                }
            \draw [thick, black] (8,0) grid[step=1.] (10,4);
            \foreach \x in {8.5,9.5}
                \foreach \y in {0.5,1.5,2.5,3.5}{
                    \fill (\x,\y) circle (0.1);
                }
        \end{scope}

    \end{tikzpicture}
}
\caption{Illustration of the buffer layers (in grey) required by CC approaches for standard and high-order lattices. Circles and squares correspond to fine and coarse nodes respectively. Red nodes are used to transfer data between coarse and fine levels through the \emph{coalescence} and \emph{explosion} steps~\cite{ROHDE_IJNMF_51_2006}.}
\label{fig:rohde_highorder}
\end{figure}

Rohde's approach originally relies on the use of a single buffer/ghost layer whose purpose is to transfer information between coarse and fine grid levels. This layer overlaps data belonging to one coarse and two fine layers of fluid cells.
As this grid refinement strategy does not require interpolations, the fine ghost-layer cells are not technically required to store data and could theoretically be skipped to save memory accesses, at the cost of a more complex implementation.
To keep our framework as simple as possible, we preferred however to keep the original overlapping layer, and further \emph{increased its size for high-order lattices}. This is illustrated in Figure~\ref{fig:rohde_highorder}.

Regardless of the grid-refinement strategy considered for LBMs, the dependency of local time step $\Delta t$ on the local space step $\Delta x$ leads to asynchronous time-stepping when running on non-uniform grids. Because of that, mismatches between dimensionless quantities (e.g., Reynolds and Mach numbers) appear between successive refinement levels. To avoid these issues, two precautions must be taken. First, one must locally adjust the time step following the so-called acoustic scaling $c=c_s (\Delta x / \Delta t)$ to enforce continuity of the speed of sound $c$, and the flow speed $\bm{u}$, between different refinement levels. Here, $c_s$ is the speed of sound in LB units, and it is either equal to the lattice constant, or $\sqrt{\gamma T_0}$, with $\gamma$ being the specific heat ratio and $T_0$ the reference temperature in LB units. Obviously, this time scaling law automatically enforces continuity of the Mach number~\cite{KRUGER_Book_2017}. To ensure continuity of the Reynolds number, one further needs to locally adjust the relaxation time $\tau$ so that the kinematic viscosity remains the same at both coarse and fine levels. In other words, one needs to make sure that $\nu = \nu_c  (\Delta x_c^2 / \Delta t_c) = \nu_f  (\Delta x_f^2 / \Delta t_f)$, with a cell size ratio of $\Delta x_c/\Delta x_f=2$, and where subscripts $c$ and $f$ stand for coarse and fine levels. Eventually, combining this condition with the acoustic scaling ($\Delta t \propto \Delta x$), one ends up with the following relationship:
$\nu_c = \nu_f/2$ and $\tau_c = (\nu_c/c_s^2 + 1/2)\Delta t_c$~\cite{FILIPPOVA_JCP_147_1998}. This eventually leads to the recursive formula: 
\begin{equation}\label{eq:tau_fine_to_coarse}
\tau_c = (\nu_f/2 c_s^2 + 1/2) \Delta t_c.
\end{equation}
When setting up the simulation, $(\nu_f,\tau_f)$ are computed from the Reynolds number and $(\Delta x_f,\Delta t_f)$, while $(\nu_c,\tau_c)$ at coarser levels are computed recursively via Eq.~(\ref{eq:tau_fine_to_coarse}). Similar relationships are used to impose the bulk and thermal diffusivity coefficients through their associated relaxation times.

Another important aspect to discuss is how LB data is transferred between refinement levels. 
In Rohde's approach, a simple conservative strategy is considered: (1) fine-to-coarse transfer is handled by merging populations, followed by, (2) collision at the coarse level, while (3) coarse-to-fine transfer simply copies the post-collision LB data. Both data transfers are conservative steps, more commonly known as the ``\emph{coalescence}'' and ``\emph{uniform explosion}'' steps. In practice, they correspond to
\begin{equation}\label{eq:coalescence}
f_{i,c}(\bm{x}_c,t)=\dfrac{1}{N_f}\sum_{\bm{x}_f \in \bm{x}_c}f_{i,f}(\bm{x}_f,t),
\end{equation}
and, 
\begin{equation}\label{eq:explosion_uniform}
\forall \bm{x}_f\in \bm{x}_c,\quad f^*_{i,f}(\bm{x}_f,t)=f^*_{i,c}(\bm{x}_c,t),
\end{equation}
respectively. $f^*_i$ stands for post-collision populations and $\bm{x}_c$ represents the coarse cell location in the buffer layer, while $\bm{x}_f$ denotes the fine cell locations within that coarse cell. $N_f$ is the number of fine cells inside a coarse cell, with $N_f \leq 4$ for 2D simulations. The fact that it does not always equal four is because several fine cells might be missing when the grid refinement interface intersects a boundary or a solid object.

In practice, \emph{coalescence}~(\ref{eq:coalescence}) is applied prior to collision on coarse ghost-layer cells to compute the values of all unknown coarse populations, i.e. all populations that did not receive a value from neighboring coarse cells during the previous streaming. Then, the coarse cell executes collision and transfers post-collision data to unknown populations on fine ghost-layer cells through \emph{uniform explosion}~(\ref{eq:explosion_uniform}). In this way, fine and coarse ghost-layer cells provide values for each other's unknown populations in a symmetric procedure. For the final details of the algorithm in the ghost layers, it is noted that the coarse cells executes the usual streaming step, while the fine cells undergo pure streaming for the next two fine-level iterations \emph{without ever colliding}. The sole purpose of the fine ghost-layer cells is to receive post-collision values directly from the coarse cells, and transfer them to the nearest fine cells in the bulk.
It is interesting to note that the \emph{coalescence} step~(\ref{eq:coalescence}) can be seen as a (rather dissipative) spatial filter. This explains why no additional filtering step is required when transferring data to the coarse level, as opposed to VC approaches~\cite{LAGRAVA_JCP_231_2012,TOUIL_JCP_256_2014}. Moreover, this \emph{coalescence} step is widely employed beyond the LB community, particularly in multigrid CFD solvers, where it serves as a conservative volume-averaged restriction step, as implemented in frameworks like PARAMESH~\cite{MACNEICE_CPC_126_2000} and Athena++~\cite{TOMIDA_AJSS_266_2023}.

As a last remark, it is important to remind that the grid refinement method proposed by Rohde et al. cannot solve accuracy or stability issues inherent to the LBM used in the bulk of the simulation. Notably, by conducting preliminary studies based on the BGK collision model with polynomial equilibria, we found out that stability issues encountered at high Reynolds numbers were still present, and sometimes amplified by the non-uniform grid. However, no such issues were encountered with the use of the collision model RR and numerical equilibria, for low and high speed flow simulations respectively. In the latter case, additional data needed to be transferred between consecutive refinement levels, i.e., the Lagrange multipliers $\lambda_{\bm\alpha}^{(n)}$ used as guess values for the Newton-Raphson algorithm that computes the numerical equilibria~(\ref{eq:exponential}). In this work, we followed the same \emph{coalescence} and \emph{uniform explosion} rules for these quantities:
\begin{equation}\label{eq:coalescence}
\lambda_{{\bm\alpha},c}^{(n)}(\bm{x}_c,t)=\dfrac{1}{N_f}\sum_{\bm{x}_f \in \bm{x}_c}\lambda_{{\bm\alpha},f}^{(n)}(\bm{x}_f,t),
\end{equation}
and, 
\begin{equation}\label{eq:explosion_uniform}
\forall \bm{x}_f\in \bm{x}_c,\quad \lambda_{{\bm\alpha},f}^{(n)}(\bm{x}_f,t)=\lambda_{{\bm\alpha},c}^{(n)}(\bm{x}_c,t).
\end{equation}
This resulted in a significant performance improvement as the Newton-Raphson step typically converges in just one iteration on average when properly initialized. In contrast, it requires 10 to 12 iterations when starting from the initial guess $\lambda_{\bm\alpha}^{(n)}=0$. More information about the compressible LBMs used in this work can be found in~\ref{app:lbm}.

\section{Implementation details and GPU acceleration based on ISO C++~\label{sec:implementation_and_c++}}

In this work, parallel algorithms, available in the Standard Template Library (STL) of C++, form the core of the implementation strategy for the collide-and-stream scheme on non-uniform grids. These shared-memory-based algorithms can run on a variety of hardware, including GPUs and many-core CPUs. Despite its general and high-level nature, this coding strategy delivers near-peak performance across various devices, assuming proper implementation choices were made. Detailed investigations on optimal choices can be found in a previous work~\cite{LATT_PLOSONE_16_2021} and in the two-part article available on the NVIDIA technical blog for developers~\cite{LATT_NVIDIA_BLOG_1_2022,LATT_NVIDIA_BLOG_2_2022}. Hereafter, we recall the key conclusions of these works before presenting the implementation details of our grid refinement strategy.

\subsection{Key aspects for efficient implementation on uniform grids and memory storage details \label{subsec:key_aspects_efficiency}}

\begin{figure}[bt!]
    \centering
    \includegraphics[width=.75\textwidth]{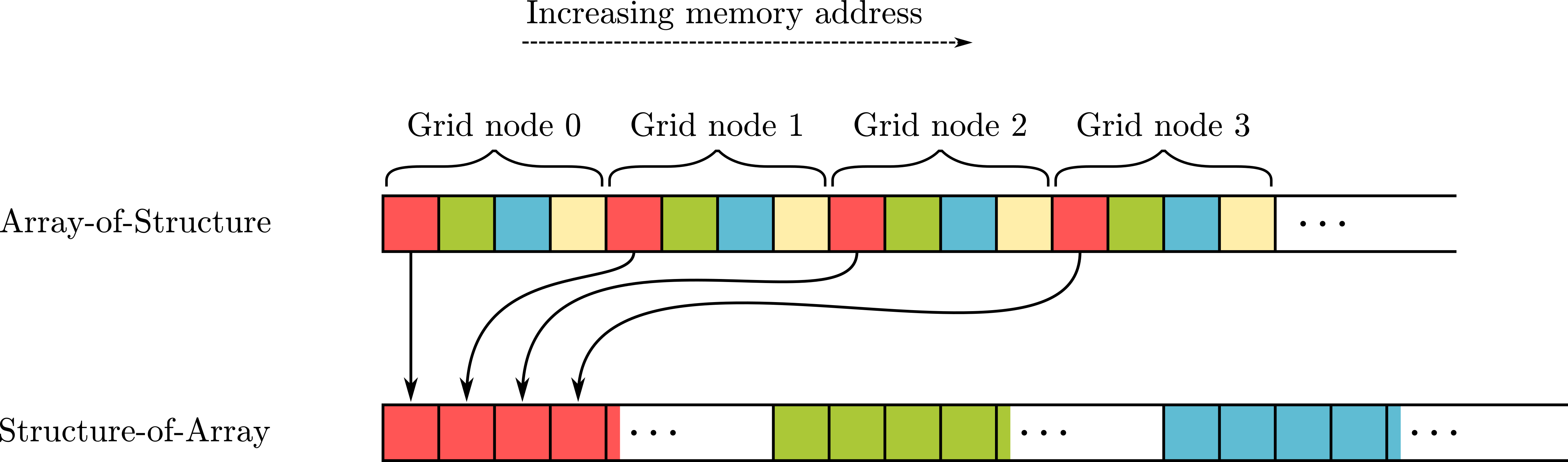}
    \caption{Illustration of the two possible data layouts used for LBM: Array-of-Strcuture (AoS) and Structure-of-Array (SoA). For the sake of illustration, only four populations are shown.}
    \label{fig:aos_vs_soa}
\end{figure}

Optimizing memory access through an efficient data layout is the first critical step to maximize performance, regardless of the GPU programming model. Since LBM solvers are memory bound on GPUs, their performance is primarily limited by memory bandwidth rather than computational power.
In the LBM literature, two widely used data layouts are Array-of-Structures (AoS) and Structure-of-Arrays (SoA), as illustrated in Figure~\ref{fig:aos_vs_soa}. The SoA layout is particularly well-suited for GPU-based LB simulations because it supports coalesced memory accesses, which are critical to access the full bandwidth of GPU memory, and which are less likely to be achieved during the streaming step using an AoS layout. 
Interestingly, the advantages of the SoA layout extend beyond LBMs as it significantly enhances GPU performance for fluid solvers based on macroscopic equations~\cite{ZIER_MNRAS_533_2024} and for particle solvers~\cite{MAGGIOAPRILE_Master_2023}.

The second step to maximize performance is to leverage parallel algorithms that are available in the C++ Standard Template Library (STL) since C++17. Efficiently parallelizing the collide-and-stream algorithm requires replacing the sequential (2D) Cartesian \texttt{for} loop 
\begin{lstlisting}[language=C++]
for (int iX=0; iX < nx; ++iX) {
  for (int iY=0; iY < ny; ++iY) {
    cell(iX, iY).collideAndStream();
  }
}
\end{lstlisting}
with an STL \texttt{for\_each} algorithm call
\begin{lstlisting}
for_each(execution::par, begin(pop), end(pop), [](Cell& cell)) {
    cell.collideAndStream();
} );
\end{lstlisting}
which runs in parallel using the \texttt{execution::par} policy. This approach is similar to parallelizing the Cartesian \texttt{for} loop with OpenMP or OpenACC directives. However, the strength of the STL lies in its wide range of efficient, ready-to-use parallel algorithms. For example, the \texttt{reduce} and \texttt{transform\_reduce} algorithms enable thread-safe parallel reduction operations, which are crucial in CFD for computing averages, minimums, and maximums of macroscopic quantities across the entire simulation domain. \texttt{sort} algorithms are also particularly useful when including the interaction between solid particles and fluid cells in particle-laden flow simulations~\cite{MAGGIOAPRILE_Master_2023}.

The present implementation uses the ``two-population'' storage formalism, where two \texttt{std::vector} are used to store populations on the heap memory. One \texttt{std::vector} is used to read populations that are modified by the collide-and-stream algorithm before being written back into the second \texttt{std::vector}, ensuring thread safety. 
This memory storage strategy, common in most CFD solvers, simplifies the implementation of various LBMs within a unified framework. However, it doubles memory requirements, which can be a significant drawback on GPUs, where available memory is typically lower than on CPUs. Fortunately, memory limitations are less of an issue when running simulations on non-uniform grids -- especially in the 2D case.

Eventually, the execution model of C++ parallel algorithms does not require explicit memory transfer instructions between CPU and GPU. Instead, GPU implementations of the STL provide a unified memory model, taking care of memory copies automatically. This enables seamless integration of CPU-bound operations, such as pre- and post-processing steps, with GPU-offloaded instructions. 
However, this convenience comes with a significant drawback: unintended memory transfers between the CPU and GPU can severely degrade performance. To prevent such performance losses, the unified memory model must be used judiciously by systematically tracking and minimizing unnecessary data movement.

\subsection{Brief comparison of multiblock and tree-based implementations of grid refinement \label{subsec:multiblock_vs_tree}}

Among the methods available for grid refinement, multiblock and tree-based approaches are the most widely used. The choice between these two strategies often depends on factors such as the specific application requirements (e.g., geometric complexity and grid adaptability), computational efficiency, implementation overhead, and the parallelization strategy.

The multiblock approach is a well-established grid-refinement implementation strategy in the field of CFD~\cite{CAMBIER_MI_14_2013,ZHANG_IJHPCA_35_2021} and LBM~\cite{LATT_CMA_81_2021,ASTOUL_PhD_2021}. It involves dividing the computational domain into distinct blocks, each of which contains a homogeneous mesh with a given refinement level. This spatial decomposition is particularly well-suited for accelerators like GPUs. First, it facilitates the contiguous alignment of data in memory which ensures coalesced read and write accesses during the streaming step. Second, memory addresses of neighboring data are easily computed through Cartesian indexing which eases the implementation of the streaming step and the computation of gradients through finite difference approximations.
Third, by maintaining a blockwise organization, multiblock methods allow developers to integrate refinement algorithms with minimal modifications to codes that were originally intended for simulations on uniform grids.
However, the multiblock method has notable limitations. Its refinement capabilities are inherently restricted to the block level, making it challenging to refine the grid to capture complex geometrical details or localized flow features effectively. This limitation often leads to over-refinement in areas that do not require high resolution, thereby increasing memory usage and simulation time. Additionally, managing complex refinement interfaces, while maintaining optimal memory access patterns, introduces significant coding challenge. 
As a result, while the multiblock approach offers an elegant and efficient framework, its practical implementation quickly becomes challenging for researchers and developers.

In contrast, tree-based grids offer a far more flexible framework to match a prescribed grid density function accurately~\cite{ZHANG_AJSS_164_2006,JUDE_JSC_2022}. Unlike multiblock methods, trees are not limited by fixed block sizes, enabling finer control over where and how the grid is refined. This flexibility reduces memory usage and computational costs by preventing unnecessary refinement in areas that do not require high resolution. Additionally, adapting the grid to complex time-dependent flow features is straightforward: leaves can be added to a branch to locally refine the mesh, or removed to coarsen it.
However, tree-based grids also come with several challenges. Without careful attention to the data memory layout, irregular memory access patterns can arise during the streaming step, which can deteriorate performance on GPUs. Furthermore, the convenient Cartesian indexing available in multiblock approaches is not available for trees. Instead, the latter rely on a list of neighbors (or connectivity table) to access non-local data, hence, increasing the number of memory accesses.
To overcome these challenges and combine the strengths of both the tree and multiblock approaches, the tree-of-blocks method can be adopted~\cite{BAUER_CMA_81_2021,HOLZER_PhD_2025}, where each leaf of the tree represents a block of cells rather than a single cell. This design significantly reduces memory accesses by storing connectivity information at the block level rather than for individual cells. In practice, however, this strategy still sacrifices fine-grained control over grid refinement, especially in highly resolved areas near boundaries or walls, where many cells within a block may exist outside the simulation domain or within the geometry. As a result, the performance benefits from reduced memory accesses can be quickly lost due to the significant number of unused cells at the highest refinement levels. This is why industry-oriented solvers typically limit block sizes to $4^3$ or $8^3$ to mitigate this issue~\cite{ASHTON_AIAA_0888_2025}.

In light of this discussion, we opted for the most flexible approach: a general tree with single-cell leaves. This reflects our vision that the computational mesh is a critical component of an efficient LBM simulation and should be implemented with minimal limitations. It also reflects our desire to demonstrate the raw performance that can be achieved in an LBM simulation without complex workarounds, and rather, by thoroughly targeting best efficiency while deploying textbook features of the LBM step by step.

\subsection{Tree construction and traversal order \label{subsec:implementation_tree_construction_and_traversal}}

Most LB mesh refinement algorithms, including Rohde's algorithm and its high-order extension, rely on a highly structured mesh layout in which cells of different level are well aligned with a 1:2 size ratio. This allows the mesh layout to be described with the help of a hierarchic tree data structure, as described in this section. The purpose of the tree data structure is to keep track of the spatial arrangement of all cells and to access the data of neighboring cells. Each node of the tree represents a spatial domain which can be subdivided, and the obtained sub-domains are represented by corresponding children/leaves of the original node. A general and straightforward choice is given by the Quadtree (in 2D) or the Octree (in 3D), where every node represents a cell of square (respectively cuboid shape), which can be subdivided recursively into 4 or 8 cells, respectively. This hierarchic decomposition proceeds until the desired level of refinement is reached, allowing to concentrate high-density portions of the mesh to areas of interest for the fluid solver. The root node thus extends over the full computational domain, while the leaves of the tree represent allocated cells. In ghost layers, where cells of different levels overlap, the parent of leave cells represent an allocated coarse ghost cell and needs to be tagged correspondingly.

Both the construction and the use of a tree rely on algorithms that are recursive in their nature. Executing this type of algorithms on GPUs is notoriously inefficient, as this hardware relies on a massive parallelization of the algorithm, and on a relatively uniform execution flow among the processing units~\cite{JABER_CPC_311_2025}. Since this work primarily investigates the accuracy, robustness, and efficiency of CC-based mesh refinement strategies on \emph{fixed} meshes, the tree is generated on the CPU. During a pre-processing step, also carried out on the CPU, the tree is parsed to create a data structure better suited for the execution of the LB algorithm on GPU.

In this context, it is useful to understand that the tree serves two purposes. The first typical use of the tree is to determine the arrangement of the cell data, especially the LB populations, in memory. A second, derived purpose of the tree is to access the populations of the neighboring cells during the execution of the streaming step for a given cell.  
For efficient data arrangement, it is crucial to map geographically close cells to nearby memory locations. 
This is done through mapping high-dimensional data (2D or 3D) into a 1D interval thanks to a space filling curve (SFC). One commonly used SFC is the Morton curve, also known as the Z-order curve, which exhibits good locally properties while being simple to create. It is created by traversing the geometrical space in a manner that first explores one quadrant/octant before moving to adjacent ones, hence, forming a characteristic ``Z'' pattern.
The locality of the SFC can be improved by changing the traversal order of a cell's children based on the position along the SFC, hence leading, for some choices, to a Hilbert curve. 
However, the increased complexity of the resulting algorithms is often considered to exhibit an unfavorable tradeoff for the performance gain. 
Moreover, when accessing data from next-to-nearest or more distant cells --as required in high-order LBMs and other CFD solvers-- the Hilbert curve offers no performance advantage over the simpler z-order curve (see Figure 14 of Ref.~\cite{HOPPE_CMAME_391_2022}). 
In this work, we therefore position the LB data along a \emph{Z-order curve}.

\subsection{Data layout for efficient GPU acceleration on trees \label{subsec:implementation_data_layout}}

Once the tree is created and the SFC selected, the data undergoes a preprocessing step on the CPU. This step reorganizes the data to optimize execution of the collide-and-stream algorithm on the GPU. Intuitively, one could think of gathering cells from all levels and arrange them according to their global SFC order. However, this method proves inefficient due to the asynchronous time-stepping of grid refinement in LB simulations, as fine-resolution cells require more frequent updates than coarse ones.
A better strategy is instead to group cells by refinement level and based on the current time iteration. This enhances memory locality at each refinement level while maintaining a good SFC ordering within each group. Additionally, the populations are stored in memory using a SoA layout to further optimize memory access performance during the streaming step, as detailed in Section~\ref{subsec:key_aspects_efficiency}. 

The final aspect requiring attention is the method for accessing neighbor data during the streaming step. One approach consists of traversing the tree data structure for every neighbor access, while storing a single tree-node index per cell as an entry point for tree traversal. However, preliminary tests revealed that the computational cost of repeatedly executing a tree-search algorithm remains prohibitively high on modern GPUs, despite their growing support for complex execution flows. 
Instead, we propose to compute the neighbor list during the preprocessing step. This involves generating the Cartesian coordinates of all neighboring cells and searching the tree for cells that match these coordinates. The SFC index of the identified cells is in a direct relationship with the storage index of the neighboring cell's data, which is subsequently stored as a component of the neighbor list. 
Thus, excluding the rest population, $V-1$ integer values are stored for every cell along with the $V$ populations. 
We store these neighbor indices as 32-bit integers, hence allowing to handle meshes with up to 4 billion cells, which is sufficient for GPUs with current memory capability. 
Given the continuous increase of performance and memory size of GPUs, it must however be expected to switch to 64-bit integers in a near future, which will automatically double the cost of these memory accesses. The neighbor lists are eventually stored in a AoS format as they are accessed locally on every cell.

In summary, the preprocessing step described above, executed on the CPU, organizes the data in memory as follows:
\begin{enumerate}
\setlength{\itemsep}{0pt} 
\setlength{\parskip}{0pt} 
\item Each cell stores $V$ real-valued populations, grouped by refinement level and ordered along the z-curve in an SoA layout to optimize memory access during the collide-and-stream algorithm.
\item Each cell also stores $V-1$ neighbor links as 32-bit integers in an AoS layout. The AoS format is chosen because neighbor lists are accessed locally for each cell during the streaming step. 
\end{enumerate}
If necessary, additional vector fields are stored in memory using the SoA layout. Depending on simulation requirements, global Cartesian coordinates are precomputed during initialization and accessed for far-field pressure relaxation~(\ref{eq:relaxation_pressure}), while the number of Newton-Raphson iterations is stored per cell and used for stability monitoring of fully compressible simulations. For fully compressible LBMs, step 1 stores twice more real-valued populations because of the second distribution used to simulation polyatomic gases.

\subsection{Parallel execution on trees\label{subsec:implementation_parallel_exe_tree}}

After data allocation, population initialization, and neighbor-list creation on the CPU, all data is automatically transferred to the GPU via the unified memory model, so that LB cycles can seamlessly be accelerated on that hardware. The sequence of events that is executed on the GPU, per time iteration \texttt{iter}, is illustrated in the following code extract:
\begin{lstlisting}[language=C++, commentstyle=\color{gray}]
for (int iLevel = tree.depth(); iLevel > 0; --iLevel) {
  if (iter % pow(2, iLevel) == 0) {
    // Refinement coupling on ghost cells at level iLevel (coarse) and iLevel-1 (fine).
    simulation.refinementCoupling(iLevel, iLevel - 1,  tau[iLevel]);
    // Collision-streaming cycle at level iLevel (coarse).
    simulation.collideAndStream(iLevel, tau[iLevel]);
  }
}
// For the finest level (iLevel=0), executes collision-streaming cycle on bulk cells.
// On ghost cells, only executes streaming.
simulation.collideAndStream(0, tau[0]);
\end{lstlisting} 

Globally speaking, the code iterates through all levels, starting from the coarsest level (equal to the tree depth, \texttt{tree.depth()}) down to the finest level (iLevel=0). It determines whether to execute the grid refinement coupling and collision-and-streaming steps based on the level's update frequency. At \texttt{iLevel=0}, the finest level, the collide-and-stream is performed at every time step. For \texttt{iLevel=1}, both the refinement coupling and collide-and-stream operations are executes, but only at even time iterations. Following the same principle, the number of time iterations between subsequent collision-streaming cycles doubles at every increase of the level \texttt{iLevel}. A modulo operation with the interval $2^\mathtt{iLevel}$ determines when computations are triggered or skipped at each level. To ensure the collision-streaming cycle has all necessary data, it is executed after the refinement coupling step, with levels processed from coarsest to finest, as shown in the loop at the beginning of the code. It iterates directly over the populations that have been regrouped for level \texttt{iLevel}, while the refinement coupling processes a pre-computed list of ghost-layer cells.

In practice, the coupling between successive coarse and fine refinement levels is managed through a function call to \texttt{simulation.refinementCoupling()} that implements the refinement algorithm described in Section~\ref{sec:grid_refinement}. Executed at the update frequency of the coarser level (\texttt{iLevel}), this step completes the unknown populations that are missing in the ghost layers at both the coarse and the fine levels (\texttt{iLevel} and \texttt{iLevel-1}). On every ghost layer, coalescence is executed to write data from the fine to the coarse level. The coarse cell is then collided, and post-collision populations are written to the corresponding fine cells during explosion. 
The \texttt{simulation.collideAndStream()} function call consists of a collision-streaming cycle, carried out on all cells of the coarse level, including those situated in the ghost layers. However, fine cells in the ghost layer only perform the streaming step without colliding.
Both the refinement coupling and collision-streaming steps are executed through a single call to the parallel algorithm \texttt{for\_each}. For the collision-streaming cycle, the algorithm iterates directly over the populations that have been regrouped for level \texttt{iLevel}. For the coupling, the algorithm processes a precomputed list of ghost-layer cells.

At this point, it is interesting to recall that the separation between refinement levels --a natural consequence of the asynchronous time-stepping in grid refinement strategies-- has a significant impact on performance due to the GPU underutilization. Indeed, GPUs are designed to execute thousands of threads concurrently while minimizing idle time to achieve maximum efficiency. However, when the workload per thread is too low, threads complete their tasks too quickly, leading to extended idle times as they wait to synchronize with others. The impact of the GPU load on performance was investigated in previous works (see Figures 7 and 15 of Refs.~\cite{LATT_PLOSONE_16_2021} and~\cite{THYAGARAJAN_PoF_35_2023}), and is also emphasized in the performance analysis of Section~\ref{sec:perfo}. As a partial workaround, the GPU could be loaded simultaneously with all levels that are active at a given time step to increase the workload per thread. This can be done thanks to the asynchronous execution capabilities of the \texttt{execution} library available since C++23 --not to be confused with the asynchronous time-stepping of the grid refinement strategy. Implementation of this performance improvement is deferred to future work.

To conclude this section, it is important to note that all GPU operations discussed here are applied to a static mesh, while mesh generation and data pre-processing are handled on the CPU for simplicity. Therefore, when using the present code in the context of adaptive mesh refinement (AMR), as discussed later in this article, mesh adaptation becomes the primary computational bottleneck. As a result, the performance analysis of our framework focuses on simulations with fixed grid refinement (see Section~\ref{sec:perfo}), while the GPU implementation of this task is left for future work. Hereafter, Section~\ref{sec:validation_weakly_compressible} presents subsonic aerodynamic and aeroacoustic validations, and Section~\ref{sec:validation_fully_compressible} extends the validation to transonic and supersonic regimes.
Finally, Section~\ref{sec:amr} provides preliminary validation of the framework’s accuracy and robustness for AMR-based simulations.

\section{Validation on weakly compressible flows~\label{sec:validation_weakly_compressible}}

The main objective of this section is to validate the proposed framework on benchmarks of increasing complexity, with a particular focus on the D2Q9-LBM. Even if higher-order lattices are more relevant for high-speed flow simulations, results for the D2Q13 and D2Q21 lattices are also included to demonstrate the universality of our refinement strategy.

Unless otherwise specified, the populations $f_i$ are initialized at equilibrium using Hermite polynomial expansions of the Maxwell-Boltzmann equilibrium distribution. Furthermore, the reference temperature $T_0$ is set to $c_s^2$. For more technical details, readers are referred to~\ref{app:lbm}.

\subsection{Double shear layer~\label{subsec:validation_weakly_compressible_dsl}}

The double shear layer benchmark is a commonly used in computational fluid dynamics to evaluate the stability and performance of standard numerical schemes~\cite{BELL_JCP_85_1989,BROWN_JCP_122_1995,MINION_JCP_138_1997,SHU_Chapter_1998}, as well as, weakly compressible~\cite{DELLAR_PRE_64_2001,MATTILA_PRE_91_2015,EZZATNESHAN_MCS_156_2019} and fully compressible lattice Boltzmann solvers~\cite{COREIXAS_PRE_96_2017,SHAN_PRE_100_2019,LI_PRE_100_2019,THYAGARAJAN_PoF_35_2023}. It involves a flow with two parallel shear layers positioned at $y/L=1/4$ and $y/L=3/4$ within a two-dimensional, periodic domain defined by $(x,y)\in [0,L]^2$. An initial small perturbation is added transversely to the flow, which leads to the development of Kelvin-Helmholtz instabilities. These instabilities cause the shear layers to roll up, resulting in the formation of two counter-rotating vortices. Thanks to this physical mechanism, it is possible to visually identify dispersion and grid refinement issues that take the form of spurious secondary vortices.

In the following simulations, the density is initialized to a constant value, $\rho_0 = 1$. The two shear layers are prescribed through the velocity field $\bm{u} = (u_x, u_y)$, where the longitudinal component is defined as
\begin{equation}
u_x = \left\{
\begin{array}{l}
u_0\tanh[k(y/L-1/4)], \quad y\leq L/2\\
u_0\tanh[k(3/4-y/L)], \quad y>L/2
\end{array} 
\right.
\end{equation}
and perturbed by a transverse velocity component  
\begin{equation}
u_y = u_0\delta \sin[2\pi(x/L+1/4)].
\end{equation}
$u_0$ represents the characteristic flow velocity, determined from the Mach number through $u_0 = c_s \mathrm{Ma}$. The parameter $k$ corresponds to the width of the shear layers, and $\delta$ controls the amplitude of the transverse perturbation. Hereafter, we consider the case of thin shear layers that are characterized by $(k, \delta) = (80, 0.05)$~\cite{BROWN_JCP_122_1995,MINION_JCP_138_1997}. 
We analyze results obtained on a non-uniform mesh featuring a horizontal refinement interface where fine cells are used for \( y > L/2 \), while coarse cells are applied for \( y \leq L/2 \). This configuration allows us to assess potential challenges that may arise when the shear layers interact with the refinement interface during their rollup.

\begin{figure}[hbt]
    \centering
    \includegraphics[width=\textwidth]{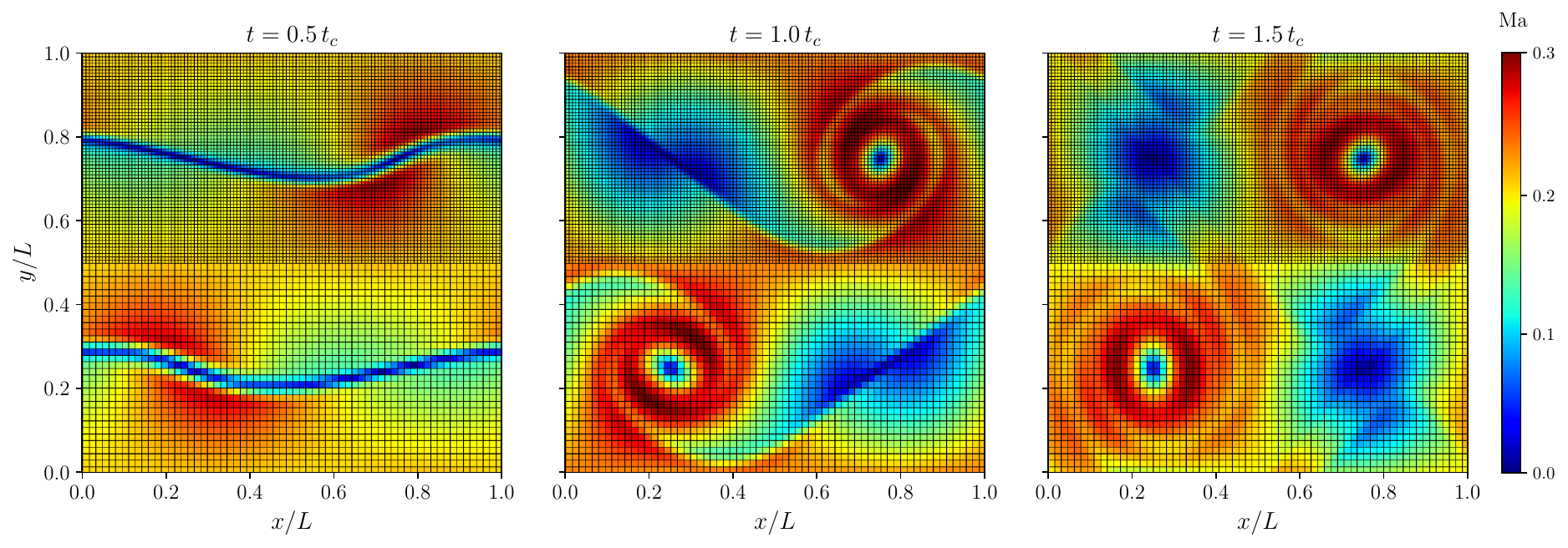}
	\captionsetup{skip=0pt}
    \caption{Impact of grid refinement on the time evolution of the double shear layer using the D2Q9-RR-LBM and for $(\mathrm{Re},\mathrm{Ma})=(10^4, 0.2)$. The convective time is defined as $t_c=L/u_0$, and the coarse space step is $\Delta x_c = L/64$. The mesh is superimposed to the local Mach number map to highlight the good accuracy of our approach at the refinement interface.}
    \label{fig:dsl_time_q9_re1e4}
\end{figure}

As a first investigation, we focus on the time evolution of the two shear layers and their interaction with the refinement interface for a moderate Reynolds number and a finite Mach number: $(\mathrm{Re},\mathrm{Ma})=(10^4, 0.2)$. Results obtained with the D2Q9-RR-LBM are presented in Figure~\ref{fig:dsl_time_q9_re1e4} for a coarse space step of $\Delta x_c=L/64$. Even at such a low resolution, the rollup is properly simulated and no spurious vortices are generated. This is thanks to both the collision model and the grid refinement algorithm, as preliminary studies conducted with the BGK collision (not shown here) led to dispersion issues that were amplified by the non-uniform mesh. Interestingly, very good results are also obtained with high-order lattices, hence, highlighting the good numerical properties of our general mesh refinement strategy (see~\ref{app:dsl}).

\begin{figure}[hbt]
    \centering
    \includegraphics[width=\textwidth]{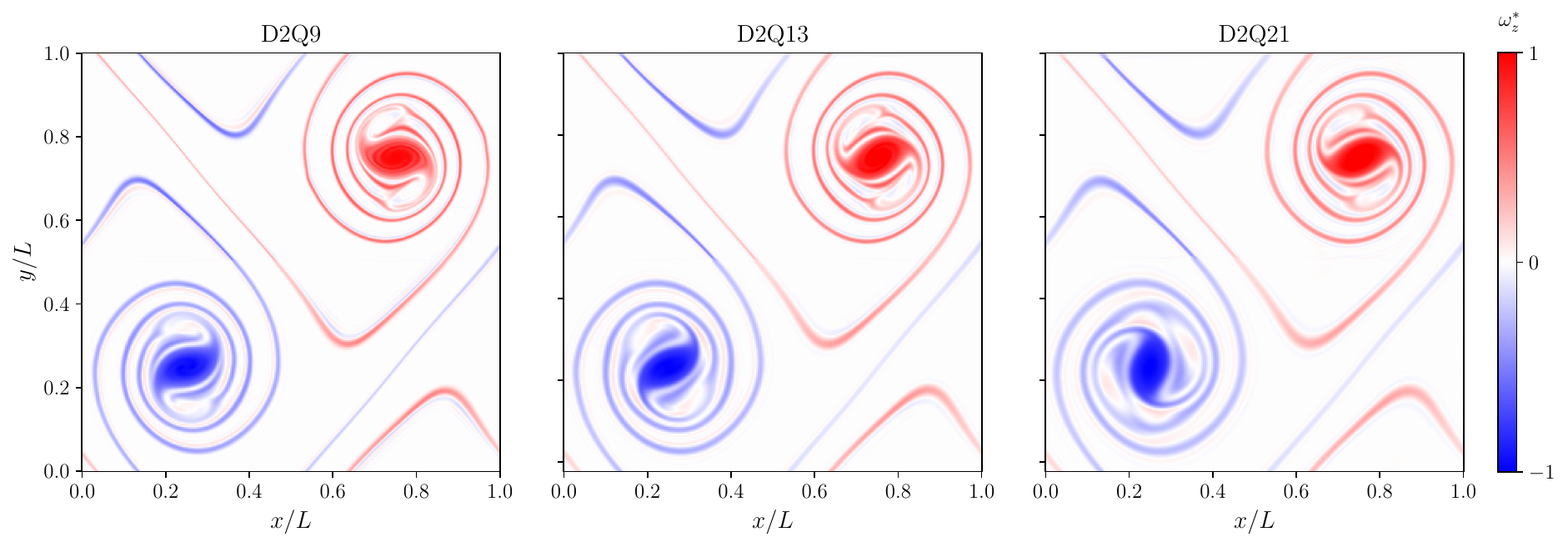}
	\captionsetup{skip=0pt}
    \caption{Impact of grid refinement on the double shear layer at $t=1.5t_c$ and $(\mathrm{Re},\mathrm{Ma})=(10^9, 0.2)$. The normalized vorticity maps ($\omega_z^* = \omega_z/\min(\omega_z)$) are reported for various lattices and the RR collision model. The convective time is defined as $t_c=L/u_0$, and the coarse space step is $\Delta x_c = L/256$. Vorticity is computed using a second-order, centered finite-difference approach.}
    \label{fig:dsl_comp_lattice_re1e9}
\end{figure}

As a more challenging configuration, we now increase the Reynolds number to $\mathrm{Re}=10^9$. The mesh resolution is kept relatively low ($\Delta x_c=L/256$) 
to highlight the robustness and physics-preserving characteristics of the grid refinement algorithm even at extremely low grid viscosity.
Figure~\ref{fig:dsl_comp_lattice_re1e9} compares the normalized vorticity field obtained with various lattices at $t=1.5\,t_c$, where the convective time is defined as $t_c=L/u_0$. Once again, no spurious vortices are generated even if the shear layers are crossing the refinement interface at several locations in space. It is worth noting that \emph{very small} visual artifacts appear within the buffer layers. However, these artefacts remain strictly confined to these regions and have no significant impact on the overall accuracy of the method.

\subsection{Steady internal aerodynamics: Lid-driven cavity}

In the lid-driven cavity test, a fluid is enclosed in a square cavity that is set in motion by a moving lid. This movement creates a shear flow and a boundary layer along the lid. As the boundary layer moves downward into the cavity, it leads to the formation of a large primary vortex centered near the middle of the cavity. This primary vortex further interacts with the stationary walls, generating smaller secondary contra-rotative vortices at the cavity corners.
The flow inside the cavity is characterized by a delicate balance between inertia and viscous forces. At moderate Reynolds numbers ($\mathrm{Re < 8000}$) the flow is steady in time~\cite{FORTIN_IJNMF_24_1997,BRUNEAU_CF_35_2006}. At higher Reynolds numbers, vortices interact with each other and with boundary layers, leading to changes in their size, shape, and location over time.

Hereafter, we validate the accuracy and robustness of our general grid refinement strategy using steady configurations. More precisely, the Mach number is fixed to $0.2$ and two Reynolds numbers are considered ($100$ and $7500$). The simulations domain size is $L \times L$ and the finest space step is $\Delta x_f = L/256$ for all grids considered below. This value $\Delta x_f$ ensures that $\Delta x_c\geq 64$ for all non-uniform meshes, hence, guaranteeing a sufficient level of accuracy in the coarsest parts of the mesh. Dirichlet no-slip and velocity boundary conditions are imposed through standard and velocity bounce-back~\cite{LADD_JFM_271_1994a,KRUGER_Book_2017}. It is worth noting that regularized velocity profiles could be employed to mitigate singularity issues that arise at the top left and right corners, where no-slip and velocity bounce-back conditions intersect~\cite{FORTIN_IJNMF_24_1997}. However, imposing a constant velocity $u_{\infty}=c_s \mathrm{Ma}$ presents a greater challenge for the LB solver, making it a more relevant test case for robustness. Eventually, macroscopic fields are initialized at $(\rho_0,\bm{u}_0) = (1,\bm{0})$ in dimensionless units.

For this benchmark, three mesh configurations are considered: a uniform mesh without refinement patches (\emph{unif}), a non-uniform mesh with vertical refinement interface at mid-span (\emph{vert}), a non-uniform mesh with three refinement layers close to walls (\emph{3 lay}), and another that further add refinement patches in corners (\emph{w{\&}c}). The \emph{vert} configuration serves as an example of simulation where the refinement interface intersects both no-slip and velocity bounce-back conditions without causing any spurious oscillations. In contrast, \emph{3 lay} and \emph{w{\&}c} configurations represent more realistic scenarios, using local refinement patches to increase mesh resolution in the regions of interest: boundary layers and cavity corners. The \emph{3 lay} is only used for $\mathrm{Re}=7500$.

\begin{figure}[bt]
    \centering
    \includegraphics[width=\textwidth]{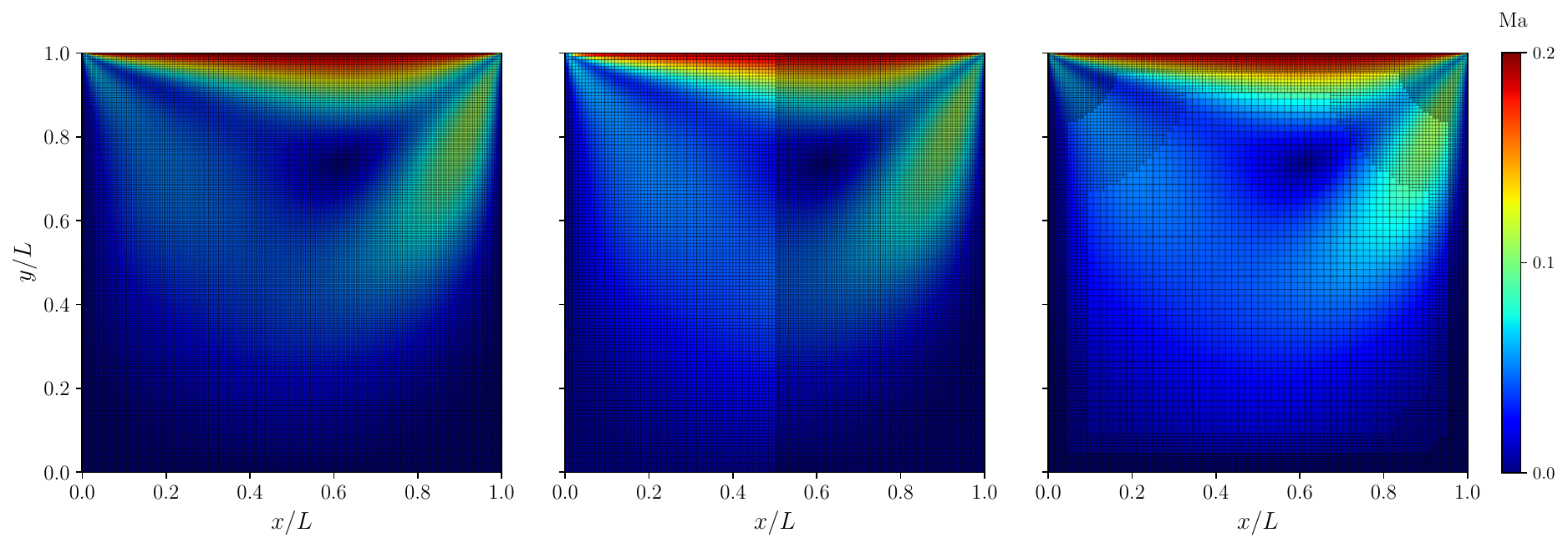}
	\captionsetup{skip=0pt}
    \caption{Impact of grid refinement on the steady state of the lid-driven cavity benchmark for $(\mathrm{Re},\mathrm{Ma})=(100, 0.2)$ and using the D2Q9-RR-LBM. The finest space step is $\Delta x_f = L/256$ for all meshes: (left) \emph{unif}, (middle) \emph{vert}, and (right) \emph{w{\&}c}. The mesh is superimposed to the local Mach number map to highlight the good accuracy of our approach at the refinement interface. Results from the \emph{3 lay} configuration are omitted as they are very similar to those from \emph{w{\&}c}.}
    \label{fig:cavity_grids_q9_re100}
\end{figure}

\begin{figure}[bt]
    \centering
    \includegraphics[width=\textwidth]{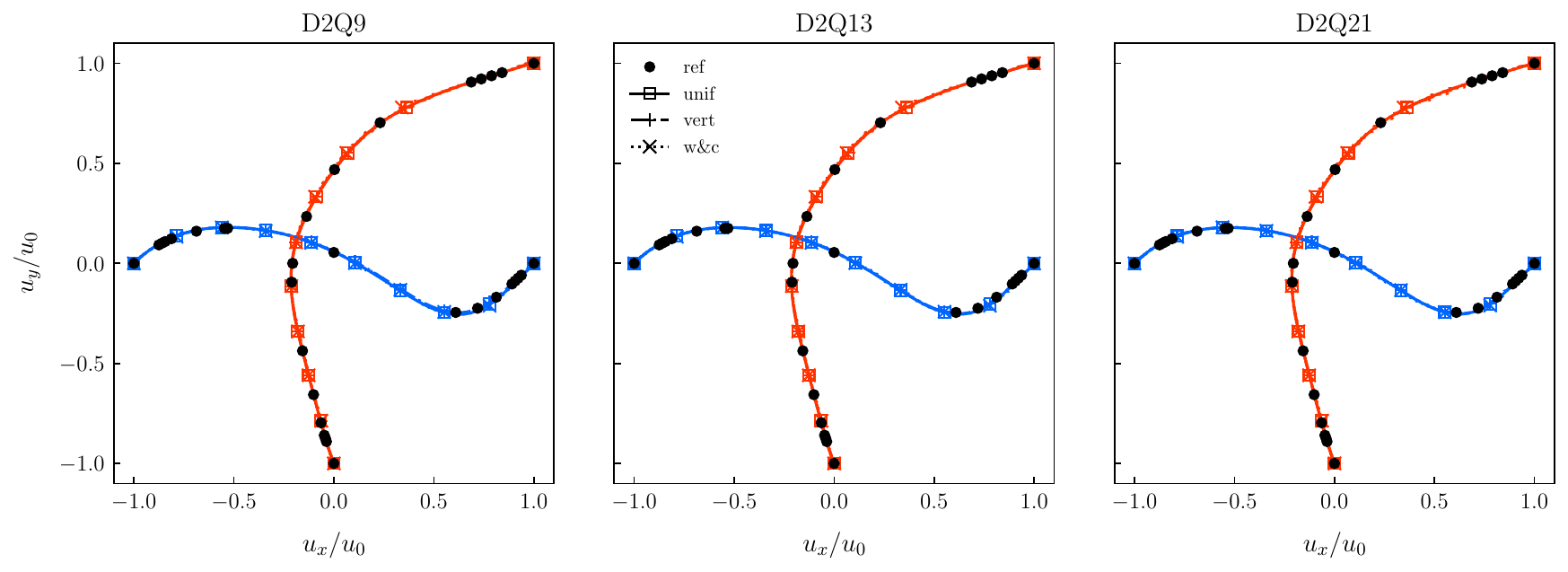}
	\captionsetup{skip=0pt}
    \caption{Velocity profiles along the centerlines of the lid-driven cavity benchmark for $(\mathrm{Re},\mathrm{Ma})=(100, 0.2)$, using various lattices and the RR collision model. Results obtained with three different meshes (\emph{unif},\emph{vert}, and \emph{w{\&}c}) are in excellent agreement with reference data from Ghia et al.~\cite{GHIA_JCP_48_1982}. Curves from \emph{3 lay} mesh are omitted as they closely resemble those from \emph{w{\&}c}.}
    \label{fig:cavity_profiles_re100}
\end{figure}

To assess convergence of the simulation, the total kinetic energy in the domain, $kin$, is monitored. When the normalized difference $\vert kin(t_1)-kin(t_0) \vert/kin(t_1)$ reaches $10^{-6}$, the flow is considered to have reached a steady state. 
This global criterion is evaluated each $t_1-t_0 = 0.1 t_c$, with the convective time being $t_c=L/u_{\infty}$. Figure~\ref{fig:cavity_grids_q9_re100} shows the local Mach fields obtained after convergence, for each mesh configuration, using the D2Q9-RR model, and for $\mathrm{Re}=100$.
Qualitatively, all configurations result in a primary recirculation of similar size and location, and this is also true for high-order lattices (not shown here). No accuracy issues are observed near refinement interfaces or at flat/corner boundary conditions. To verify the accuracy of our refinement strategy, we compare velocity profiles along the centerlines with reference data from Ghia et al.\cite{GHIA_JCP_48_1982}. Remarkably, all lattices and mesh configurations show excellent agreement with the reference data (see Figure~\ref{fig:cavity_profiles_re100}). In particular, the vertical velocity profiles \emph{measured within the buffer layers} for configuration \CC{\emph{vert}} confirm the accurate reconstruction of missing information, even when using two (D2Q13) or three (D2Q21) consecutive buffer layers.

For the sake of completeness, results are also provided at $\mathrm{Re}=7500$ in~\ref{app:ldc}. This Reynolds number was selected because it is the highest for which steady-state data is available in the literature~\cite{GHIA_JCP_48_1982}. Corresponding results further confirm the accuracy and robustness of our grid-refinement algorithm in the context of internal flows with steady and moving walls.

\subsection{Aeolian aeroacoustics: Flow past a cylinder \label{subsec:aeolian_noise}}

Aeolian noise simulations study the noise generated by the interaction between a fluid flow and a cylinder, specifically for Reynolds numbers where \emph{periodic} vortex shedding occurs behind the cylinder. In 1878, experiments conducted by Strouhal~\cite{STROUHAL_AP_241_1878} showed that Aeolian tones were generated at a nearly constant frequency $f$, over a wide range of Reynolds numbers ($300$ to $10000$). Furthermore, the frequency was proportional to the cylinder diameter $D$ and the freestream velocity $u_{\infty}$, and the proportionality constant has since become the Strouhal number $\mathrm{St} = fD/u_{\infty}$. Later research found that this frequency is actually related to force fluctuations on the cylinder, particularly, the lift fluctuations associated with vortex shedding~\cite{GERRARD_PPSSB_68_1955}.

\begin{figure}[hbt]
    \centering
    \includegraphics[width=\textwidth]{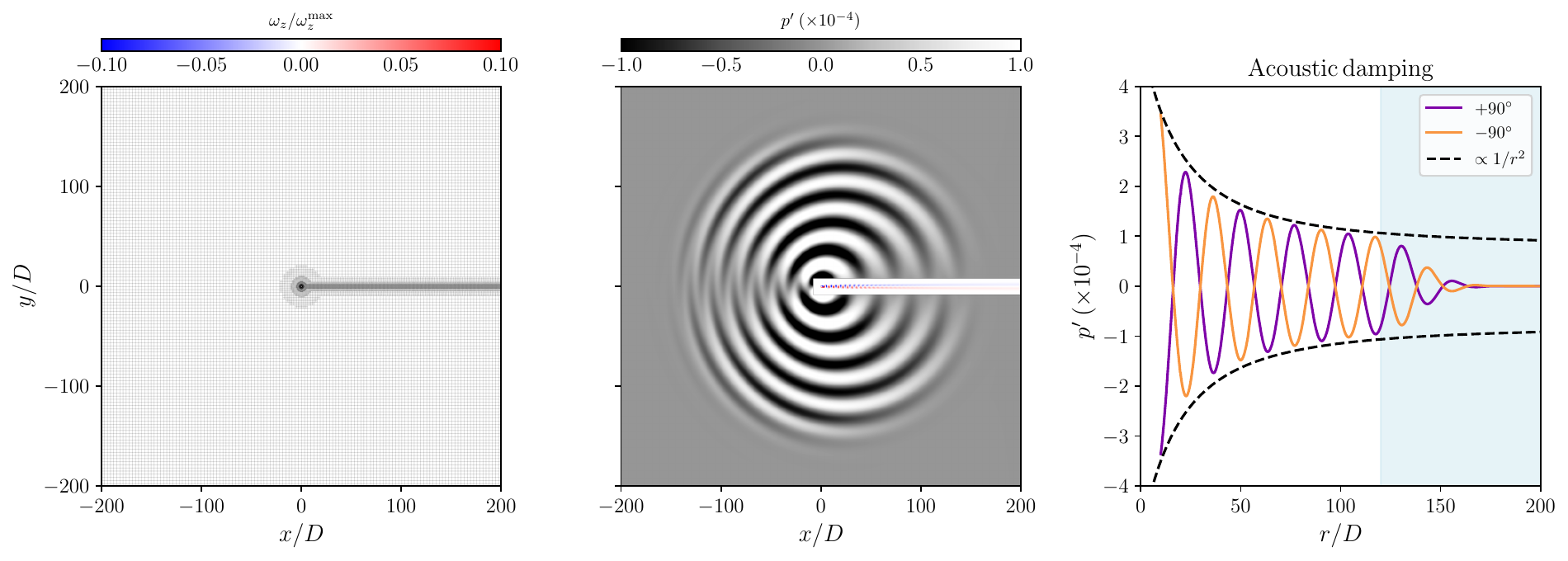}
	\captionsetup{skip=0pt}
    \caption{Far-field data from Aeolian noise simulations at $(\mathrm{Re},\mathrm{Ma},\Delta x_f)=(150,0.2,D/160)$ with the D2Q9-RR-LBM. Left panel: Grid composed of four circular and three rectangular refinement patches of normalized diameters $d/D\in\{2.5, 5, 10, 20\}$ and heights $h/D\in\{5, 10, 20\}$, respectively. Only one cell over eight is plotted for the sake of clarity. Middle panel: Far-field pressure fluctuations and vorticity overlay. Right panel: Decay of pressure fluctuations with the distance to the cylinder $r$. The transparent blue area indicates the region where the relaxation of pressure is activated.}
    \label{fig:aeolian_noise_farfield}
\end{figure}

Hereafter, we investigate the ability of the present framework to simulate Aeolian noise through comparisons against direct numerical simulations (DNS) of the Navier-Stokes equations~\cite{INOUE_JFM_471_2002,INOUE_AIAA_2132_2001}. Specifically, we simulate the flow past a circular cylinder using the D2Q9-RR-LBM for various Reynolds and Mach numbers, starting with  $(\mathrm{Re},\mathrm{Ma})=(150,0.2)$. The simulation domain size is $L\times L$ with $L=400D$, and it is centred about a cylinder of diameter $D$. This ensures that (1) boundary conditions have a reduced impact on the flow  physics as the cylinder remains $200D$ away from them, and (2) direct far-field observations can be made without relying on acoustic analogies. The left boundary imposes the freestream velocity through a Dirichlet condition (velocity bounce-back), while the other boundaries are Neumann conditions (zero-gradient of all populations). 
To avoid the reflection of acoustic waves over boundary conditions, a relaxation approach is adopted for the pressure $p$. In practice, $p$ progressively tends towards its free-stream value $p_{\infty}$ as the distance from the cylinder increases. In the context of weakly compressible LBMs, for which temperature is constant ($T_0$), the relaxation of pressure or density is equivalent. Here, it is implemented through:
\begin{equation}
p = p(1-\sigma) + \sigma p_{\infty}
\end{equation}
with
\begin{equation}
\sigma = \left\{
\begin{array}{l  l}
\sigma_{max}(r-130D)/(180D-130D) , & r \geq 130D \\[0.1cm]
\sigma_{max}, & r \geq 180D
\end{array} 
\right.
\label{eq:relaxation_pressure}
\end{equation}
where $r$ is the distance to the cylinder center, $p=\rho T_0$ and $p_{\infty}=\rho_{\infty} T_0$. $\sigma_{max}$ is kept relatively low ($0.1$) to avoid confinement effects that would hinder the correct propagation of acoustic waves in the far-field. The relaxation~(\ref{eq:relaxation_pressure}) is applied before the collision step so that it is compatible with any collision model and velocity discretization. Obviously, this approach is not as accurate as Navier-Stokes characteristic boundary conditions~\cite{POINSOT_JCP_101_1992,WISSOCQ_JCP_331_2017,FENG_PF_31_2019,CHEN_JCP_490_2023}, but it will be shown to meet the accuracy requirements of direct aeroacoustics simulations when the domain is large enough.

Simulations are performed over $800$ convective times $(t_c=D/u_{\infty})$ so that the transient time --approximately 200 to 600 $t_c$ depending on the Mach and Reynolds numbers-- is properly skipped before recording data. This long transient time is explained by the large size of the simulation domain ($t_c^{\mathrm{domain}}=400 D/u_{\infty}=400 t_c$), and by the symmetry of the initial condition (uniform density $\rho_0=1$ and velocity $\bm{u}=(u_{\infty},0)$) that usually requires an asymmetric perturbation to speedup the onset of the vortex shedding behind the cylinder~\cite{INOUE_JFM_471_2002,INOUE_AIAA_2132_2001}. 

Due to the large domain size, several grid refinement levels are required to minimize the number of cells in the whole domain. As we are interested in reproducing the mechanisms behind the noise radiated from the cylinder, and its propagation in the far-field, we use several circular refinement patches centered on the cylinder. Additionally, we include rectangular patches extending downstream the cylinder to accurately capture and propagate its vortex shedding. This setup is similar to the one proposed by Ishida~\cite{ISHIDA_AIAA_0259_2016} using the lattice Boltzmann code developed by the Japan Aerospace and Exploration Agency (JAXA). Based on a preliminary mesh convergence analysis performed over a wide range of Reynolds and Mach numbers, the diameter resolution is eventually fixed to $\Delta x_f=D/160$, and data is averaged at every coarsest time step during the last $100t_c$. 

Figure~\ref{fig:aeolian_noise_farfield} depicts the computational setup and presents far-field data obtained using the D2Q9-RR-LBM. The middle panel highlights the vortex shedding behind the cylinder, and the resulting acoustic wave generation and propagation in the far-field. As predicted by theoretical models~\cite{LANDAU_Book_2nd_1987} and corroborated by numerical simulations~\cite{INOUE_JFM_471_2002,INOUE_AIAA_2132_2001,ISHIDA_AIAA_0259_2016}, the amplitude of the sound waves decreases proportionally to the inverse square of the distance from their source as shown on the right panel. Moreover, pressure fluctuations around $p_{\infty}$ exhibit symmetry in propagation directions perpendicular to the flow, providing initial evidence that the emitted noise is of dipolar nature. 
Eventually, the effectiveness of the simple relaxation procedure~(\ref{eq:relaxation_pressure}) is confirmed as no wave reflections are observed near the boundary conditions.

\begin{figure}[hbt]
    \centering
    \includegraphics[width=\textwidth]{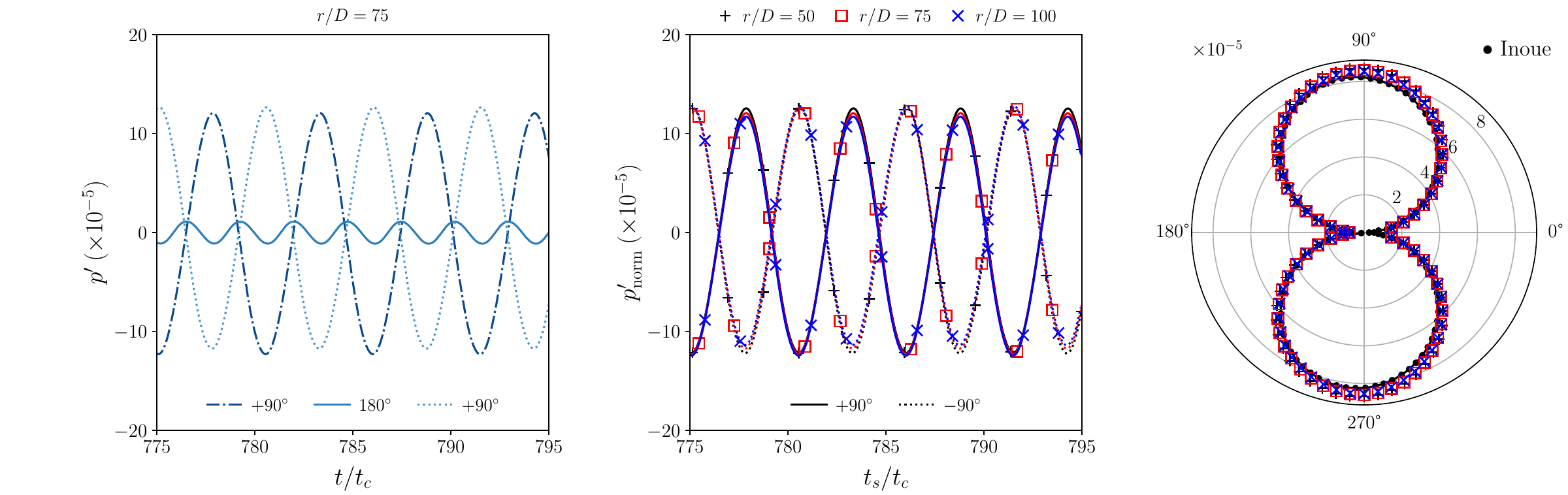}
	\captionsetup{skip=0pt}
    \caption{
Aeolian noise characteristics at $(\mathrm{Re}, \mathrm{Ma}, \Delta x_f) = (150, 0.2, D/160)$ using the D2Q9-RR-LBM.
Left panel: Time evolution of pressure fluctuations, $p' = (p - p_{\mathrm{mean}})/p_{\infty}$, at $r/D=75$ and for $\theta\in \{+90^{\mathrm{o}}, 180^{\mathrm{o}}, -90^{\mathrm{o}}\}$.
Middle panel: Time evolution of normalized pressure fluctuations, $p'_{\mathrm{norm}} = p' / p_{\infty}(r/r_{\mathrm{ref}})^{1/2}$, for various locations $r/D \in \{50,75,100\}$ and $\theta = \pm 90^{\mathrm{o}}$.
Right panel: Directivity of the normalized root-mean-square (rms) pressure, $p^{\mathrm{rms}}_{\mathrm{norm}} = p^{\mathrm{rms}} / p_{\infty}(r/r_{\mathrm{ref}})^{1/2}$. 
The reference distance, $r_{\mathrm{ref}} / D = 75$, is chosen to facilitate comparison with DNS data presented in Figure 12 of Ref.~\cite{INOUE_JFM_471_2002}. Mean and rms values are calculated over the final $100 t_c$. $t_s$ accounts for the time shift between different locations, assuming sound waves propagate at the speed of sound $c_s$. \CC{Reference $p^{\mathrm{rms}}$ values are extracted from Figure 12 (a) of Ref.~\cite{INOUE_JFM_471_2002} using \href{https://automeris.io/wpd/}{WebPlotDigitizer}.}
    }
    \label{fig:aeolian_noise_pressure_fluct_norm_polar}
\end{figure}

Figure~\ref{fig:aeolian_noise_pressure_fluct_norm_polar} gathers more quantitative data about the dipolar nature of the sound radiated from the cylinder. For such configuration, lift and drag fluctuations are expected to lead to pressure waves of different amplitudes and frequency~\cite{INOUE_JFM_471_2002,INOUE_AIAA_2132_2001}. These fluctuations are captured by the pressure measurements at $r/D=75$ for various angles $\theta\in \{+90^{\mathrm{o}}, 180^{\mathrm{o}}, -90^{\mathrm{o}}\}$. More precisely, lift force fluctuations in the $\pm 90^{\mathrm{o}}$ directions produce acoustic waves with amplitudes approximately 10 times larger than those generated by drag fluctuations in the $180^{\mathrm{o}}$ direction.
In our simulations, the lift dipole oscillates at half the frequency of the drag dipole. The Strouhal number was determined via FFT analysis of the time signal recorded over $100t_c$, utilizing Hamming windowing and zero-padding to enhance resolution~\cite{PRESS_Book_3rd_2007}. This method yielded $\mathrm{St} = 0.1824$, which aligns closely with DNS results~\cite{INOUE_JFM_471_2002} and experimental data~\cite{WILLIAMSON_JFM_206_1989}: $\mathrm{St} = 0.183$ and $\mathrm{St} = 0.179 - 0.185$, respectively. These findings remain consistent across different observation points ($r/D = 50$, $r/D = 75$, and $r/D = 100$), as illustrated in the middle panel of Figure~\ref{fig:aeolian_noise_pressure_fluct_norm_polar}. Interestingly, the pressure fluctuations exhibit a form of ``self-similar'' behavior as all curves almost perfectly coincide when (1) their amplitude is normalized by the decay rate of $1/r^2$, and (2) their time evolution is shifted according to the speed of sound and the distance between the two locations. 
The right panel of Figure~\ref{fig:aeolian_noise_pressure_fluct_norm_polar} further quantifies the noise directivity using the normalized root-mean-square (rms) pressure values, $p^{\mathrm{rms}}_{\mathrm{norm}}$. The observed rms values are in \CC{very good} agreement with DNS data from Inoue et al.~\cite{INOUE_JFM_471_2002}. This polar plot confirms that, regardless of the measurement distance, the emitted noise is predominantly dipolar in nature, with the lift dipole serving as the primary contributor to the radiated acoustic field. 
In fact, this holds true for a wide range of Reynolds and Mach numbers as shown in~\ref{app:aeolian_noise}.

\subsection{External aerodynamics at realistic conditions: Flow past the 30P30N three-element airfoil \label{subsec:30p30n}}

\begin{figure}[hbt]
    \centering
    \includegraphics[width=\textwidth]{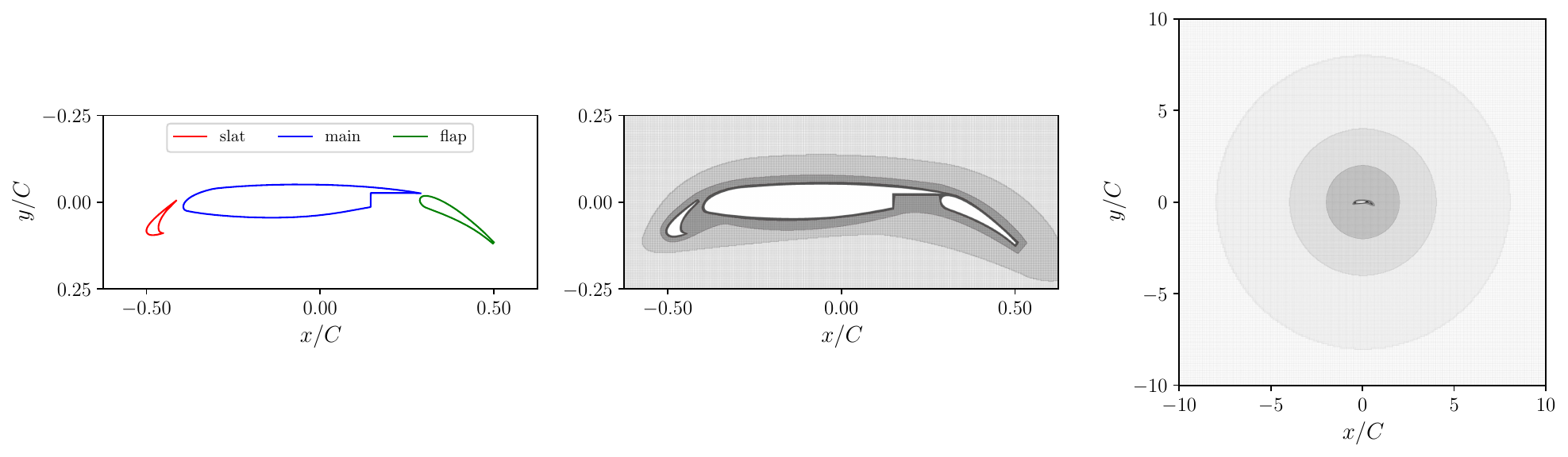}
	\captionsetup{skip=0pt}
    \caption{Key features of the 30P30N three-element airfoil and the grid used for the simulations. Left panel: Each element of the airfoil is identified. Middle panel: Focus on the three local refinement levels. Right panel: Global view of the grid composed of seven levels with three circular and three local refinement levels. The geometry was obtained through a 2D slice of the STL file available on the website of \href{https://www.wolfdynamics.com/tutorials.html?id=119}{Wolf Dynamics}. For visualization purposes, the figure merges groups of 16 cells into a single one in the global view.}
    \label{fig:30p30n_geo_mesh}
\end{figure}

In the early 1990s, the Douglas Aircraft Company (now part of Boeing) and the NASA's Langley Research Center collaborated on an experimental program to create a high-quality database for advancing and validating computational methods~\cite{VALAREZO_JA_30_1993}. This database was developed through an in-depth study of the aerodynamic properties of the 30P30N three-element wing configuration under realistic flow conditions.
This configuration consists of the main wing, a leading-edge slat, and a trailing-edge flap. The main wing provides primary lift, the leading-edge slat is added to delay airflow separation, and the trailing-edge flap further increases camber and lift during takeoff and landing phases (see left panel of Figure~\ref{fig:30p30n_geo_mesh}). These components together offer increased lift and improved stall characteristics, enhancing control and overall aircraft performance at high angles of attack.

This three-element airfoil has been widely studied over the last three decades through experiments~\cite{PASCIONI_AIAA_3062_2014,MURAYAMA_AIAA_2080_2014,MURAYAMA_AIAA_3460_2018} and numerical simulations based on Navier-Stokes solvers~\cite{VALAREZO_JA_32_1995,RUMSEY_PAS_38_2002,KHORRAMI_AIAA_2802_2004,CHOUDHARI_AIAA_2844_2015,HOUSMAN_AIAA_2963_2016,MONTALA_PoF_36_2024} and LBMs~\cite{ISHIDA_AIAA_0259_2016,ISHIDA_AIAA_2306_2019,RUMSEY_JA_56_2018,DEGRIGNY_PhD_2021,MAEYAMA_CF_233_2022}. However, most 2D simulations are based on steady Reynolds-Averaged Navier-Stokes (RANS) solvers~\cite{VALAREZO_JA_32_1995,RUMSEY_PAS_38_2002,KHORRAMI_AIAA_2802_2004,MURAYAMA_AIAA_2080_2014,MURAYAMA_AIAA_3460_2018}, with the exception of more recent unsteady LB simulations by Ishida et al.~\cite{ISHIDA_AIAA_0259_2016,ISHIDA_AIAA_2306_2019} from JAXA. Hereafter, studies by Ishida et al. are then used for both qualitative and quantitative validations of our grid-refinement strategy at realistic Reynolds and Mach number conditions.
Hereafter, studies by Ishida et al. are then for both qualitative comparisons of 2D fields and quantitative validation of the wall pressure coefficient, with RANS simulations serving as the reference for the wall pressure data.

To investigate the ability of our framework in simulating high-lift configurations under realistic conditions, we conducted simulations of the flow past a 30P30N three-element airfoil at $(\mathrm{Re}, \mathrm{Ma}) = (1.7 \times 10^6, 0.17)$ using the D2Q9-RR-LBM. For angles of attack (AoAs) considered in this work, $\mathrm{AoA}\in\{5.5^{\circ},9.5^{\circ},14^{\circ},20^{\circ}\}$, the local Mach number reaches values of 0.5-0.6 in between the slat and main wing, and this pushes the D2Q9-LBM to its stability limit. For the highest AoA, the boundary layer becomes increasingly sensitive to no-slip boundary conditions, often leading to early detachment in case of inaccurate wall treatment. This provides valuable insights into scenarios where stair-cased boundaries, such as bounce-back conditions, may require improvements (e.g., interpolation or wall modeling) to capture the correct physics. In the end, these extreme conditions are a formidable opportunity to assess the stability and accuracy limits of choices made in this work, to provide best practices to the community, and to guide future research.

Hereafter, the domain size is set to $20C \times 20C$, where $C$ represents the chord length of the airfoil in its high-lift configuration. The airfoil was kept horizontal, and the AoA was imposed through Dirichlet boundary conditions at the left and bottom boundaries (as in Section 7.3 of Ref~\cite{DEGRIGNY_PhD_2021}) while Neumann boundary conditions were applied to the remaining boundaries.
As for Aeolian noise simulations, the pressure relaxation approach~(\ref{eq:relaxation_pressure}) was employed. However, the maximum damping coefficient was reduced to $0.05$ to avoid confinement effects resulting from the smaller domain size (compared to Section~\ref{subsec:aeolian_noise}). 
To attenuate the strength of the initial gradients  that appear near the geometry because of the homogeneous initialization at $\mathrm{Ma}=0.17$, a viscosity ramp was applied at the beginning of the simulation. Specifically, the kinematic viscosity was initially set to $\nu_{\mathrm{ini}} = 100 \nu$ and then decreased linearly to $\nu$ over a period of $20 t_c$, where the convective time scale is defined as $t_c = C/(c_s \mathrm{Ma})$. This smoothing technique is commonly encountered in industrial solvers and was notably used in ProLB for the simulation of landing gear noise under realistic conditions (see Section 3.3.5.3 of Ref.~\cite{COREIXAS_Master_2014}). 

To further improve the stability of the D2Q9-RR-LBM, we chose to equilibrate the trace of the second-order moments \CC{--a method similar to the bulk viscosity modification proposed by Dellar~\cite{DELLAR_PRE_64_2001}--} which is widely used in multi-relaxation-time and central-moment-based collision models for single-phase~\cite{GEIER_PRE_73_2006,ASINARI_PRE_78_2008,FEI_PRE_97_2018}, multiphase~\cite{FEI_PoF_31_2019,SAITO_PRE_98_2018}, and magnetohydrodynamic flow simulations~\cite{DEROSIS_PoF_31_2019,DEROSIS_PoF_32_2020}. This introduces an additional bulk viscosity proportional to $\Delta t$~\cite{WISSOCQ_JCP_450_2022}. As such it adapts to the local mesh resolution, hence being smaller in well-resolved regions and larger in more critical under-resolved areas. The method is generalizable to any collision model~\cite{COREIXAS_PRE_100_2019}, with implementations available in open-source solvers like Palabos~\cite{LATT_CMA_81_2021,LATT_ARXIV_09242_2025} and STLBM~\cite{LATT_PLOSONE_16_2021}. 

\begin{figure}[hbt]
    \centering
    \includegraphics[width=\textwidth]{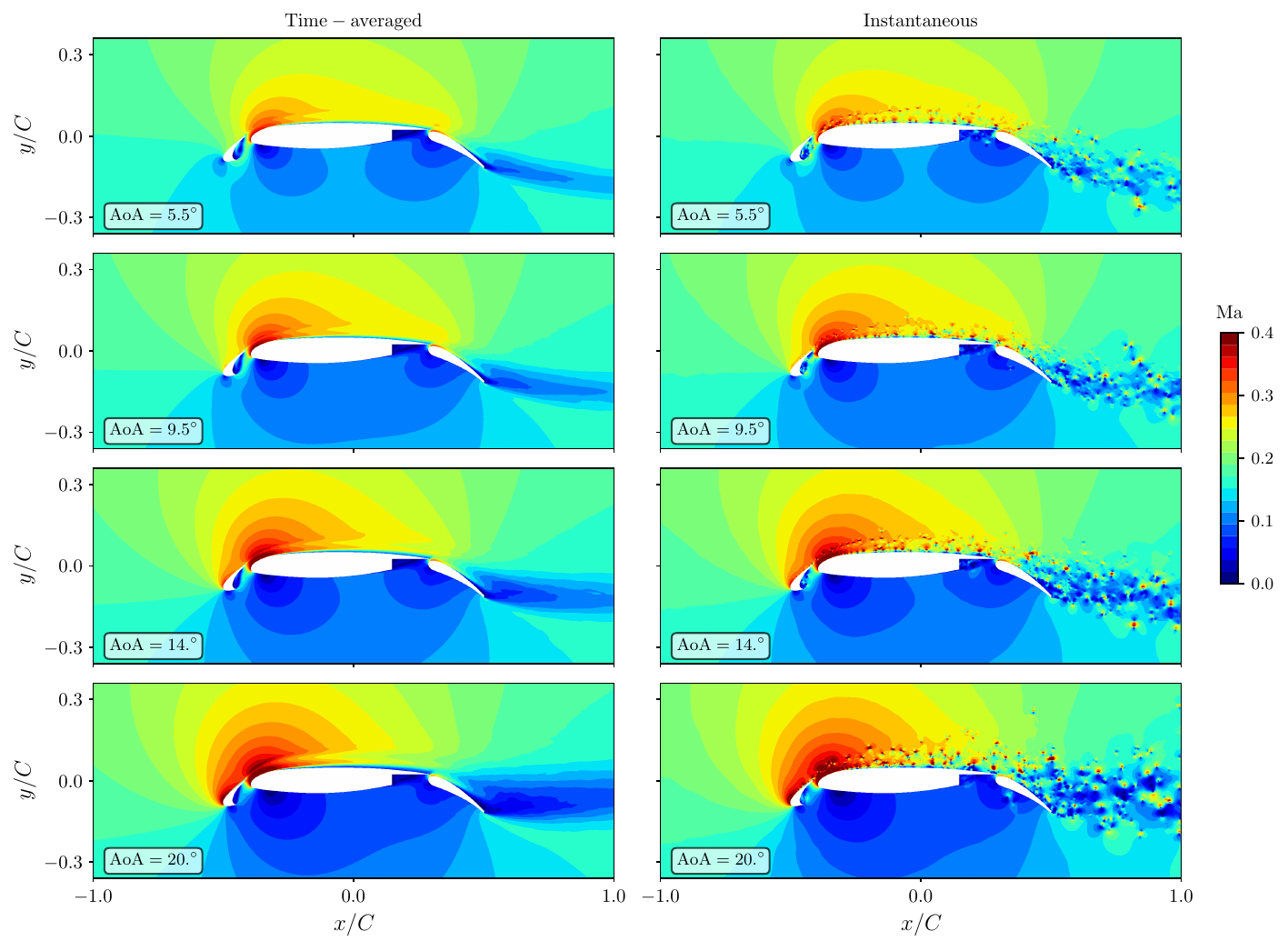}
	\captionsetup{skip=0pt}
    \caption{Simulations of the flow past the 30P30N three-element airfoil at $(\mathrm{Re},\mathrm{Ma},\Delta x_f)=(1.7\times 10^6 ,0.17,C/6400)$ with the D2Q9-RR-LBM. Time-averaged (left) and instantaneous (right) Mach fields are shown for several angles of attack (AoAs). Time averaging is performed at every coarsest time step and for $20t_c$, while instantaneous data is outputted at the end of the simulation ($t=40 t_c$). The colormap range follows that of Figure 4 in Ref.~\cite{ISHIDA_AIAA_2306_2019}. Across all AoAs, the simulation results align closely with those of JAXA.}
    \label{fig:30p30n_avg_inst_mach_impact_aoa}
\end{figure}

The simulations are carried out over $40 t_c$ for a fine resolution of about $\Delta x_f = C/6400$, which corresponds to a bit more than 2.6 millions of time iterations. 
Here, about $20 t_c$ are required to fully evacuate the initial transient from the domain, and for the different flow features to develop over the entire airfoil (boundary and shear layers, slat cove recirculation, etc). The last $20 t_c$ are dedicated to data processing, including mean and rms computations, as well as flow visualizations. To prevent spurious reflections at the inlet and outlet, pressure relaxation~(\ref{eq:relaxation_pressure}) is applied. $\sigma_{max}$ is reduced from $0.1$ to $0.05$ to avoid confinement effects. Eventually, wall pressure coefficient ($C_p$) values are outputted at the nearest fluid cells.

Figure~\ref{fig:30p30n_avg_inst_mach_impact_aoa} compares the time-averaged and instantaneous Mach number fields obtained for various AoAs. 
For $\mathrm{AoA}=5.5^{\circ}-14^{\circ}$, the flow remains mostly attached across all three elements. It accelerates while being compressed in the narrow regions between the airfoil elements, which significantly raise the local Mach number (up to 0.50-0.55). This interaction re-energizes the boundary layer and delays separation, a role expected from the slat and flap. At higher AoA ($20^{\circ}$), adverse pressure gradients intensify, and an early boundary layer separation seems to develop on the flap as we are getting closer to stall conditions.

Looking into more detail, several key trends are also accurately captured by our simulations. They include the downstream shift of the stagnation point on the slat, and the growth of the high-speed bubble at the leading edge of the main wing as AoA increases. All of this matches very well the 2D unsteady simulations by Ishida~\cite{ISHIDA_AIAA_0259_2016,ISHIDA_AIAA_2306_2019}, as well as 2D RANS~\cite{VALAREZO_JA_32_1995,RUMSEY_PAS_38_2002,MURAYAMA_AIAA_2080_2014,MURAYAMA_AIAA_3460_2018} and 3D unsteady NS simulations~\cite{CHOUDHARI_AIAA_2844_2015,HOUSMAN_AIAA_2963_2016,ISHIDA_AIAA_0259_2016,ISHIDA_AIAA_2306_2019,
RUMSEY_JA_56_2018,DEGRIGNY_PhD_2021,MAEYAMA_CF_233_2022,MONTALA_PoF_36_2024}. 

\begin{figure}[hbt]
    \centering
    \includegraphics[width=\textwidth]{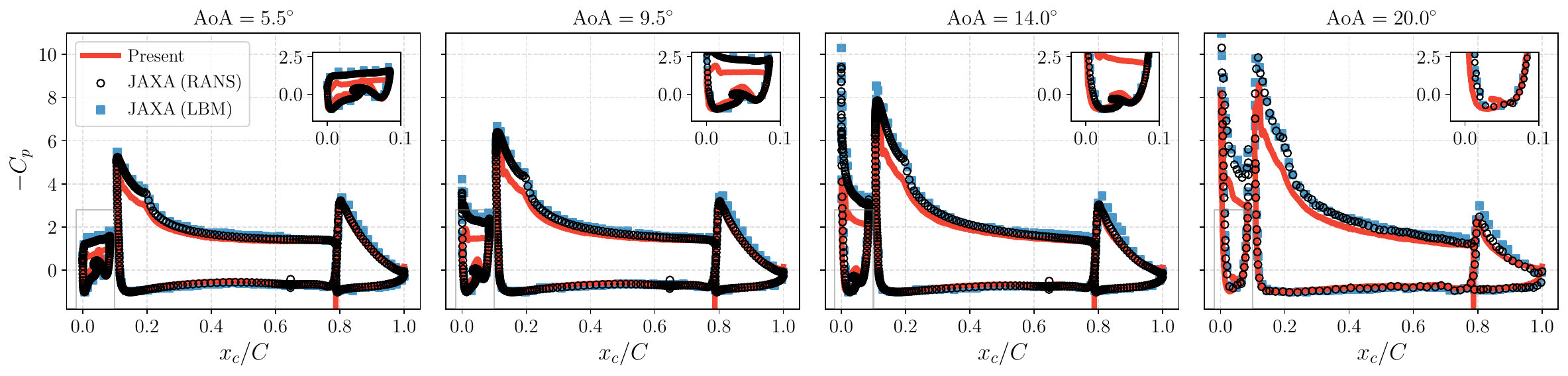}
	\captionsetup{skip=0pt}
    \caption{Simulation of flow past the 30P30N three-element airfoil at  $(\mathrm{Re}, \mathrm{Ma}, \Delta x_f) = (1.7 \times 10^6, 0.17, C/6400)$  using the D2Q9-RR-LBM. Time-averaged $C_p$ profiles are compared for all angles of attack, $\mathrm{AoA} \in \{5.5^\circ, 9.5^\circ, 14^\circ, 20^\circ\}$, against reference results from steady RANS~\cite{MURAYAMA_AIAA_2080_2014,MURAYAMA_AIAA_3460_2018} and unsteady LB simulations~\cite{ISHIDA_AIAA_2306_2019} conducted by JAXA. The coordinate $x_s$ is centered about the leading edge of the slat. Time averaging is performed every coarse step over $20 t_c$. Most of the RANS data are publicly available on the \href{https://cfdws.chofu.jaxa.jp/apc/apc5/rans.html}{APC5 website}, while RANS data at $\mathrm{AoA} = 20^\circ$ and all LB data are digitized from Figure 5 of Ref.~\cite{ISHIDA_AIAA_2306_2019} using \href{https://automeris.io/wpd/}{WebPlotDigitizer}, resulting in lower resolution, especially on the slat.}
    \label{fig:30p30n_cp_profile_impact_aoa}
\end{figure}

Further analysis confirms that the simple bounce back captures reasonably well the wall pressure evolution across the airfoil at the lowest AoA (see Figure~\ref{fig:30p30n_cp_profile_impact_aoa}). 
As expected, the under-resolution of the slat (about 500 points) leads to an underestimation of the shear layer strength at its trailing edge, which subsequently affect the wall pressure at the main wing’s leading edge, especially at higher AoA.
Adding local refinement patches in this critical region could mitigate the issue (see Figure 7.3 of Ref~\cite{DEGRIGNY_PhD_2021}). However, it is more likely that improving the treatment of solid boundaries would enhance accuracy under near-stall conditions, as Ishida achieved accurate $C_p$ profiles without requiring such refinement patches~\cite{ISHIDA_AIAA_2306_2019}. This will be investigated in future work.


\subsection{Partial conclusions~\label{sec:partial_conclusion_weakly}}


This initial validation demonstrates that the straightforward combination of (1) conservative CC grid refinement, (2) stair-cased bounce-back, and (3) the RR collision model is more than adequate for accurate and robust simulations of subsonic aerodynamics and aeroacoustics across a wide range of Reynolds and Mach numbers.

We were particularly surprised to find that this combination enables stable simulations of Aeolian noise with local Mach numbers reaching up to 0.72 in the low Reynolds regime ($\mathrm{Re} \leq 500$), and up to 0.55 for realistic simulations of flow past the 30P30N airfoil ($\mathrm{Re} > 10^6$). These Reynolds and Mach numbers significantly exceed what is typically reported in the literature for simulations performed using the D2Q9-LBM~\cite{KRUGER_Book_2017,SUCCI_Book_2018}, and particularly on non-uniform grids. 
Usually, one would need to employ subgrid-scale models~\cite{SAGAUT_CMA_59_2010} or correction terms~\cite{DELLAR_JCP_259_2014,WISSOCQ_JCP_450_2022} to achieve similar levels of accuracy and robustness. 

Although high-order lattices (e.g., D2Q13 and D2Q21) are typically reserved for high-speed flow simulations, we also conducted preliminary validations of our framework using these lattices. Notably, our extension of the original CC grid refinement method by Rohde et al.~\cite{ROHDE_IJNMF_51_2006} demonstrated both accuracy and robustness for low-speed flow simulations, even when the grid refinement interface intersects boundary conditions across multiple ghost layers (e.g., in lid-driven cavity simulations). To the best of our knowledge, this marks the first time such an extension has been proposed within the LB community.

Benchmarks in this section reaffirm the recent findings of Schukmann et al.~\cite{SCHUKMANN_FLUIDS_8_2023,SCHUKMANN_FLUIDS_10_2025}, and to some extent those of Astoul et al.~\cite{ASTOUL_JCP_418_2020,ASTOUL_JCP_447_2021,ASTOUL_PhD_2021}, that \emph{ensuring conservation at the grid refinement interface is the most critical requirement for a grid refinement algorithm}. This is even more important than the standard rescaling of the non-equilibrium part of populations. From an LB perspective, this is explained by the fact that conservation errors affect macroscopic quantities at the interface, hence introducing errors in equilibrium populations (zeroth-order terms in the Chapman-Enskog expansion). However, the non-rescaling introduces errors in the gradients of macroscopic quantities which are first-order terms in the Chapman-Enskog expansion. As such, the latter errors are of smaller magnitude and should be addressed only after conservation laws at the interface are rigorously satisfied.

To be fair, recent works on grid refinement~\cite{ASTOUL_JCP_418_2020,ASTOUL_JCP_447_2021,ASTOUL_PhD_2021,SCHUKMANN_FLUIDS_8_2023,SCHUKMANN_FLUIDS_10_2025} have also highlighted the critical impact of the collision model on the accuracy and stability of simulations performed on non-uniform grids. This emphasizes the need to reassess the impact of recent collision models~\cite{COREIXAS_PRE_100_2019,COREIXAS_RSTA_378_2020} on grid-refinement strategies so that fair and reliable best practices can be established for the design of new LB solvers. This will be considered in future work.

\section{Validation on fully compressible flows~\label{sec:validation_fully_compressible}}

With the successful validation of our framework for weakly compressible flow simulations, we now tackle the simulation of fully compressible flows. High-order lattices (D2Q13, D2Q21, and D2Q37) will be utilized to simulate both viscous and inviscid conditions, with a particular emphasis on the D2Q21 lattice. To enhance the robustness and accuracy of these high-order LBMs, a numerical equilibrium strategy is applied and a shock capturing technique based on a kinetic sensor is adopted. The double distribution function (DDF) approach is used for simulating polyatomic gases, where the second set of distribution functions, $g_i$, is closely related to the energy contained in internal degrees of freedom of the molecules. Both sets of populations ($f_i$ and $g_i$) are initialized at equilibrium (\ref{eq:exponential} and \ref{eq:gEq}), by enforcing the constraints (\ref{eq:G0_8mom})-(\ref{eq:G3yaa_8mom}) obtained from the initial state of macroscopic fields. This section focuses on diatomic gases, such as air, with a specific ratio $\gamma_r = 7/5$. 
The reference temperature $T_0$ plays a critical role as it influences macroscopic errors, dispersion properties, and the stability domains of the current compressible LBMs. Therefore, it must be chosen with care~\cite{LATT_RSTA_378_2020}. For inviscid simulations, $T_0$ is set to $0.35$, $0.9$, and $1.0$ for the D2Q13, D2Q21, and D2Q37 lattices, respectively. However, for viscous simulations, the values are decreased to $0.65$ and $0.4$ for the D2Q21 and D2Q37 lattices to reduce dispersion issues that arise near no-slip boundary conditions.
Further technical details about the present compressible DDF-LBMs can be found in~\ref{app:lbm}.

\subsection{Inviscid Sod shock tube \label{subsec:sod}}

Hereafter, we consider the 1D Riemann problem which is commonly referred to as the Sod shock tube~\cite{SOD_JCP_27_1978}. This setup consists of a closed tube divided into two subdomains by a thin membrane. Both regions contain the same gas but differ in their thermodynamic properties: density $\rho$, temperature $T$, pressure $P$ and velocity $u$: 
\begin{equation}\label{eq:sod_init}
\left(p/p_0,\rho/\rho_0,u\right) = \left\{
\begin{array}{l}
(10,8,0), \quad \mathrm{if}\:\: x/L\leq 1/2,\\
(\phantom{0}1,1,0), \quad \mathrm{if}\:\: x/L>1/2,
\end{array}
\right.
\end{equation}
where the subscript ``0'' stands for right state values, and $L$ is the length of the tube. At the initial instant, the membrane is removed, causing a rapid flow acceleration from the high-pressure region to the low-pressure one, aiming to equalize pressure within the tube. This triggers the formation and propagation of three characteristic waves. First, the gas compression generates a shock wave that moves toward the low-pressure side. Second, the expansion of the gas toward the high-pressure side produces an expansion wave. Finally, the contact discontinuity, a slip line between the two waves, travels at a constant speed toward the low-pressure side.

In this paper, we consider the above canonical configuration that was used to validate numerous numerical methods for Euler/Navier-Stokes-Fourier equations~\cite{SOD_JCP_27_1978,JIANG_JCP_126_1996,BORGES_JCP_227_2008,GEROLYMOS_JCP_228_2009,FU_CiCP_26_2019,DIRENZO_CPC_255_2020,LUSHER_CPC_267_2021,HOPPE_CMAME_391_2022}, discrete velocity methods that do not rely on the collide-and-stream algorithm~\cite{NADIGA_JSP_81_1995,LIa_PRE_76_2007,CHEN_CTP_52_2009,WANG_IJMPC_21_2010,YANG_CF_79_2013,GUO_PRE_91_2015,LIU_PRE_98_2018,DZANIC_JCP_486_2023}, LBMs based on fixed~\cite{COREIXAS_PRE_96_2017,LATT_RSTA_378_2020,THYAGARAJAN_PoF_35_2023,MATTILA_PF_29_2017,JAMMALAMADAKA_AIAA_3055_2020} or adaptive lattices~\cite{SUN_PRE_58_1998,SUN_TST_5_2000,COREIXAS_PoF_32_2020}, and hybrid LBMs~\cite{LI_AIAA_4128_2015,KOPRIVA_AIAA_3929_2019,WISSOCQ_PoF_34_2022}. 
To highlight the accuracy and robustness of our meshing strategy, several grids are used to discretize the $L\times L$ simulation domain: \emph{band} with a refinement patch centered around the initial discontinuity and located between $x/L=0.25$ and $0.75$, \emph{sqr} with a square-shape refinement patch centered about the domain center and of size $L/2$, \emph{hori} with a horizontal refinement interface located at $y/L=0.5$. \emph{band} is ideal as it properly captures the generation of the three waves at the beginning of the simulation. \emph{hori} is the worst configuration as the refinement interface is normal to the propagation of all three waves for the entire domain. \emph{sqr} is an intermediate configuration, less adequate than \emph{band} as it only partially covers the initial discontinuity, but better than \emph{hori} since its refinement interface is normal to the propagation direction of waves on a restricted part of the domain. Neumann boundary conditions are imposed at $x/L=0$ and $1$ on both $f_i$ and $g_i$, while periodic conditions are used at $y/L=0$ and $1$. To increase the complexity of this benchmark, \emph{inviscid conditions} are enforced by setting viscosity to zero, and a shock-capturing technique based on a kinetic sensor~\cite{LATT_RSTA_378_2020,COREIXAS_PoF_32_2020,THORIMBERT_JoCS_64_2022} is employed to ensure robustness near discontinuities (see Eq.~(\ref{eq:shock_function})). 

\begin{figure}[bt]
    \centering
    \includegraphics[width=\textwidth]{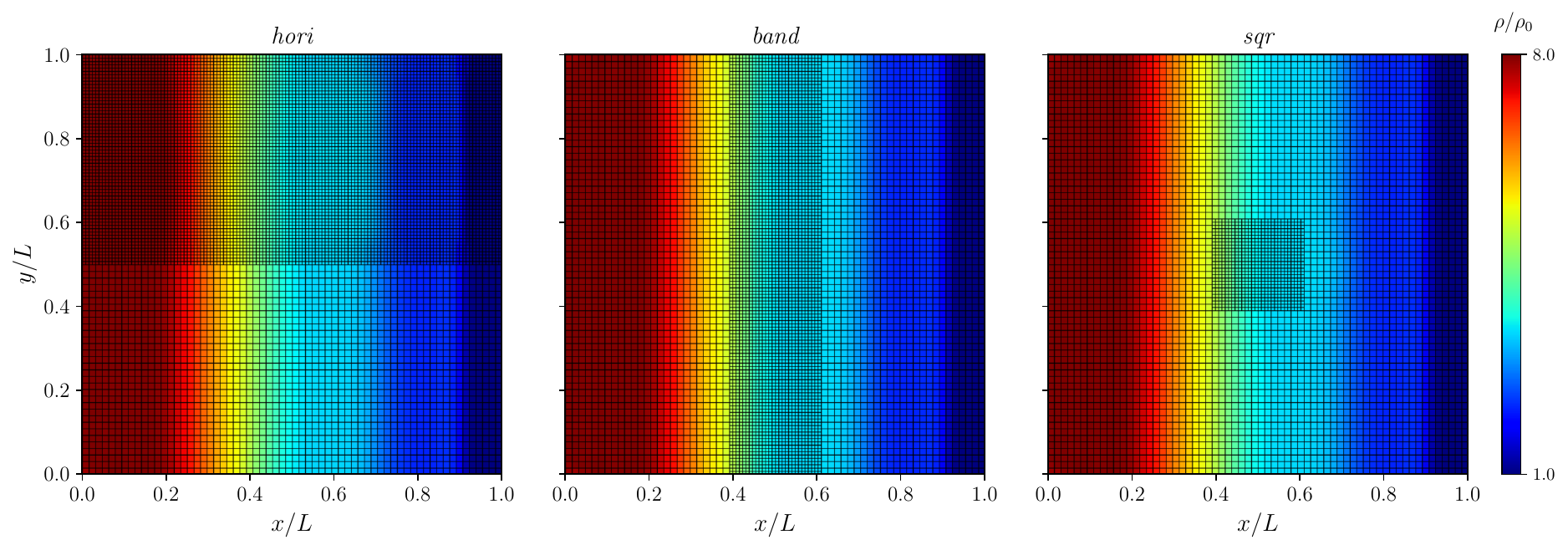}
	\captionsetup{skip=0pt}
    \caption{Interaction between canonical compressible waves (rarefaction, contact discontinuity and shock) and grid refinement interfaces for the Sod shock tube benchmark. The D2Q21-LBM is used with a numerical equilibrium (8 moments) and for inviscid conditions ($\nu=\nu_T=0$). The coarse space step is $\Delta x_c = L/64$ for all meshes: (left) \emph{hori}, (middle) \emph{band}, and (right) \emph{sqr}. The mesh is superimposed to the normalized density field ($\rho/\rho_0$) to highlight the good accuracy and robustness of our approach in under-resolved conditions. Data is outputted at $t=0.25 t_c$ with the characteristic time being $t_c=L/\sqrt{\gamma_r T_0}$.}
    \label{fig:riemann_grids_q21}
\end{figure}

\begin{figure}[bt]
    \centering
    \includegraphics[width=\textwidth]{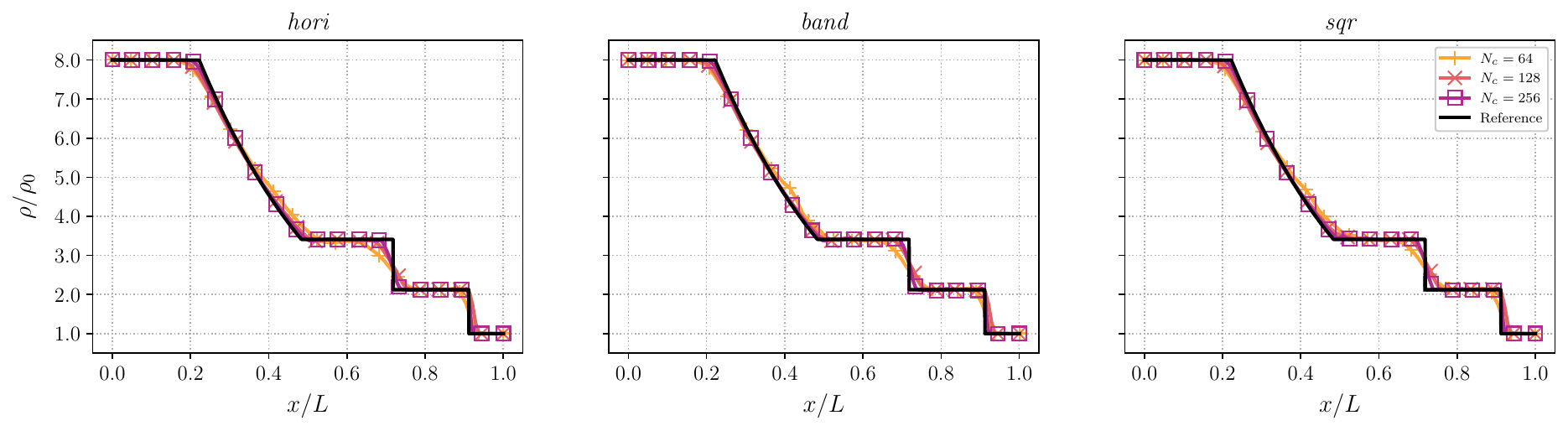}
	\captionsetup{skip=0pt}
    \caption{Normalized density profiles along the centerline ($y/L=0.5$) of the Sod shock tube benchmark for inviscid conditions ($\nu=\nu_T=0$), using the D2Q21-LBM is used with a numerical equilibrium (8 moments). Results obtained with three different mesh configurations (\emph{hori}, \emph{band}, and \emph{sqr}) and mesh resolutions ($N_c=L/\Delta x_c$) are in very good agreement with the analytical solution. Data is outputted at $t=0.25 t_c$ with the characteristic time being $t_c=L/\sqrt{\gamma_r T_0}$.}
    \label{fig:riemann_profiles_q21}
\end{figure}

Figure~\ref{fig:riemann_grids_q21} gathers the results obtained with the D2Q21 lattice for the three aforementioned mesh configurations. Here, a relatively coarse mesh resolution ($N_c=L/\Delta x_c= 64$) is used to assess the accuracy and robustness of our meshing strategy in under-resolved conditions. Despite the significant differences in mesh configurations, all simulations yield similar density fields. No numerical artefacts are observed at the interface between coarse and fine grids, despite the three-layer buffer zone required for data transfer by the D2Q21 lattice. We expected dispersion errors to dominate in the buffer region, especially in the \emph{hori} configuration, due to differences in wave propagation speeds between consecutive mesh levels. However, this is clearly not the case.

Going into more details, Figure~\ref{fig:riemann_profiles_q21} shows that all density profiles match very well the analytical solution.
As usual, overdissipation is observed close to the contact discontinuity ($x/L\approx 0.71$) because this wave requires particular treatments (local anti-diffusion, compression strategy, etc) to remain as sharp as possible~\cite{SHU_SIAM_51_2009}. While discrepancies are evident for $N_c = 64$ and $0.2 \leq x/L \leq 0.5$, the results improve with increased mesh resolution.
Interestingly, these impressive results are not limited to the D2Q21 lattice since LBMs based on both smaller (D2Q13) and larger (D2Q37) lattices also achieve similar accuracy (see Figure~\ref{fig:riemann_profiles_q13_q37}). 
The only noticeable difference between all three lattices is their respective robustness, as the D2Q13-LBM is very close to its stability limit in terms of temperature variations and low viscosity. 

To the best of our knowledge, this is the first time a grid refinement algorithm for LBMs demonstrates such a high level of accuracy and robustness for the Sod shock tube despite \emph{inviscid} and \emph{under-resolved} conditions.

\subsection{Inviscid 2D Riemann problem \label{subsec:riemann2d}}

The following 2D Riemann problem can be broken down into four 1D Riemann problems (or Sod shock tubes), each initialized in one of the four quadrants of the simulation domain. Depending on the density, pressure, and velocity conditions in each quadrant, various compressible phenomena emerge at the quadrant interfaces, particularly near the center of the simulation domain. These phenomena have been thoroughly investigated by Schulz-Rinne et al.~\cite{SCHULZRINNE_SIAM_14_1993}, Lax \& Lui~\cite{LAX_SIAM_19_1998}, and Kurganov \& Tadmor~\cite{KURGANOV_NMPDE_18_2002}. Building on these studies, we focus on the configuration commonly referred to as ``F'' or ``12'':

\begin{table}[htbp]
\centering
\small
\renewcommand{\arraystretch}{1.2} 
\begin{tabular}{l|l}
$\rho^{[2]}/\rho_0 = 1$         & $\rho^{[1]}/\rho_0 = 0.513$         \\
$u_x^{[2]}/u_0 = 0.7276$        & $u_x^{[1]}/u_0 = 0$                \\
$u_y^{[2]}/u_0 = 0$             & $u_y^{[1]}/u_0 = 0$                \\
$p^{[2]}/p_0 = 1$               & $p^{[1]}/p_0 = 0.4$                \\[0.2cm]
\hline \\[-0.3cm] 
$\rho^{[3]}/\rho_0 = 0.8$       & $\rho^{[4]}/\rho_0 = 1$            \\
$u_x^{[3]}/u_0 = 0$             & $u_x^{[4]}/u_0 = 0$                \\
$u_y^{[3]}/u_0 = 0$             & $u_y^{[4]}/u_0 = 0.7276$           \\
$p^{[3]}/p_0 = 1$               & $p^{[4]}/p_0 = 1$                  \\
\end{tabular}
\end{table}

\noindent with reference quantities being $p_0=\rho_0 T_0$, $\rho_0=1$, $u_0=\sqrt{\gamma_r T_0}$ in LB units. 
 
For the above configuration, several key phenomena are known to occur during the simulation. 
First, the initial state is symmetric along the diagonal axis ($x=y$), so the macroscopic fields are expected to preserve this symmetry.
Second, all quadrants (except [1]) share the same pressure field and longitudinal velocities along the interfaces of [3], thus leading to contact discontinuities at its boundaries.
Third, two shock waves will form at the interfaces with quadrant [1] and interact, hence, creating a complex pattern in this area.
The latter pattern will eventually cause the contact discontinuities to roll up into a pair of vortices in quadrant [3].

To further investigate the ability of our refinement strategy to accurately and robustly simulate the above phenomena in \emph{inviscid} conditions ($\nu=\nu_T=0$), we consider three mesh configurations to discretize the $[-L,L]\times [-L,L]$ simulation domain: \emph{diag}, \emph{axes}, and \emph{optim}. 
The \emph{diag} configuration features a grid refinement interface along the diagonal $x=y$, hence, aligning with the symmetry plane of the initial state. This setup was chosen because it poses significant challenges in terms of accuracy and robustness, as various physical phenomena evolve along this diagonal. The configuration \emph{axes} employs refinement patches along the mesh axes, which enhances the capture of the initial evolution of discontinuities. However, this configuration is less effective at accurately capturing the interaction and evolution of waves along the diagonal axis ($x=y$). Lastly, \emph{optim} was designed to represent an optimal configuration that successfully captures the initial evolution of discontinuities, the complex patterns emerging in quadrant [1], as well as the mushroom-like vortices forming in quadrant [3].

Neumann boundary conditions are unsuitable here because the \emph{diag} configuration has a refinement interface that intersects the boundaries in a non-axis-aligned manner, which the current version of the code cannot perfectly handle. Instead, periodic boundary conditions are applied at all boundaries and data is outputted on a reduced domain $[-L/2, L/2] \times [-L/2, L/2]$ at $t = 0.30 t_c$, where the characteristic time is $t_c = L/\sqrt{\gamma_r T_0}$. On top of the three mesh configurations, three mesh resolutions are considered to better quantify the accuracy and robustness of our approach in severely under-resolved ($N_c=32$) and well-resolved ($N_c=512$) conditions.

\begin{figure}[bt!]
    \centering
    \includegraphics[width=0.9\textwidth]{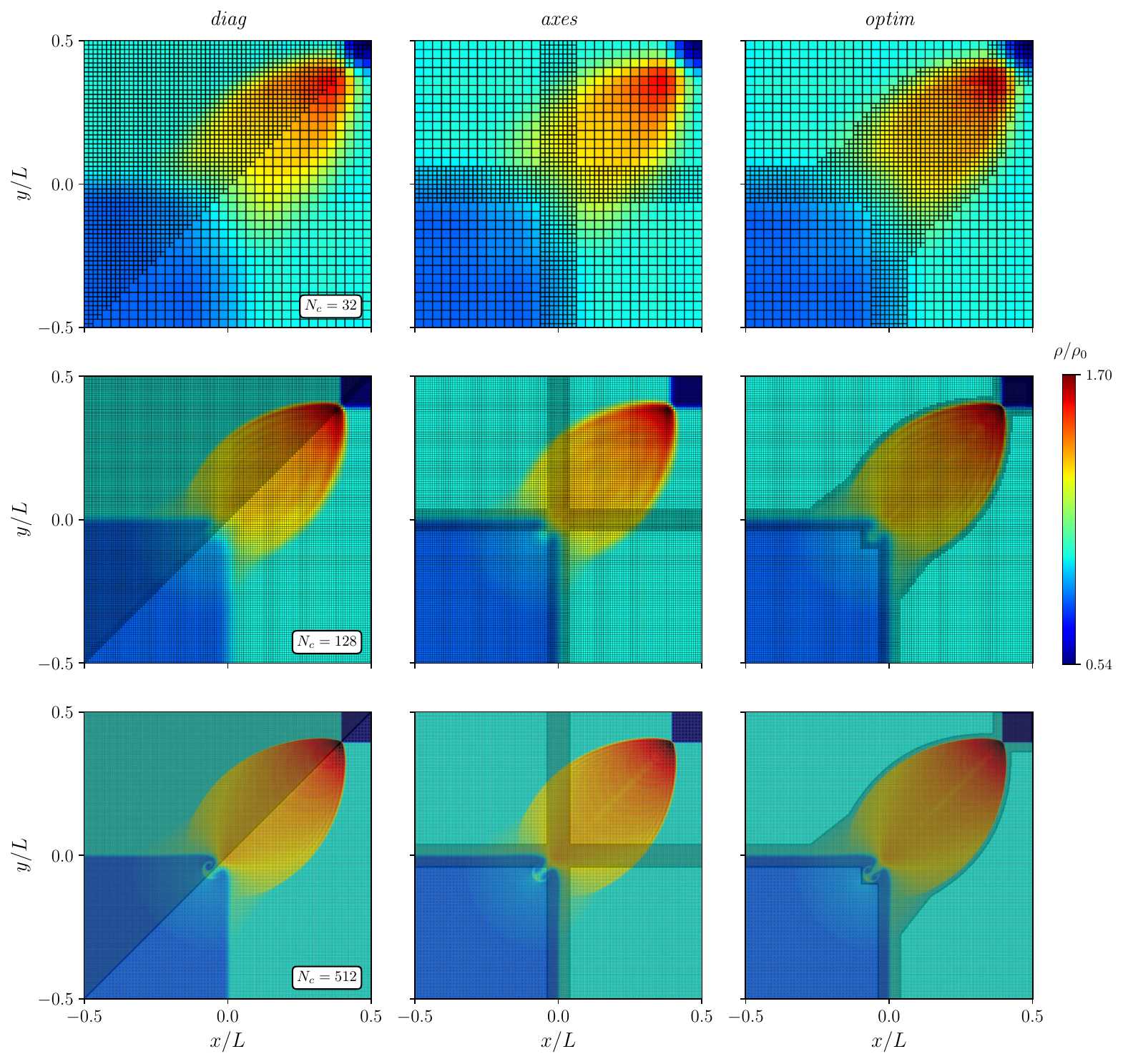}
	\captionsetup{skip=0pt}
    \caption{Normalized density fields of the 2D Riemann problem for inviscid conditions ($\nu=\nu_T=0$), using the D2Q21-LBM with a numerical equilibrium (8 moments). Each column corresponds to a mesh configuration (\emph{diag}, \emph{axes}, and \emph{optim}), while each row stands for a given mesh resolution ($N_c=L/\Delta x_c$). Data is outputted at $t=0.30 t_c$ with the characteristic time being $t_c=L/\sqrt{\gamma_r T_0}$.}
    \label{fig:riemann2d_grids_q21}
\end{figure}

Figure~\ref{fig:riemann2d_grids_q21} gathers density fields obtained with the D2Q21-LBM for each mesh configuration (\emph{diag}, \emph{axes}, and \emph{optim}) and coarse resolution ($N_c\in\{32, 128, 512\}$). 
Globally speaking, it is remarkable that all simulations remain stable, even in very challenging and severely under-resolved conditions. They produce --with various degrees of accuracy-- the expected symmetric flow features, such as the complex interaction between shock waves in quadrant [1] and the formation of mushroom-like vortices in quadrant [3].
Notably, no spurious vortices or significant dispersion issues arise at the interface between refinement levels, despite the simulations being run under inviscid conditions. 
Going into more details, the \emph{optim} configuration is, as expected, the only one to accurately capture all relevant physical phenomena, including the vortex pair in quadrant [1]. Interestingly, despite being the most challenging, the \emph{diag} configuration maintains a high degree of symmetry for most flow features. The \emph{axes} configuration also produces excellent results without disrupting the symmetric formation of vortices. 
While the D2Q13-LBM only provides stable results in under-resolved conditions due to numerical dissipation, the D2Q37-LBM produces similar results to the D2Q21-LBM. For brevity, these findings are summarized in~\ref{app:riemann}.

\subsection{Viscous flow past a NACA0012 airfoil at transonic speeds\label{subsec:naca_transonic_compressible}}

The simulation of viscous flows past airfoils is crucial for optimizing the design of commercial and military aircraft. One of the critical aspects of wing design is the ability to predict the evolution of airflow around the wing in transonic cruise conditions. In such conditions, shock waves can form on the airfoil surface as the flow locally reaches supersonic speeds. These shock waves cause a sudden deceleration of the airflow, resulting in a sharp pressure rise that can lead to shock-induced boundary layer separation. This interaction can create unstable flow conditions, such as buffeting, where periodic flow separation and reattachment produce fluctuating aerodynamic forces. In addition to significantly increasing drag and deteriorating aircraft performance, buffeting can lead to structural vibrations, which, if severe, can cause fatigue or even failure of the wing structure.

To assess the effectiveness of our grid refinement strategy in accurately capturing the complex interaction between shock waves and boundary layers, we simulate the flow past a NACA0012 airfoil under viscous conditions ($\mathrm{Re}=10^4$), at transonic speeds ($\mathrm{Ma}=0.85$) and a zero-degree angle of attack. 
Satofuka et al. were among the first to examine this setup using a 2D Navier-Stokes-Fourier solver, demonstrating its highly unsteady behavior near the trailing edge, as shown by the six instantaneous fields and $C_p$ profiles in Figure 17 of Ref.~\cite{SATOFUKA_GAMM_1987}. This configuration has been used both for validating academic codes on uniform meshes~\cite{FRAPOLLI_PhD_2017} and for assessing the grid refinement strategy implemented in the industrial solver ProLB~\cite{FENG_PRE_101_2020}. However, neither study discussed the unsteady nature of the flow in detail, nor specified which of the six curves served as the reference data for the $C_p$ profile comparison.

\begin{figure}[bt!]
    \centering
    \includegraphics[width=\textwidth]{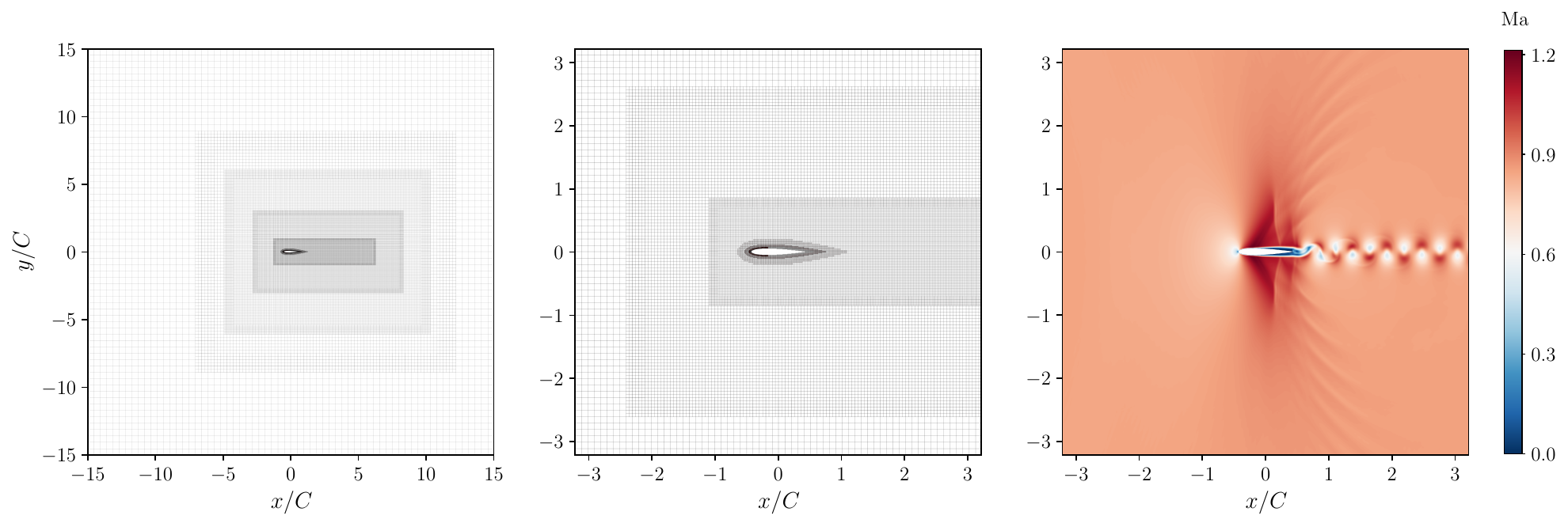}
	\captionsetup{skip=0pt}
    \caption{Viscous flow past a NACA0012 airfoil at transonic speeds using the D2Q21-LBM based on a numerical equilibrium (8 moments), and with $(\mathrm{Re},\mathrm{Ma})=(10^4,0.85)$. Left panel: Mesh for the full simulation domain. Middle panel: Mesh zoom-in around the airfoil. Right panel: Mach number evolution around the airfoil. The refinement patch at the leading edge has a fine mesh resolution of $N_f=C/\Delta x_f\approx 1100$, while the rest of the airfoil has a resolution twice coarser. For the sake of clarity, only one cell over 16 are shown.}
    \label{fig:naca_grid_zoom_mach}
\end{figure}

To tackle this configuration, we use a simulation domain of size $\left[-15C, 15C\right]^2$ -with $C$ the airfoil chord- that is bounded by three equilibrium Dirichlet boundary conditions (left, top, and bottom) and one Neumann condition (right). As a best practice from previous works~\cite{LATT_RSTA_378_2020,THYAGARAJAN_PoF_35_2023}, we start with a uniform flow at a freestream Mach number lower than the targeted one. The freestream Mach number imposed by the Dirichlet boundary conditions is then rapidly increased (over one convective time) to its nominal value. This notably allows to avoid convergence issues that were encountered with the Newton-Raphson algorithm at the beginning of the simulation for the D2Q13-LBM. When it comes to the mesh generation, four rectangular and three NACA-shaped refinement patches surround the airfoil. The finest refinement patch aims at (1) better capturing the boundary layer development at the leading edge of the airfoil, and (2) reducing pressure oscillations at the wall that are partly induced by the staircase nature of the bounce-back approach. Hereafter, simulation data are outputted at $t=5 t_c^{\mathrm{domain}}$, so that the transient flow is \emph{completely} evacuated from the simulation domain, and with the characteristic time being $t_c^{\mathrm{domain}}=30C/(\mathrm{Ma}\sqrt{\gamma_r T_0})$. If one is only interested in what is happening close to the airfoil, the output time can be reduced by a factor two to three which reduces the simulation time by approximately the same factor. Eventually, $C_p$ values are obtained at the wall using the inverse distance weighting (order two) of values available on the nearest and second nearest fluid cells.

Figure~\ref{fig:naca_grid_zoom_mach} shows snapshots of the mesh and the instantaneous Mach field obtained using the D2Q21-LBM. This 2D field clearly illustrates the interaction between the boundary layer and the primary shock wave, which manifests as a $\lambda$-shaped shock structure. This interaction causes early boundary layer separation and the formation of a secondary $\lambda$-shaped shock wave. Interestingly, this secondary shock wave  periodically merges with and detaches from the primary shock wave. Additionally, strong pressure waves -generated by the vortex shedding occurring at the trailing edge- start accumulating on a plane perpendicular to the meanflow direction.
These results align well with the observations by Satofuka et al.~\cite{SATOFUKA_GAMM_1987} and confirm that, despite crossing several grid refinement interfaces, the time evolution of the boundary layer and shock waves remains unaffected.

\begin{figure}[bt!]
    \centering
    \includegraphics[width=\textwidth]{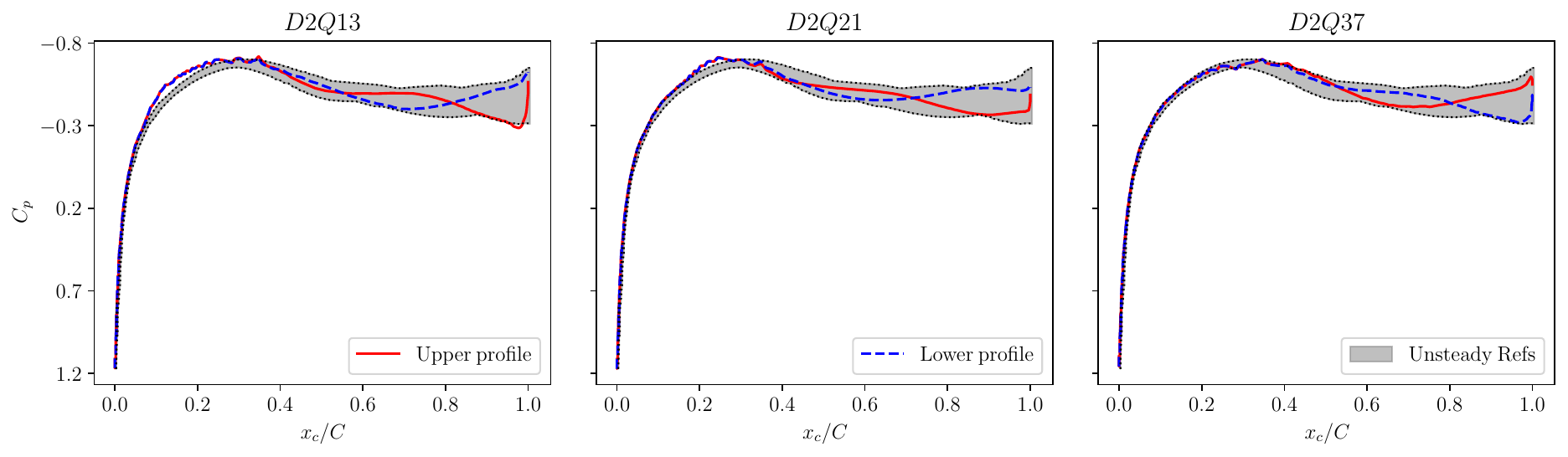}    
	\captionsetup{skip=0pt}
    \caption{Instantaneous wall $C_p$ profile comparisons for the viscous flow past a NACA0012 airfoil at transonic speeds with $(\mathrm{Re},\mathrm{Ma})=(10^4,0.85)$. The longitudinal coordinate $x_c$ is centered about the leading edge. The numerical equilibrium (8 moments) is used for all simulations. Unsteady reference data is gathered from the work by Satofuka et al.~\cite{SATOFUKA_GAMM_1987} and merged into a single grey area that highlights the unsteadiness of the flow.}
    \label{fig:naca_cp_profiles_ma085_LocRef}
\end{figure}

Further comparisons with reference wall $C_p$ profiles (Figure~\ref{fig:naca_cp_profiles_ma085_LocRef}) show that all LBMs provide results in good agreement with the merged profiles extracted from Satofuka et al.~\cite{SATOFUKA_GAMM_1987} using WebPlotDigitizer (https://automeris.io/). 
Interestingly, while all simulations are run for the same physical time, the profiles differ as we get closer to the trailing edge. This is likely due to the inherent numerical dissipation of each LBM that causes variations in the timing of boundary layer and shock wave interactions earlier in the simulation.
Spurious oscillations can be seen near the leading edge ($x/C \leq 0.3$), where the primary shock waves interact with the thin boundary layer, which is consistent with results reported in Figure 15 of Ref.~\cite{FENG_PRE_101_2020}.
This region is also where staircase nature of the bounce-back approach could be problematic, as already seen for the flow past the three-element 30P30N airfoil in Section~\ref{subsec:30p30n}. 
In fact, preliminary simulations without the local refinement at the leading edge showed oscillations with a higher amplitude (see Figure~\ref{fig:naca_cp_profiles_ma085} in~\ref{app:naca}). Nonetheless, the present results are very promising and further confirm the viability of our refinement strategy for the simulation of complex interactions, such as buffet phenomena that occur at transonic speeds.

\subsection{Viscous flow past a NACA0012 airfoil at supersonic speeds\label{subsec:naca_supersonic_compressible}}

In addition to transonic conditions, simulating an airflow past a wing in supersonic conditions is critical for the design of high-speed aircraft, particularly supersonic commercial aircrafts, rockets, military jets, and spacecrafts during re-entry. At supersonic speeds, the flow dynamics change significantly due to the presence of strong shock waves that form in front of and along the airfoil. Similarly to the transonic case, these shock waves will cause drastic increases in drag, and can result in flow separation. However, heating effects are more significant in the supersonic regime, notably because of the intense compression of air close to the airfoil surface. 

\begin{figure}[bt!]
    \centering
\includegraphics[width=\textwidth]{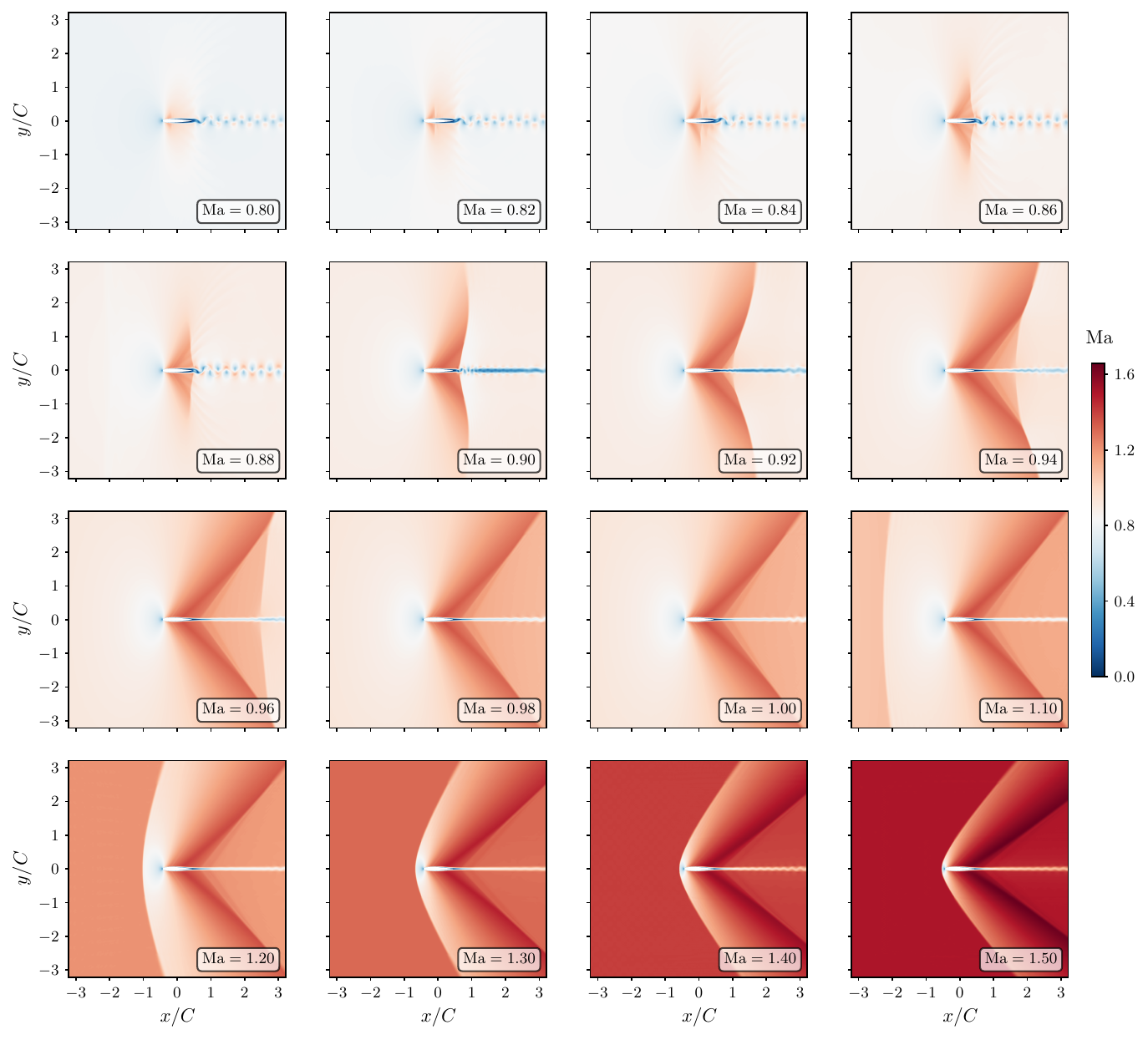}
	\captionsetup{skip=0pt}
    \caption{Evolution of the local Mach number close to the airfoil for freestream Mach numbers ranging from transonic to supersonic regimes. The colormap of the simulation performed at $\mathrm{Ma=1.5}$ is used as a reference for all data. The 16 simulations are based on high-order lattices with a numerical equilibrium (8 moments). Data are plotted at $t = 5 t_c^{\mathrm{domain}}$, with $t_c^{\mathrm{domain}}=30C/(\mathrm{Ma}\sqrt{\gamma_r T_0})$. The lattice size is chosen based on the freestream Mach number: D2Q13 for $\mathrm{Ma} \leq 0.98$, and D2Q21 for $1.0 \leq \mathrm{Ma} \leq 1.5$.}
    \label{fig:naca_comp_ma_q21}
\end{figure}

\begin{figure}[bt!]
    \centering
\includegraphics[width=\textwidth]{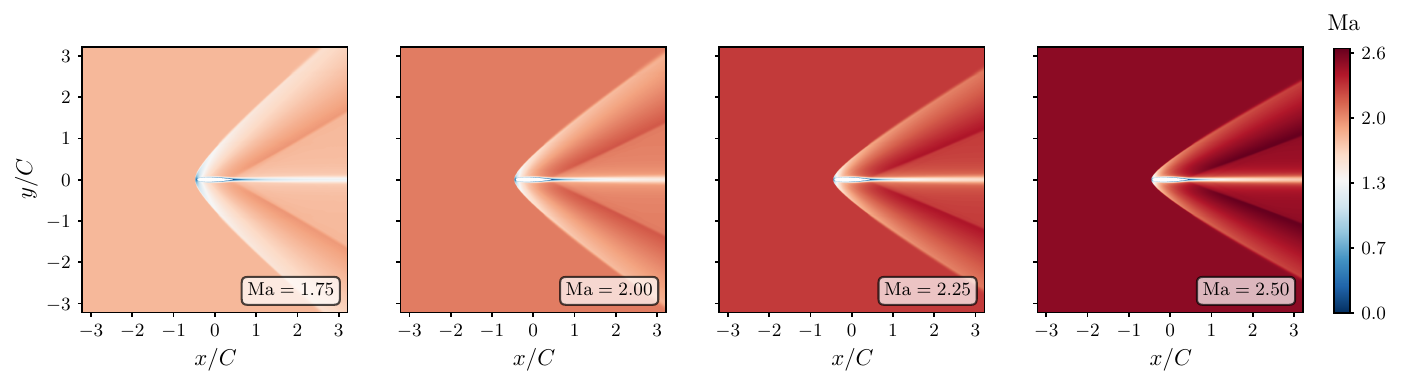}
	\captionsetup{skip=0pt}
    \caption{Evolution of the local Mach number close to the airfoil for freestream Mach numbers ranging from $1.75$ to $2.5$. The colormap of the simulation performed at $\mathrm{Ma=2.5}$ is used as a reference for all data. These simulations are based on the D2Q37-LBM with a numerical equilibrium (8 moments). Data are plotted at $t = 5 t_c^{\mathrm{domain}}$, with $t_c^{\mathrm{domain}}=30C/(\mathrm{Ma}\sqrt{\gamma_r T_0})$.}
    \label{fig:naca_comp_ma_q37}
\end{figure}

Hereafter, we use the same setup of Section~\ref{subsec:naca_transonic_compressible} to simulate the viscous flow ($\mathrm{Re}=10^4$) past a NACA0012 airfoil at supersonic speeds. We first increase the freestream Mach number up to 1.5, and move from subsonic to supersonic inlet boundary conditions by imposing all macroscopic variables. Figure~\ref{fig:naca_comp_ma_q21} gathers 16 simulations, performed with the D2Q13- and D2Q21-LBM, for increasing values of the freestream Mach number ($0.8 \leq \mathrm{Ma}\leq 1.5$). These simulations highlight several physical phenomena that appear when transitioning from the transonic to the supersonic regime. Notably, as the primary shock wave gains in strength, it merges with the pressure waves that were accumulating on a perpendicular plane located at the trailing edge of the airfoil. By further increasing the Mach number, this plane starts to bend and compress the wake, hence, increasing the frequency of vortex shedding. 
In the supersonic regime $\mathrm{Ma}>1.0$, a strong bow shock wave appears upstream the airfoil. As $\mathrm{Ma}$ is progressively increased to 1.5, the bow shock wave gets closer to the airfoil and starts bending. This induces a strong compression of the flow between the shock and the leading edge, which, in turn, locally increases wall pressure, temperature and heat transfer. The boundary layer is now steady and remains attached over the entire airfoil. Complex interactions between the secondary $\lambda$-shaped shock waves, emerging at the trailing edge, and the wake impact the exact location at which the vortex shedding is being triggered. 
Additional simulations confirm that compression effects intensify when the Mach number is increased to 2.5 (see Figure~\ref{fig:naca_comp_ma_q37}) --value beyond which the D2Q37-LBM begins to encounter stability issues due to the velocity norm locally exceeding $2.0$ (in mesh/lattice units). At these high Mach numbers, the airflow is further compressed against the airfoil surface because of the reduced shock standoff distance, resulting in a significant increase in wall pressure. 
The wake stabilizes under the intense compression of the secondary shock, preventing any instabilities from forming --at least within the limits of the present mesh resolution.

\begin{figure}[bt!]
    \centering
    \includegraphics[width=0.7\textwidth]{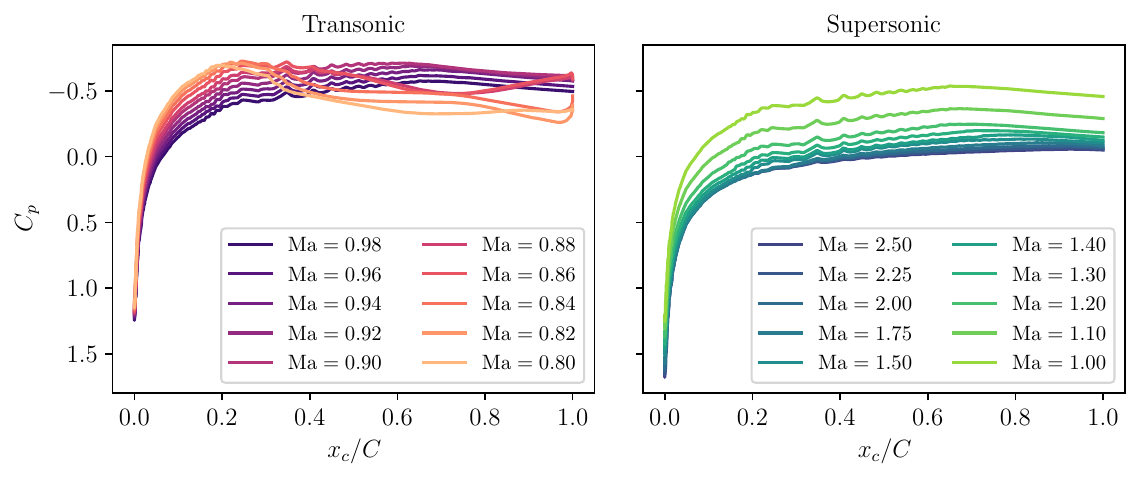}    
	\captionsetup{skip=0pt}
    \caption{Instantaneous wall $C_p$ profile comparisons for the viscous flow past a NACA0012 airfoil at $\mathrm{Re}=10^4$, and for a wide range of freestream Mach numbers ($0.80 \leq \mathrm{Ma}\leq 2.5$). The longitudinal coordinate $x_c$ is centered about the leading edge. The lattice size is adjusted based on the freestream Mach number: D2Q13 for $\mathrm{Ma} \leq 0.98$, D2Q21 for $1.0 \leq \mathrm{Ma} \leq 1.5$ and D2Q37 for $1.75 \leq \mathrm{Ma} \leq 2.5$. The numerical equilibrium (8 moments) is used for all simulations.}
    \label{fig:naca_cp_profiles_shock_standoff_distance}
\end{figure}

Globally speaking, all physical phenomena cross refinement interfaces without any issue. Notably, the bow and secondary shock waves cross \emph{four} refinement interfaces with only some noticeable thickening. Remarkably, a vortex shedding is formed in the wake of the airfoil at $\mathrm{Ma}\leq 1.5$, despite crossing \emph{two} refinement interfaces over a distance of less than $C$. 
Figure~\ref{fig:naca_cp_profiles_shock_standoff_distance} compiles $C_p$ profiles obtained for the full range of freestream Mach numbers. These profiles provide further insight into the increased pressure exerted over the airfoil by the surrounding airflow for $\mathrm{Ma} > 0.85$. Notably, for $\mathrm{Ma}\geq 0.90$, $C_p$ profiles become steady and symmetric because the shock/boundary-layer interaction is pushed away from the airfoil surface by the increased freestream velocity. As expected, the profiles are shifted downwards as pressure increases over the entire airfoil for supersonic speeds. For $\mathrm{Ma} \leq 2.00$, the $C_p$ profiles are positive, indicating that the wall pressure exceeds the freestream pressure. Further increases in $\mathrm{Ma}$ do not lead to significant changes in $C_p$. 

\subsection{Partial conclusions~\label{sec:partial_conclusion_fully}}

This validation has further confirmed that our approach, which combines (1) a high-order formulation of conservative CC grid refinement, (2) stair-cased bounce-back techniques, (3) numerical equilibria, and (4) kinetic-based shock sensors is both accurate and robust for fully compressible flow simulations. The generality of our framework also facilitated the implementation of the grid refinement algorithm for several high-order lattices (D2Q13, D2Q21, and D2Q37). This significantly broadens the operational range of the initial low-speed solver so that transonic, supersonic, and potentially higher-speed regimes can be simulated.

Fully compressible simulations often pose challenges on non-uniform grids due to sharp gradients in macroscopic fields created by shock waves and discontinuities. Nonetheless, our framework has shown an excellent ability to handle these daunting discontinuities, ensuring high accuracy and robustness even for the most challenging configurations --such as those where shock waves propagate along the refinement interface.

Additionally, our framework demonstrated the ability to handle refinement interfaces that intersect boundary conditions without introducing numerical artefacts. This was notably the case for the transonic and supersonic simulations of flow past a NACA0012 airfoil, where the grid was locally refined at the leading edge to improve the capture of boundary layer development. Even in this complex scenario, where multiple ghost layers overlapped with bounce-back conditions, our method maintained a high level of accuracy by allowing the flow physics to smoothly cross several refinement interfaces.

Eventually, this set of compressible benchmarks extend the recent findings of Schukmann et al.~\cite{SCHUKMANN_FLUIDS_8_2023,SCHUKMANN_FLUIDS_10_2025} and Astoul et al.~\cite{ASTOUL_JCP_418_2020,ASTOUL_JCP_447_2021,ASTOUL_PhD_2021}, that \emph{ensuring conservation rules} is of paramount importance for fully compressible LB simulations on non-uniform grids. As for the weakly compressible case, the collision model plays an important role in the entire framework since the grid refinement strategy cannot compensate for accuracy or robustness issues encountered in the bulk of the simulation domain. The question then arises as to whether the present strategy also maintains the computational efficiency of LBMs, particularly on GPUs, and how the grid refinement strategy impacts the overall performance of the framework.

\section{Performance analysis~\label{sec:perfo}}

Building on the excellent results presented in Sections~\ref{sec:validation_weakly_compressible} and~\ref{sec:validation_fully_compressible}, we now turn our attention to the performance analysis of both the weakly and fully compressible frameworks. This section will primarily focus on (1) establishing performance models for weakly and fully compressible LBMs, (2) quantifying the impact of tree-based grid refinement strategies, (3) understanding how the time-asynchronous nature of the grid refinement strategy affects solver efficiency, (4) comparing measured performance with theoretical peak values for memory-bound solvers, and (5) quantifying the GPU speedup for compute-intensive solvers.

To ensure a fair and meaningful comparison across different configurations, we designed our framework with a strong emphasis on generality rather than full optimizations tailored to a specific lattice or collision model. However, to provide a relevant performance baseline, preliminary optimizations were applied to the reference LBMs used in the validation sections: the D2Q9 lattice (polynomial equilibrium) for low-speed flows and the D2Q21 model (numerical equilibrium) for high-speed flows. 
Thanks to this, the results presented below demonstrate that our approach achieves excellent raw performance for memory-bound solvers, and substantial GPU speedup for compute-bound ones, highlighting the efficiency of our general framework.

\subsection{Compilation flags, optimization levels and performance measurements~\label{subsec:perfo_flags_optim}}

The performance analysis is conducted on NVIDIA GPU cards, using the \texttt{nvc++} compiler (version \href{https://developer.nvidia.com/nvidia-hpc-sdk-241-downloads}{24.1}). The code was compiled on an A100 SXM4 GPU (memory size of 80GB, and bandwidth of $bw=2.039$ GB/s) with the following compilation flags:
\begin{center} 
\texttt{nvc++ -std=c++17 -stdpar=gpu -Msingle -Mfcon -O1 lbm.cpp -o lbm} 
\end{center}
The \texttt{-std=c++17} flag enables parallel algorithms from the C++17 standard library. The \texttt{-stdpar=gpu} flag instructs the compiler to target GPU offloading.

The \texttt{-M<nvflag>} options are specific to the \texttt{nvc++} compiler, with a full list available in the \href{https://docs.nvidia.com/hpc-sdk/compilers/hpc-compilers-ref-guide/index.html}{NVIDIA HPC Compilers Reference Guide}. Specifically, the \texttt{-Mfcon} option ensures floating-point calculations are performed in single precision without implicit conversion from the default double precision. 
The \texttt{-Msingle} flag prevents implicit float-to-double conversions. While it may seem redundant with \texttt{-Mfcon}, using both flags together was found to yield more consistent performance across a range of GPUs, including both gaming and cluster-class models. When performance analysis is conducted with double precision arithmetic, both flags are omitted.

The \texttt{-O1} optimization level was chosen after comparison with higher levels (\texttt{-O2}, \texttt{-O3}, \texttt{-Ofast}, and \texttt{-O4}). While these higher levels may improve performance in some cases, they often led to performance degradation. Notably, we found out that the more aggressive optimizations (\texttt{-O3} and above) could cause up to $20\%$ performance losses for specific versions of the compiler. For the 24.1 version, \texttt{-O1} was chosen as a good tradeoff between code optimization and performance. Given the frequent updates to the \texttt{nvc++} compiler (\href{https://developer.nvidia.com/nvidia-hpc-sdk-releases}{6 to 7 releases per year}), it is recommended to regularly run performance analyses to prevent regressions when upgrading to newer versions.

For compute-bound solvers, the speedup achieved on GPU is quantified thanks to additional performance benchmarks conducted on a 48-core Intel Xeon Gold 6240R CPU. Although \texttt{nvc++} enables seamless compilation for both GPU and CPU using the flags \texttt{-stdpar=gpu} and \texttt{-stdpar=multicore}, preliminary benchmarks indicate that \texttt{g++} delivers superior performance on the CPU, as previously noted in Ref.~\cite{LATT_PLOSONE_16_2021}. Therefore, all CPU performance results presented in this section are based on the code compiled with \texttt{g++} using the following command:
\begin{center}
\texttt{g++ -std=c++17 -O3 -march=native -ltbb lbm.cpp -o lbm}
\end{center}
Here, the \texttt{-march=native} flag enables architecture-specific optimizations, which ensures the best possible performance on the target CPU, while the \texttt{-ltbb} flag links to the Intel Threading Building Blocks (TBB) library to leverage efficient and automatic multi-threading capabilities~\cite{KIM_IEEE_2011}. 

For all benchmarks, performance is averaged over three consecutive full simulations, and it is computed as
\begin{equation}
P_{\mathrm{meas}} = \dfrac{n_{\mathrm{cells}} \times n_{\mathrm{ite}}}{t_{\mathrm{sim}}},
\end{equation}
with $n_{\mathrm{cells}}$ being the total number of cells in the grid. The simulation time $t_{\mathrm{sim}}$ is measured in seconds, and $n_{\mathrm{ite}}$ is the number of time iterations. $P_{\mathrm{meas}}$ corresponds to the total number of lattices (or cells) updated per second (LUPS), commonly referred to as MLUPS or GLUPS since $P_{\mathrm{meas}}$ is typically of the order of millions or billions of LUPS.

\subsection{Performance on matrix-like grids~\label{subsec:perfo_matrix_uniform}}

The performance of weakly compressible LBMs (Section~\ref{sec:validation_weakly_compressible}) --which use a polynomial equilibrium-- is primarily limited by the hardware memory bandwidth~\cite{BAUER_JCS_49_2021,LATT_PLOSONE_16_2021,LATT_ARXIV_09242_2025}. This condition, referred to as being memory-bound, implies that the execution time is mostly spent transferring data between memory and processing units, while computations are (almost) free in cost.
In contrast, the performance of fully compressible LBMs (Section~\ref{sec:validation_weakly_compressible}) depends more on computational power than memory bandwidth, as these methods rely on iterative calculations of equilibrium populations. However, with the computational power of modern GPUs that keeps improving, bandwidth-based performance models remain a good way to quantify the performance of these solvers~\cite{THYAGARAJAN_PoF_35_2023}.
Additionally, another exception applies to machine-learning- and gaming-dedicated GPU cards, where double-precision computational power is significantly lower than single-precision capabilities. This design choice prioritizes higher throughput for workloads that predominantly rely on single-precision arithmetic, such as deep learning and real-time graphics rendering. For such GPUs, simulations must be performed using single precision to satisfy the memory-bound assumption.

By definition, the theoretical peak performance of a memory-bound solver can be computed as the ratio of the available memory bandwidth, $bw$, to the amount of data that needs to be transferred per time step. For LBMs that rely on uniform and matrix-like grids, this relationship is expressed as:
\begin{equation}
P_{\mathrm{peak}}^{\mathrm{matrix}} = bw / (2 n_{\mathrm{LBM}} s_{\mathrm{data}} + s_{\mathrm{index}} + s_{\mathrm{type}}).
\label{eq:perfo_model_matrix}
\end{equation}
Here, $n_{\mathrm{LBM}}$ represents the number of variables stored in memory for a given LB scheme. Specifically, $n_{\mathrm{LBM}}=Q$ and $2Q + n_{\lambda}$ for weakly and fully compressible models, respectively. Each variable requires two memory accesses per time step --one for reading it and one for writing it back to memory. In this work, four different lattices are used ($Q\in\{9,13,21,37\}$), and the number of Lagrange multipliers is $n_{\lambda}=8$. $s_{\mathrm{data}} = 4$ or $8$ bytes depending on the arithmetic precision used for the simulations, and two flags of $4$ bytes each ($s_{\mathrm{index}}$ and $s_{\mathrm{type}}$) are used for the cell index and its type (fluid, solid, etc). In practice, one can expect weakly compressible LBMs to reach 60 to 80\% of the peak performance on GPUs using modern C++~\cite{LATT_PLOSONE_16_2021,LATT_ARXIV_09242_2025}, while fully compressible LBMs used in this work perform at about 40\% for the most powerful GPUs~\cite{THYAGARAJAN_PoF_35_2023}. Achieving this level of performance requires extensive code optimization, including techniques such as loop unrolling and precomputing redundant terms. Additionally, it necessitates sufficiently high memory utilization, typically ranging from a few percent to up to 20\% of the total memory for server-based and gaming-class GPUs, respectively.

\begin{figure}[bt!]
    \centering
    \includegraphics[width=\textwidth]{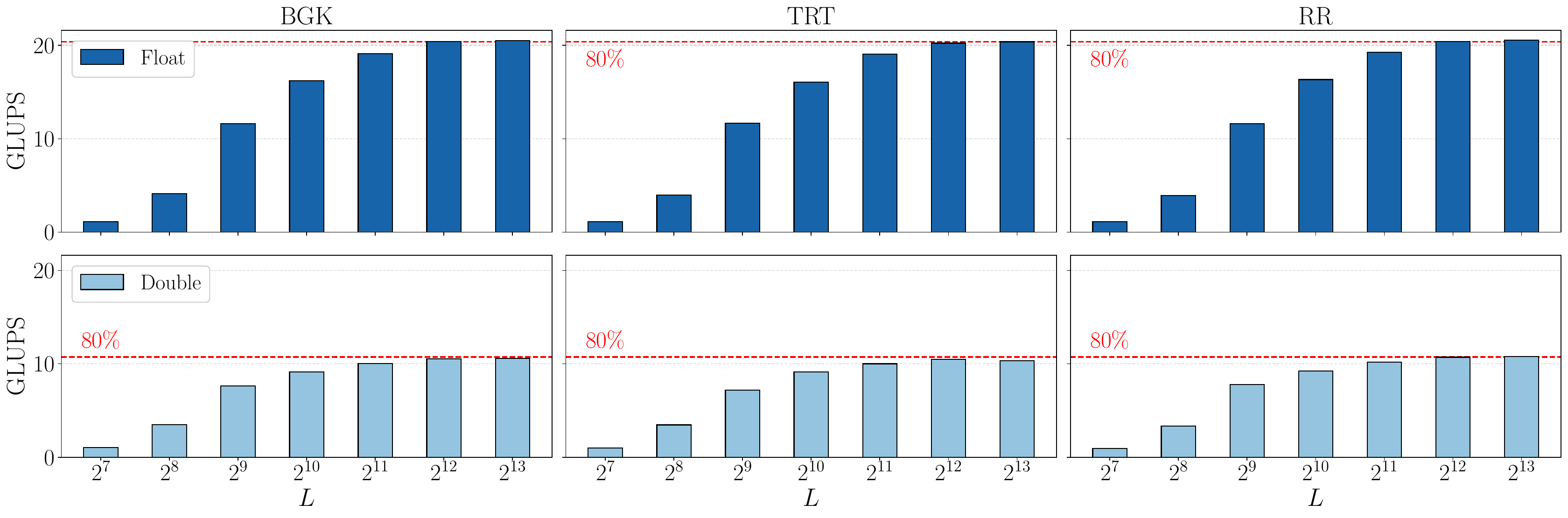}    
	\captionsetup{skip=0pt}
    \caption{Performance analysis of the lid-driven cavity simulation using a \emph{matrix-like} SoA implementation of the D2Q9-LBM with various collision models and $(\mathrm{Re}$,$\mathrm{Ma})=(100,=0.2)$. For the lowest cavity resolution, $L=2^7=128$ cells per direction, the performance is strongly impacted by thread over-synchronization. As $L$ increases, all models progressively reach 80\% of the peak performance of the A100 SXM4 (80GB) that is computed using the performance model~(\ref{eq:perfo_model_matrix}). The Reynolds number is increased to 10'000 when $L \geq 2^{10}=1024$ to reduce the value of $\tau$ and avoid stability issues.}
    \label{fig:perfo_matrix_comp_lattices_coll}
\end{figure}

To support our arguments, we included in our framework an additional implementation of the lid-driven cavity benchmark using matrix-like data structures with an SoA memory layout and ``two-population'' storage formalism (see Section~\ref{subsec:key_aspects_efficiency}). Performance data obtained with the D2Q9 lattice and several collision models (BGK, TRT and RR) are reported on Figure~\ref{fig:perfo_matrix_comp_lattices_coll}. As previously discussed, the performance barely depends on the collision model, and it is an increasing function of the cavity length resolution $L$. More precisely, it reaches a plateau (about $80$\% of the peak performance), when $L$ is greater than $2^{12}=4096$ cells per direction, i.e., a few percent of the memory size of the A100 SXM4 (80GB). This benchmark also confirms that there is no need to rely on low-level coding languages, like CUDA, to achieve near-peak performance on GPUs with matrix-like LBMs.

\subsection{\REV{Performance on tree-like structures with uniform grids: Quantifying the impact of the neighbor-list}~\label{subsec:perfo_tree_uniform}}

While tree-based structures retain the core principles of matrix-like LBMs, they suffer from additional overhead due to the unstructured nature of the grid. Starting from a uniform mesh constructed in a tree-like manner, the primary source of performance loss stems from the extra memory accesses required to access indexes stored in the list of neighbors (LoN). This adjustment leads to the following performance model for tree-based LBMs: 
\begin{equation}
P_{\mathrm{peak}}^{\mathrm{tree}} = bw / (2 n_{\mathrm{LBM}} s_{\mathrm{data}} +  n_{\mathrm{LoN}} s_{\mathrm{LoN}} + s_{\mathrm{index}} + s_{\mathrm{type}}).
\label{eq:perfo_model_tree}
\end{equation}
Here, $n_{\mathrm{LoN}}=(Q-1)/B_{\mathrm{size}}$ represents the number of additional memory accesses to the LoN, with each access requiring $s_{\mathrm{LoN}}=4$ bytes, and normalized by the size of blocks $B_{\mathrm{size}}$ that serve as leaves in the tree. Compared to the previous performance model~(\ref{eq:perfo_model_matrix}), the additional overhead is quantified by the relative performance ratio $P_{\mathrm{ratio}}=P_{\mathrm{peak}}^{\mathrm{tree}}/P_{\mathrm{peak}}^{\mathrm{matrix}}$ that reads:
\begin{equation}\label{eq:perfo_ratio}
P_{\mathrm{ratio}} = 1 - (n_{\mathrm{LoN}} s_{\mathrm{LoN}}) / (2 n_{\mathrm{LBM}} s_{\mathrm{data}} + n_{\mathrm{LoN}} s_{\mathrm{LoN}} + s_{\mathrm{index}} + s_{\mathrm{type}}).
\end{equation}
This ratio~(\ref{eq:perfo_ratio}) is very useful to derive corner cases before running any simulation. As an example, the performance of matrix-like codes is quickly recovered as the size of blocks $B_{\mathrm{size}}$ is increased. In the present framework, for which $B_{\mathrm{size}}=1$, the worst case scenario is obtained for weakly compressible LBMs ($n_{\mathrm{LBM}}=Q$) and single precision arithmetics ($s_{\mathrm{data}}=4$ bytes):
\begin{equation}
P_{\mathrm{ratio}}^{\mathrm{weakly,sp}} = 1 - (Q-1) / (3Q+1) \geq 2/3 \approx 0.67.
\end{equation}
In contrast, fully compressible LBMs ($n_{\mathrm{LBM}}=2Q + 8$) and double precision arithmetics ($s_{\mathrm{data}}=8$ bytes) result in a more favorable performance ratios:
\begin{equation}
P_{\mathrm{ratio}}^{\mathrm{fully,dp}} = 1 - (Q-1) / (9Q+33) \geq 8/9 \approx 0.89.
\end{equation}

\begin{table}[hbt]
\centering
\begin{tabular}{l|ccccc}
\toprule
$Q$ & $9$ & $13$ & $21$ & $37$ & $\infty$ \\
\midrule
Weakly (sp) & 0.71 & 0.70 & 0.69 & 0.68 & 0.67\\
Weakly (dp) & 0.83 & 0.82 & 0.81 & 0.81 & 0.80\\
\midrule
Fully  (sp) & 0.87 & 0.85 & 0.84 & 0.82 & 0.80\\
Fully  (dp) & 0.93 & 0.92 & 0.91 & 0.90 & 0.89\\
\bottomrule
\end{tabular}
\caption{\REV{Theoretical} performance ratios $P_{\mathrm{ratio}}$~(\ref{eq:perfo_ratio}) computed for various LBMs and arithmetic precisions. sp and dp stands for single precision and double precision, respectively.}
\label{tab:perfo_ratios}
\end{table} 

Performance ratios are provided in Table~\ref{tab:perfo_ratios} for all configurations of interest in this work. They allow us to draw several conclusions regarding the impact of the LoN on the performance of LBMs:
\begin{itemize}
\item The performance loss is directly influenced by the size of the lattice, and LBMs with larger lattices are more affected than those with smaller ones. This is because the fixed contributions, such as the memory accesses to flags and Lagrange multipliers, become negligible as $Q$ increases.
\item Simulations using single precision arithmetic are more significantly impacted by memory accesses to the LoN. This can be attributed to the fact that the data size, $s_{\mathrm{data}}$, appears in the denominator of the performance ratio $P_{\mathrm{ratio}}$~(\ref{eq:perfo_ratio}), making performance more sensitive to extra memory accesses when using single precision.
\item The performance loss can reach up to 33\% for weakly compressible LBMs and 20\% for fully compressible LBMs. This demonstrates that weakly compressible LBMs are more sensitive to the overhead introduced by the LoN than fully compressible LBMs. However, this is only true if the latter LBMs are memory-bound, which is not the case in this work (see Section~\ref{subsec:perfo_tree_fully_compressible}).
\end{itemize}

\begin{figure}[bt!]
    \centering
    \includegraphics[width=\textwidth]{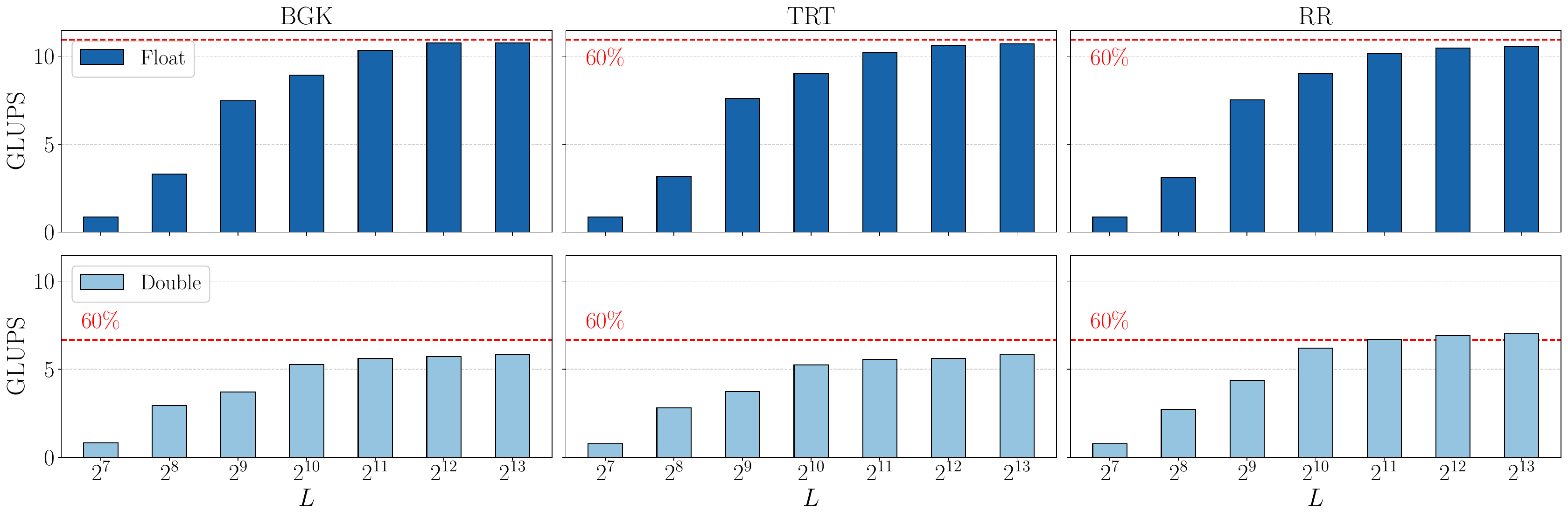}    
	\captionsetup{skip=0pt}
    \caption{Performance analysis of the lid-driven cavity simulation using a \emph{tree-like} SoA implementation of the D2Q9-LBM with various collision models and $(\mathrm{Re}$,$\mathrm{Ma})=(100,=0.2)$. For the lowest cavity resolution, $L=2^7=128$ cells per direction, the performance is strongly impacted by thread over-synchronization. As $L$ increases, all models progressively reach 60\% of the peak performance of the A100 SXM4 (80GB) that is computed using the performance model~(\ref{eq:perfo_model_tree}). The Reynolds number is increased to 10'000 when $L \geq 2^{10}=1024$ to reduce the value of $\tau$ and avoid stability issues.}
    \label{fig:perfo_tree_comp_lattices_coll}
\end{figure}

Building on this theoretical performance analysis, we further evaluate the performance of our framework using the lid-driven cavity benchmark with a uniform grid stored in a tree-like structure. As in Section~\ref{subsec:perfo_matrix_uniform}, the simulations are conducted with the D2Q9-LBM and various collision models (BGK, TRT, RR). Figure~\ref{fig:perfo_tree_comp_lattices_coll} presents results for both single-precision and double-precision arithmetic.
Consistent with matrix-like grids, (1) performance remains rather independent of the collision model, and (2) it increases with memory usage, reaching a plateau for $L \geq 2^{12}$.
However, the peak performance reaches only 60\% of the theoretical maximum. This \CC{value} and the difference between collision models are attributed to the partial optimization of our framework, which prioritizes the validation of various LBMs rather than optimization for a specific model.
Nevertheless, the performance of the D2Q9-RR ranks among the best reported in the literature, notably those obtained with highly optimized approaches based on trees of blocks and GPU-specific coding languages like CUDA and HIP~\cite{HOLZER_PhD_2025}. \REV{In the latter case, the performance is roughly halved, compared to simpler matrix-based formulations. This is in agreement with the performance ratios measured with our framework (see Table~\ref{tab:perfo_loss_neighbor_list}), even if they are approximately 20\% lower than our theoretical estimates (Table~\ref{tab:perfo_ratios}). Nevertheless, they remain sufficiently close to lend confidence to the validity of our performance model.}
\begin{table}[bt!]
\centering
\begin{tabular}{C{1.25cm} *{6}{C{1.25cm}}}
\toprule
& \multicolumn{3}{c}{Single Precision (GLUPS)} & \multicolumn{3}{c}{Double Precision (GLUPS)} \\
\cmidrule(lr){2-4}\cmidrule(lr){5-7}
Collision & Matrix & Tree & $P_{\mathrm{ratio}}$ & Matrix & Tree & $P_{\mathrm{ratio}}$ \\
\midrule
BGK & 20.503 & 10.739 & 0.52 & 10.598 & 5.824 & 0.55 \\
TRT & 20.390 & 10.688 & 0.52 & 10.178 & 5.847 & 0.57 \\
RR  & 20.584 & 10.541 & 0.51 & 10.795 & 7.041 & 0.65 \\
\bottomrule
\end{tabular}
\caption{Measured performance ratios $P_{\mathrm{ratio}}$ for D2Q9-LBMs. Data is gathered from lid-driven cavity simulations using $L=2^{13}$.}
\label{tab:perfo_loss_neighbor_list}
\end{table}

\REV{Similar trends in peak performance and impact of neighbor-list are observed for other weakly compressible LBMs used in this study, such as the D2Q13-RR and D2Q21-RR. For these high-order LBMs, the performance results remain promising and suggest clear opportunities for further optimization (see~\ref{app:perfo_weakly_tree_q13_q21}).}

\subsection{\REV{Performance on tree-like structures with non-uniform grids: Theoretical impact of asynchronous time-stepping \label{subsec:perfo_tree_nonuniform}}}

Moving on to grids composed of several refinement levels, the extra memory accesses required for buffer layer cells can be neglected as a first approximation. This is because the number of cells in the buffer layers is usually small compared to the number of cells in the bulk. 

Including the impact of the number of refinement levels, as well as the asynchronous time-stepping of the grid refinement strategy, into the performance model~(\ref{eq:perfo_model_tree}) requires several adjustments. Notably, the overall peak performance $P_{\mathrm{peak}}^{\mathrm{tree},*}$ should be computed through a \emph{weighted time average} of peak performances per level $i$, noted $P_{\mathrm{peak}}^{\mathrm{tree}}(i)$. 
For the performance model to be more realistic, $P_{\mathrm{peak}}^{\mathrm{tree}}(i)$ includes performance losses related to the hardware (see Figure~\ref{fig:perfo_tree_comp_lattices_coll}), and which can be amplified by the distribution of cells over all refinement levels. To ease the derivation of the performance model, we first assume that the peak performance is averaged over one coarsest time step $\Delta t_c$.
Accounting for all these modifications, the peak performance reads
\begin{equation}\label{eq:perfo_model_tree_nonuniform_time}
P_{\mathrm{peak}}^{\mathrm{tree},*}= \dfrac{1}{\Delta t_c}\sum_{t=0}^{\Delta t_c} \left[\sum_{i=0}^{i_{max}} P_{\mathrm{peak}}^{\mathrm{tree}}(i) \dfrac{\delta_{2^i}(t)}{\vert\vert \delta_{2^i}(t) \vert\vert}\right],
\end{equation}
with $\delta_{2^i}(t)$ being the Kronecker function that returns one when level $i$ requires an update (every $2^i$ time iterations), and zero otherwise. $\vert\vert \delta_{2^i}(t) \vert\vert $ is defined as the number of times $\delta_{2^i}(t) = 1$ when looping over all refinement levels during $\Delta t_c$. 

Because of the time dependency, the performance model~(\ref{eq:perfo_model_tree_nonuniform_time}) is too complex to provide valuable information in an a priori manner. To obtain a more convenient performance model, we propose to replace the time dependency by a level-wise dependency thanks to the convective scaling ($\Delta t \propto \Delta x$). Indeed, under this scaling, the number of finest time iterations spent in a refinement level $i$ during one coarsest time step is $n_t(i) = 2^{i_{max}-i}$, where $i_{max}$ is the index of the coarsest refinement level. Normalizing $n_t(i)$ by $\sum_{i=0}^{i_{max}} n_t(i)$, we get the level-wise weight 
\begin{equation*}
w(i) = \dfrac{n_t(i)}{\sum_{n=0}^{i_{max}}  n_t(i)} = \dfrac{2^{i_{max}-i}}{2^{i_{max}+1} -1}
\end{equation*}
which lies between 0 and 1, and satisfies the property $\sum_i w(i) = 1$. Based on this, the performance model for tree-like structures on non-uniform grids is expressed as:
\begin{equation}\label{eq:perfo_model_tree_nonuniform_general}
P_{\mathrm{peak}}^{\mathrm{tree},*} = \sum_{i=0}^{i_{max}} w(i) P_{\mathrm{peak}}^{\mathrm{tree}}(i).
\end{equation}

To validate the performance model~(\ref{eq:perfo_model_tree_nonuniform_general}), we can analyze several corner cases. First, in the absence of performance losses, we have $P_{\mathrm{peak}}^{\mathrm{tree}}(i) = P_{\mathrm{peak}}^{\mathrm{tree}}$ for all $i$. Therefore,
\begin{equation*}
P_{\mathrm{peak}}^{\mathrm{tree},*} = \sum_{i=0}^{i_{max}} w(i) P_{\mathrm{peak}}^{\mathrm{tree}}(i) = P_{\mathrm{peak}}^{\mathrm{tree}} \sum_{i=0}^{i_{max}} w(i) = P_{\mathrm{peak}}^{\mathrm{tree}}.
\end{equation*}
This confirms our decision to incorporate hardware-dependent performance losses into the study, ensuring that the impact of grid refinement on performance is properly captured.
Second, for a uniform grid, where $i_{max} = i = 0$, we find $w(0) = 1$, which leads to the recovery of the performance model~(\ref{eq:perfo_model_tree}).
Third, for more realistic non-uniform grids, where the number of refinement levels can be large ($i_{max} \gg 1$), we observe the asymptotic behavior:
\begin{equation}
P_{\mathrm{peak}}^{\mathrm{tree},*} \overset{i_{max}\gg 1}{\longrightarrow} \sum_{i=0}^{i_{max}} \dfrac{1}{2^{i+1}}P_{\mathrm{peak}}^{\mathrm{tree}}(i).
\end{equation}
In practice, $w(i)$ rapidly converges toward $1/2^{i+1}$, as demonstrated by the cases with three and four refinement levels:
\begin{equation*}
P_{\mathrm{peak}}^{\mathrm{tree},*} = \dfrac{4}{7} P_{\mathrm{peak}}^{\mathrm{tree}}(0) + \dfrac{2}{7} P_{\mathrm{peak}}^{\mathrm{tree}}(1) + \dfrac{1}{7} P_{\mathrm{peak}}^{\mathrm{tree}}(2),
\end{equation*}
and
\begin{equation*}
P_{\mathrm{peak}}^{\mathrm{tree},*} = \dfrac{8}{15} P_{\mathrm{peak}}^{\mathrm{tree}}(0) + \dfrac{4}{15} P_{\mathrm{peak}}^{\mathrm{tree}}(1) + \dfrac{2}{15} P_{\mathrm{peak}}^{\mathrm{tree}}(2) + \dfrac{1}{15} P_{\mathrm{peak}}^{\mathrm{tree}}(3),
\end{equation*}
respectively.

After validating the performance model~(\ref{eq:perfo_model_tree_nonuniform_general}) for general performance losses per level $i$, we now turn our attention to the specific form of $P_{\mathrm{peak}}^{\mathrm{tree},*}(i)$. As illustrated in Figures~\ref{fig:perfo_matrix_comp_lattices_coll} and~\ref{fig:perfo_tree_comp_lattices_coll}, memory usage significantly influences performance. In the context of non-uniform grids, this means that the distribution of cells across all levels must be factored into the performance model to provide an estimate of the memory usage at each level. This can be done via the simple modification
\begin{equation}\label{eq:perfo_model_tree_nonuniform_mem_usage}
P_{\mathrm{peak}}^{\mathrm{tree},*}(i)= A(i) P_{\mathrm{peak}}^{\mathrm{solver}}.
\end{equation}
where $A(i)$ is a monotonically increasing function of the memory usage that goes from 0 to 1 as the number of cells becomes high enough for the solver peak performance to be achieved at level $i$. As already said, this function is hardware- and precision-dependent and can be evaluated, e.g., through fitting of benchmark data gathered in Figure~\ref{fig:perfo_tree_comp_lattices_coll} for which $P_{\mathrm{peak}}^{\mathrm{solver}}$ would be close to $0.6$. Interestingly, the performance model~(\ref{eq:perfo_model_tree_nonuniform_general}) that accounts for level-wise memory usage~(\ref{eq:perfo_model_tree_nonuniform_mem_usage}) can be used to provide simulation time estimations as soon as the grid has been generated and prior to running the actual simulation. 

\subsection{Concrete impact of the grid hierarchy on performance: Aeolian noise simulations~\label{subsec:perfo_tree_aeolian}}

On top of the theoretical analysis that led to the performance model~(\ref{eq:perfo_model_tree_nonuniform_general})-(\ref{eq:perfo_model_tree_nonuniform_mem_usage}), we propose to conduct a performance analysis based on Aeolian noise simulations. This will allow us to better grasp what is at stake when running simulations on non-uniform grids (see Table~\ref{tab:perfo_configs_details}). Notably, we examine how the number of refinement levels, along with the distribution of cells across these levels, influences both performance and simulation time.
Starting from the mesh composed of six grid levels that was used for simulations performed in Section~\ref{subsec:aeolian_noise} (\#{1}), we conducted two parametric studies at \emph{fixed finest resolution}, and which both converges towards a uniform grid (\#{12}). In the first study, grid levels are merged sequentially starting from the coarsest level (\#{2} to \#{6}), while in the second study, merging begins at the finest level (\#{7} to \#{11}). The first strategy is typically used to derive best practices for physics in the far-field, whereas the second strategy focuses on physics closer to the cylinder.

In Table~\ref{tab:perfo_configs_details}, two key metrics are the \emph{total number of cells} in the grid, and the \emph{equivalent fine cells}. The former quantifies the memory footprint of the simulation, whereas the latter quantifies the computational savings achieved by the asynchronous grid refinement strategy. Specifically, it is derived by summing all cells and weighting them by their update frequency (\(1/2^i\)). By doing so, one can measure how much of the computational load is concentrated in finer grid areas, while coarser areas are updated less frequently. 
Additionally, the table includes a configuration with a uniform grid consisting solely of fine cells. This grid requires $2^{\text{depth}}$ cells per direction, corresponding to a total cell count of 67'108'864. This configuration is mainly used to quantify the differences in computational load and memory requirements between grids with and without refinement. 
\CC{Eventually, the ratio between the total number of cells and the equivalent fine cells can be interpreted as the theoretical speedup provided by asynchronous time stepping, under the \emph{ideal assumption} that performance is independent of memory usage. Specifically, under iso-LUPS conditions, configuration \#{1} achieves a speedup of approximately $23114/1338\approx 17$, thanks to asynchronous time stepping.}
Additional discussions about the parametric studies can be found in~\ref{app:perfo_aeolian}.
\begin{table}[btp!]
\centering
\scriptsize
\begin{tabular}{crrrrrrrrrrr}
\specialrule{0.15em}{0em}{0.1em} 
\# & Depth & Level 0 & Level 1 & Level 2 & Level 3 & Level 4 & Level 5 & Level 6 & Tot Cells & Eq Fine Cells & \textpertenthousand \\
\toprule
1 & 13 & 260 & 236 & 288 & 4'296 & 1'246 & 734 & 16'054 & 23'114 & 1'338 & 3 \\
\midrule
2 &  & 260 & 236 & 288 & 4'296 & 1'246 & 64'950 & - & 71'276 & 3'094 & 11 \\
3 &  & 260 & 236 & 288 & 4'296 & 261'046 & - & - & 266'126 & 17'302 & 40 \\
4 &  & 260 & 236 & 288 & 1'048'480 & - & - & - & 1'049'264 & 131'510 & 156 \\
5 &  & 260 & 236 & 4'194'208 & - & - & - & - & 4'194'704 & 1'048'930 & 625 \\
6 &  & 260 & 16'777'068 & - & - & - & - & - & 16'777'328 & 8'388'794 & 2500 \\
\midrule
7 &  & 1'428 & 1'320 & 25'352 & 7'056 & 3'984 & 63'696 & - & 102'836 & 11'547 & 15 \\
8 &  & 7'300 & 101'756 & 28'596 & 16'308 & 254'660 & - & - & 408'620 & 83'281 & 61 \\
9 &  & 431'292 & 118'918 & 67'718 & 1'017'470 & - & - & - & 1'635'398 & 634'864 & 244 \\
10 &  & 909'316 & 272'636 & 4'069'292 & - & - & - & - & 5'251'244 & 2'062'957 & 782 \\
11 &  & 2'033'916 & 16'268'654 & - & - & - & - & - & 18'302'570 & 10'168'243 & 2727 \\
\midrule
12 &  & 67'108'532 & - & - & - & - & - & - & 67'108'532 & 67'108'532 & 10000 \\
\specialrule{0.15em}{0.1em}{0em}
\end{tabular}
\caption{Overview of the mesh configurations for the performance analysis of Aeolian noise simulations with the D2Q9-RR-LBM. The finest resolution (level 0) uses fine steps \(\Delta x_f\) and \(\Delta t_f\), while coarser levels scale both of them by \(2^i\). The number of cells per refinement level composes the left part of the table. The sum of all cell counts (`Tot cells') provide a glimpse of how much memory is saved compared to the uniform mesh (Grid \# 12). The `equivalent fine cells' metric highlights computational savings through asynchronous refinement, and the final column (\textpertenthousand) compares the tree-like grid size to the one of a uniform grid composed of $2^{\mathrm{depth}}$ fine cells in each direction. Two parametric studies are conducted, each exploring different grid merging strategies, i.e., one merging grid levels from the coarsest while the other does it from the finest. The tree depth is $13$, with a corresponding finest grid resolutions of $\Delta x_f \approx D/20$, and total number of time iterations of $n_{\mathrm{ite}}=$141'952.}
\label{tab:perfo_configs_details}
\end{table}

\begin{figure}[hbt!]
    \centering
    \includegraphics[width=\textwidth]{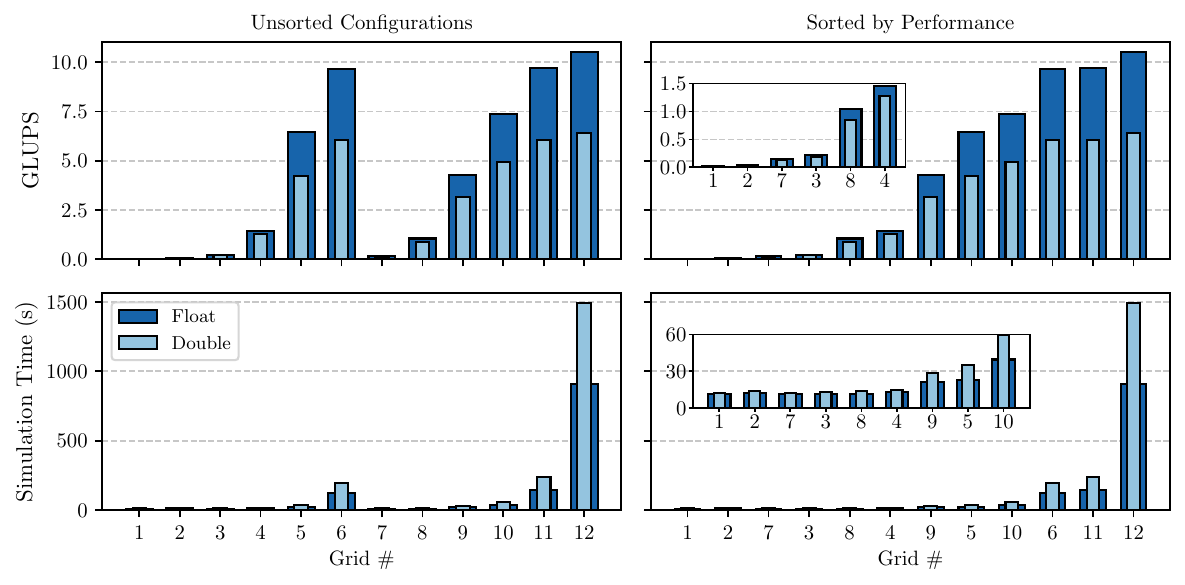}    
	\captionsetup{skip=0pt}
    \caption{Performance data (top) and simulation times (bottom) measured for Aeolian noise simulations based on the D2Q9-RR and with a finest resolution of $\Delta x_f = D/20$. To better identify relationships between performance and simulation time, data sorted by increasing performance is also plotted.}
    \label{fig:perfo_tree_eolian_depth13}
\end{figure}

Figure~\ref{fig:perfo_tree_eolian_depth13} presents performance metrics and simulation times for 12 different grid configurations. Several key trends can be observed:
\begin{enumerate}
\item As the grid size increases, performance gradually converges to approximately 60\% of the peak performance, that was previously achieved in lid-driven cavity simulations on a uniform grid (see Figure~\ref{fig:perfo_tree_comp_lattices_coll}). This supports the hypothesis that neither the additional memory accesses on the buffer layer, nor Rohde's algorithm itself, significantly impact performance.
\item For smaller grids, performance becomes independent of arithmetic precision (single and double precision yield similar results) and remains significantly below its expected peak value. This is because CUDA cores spend the majority of their time waiting for synchronization between them.
\item Configurations with the highest update frequencies do not correspond to those with the shortest simulation times. Instead, near-peak performance is observed for the largest grids \CC{(\#{6}, \#{11}, and \#{12})}, which, conversely, result in the longest simulation times. Opposite conclusions are reported for the smallest grids \CC{(\#{1}, \#{2}, and \#{7})}.
\end{enumerate}

The third observation is particularly important as it highlights the tradeoff required when performing simulations on non-uniform grids using GPUs. \CC{To minimize simulation time, two key factors must be considered: (1) reducing the number of cells at the finest refinement levels, as these are updated most frequently, and (2) maintaining a sufficiently high global update rate (LUPS). Based on these principles, configurations \#{4} and \#{5} provide optimal far-field acoustic propagation. They offer the best performance-to-simulation-time trade-off, while also preserving high mesh resolution in the far field, which is critical for accurately capturing acoustic wave propagation. Conversely, configurations \#{9} and \#{10} lead to optimal near-field aerodynamics with good far-field accuracy, since these grids keep a high mesh resolution close to the cylinder while coarsening the mesh far away from the cylinder.} Yet, this comes at the cost of larger memory footprints compared to \#{4} and \#{5}. \CC{Similar trends are observed for grids with higher mesh resolutions, except for the normalized performance-to-simulation-time ratio, which consistently improves with increasing resolution due to the higher workload across all refinement levels (see Appendix~\ref{app:perfo_aeolian}).}

In conclusion, adopting best practices for the grid generation is essential to balancing accuracy, performance, and simulation time effectively. \CC{These decisions are inherently case-dependent, as the priority between minimal simulation time, memory footprint, accurate near-field aerodynamics, or accurate far-field acoustics dictates the appropriate grid configuration.} The performance model, and the insights from this case study, provide a solid foundation for understanding all the aspects of such problems and for deriving best practices in a semi-automated manner.

\subsection{Preliminary performance results on fully compressible models~\label{subsec:perfo_tree_fully_compressible}}

In this work, we employ a double-distribution-function (DDF) LBM with numerical equilibria to simulate fully compressible flows in transonic and supersonic regimes, with local Mach numbers approaching $\mathrm{Ma}=3$. The same lattice is used for both the monatomic population, $f_i$, and the second population that represents extra degrees of freedom of molecules, $g_i$, thus providing a straightforward way to adjust the specific heat ratio. However, this simple strategy results in a large number of equations that must be solved at each time step for every cell. Additionally, the numerical equilibrium approach requires solving an optimization problem using a computationally expensive Newton-Raphson algorithm to achieve good accuracy and stability. These factors make the present DDF-LBM partially compute-bound. Despite this, previous studies have shown that these choices are a solid foundation for efficient simulation of fully compressible flows using purely LBM-based methods, notably due to the great speedup achievable on GPUs compared to CPUs~\cite{LATT_PLOSONE_16_2021,THYAGARAJAN_PoF_35_2023}.

\begin{figure}[hbt!]
    \centering
    \includegraphics[width=\textwidth]{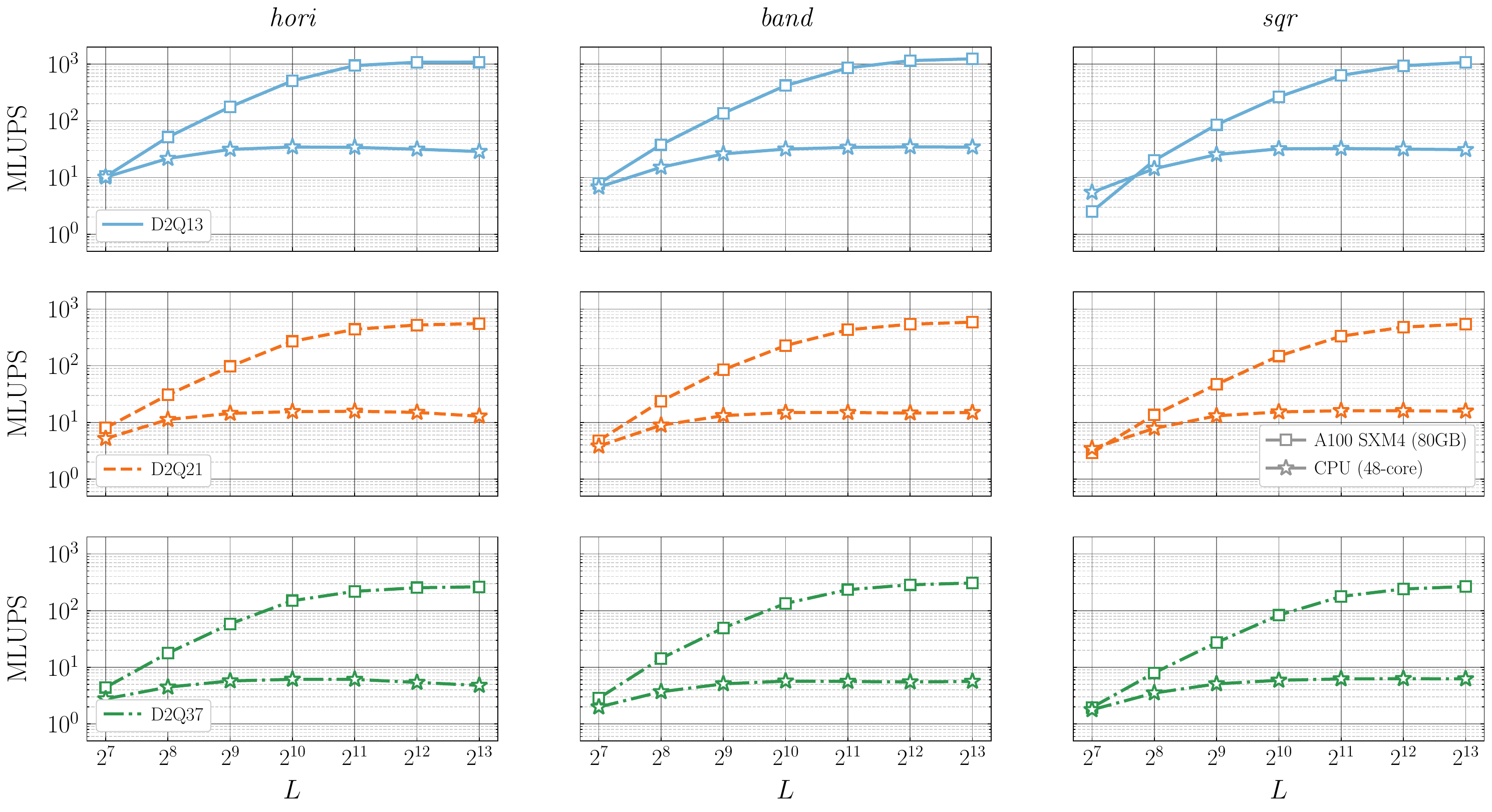}    
	\captionsetup{skip=0pt}
    \caption{Performance analysis of fully compressible LBMs using the inviscid Sod shock tube benchmark. From left to right: Impact of the grid configuration. From top to bottom: Impact of the lattice size.}
    \label{fig:perfo_riemann_1D_comp_gpu_cpu_lattices}
\end{figure}

\begin{table}[ht]
\centering
\begin{minipage}{0.45\textwidth}
\centering
\begin{tabular}{lcccc}
\multicolumn{5}{c}{(a) GPU Perfo in MLUPS} \\
\toprule
Lattice & \emph{unif} & \emph{hori} & \emph{band} & \emph{sqr} \\
\midrule
D2Q13 & 1259.1 & 1088.4 & 1244.8 & 1072.7 \\
D2Q21 & \phantom{1}618.5 & \phantom{1}555.9 & \phantom{1}588.2 & \phantom{1}544.4 \\
D2Q37 & \phantom{1}286.9 & \phantom{1}263.0 & \phantom{1}307.4 & \phantom{1}266.1 \\
\bottomrule
\end{tabular}
\end{minipage}
\hspace*{0cm}
\begin{minipage}{0.45\textwidth}
\centering
\begin{tabular}{lcccc}
\multicolumn{5}{c}{(b) GPU/CPU speedup} \\
\toprule
Lattice & \emph{unif} & \emph{hori} & \emph{band} & \emph{sqr} \\
\midrule
D2Q13 & 37.9 & 37.9 & 36.2 & 34.6 \\
D2Q21 & 48.9 & 43.3 & 39.4 & 34.6 \\
D2Q37 & 63.5 & 54.9 & 54.5 & 42.7 \\
\bottomrule
\end{tabular}
\end{minipage}
\caption{Key performance metrics for fully compressible LBMs on the Sod shock tube benchmark with $L=2^{13}$ and single precision arithmetics. (a) Raw performance obtained on the A100-SXM4 (80GB) GPU. (b) Performance speedup $P_{\text{GPU}}/P_{\text{CPU}}$. Results include non-uniform (\emph{hori}, \emph{band}, \emph{sqr}), as well as uniform (\emph{unif}) grid configurations when possible. For the D2Q37 lattice on a uniform grid, results are reported for $L=2^{12}$ due to int32 storage limitations encountered by the list of neighbors for $L=2^{13}$.}
\label{tab:perfo_fully_gpu_and_speedup}
\end{table}

With uniform grids, Cartesian indexing, SoA memory layout, and a ``two-population'' storage implementation, around 40\% of the peak performance can be achieved when using single precision arithmetics~\cite{THYAGARAJAN_PoF_35_2023}. Compared to a DDF-LBM with polynomial equilibria that would reach 80\% of $P_{\mathrm{peak}}^{\mathrm{tree}}$, the performance loss due to the iterative equilibrium computation is approximately 40\%. One may then wonder whether the performance loss induced by the Newton-Raphson step adds up with the loss induced by the neighbor list. To investigate this, we evaluate the performance of DDF-LBMs with numerical equilibria (eight constraints~(\ref{eq:G0_8mom})-(\ref{eq:G3yaa_8mom})) through the simulation of the Sod shock tube studied in Section~\ref{subsec:sod}. Figure~\ref{fig:perfo_riemann_1D_comp_gpu_cpu_lattices} gathers data obtained with several lattices (D2Q13, D2Q21 and D2Q37), and different grids (\emph{hori}, \emph{band}, and \emph{sqr}). Whatever the grid configuration, similar trends can be identified:
\begin{enumerate}
\item The performance is indeed impacted by both the compute-intensive nature of the numerical equilibrium, and the list of neighbors. As for weakly compressible LBMs, when the grid resolution is too low, GPU performance drops significantly and can even fall below CPU performance, as observed for the D2Q13 lattice with the \emph{sqr} configuration for $ L = 2^7 = 128 $. When the hardware is sufficiently loaded, the GPU consistently outperforms the CPU, regardless of lattice size or mesh configuration, thanks to its superior computational power. In particular, the GPU achieves a performance speedup ranging from 35 to 64 for single precision arithmetics (see Table~\ref{tab:perfo_fully_gpu_and_speedup}).
\item Due to the compute-bound nature of our compressible LBMs, increasing the lattice size leads to additional performance loss compared to memory-bound LBMs. For instance, both matrix-like and tree-like implementations of the D2Q13 and D2Q21 RR-LBM exhibit a performance ratio of
$ P_{Q21}/P_{Q13} \approx 13/21 \approx 62\%$ (see Figures~\ref{fig:perfo_matrix_comp_lattices_coll_merged} and~\ref{fig:perfo_tree_comp_lattices_coll_merged}). Hence, moving from 13 to 21 velocities leads to a performance loss is 38\%. However, in the fully compressible case, the performance loss reaches about 50\%, indicating an additional 10\% performance loss. Interestingly, this 10\% loss remains consistent across all lattices and grid configurations considered in this work.
\item On the CPU, performance initially improves with increasing grid resolution but declines once the grid becomes too dense, as the computational workload largely exceeds the available computational power. This issue is more pronounced for the D2Q37 lattice, as this LBM requires the highest number of operations to construct the Jacobian matrix required by the Newton-Raphson step. While additional lattice-specific optimizations could mitigate this effect, they would not alter the overall conclusions of this study.
\end{enumerate}

In summary, the present fully compressible LBMs benefit significantly from GPU acceleration, even on non-uniform grids. While their performance is impacted by the iterative computation of equilibrium populations, they leverage GPU computational power more effectively than memory-bound LBMs, which results in substantially higher speedups.
Obviously, their raw performance could be further improved through lattice-specific optimizations, a more efficient root-finding solvers, a smaller lattice for the second population, or even a finite-difference approximation of the energy equation. Nonetheless, the continuously increasing computational power and memory bandwidth of modern GPUs make these optimizations less critical for 2D applications. Consequently, these studies will be considered in future work on 3D LBMs with Octree-based grid refinement strategies.

\subsection{Partial conclusions}

The present performance analysis rigorously decomposed, modelled, and, where possible, quantified performance losses induced by each component of our framework.

First, by deriving performance models based on memory bandwidth limitations, we extended existing models for matrix-like grids (Section~\ref{subsec:perfo_matrix_uniform}) by incorporating the impact of the neighbor list (Section~\ref{subsec:perfo_tree_uniform}) and of the asynchronous time-stepping (Section~\ref{subsec:perfo_tree_nonuniform}). By identifying corner cases, we determined which LBMs, lattice sizes, and arithmetic precisions are most and least affected by the additional memory accesses required for the neighbor list (see Table~\ref{tab:perfo_ratios}). Parametric studies on Aeolian noise simulations further allowed us to establish best practices for benefiting from the asynchronous time-stepping of the solver on non-uniform grids (Section~\ref{subsec:perfo_tree_aeolian}). To the best of our knowledge, this is the first time these grid refinement components have been integrated into a performance model and examined through in-depth parametric studies. This paves the way for semi-automatic mesh generation strategies that incorporate performance estimation before running simulations.

Second, while the in-depth study on weakly compressible LBMs clarified the impact of grid refinement on the performance of memory-bound solvers, preliminary results for fully compressible LBMs provided valuable insights into its effects on the performance of compute-bound solvers.
Notably, weakly compressible LBMs achieve between 45\% and 60\% of theoretical peak performance on an A100-SXM4 (80GB), comparable to performance reported by Holzer using CUDA and HIP in waLBerla for the same refinement strategy~\cite{HOLZER_PhD_2025}. For fully compressible LBMs, performance losses caused by the neighbor list were shown to add up with those from the iterative computation of equilibrium populations. However, these LBMs benefit significantly from GPU acceleration over high-end CPUs due to their compute-intensive nature. For single precision arithmetic, performance speedups of up to 64 are observed, which is far beyond those achieved by memory-bound LBMs on non-uniform grids using similar hardware~\cite{HOLZER_PhD_2025}. While these results are somewhat optimistic, as the framework is not yet fully optimized --particularly for CPUs-- the overall conclusions remain valid.

Finally, both physical validation and performance analysis demonstrate that modern C++ can be used as a quick prototyping tool for identifying key components in the design of accurate, robust, and efficient LBMs on non-uniform grids. 
While ISO languages have primarily enabled GPU offloading on NVIDIA hardware~\cite{LARKIN_GTC_2022,LARKIN_GTC_2024}, recent advancements indicate they are becoming viable alternatives to low-level languages on AMD and Intel GPUs as well. Although performance on these GPUs was previously suboptimal for ISO languages~\cite{LIN_ARXIV_02680_2024}, this gap is narrowing~\cite{CAPLAN_SC_2025} and should continue to do so in the coming years, hence mirroring the improvements observed between the C++17 and upcoming C++26 versions of Parallel Algorithms.

\section{Preliminary results on adaptive mesh refinement for subsonic and supersonic applications~\label{sec:amr}}

Before moving to the general conclusion of our work, we examine the straightforward extension of our framework to adaptive mesh refinement (AMR). By definition, AMR dynamically adjusts mesh resolution based on either prescribed constraints or local flow features, thus optimizing computational resources by refining only where necessary.
Given the critical role of conservation at refinement interfaces and the impact of collision models on stability, it is also essential to assess how AMR influences these aspects across different flow regimes.

\subsection{Brief literature overview and methodology~\label{subsec:amr_overview}}

The concept of Adaptive Mesh Refinement (AMR) was introduced by Berger in the early 1980s for finite difference solvers of hyperbolic partial differential equations~\cite{BERGER_PhD_1982,BERGER_JCP_53_1984}. Since then, AMR has been widely adopted across numerous fields and numerical methods. Its applications range from shock hydrodynamics~\cite{BERGER_JCP_82_1989}, wave propagation~\cite{BERGER_SIAM_35_1998}, and magnetohydrodynamics~\cite{BALSARA_JCP_174_2001} to relativistic flows~\cite{ZHANG_AJSS_164_2006}, multiphase flows~\cite{YU_JCP_228_2009}, and wind turbine performance analysis~\cite{DEITERDING_JPCS_753_2016}. AMR has also played a key role in simulating radiative transfer in astrophysical flows~\cite{ZIER_MNRAS_533_2024}, hypersonic rarefied flows~\cite{ARSLANBEKOV_PRE_88_2013,BARANGER_JCP_257_2014,BRULL_JCP_266_2014}, laminar-to-turbulent transitions in hypersonic boundary layers~\cite{SHARMA_AST_141_2023}, and deflagration-to-detonation transitions in reactive flows~\cite{ZHAO_AST_136_2023}, among many others.
In terms of numerical methods, AMR has been successfully integrated into general partial differential equation solvers based on finite-difference~\cite{BERGER_JCP_53_1984}, finite-element~\cite{LOHNER_CMAME_61_1987,BALSARA_JCP_174_2001,HARTMANN_SIAM_24_2003}, finite-volume schemes~\cite{BERGER_SIAM_35_1998,SHARMA_AST_141_2023,ZHAO_AST_136_2023,ZIER_MNRAS_533_2024}, and coupled approaches~\cite{MOSSIER_JoCS_97_2023,MOSSIER_JCP_520_2025}. It has also been employed to solvers based on the kinetic theory of gases, such as lattice Boltzmann methods~\cite{YU_JCP_228_2009,DEITERDING_JPCS_753_2016,SCHORNBAUM_PhD_2018,BAUER_CMA_81_2021}, discrete velocity methods~\cite{ARSLANBEKOV_PRE_88_2013,BARANGER_JCP_257_2014,BRULL_JCP_266_2014}, and gas kinetic schemes~\cite{CHEN_JCP_231_2012b,XIAO_JCP_415_2020}.
Due to its ability to significantly reduce computational costs while maintaining accuracy, AMR has been incorporated into both open-source frameworks (e.g., AMROC~\cite{DEITERDING_HPCC_2005,DEITERDING_JPCS_753_2016,SHARMA_AST_141_2023,ZHAO_AST_136_2023}, waLBerla~\cite{SCHORNBAUM_PhD_2018,BAUER_CMA_81_2021}, AMReX~\cite{ZHANG_IJHPCA_35_2021,ZIER_MNRAS_533_2024}, PARAMESH~\cite{MACNEICE_CPC_126_2000}, SAMRAI~\cite{WISSINK_SC_2001,VINOD_AIAA_0028_2018}, Athena++\cite{TOMIDA_AJSS_266_2023}) and industrial solvers, including Converge~\cite{CONVERGE_WEBSITE} and XFlow~\cite{XFLOW_WEBSITE}.

A key aspect of AMR is the dynamic adaptation of the computational mesh at selected time intervals, and according to a given criterion. The refinement process begins by generating a 2D mesh density field, where each cell is assigned an integer value ranging from 0 (finest mesh level) to the tree depth (coarsest mesh level), based on either a prescribed criterion (e.g., fixed rotation or translation speed) or local flow characteristics (e.g., vorticity, density gradient, or deviation from equilibrium). Once the new mesh structure is determined, LB populations are transferred whenever possible or reconstructed otherwise. This involves iterating over all cells in the updated tree and checking if corresponding LB data exists at the same location in the previous tree. If a match is found, populations are directly copied; otherwise, they are recomputed using conservative \emph{coalescence} (for cell coarsening) or \emph{uniform explosion} (for cell refinement) steps. After these modifications, the standard collide-and-stream cycles resume until the next mesh adaptation step is triggered.

As discussed in Section~\ref{subsec:implementation_parallel_exe_tree}, our current framework executes LB computations on the GPU using a static non-uniform mesh, while mesh adaptation is managed on the CPU. In that context, AMR introduces a new computational bottleneck, shifting the performance constraints from LBM tasks to mesh reconstruction itself. While the porting of the mesh adaptation to the GPU is still on-going, we provide below preliminary validation of the framework’s accuracy and robustness in AMR-based simulations. For this, we consider three test cases that are simulated with D2Q21-LBMs: convection of a vortex in a periodic domain, double shear layer, and flow past a NACA airfoil at supersonic speeds.

\subsection{Testcase presentation, motivation, and results~\label{subsec:amr_testcase_presentation}}

\begin{figure}[hbt!]
    \centering
    \includegraphics[width=\textwidth,trim=0 1.3cm 0 0,clip]{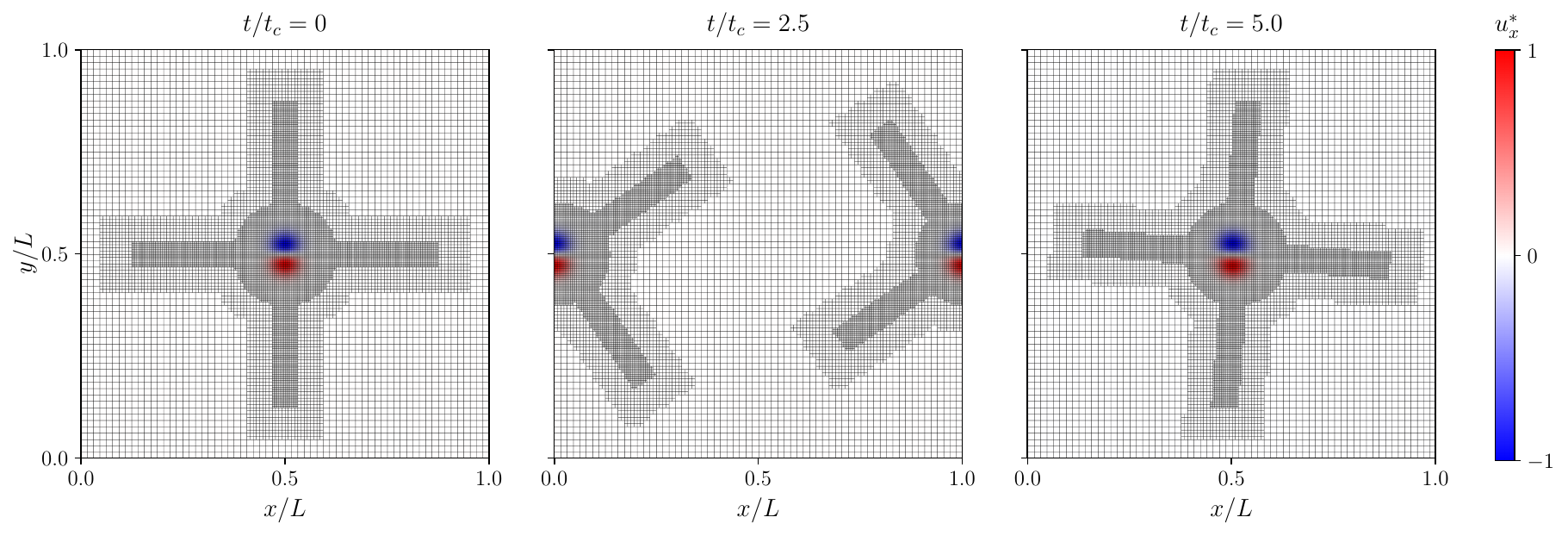}
    \includegraphics[width=\textwidth]{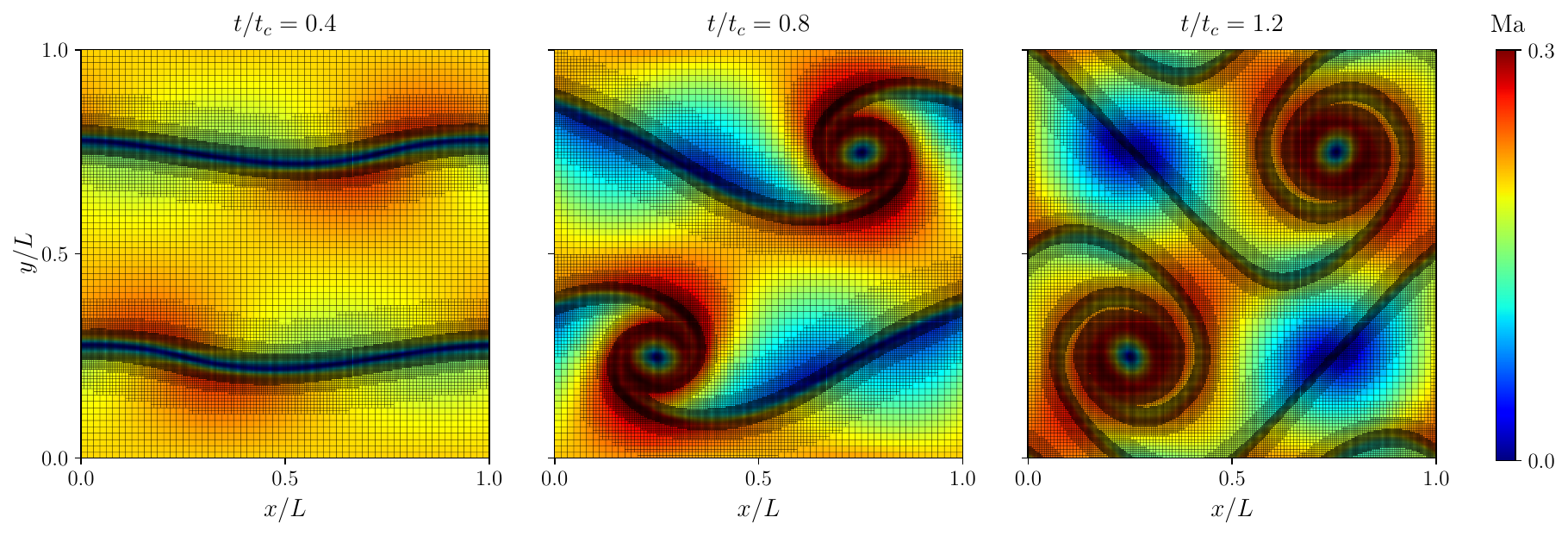}    
	\captionsetup{skip=0pt}
    \caption{Illustration of AMR for subsonic applications with D2Q21-LBM: (top) vortex convection and (bottom) double shear layer. Normalized $x$-component of the velocity field $u_x^*=u_x/u_{x,max}^{t=0}$, and local Mach number $\mathrm{Ma}$ are reported. The coarsest mesh resolution is $\Delta x_c = L/64$ for both testcases. The vortex convection test demonstrates the ability of AMR to track a translating and rotating vortex with synchronized refinement patches without generating spurious oscillations, while the double shear layer test showcases how AMR adapts to regions of high vorticity without generating spurious secondary vortices.}
    \label{fig:amr_covo_dsl}
\end{figure}

\begin{figure}[hbt!]
    \centering
    \includegraphics[width=\textwidth]{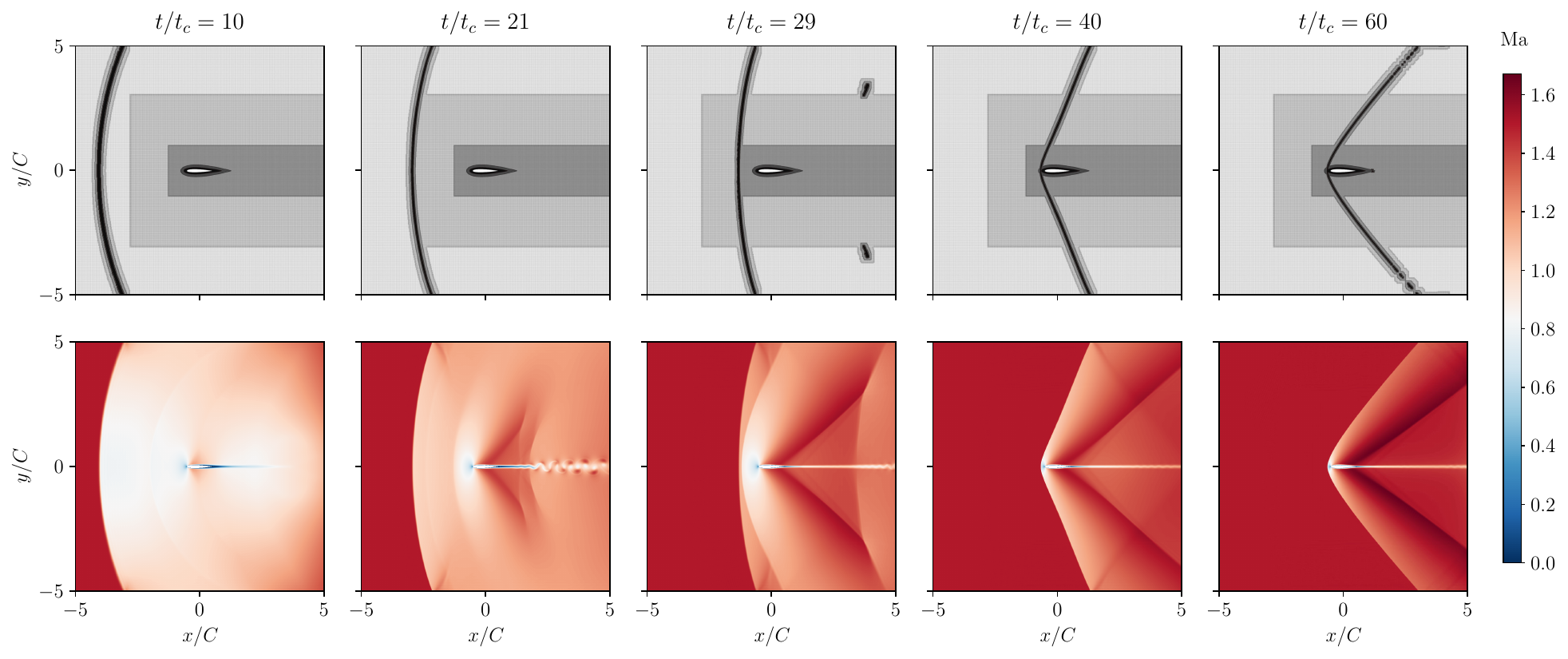}
	\captionsetup{skip=0pt}
    \caption{Illustration of AMR for the supersonic flow past a NACA airfoil with D2Q21-LBM. Time evolution of the grid (top), and Mach field (bottom) highlight how adaptive refinement patches track the primary bow shock while seamlessly merging with fixed patches.}
    \label{fig:amr_naca}
\end{figure}

The convected vortex benchmark evaluates the accuracy of AMR when refinement patches follow a translating and rotating vortex. 
A circular refinement region encloses the vortex, with two perpendicular rectangular patches forming a `propeller-like' structure that moves and rotates synchronously with the vortex. 
As the mesh motion continuously refines and coarsens cells upstream and downstream of the vortex, this test helps quantifying whether the basic \emph{coalescence} and \emph{uniform explosion} steps introduce round-off errors that could significantly distort vortex evolution. \CC{It also provides insight into whether this strategy could be used in the future to simulate scenarios involving moving/rotating body forces, such as those in actuator models commonly used to replicate the effects of drone or wind turbine blades without explicitly modeling them (see, e.g., Ref.~\cite{ASMUTH_WES_5_2020}).}
In this work, the vortex is initialized in a periodic domain of size $L$, using a barotropic formulation (see Eq. (29) in Ref.~\cite{WISSOCQ_PRE_101_2020}) with a radius $R=L/40$. The Mach number is 0.2, and the Reynolds number is chosen very high ($\mathrm{Re}=10^9$) so that possible errors originating from coarsening and refining cannot be attenuated by viscous effects. Velocity fields are reported on the top panels of Figure~\ref{fig:amr_covo_dsl} for different dimensionless times. They show that the motion of the vortex is not impacted by the dynamic remeshing happening in front and behind it, as it crosses the full domain five times.

The double shear layer test is used to assess the ability of the present AMR strategy to dynamically track the emergence and evolution of the counter-rotating vortices. In particular, the AMR is expected to enhance accuracy in regions of high vorticity while avoiding excessive computational overhead in less dynamic areas. The finest mesh level adapts dynamically based on local vorticity, while coarser levels are recursively constructed through a simple cell-flagging approach that imposes a fixed number of cell layers.
Here, we reused the simulation setup described in Section~\ref{subsec:validation_weakly_compressible_dsl}, and we fix $(\mathrm{Re},\mathrm{Ma})=(10^4,0.2)$. The Reynolds is kept moderate to avoid dispersion issues, already present on uniform grids with $\Delta x=L/64$, and that would prevent the proper accuracy evaluation of the AMR. Bottom panels in Figure~\ref{fig:amr_covo_dsl} highlight how the four mesh levels adapt to the local changes in vorticity (roll-up of the shear layers and vortex formations) without leading to the generation of spurious vortices.

For the flow past a NACA airfoil, we assess the ability of our AMR framework to combine fixed and dynamic refinement patches while simulating the flow past an bluff body. In this configuration, static refinement regions ensure adequate resolution around the airfoil, while additional patches dynamically adapt to the shock wave formation upstream the airfoil and its motion towards the leading edge. For this benchmark, we rely on the kinetic sensor introduced in a previous work~\cite{THORIMBERT_JoCS_64_2022}, and which basically evaluates the local departure to equilibrium.  
Beyond tracking high-speed phenomena like shock waves, this test highlights the framework’s accuracy and robustness, especially when dynamic patches interact with, and merge into, fixed ones.
In this study, we employed a configuration similar to the one discussed in Section~\ref{subsec:naca_supersonic_compressible}, specifically with $(\mathrm{Re},\mathrm{Ma})=(10^4,1.5)$. However, we used a smaller computational domain of size $10C \times 10C$, incorporating two fixed NACA-shaped refinement patches near the airfoil, along with two additional fixed rectangular-shaped patches. 
A refinement patch of finest cells is used to dynamically track the location of the primary bow shock by flagging cells where the sensor exceeds a threshold value of 0.2.
Additionally, two coarser refinement patches of predefined thickness are automatically added around the patch of finest cells that tracks the bow shock, hence ensuring a smooth transition from the finest to the coarsest grid level, with resolutions of $\Delta x_f \approx C/800$ and $\Delta x_c \approx C/50$, respectively.
Figure~\ref{fig:amr_naca} illustrates the time evolution of both the Mach field and the adaptive grid.
Our AMR strategy once again demonstrates high levels of accuracy and robustness, by effectively tracking the primary shock wave as it moves toward the airfoil’s leading edge, partially blending with fixed refinement patches along the way. Further optimization could reduce discontinuities in the mesh structure, such as those observed near the top and bottom boundary conditions at $t/t_c=60$, or eliminate unnecessary refinement when waves interacts between each others near the outflow boundary at $t/t_c=29$. Nevertheless, these occurrences further validate the robustness of our AMR-based D2Q21-LBM approach, which maintains accuracy and stability even in the presence of mesh discontinuities. 

These preliminary results confirm the high accuracy of our high-order CC AMR strategy for both convection-dominated and viscous phenomena in the subsonic regime while also demonstrating its robustness in simulating supersonic flows. To the best of our knowledge, this marks the first successful proposal and validation of AMR for high-order LBMs in both subsonic and supersonic regimes.
Future work will focus on a deeper investigation of CC refinement strategies for AMR-based high-order LBMs, aiming to define their operability range and establish best practices for remeshing frequency, refinement criteria, and smoothing techniques to optimize grid size and overall efficiency.

{
\color{black}{
\section{Synthesis of findings\label{sec:findings}}
In this work, we propose a strategy that couples simple, yet effective ideas to achieve (1) seamless hardware acceleration, (2) accurate and robust simulations across a wide range of Reynolds and Mach numbers, and (3) easy handling of complex geometries encountered in industrial applications.

Going into more details, our framework includes four lattices (D2Q9, D2Q13, D2Q21, D2Q37), four collision models (BGK, TRT, RR, and numerical equilibria), simple boundary conditions (staircase bounce-back, equilibrium, Neumann), and a high-order extension of the conservative cell-centered (CC) grid refinement strategy by Rohde et al.~\cite{ROHDE_IJNMF_51_2006}. Additionally, a double distribution function (DDF) formalism is used for simulation of fully compressible polyatomic flows. For efficient execution on CPUs and GPUs, our framework is based on C++ parallel algorithms with: (1) a ``two-population'' strategy to ensure thread-safety and ease of implementation, (2) a structure-of-array (SoA) data layout for optimal efficiency on GPUs, and (3) a grid generation based on a tree-structure that easily and efficiently handles grid refinement around realistic geometries. 

\paragraph{\textbf{Physical validation}}
Our framework is validated through an extensive set of \CC{2D} benchmarks, covering a wide range of flow regimes from low-speed aerodynamics and aeroacoustic to supersonic polyatomic flows. These benchmarks include simulations of lid-driven cavity flows, Aeolian noise generated by the flow past a circular cylinder, external aerodynamics past the 30P30N airfoil at realistic conditions, inviscid Riemann problems, as well as, the viscous flow past a NACA airfoil in both transonic and supersonic regimes. These validations were designed to address two central questions:
\begin{enumerate}
\setlength{\itemsep}{0pt} 
\setlength{\parskip}{0pt} 
\item Can standard LBMs based on appropriate collision models, CC grid refinement strategy, and simple boundary conditions, accurately simulate fluid flows across Reynolds numbers from 100 to over $10^6$, and for local Mach numbers up to 0.5-0.6?
\item Can our extension of the grid refinement strategy to high-order LBMs achieve similar levels of accuracy and robustness for transonic and supersonic regimes in both inviscid and viscous conditions?
\end{enumerate}
The results provide affirmative answers to both questions and further reveal that:
\begin{itemize}
\setlength{\itemsep}{0pt} 
\setlength{\parskip}{0pt} 
\item The inherent conservation properties of CC-based grid refinement strategies offer a clear advantage over the more widely used vertex-centered (VC) approaches, especially in the context of aeroacoustics simulations.
\item The refinement strategy naturally handles complex geometries, including interfaces intersecting inlets, outlets, and no-slip boundaries, allowing for flexible mesh designs with minimal memory and computational overhead.
\item Its versatility also allows for seamless extension to adaptive mesh refinement (AMR), with preliminary validations conducted on convected vortices, double shear layers, and supersonic flows past a NACA airfoil using D2Q21-LBMs.
\end{itemize}
Overall, the proposed approach combines simplicity, generality, and robustness which allows for accurate LB simulations on non-uniform grids across a wide spectrum of flow conditions.

\paragraph{\textbf{Performance study}} 
To evaluate the computational efficiency of our framework on non-uniform grids, we focused on answering the following two questions:
\begin{enumerate}
\setlength{\itemsep}{0pt} 
\setlength{\parskip}{0pt} 
\item How can performance models be derived to account for the impact of the grid refinement strategy (e.g., neighbor lists and time asynchronicity)? What is the corresponding theoretical performance loss compared to LBMs on matrix-like grids?
\item How does the performance of GPU-accelerated LBMs implemented with modern C++ compare to GPU-native CUDA implementations? How close is our framework to achieving peak GPU performance? What speedups can be achieved relative to CPU-based implementations?
\end{enumerate}
To answer the first question, we derived analytical performance models that incorporate extra memory accesses to the neighbor-list, as well as, the asynchronous time-stepping though a level-wise weighted performance model. Where possible, theoretical performance models were compared to practical measurements from lid-driven cavity, Aeolian noise, and Sod shock tube simulations. Notably, we found out that:
\begin{itemize}
\setlength{\itemsep}{0pt} 
\setlength{\parskip}{0pt} 
\item The overhead induced by the neighbor-list cuts performance approximately in half for both weakly and fully compressible LBMs. 
\item If the update frequency is similar across all refinement levels, the speedup achieved through asynchronous time-stepping is equivalent to the ratio of the total number of cells to the number of equivalent fine cells.
\item If not properly anticipated, asynchronous time-stepping can significantly degrade raw performance due to GPU memory underuse or imbalanced memory accesses between levels. Nevertheless, the resulting reduction in simulation time often compensates for the raw performance drop, yielding substantial overall speedups.
\item Our C++ implementation of weakly compressible LBMs with tree-based refinement achieves raw performance comparable to state-of-the-art CUDA-native solvers.
\item Fully compressible LBMs with numerical equilibria have lower raw performance due to their higher computational intensity. However, they achieve better GPU speedups, even matching the performance of 1000 CPU cores in the most compute-intensive scenarios.
\end{itemize}
In summary, our performance analysis is able to properly isolate the contribution of each element of the grid refinement algorithm. Furthermore, it shows that ISO-standard languages like C++ can rival hardware-specific approaches in raw efficiency while offering superior ease of implementation, modularity and maintainability. This positions modern C++ as a compelling choice not only for rapid prototyping but also for more advanced design of accurate, robust, and efficient LBMs on non-uniform grids.

\paragraph{\textbf{General conclusion}}
This work shows that the combination of simple LBMs, boundary conditions, and grid refinement strategies can lead to excellent results in the 2D context. Furthermore, it repositions Rohde's refinement approach, and similar CC strategies, at the forefront of discussions on the design of industry-oriented LBMs. Thanks to modern C++, the GPU port of LBMs is rather straightforward while showing performance comparable to GPU-native solvers. In the end, this study establishes a new baseline for developing GPU-accelerated, tree-based LBMs for weakly and fully compressible flows.

\section{Current and future directions \label{sec:perspectives}}

Looking ahead, we plan to enhance the current framework to meet the rigorous demands of 3D industrial solvers. To achieve this, we are investigating several key aspects:

\begin{itemize}
    \item \emph{Memory footprint:} As grids composed of billions of fluid cells become more common in the industry, memory constraints on affordable GPU cards pose a significant challenge. These cards often struggle to store 3D data for more than half a billion cells using single precision. Thus, minimizing memory usage is crucial. While our current ``two-population'' strategy simplifies implementation, it also doubles memory consumption. We are exploring more memory-efficient alternatives, such as the AA pattern~\cite{BAILEY_ICPP_2009}, which can potentially halve the memory footprint with minimal added complexity.

    \item \emph{Multi-GPU capabilities:} To scale simulations across multiple GPUs, efficient domain decomposition and communication schemes are essential, as they enable handling larger and more complex 3D cases. Although the present study focuses on single-GPU performance, recent advancements reported in our GPU port of the Palabos fluid solver show promising results for multi-GPU scalability, albeit currently limited to uniform grids~\cite{LATT_ARXIV_09242_2025}. Extending these strategies to support grid refinement and adaptive meshes remains to be done.

    \item \emph{Adaptive mesh refinement:} While initial CPU-based results have shown the potential of our AMR strategy, developing GPU-efficient implementations remains a priority. We aim to enable dynamic mesh adaptation on-the-fly using flow-dependent or kinetic criteria~\cite{THORIMBERT_JoCS_64_2022}, thereby improving accuracy and efficiency for multi-scale problems.

    \item \emph{Boundary conditions:} Staircased boundary conditions, such as the current bounce-back approach, are known to suffer from accuracy issues in under-resolved conditions~\cite{KRUGER_Book_2017}. Moving to curved boundary conditions that account for the distance to the wall (e.g., Bouzidi et al.~\cite{BOUZIDI_PoF_13_2001} or Filippova and H\"{a}nel~\cite{FILIPPOVA_JCP_147_1998}) is a necessary step to ensure accurate simulations in under-resolved conditions while maintaining a high level of accuracy near the wall.

    \item \emph{Wall modeling:} Combining curved boundary conditions with wall models, such as algebraic wall functions or more advanced approaches, is crucial for accurately capturing near-wall effects in high-Reynolds-number flows~\cite{MALASPINAS_JCP_275_2014,DEGRIGNY_PhD_2021}. Integrating such models is particularly important for achieving the required accuracy without fully resolving the boundary layer, ensuring computational efficiency and suitability for industrial applications~\cite{SENGISSEN_AIAA_2993_2015}.

    \item \emph{Turbulence modeling:} While the numerical dissipation introduced by the collision model can stabilize simulations at high Reynolds numbers, explicit turbulence models are essential for accurately accounting for dissipation mechanisms that cannot be fully resolved by coarse grids. Implementing subgrid-scale models (Smagorinsky, Vreman, or WALE models) would enable more accurate simulations at coarse grid resolutions, while allowing for better control over the amount of turbulent dissipation introduced in the simulation domain~\cite{POPE_Book_2000, SAGAUT_CMA_59_2010}.
\end{itemize}

These enhancements are being progressively integrated into a 3D version of the framework. Figure~\ref{fig:amr_drag_crisis} gathers preliminary results about the flow past a sphere that is simulated with the D3Q27-RR in the super-critical regime using AMR and the kinetic criterion. More details will be presented in future work dedicated to multi-GPU, industry-oriented 3D-LBM with AMR capabilities.

\begin{figure}[hbt!]
    \centering
    \includegraphics[width=\textwidth]{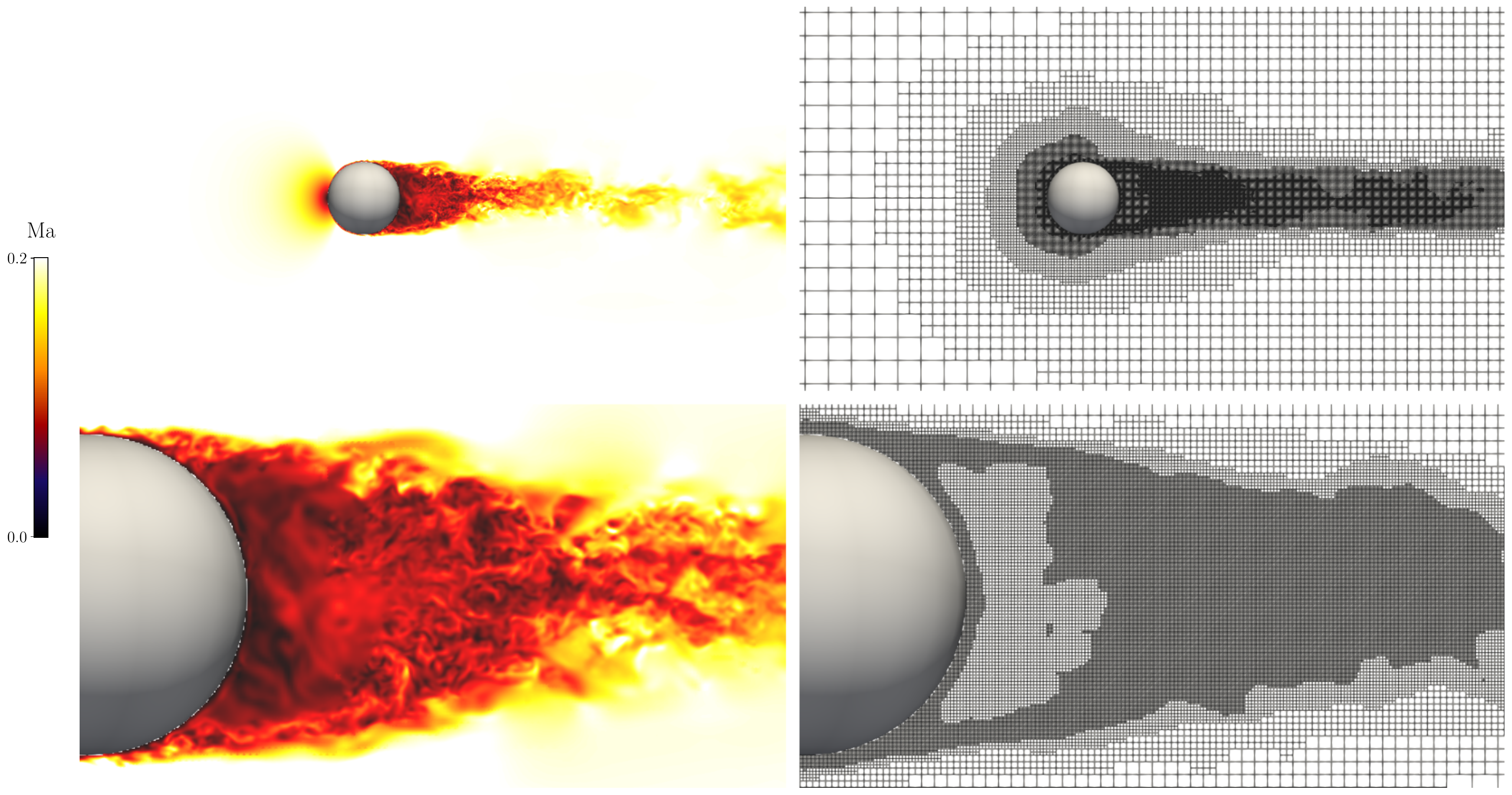}
	\captionsetup{skip=0pt}
    \caption{
\CC{Illustration of AMR for the flow past a sphere with the D3Q27-RR-LBM in the super-critical regime $(\mathrm{Re},\mathrm{Ma})=(10^6,0.2)$. Left: Local Mach number. Right: Dynamic grid. Seven refinement levels are used to discretize the $48D \times 32D \times 32D$ simulation domain with $D$ being the sphere diameter. Equilibration of the trace of second-order moments is used for improved robustness. The finest mesh resolution is set to $\Delta x_f = D/200$, a threshold value required to get the flow separation past the sphere’s maximum diameter --which is characteristic of the supercritical regime.}
    }
    \label{fig:amr_drag_crisis}
\end{figure}
}
}

\section*{Acknowledgements}
C.C. is grateful to Alexander Schukmann for insightful discussions on grid refinement strategies and for sharing an early version of Ref.~\cite{SCHUKMANN_FLUIDS_10_2025}. 
Additionally, C.C. thanks Markus Holzer for providing valuable insights into the impact of grid refinement on the GPU performance of LBMs and for discussing strategies to mitigate these effects.
\CC{Eventually, the authors would also like to thank the reviewers for their pertinent comments that help us improving the quality of the manuscript.}
This work received financial support from BNBU Research Fund $\mathrm{n^o UICR0700094}$-$24$ entitled ``GPU-accelerated multidisciplinary methods for industrial computational fluid dynamics''.

\appendix
\section{Details on lattice Boltzmann methods used in this work\label{app:lbm}}
\subsection{General knowledge}
Lattice Boltzmann methods are numerical schemes dedicated to the solving of the discrete velocity Boltzmann equations~\cite{KRUGER_Book_2017}:
\begin{equation}
\partial_t f_i + \bm\xi_i\cdot\bm\nabla f_i = \Omega_i,
\label{eq:GeneralLBE} 
\end{equation} 
where external forces have been ignored, and a general collision model $\Omega_i$ was considered.
These equations consist of a linear convection term and a non-linear collision term. The convection term is integrated exactly using the method of characteristics, while the collision term is integrated with second-order accuracy using the trapezoidal rule. This combination of numerical discretizations leads to:
\begin{equation}
f_i\left(\bm{x} + \bm\xi_i\Delta t, t + \Delta t\right) - f_i\left(\bm{x}, t\right) = \dfrac{ \Omega_i\left(t + \Delta t\right)+ \Omega_i(t)}{2}\Delta t + \mathcal{O}(\Delta t^2, \Delta x^2),
\label{eq:implicitTimeDis}
\end{equation}
where $\Delta t$ and $\Delta x$ are the time and space steps respectively.
This space/time discretization results in an implicit formulation since $\Omega_i\left(t + \Delta t\right)$ depends on $f_i\left(\bm{x} + \bm\xi_i\Delta t, t + \Delta t\right)$. This is why a change of variables, compliant with conservation laws, is commonly used to get around it~\cite{HE_JCP_146_1998,DELLAR_PRE_64_2001}:
\begin{equation}
\overline{f_i}\left(\bm{x}, t\right) =  f_i\left(\bm{x}, t\right) - \dfrac{\Delta t}{2}\overline{\Omega_i}(t).
\label{eq:changeOfVariables}
\end{equation}
This leads to a numerical scheme that is explicit in time:
\begin{equation}
\overline{f_i}\left(\bm{x} + \bm\xi_i\Delta t, t + \Delta t\right) - \overline{f_i}\left(\bm{x}, t\right) = \Delta t \,\Omega_i(t) .
\end{equation}
Assuming a relaxation of populations $f_i$ towards their equilibrium $f_i^{eq}$, the general collision model $\Omega_i(t)$ can be replaced by the BGK operator~\cite{BHATNAGAR_PR_94_1954}
\begin{equation}
\Delta t\,\Omega_i(t) = -\dfrac{\Delta t}{\tau}\left[ f_i\left(\bm{x}, t\right) -  f^{eq}_i\left(\bm{x}, t\right)\right].
\end{equation}
By including all the modifications induced by the change of variables, it then reads
\begin{equation}
\Delta t\,\overline{\Omega_i}(t) = -\dfrac{\Delta t}{\overline{\tau}}\left[ \overline{f_i}\left(\bm{x}, t\right) -  f^{eq}_i\left(\bm{x}, t\right)\right]
\end{equation}
where, notably, $\overline{f_i^{eq}}=f_i^{eq}$ by construction of the change of variables. Additionally, the relaxation time absorbed part of the numerical error introduced by the trapezium time integration: $\overline{\tau} = \tau + \Delta t /2$. Dropping the overline notation, we end up with the explicit and second-order accurate in space/time scheme: 
\begin{equation}
f_i\left(\bm{x} +1, t + 1\right) - f_i\left(\bm{x}, t\right) = -\dfrac{1}{\tau}\left[ f_i\left(\bm{x}, t\right) -  f^{eq}_i\left(\bm{x}, t\right)\right],
\end{equation}
where all quantities have been normalized by $\Delta x$ and $\Delta t$ to end up with quantities based on so-called mesh/lattice units. In practice, this numerical scheme is decomposed into two successive steps:
\begin{align}
f^*_i(\bm{x}, t) &= f_i(\bm{x}, t) -\dfrac{1}{\tau}\left[ f_i\left(\bm{x}, t\right) -  f^{eq}_i\left(\bm{x}, t\right)\right],\label{eq:CollisionStep} \\ 
f_i(\bm{x} +\bm{\xi}_i, t+1) &= f_i^*(\bm{x}, t), \label{eq:StreamingStep}
\end{align}
where $f^*_i$ are post collision populations. Eventually, we recall in Tables~\ref{tab:lattices_structure} and~\ref{tab:lattices_hermite} the properties of the velocity discretizations used in this work.

\renewcommand{\arraystretch}{1.2}
\begin{table}[btp!]
\centering
\begin{tabular}{c @{\quad} c @{\quad\quad} c c c c  @{\quad\quad} c @{\quad\quad\quad} c @{\quad\quad\quad} c @{\quad\quad\quad} c}
\hline
\hline
$\text{Group}$ & $\boldsymbol\xi_i$ & $p$ & & & & $ D2Q9 $ & $ D2Q13 $ & $ D2Q21 $ & $ D2Q37 $\\
\cline{1-4} \cline{6-10}
$1$ & $(0,0)$ & $1$ & & & & $4/9$ & $36/96$ & $ 91/324$ & $0.233150669132352$\\

$2$ & $(1,0)$ & $4$ & & & & $1/9$ & $8/96$ & $1/12$ & $0.107306091542219$\\
$3$ & $(1,1)$ & $4$ & & & & $1/36$ & $6/96$ & $2/27$ & $0.057667859888794$\\

$4$ & $(2,0)$ & $4$ & & & & & $1/96$ & $7/360$ &     $0.014208216158450$\\
$5$ & $(2,1)$ & $8$ & & & &  & &   &                 $0.005353049000513$\\
$6$ & $(2,2)$ & $4$ & & & &  & & $1/432$ &           $0.001011937592673$\\

$7$ & $(3,0)$ & $4$ & & & &  & & $1/1620$ &          $0.000245301027757$\\
$8$ & $(3,1)$ & $8$ & & & &  & &  &                  $0.000283414252994$\\
\cline{1-4} \cline{6-10}
 & $c_s^2$ & & & & & $1/3$ & $1/2$ & $2/3$ & $0.697953322019683$ \\
\hline
\hline
\end{tabular}
\caption{Key features of lattices considered in this work. The parameter $p$ represents the number of discrete velocities in each velocity set, with the corresponding weights $w_i$ listed on the right side of the table. Data is compiled from Refs~\cite{WEIMAR_PA_224_1996,SHAN_JFM_550_2006,PHILIPPI_PRE_73_2006}.}
\label{tab:lattices_structure}
\end{table}

\renewcommand{\arraystretch}{1.4}
\begin{table}[btp!]
\centering
\begin{tabular}{ c c c c c c c c c c c c }
\hline
\hline
Lattices & & $\mathcal{H}^{(0)}$ & $\mathcal{H}_x^{(1)}$ & $\mathcal{H}_{xx}^{(2)}$ & $\mathcal{H}_{xy}^{(2)}$ & $\mathcal{H}_{xxx}^{(3)}$ & $\mathcal{H}_{xxy}^{(3)}$ & $\mathcal{H}_{xxxx}^{(4)}$ & $\mathcal{H}_{xxxy}^{(4)}$ & $\mathcal{H}_{xxyy}^{(4)}$  \\
\cline{1-1} \cline{3-11}
$D2Q9\phantom{3}$ & & \ding{109} & \ding{109} & \ding{109} & \ding{109} & \ding{55} & \ding{109} & \ding{55} & \ding{55} & \ding{109} \\
$D2Q13$ & & \ding{109} & \ding{109} & \ding{109} & \ding{109} & \ding{109} & \ding{109} & \ding{55} & \ding{55} & \ding{55} \\
$D2Q21$ & & \ding{109} & \ding{109} & \ding{109} & \ding{109} & \ding{109} & \ding{109} & \ding{55} & \ding{55} & \ding{55} \\
$D2Q37$ & & \ding{109} & \ding{109} & \ding{109} & \ding{109} & \ding{109} & \ding{109} & \ding{109} & \ding{109} & \ding{109} \\
\hline
\hline
\end{tabular}
\caption{Hermite tensor bases used for lattices considered in this work. The tensors are classified into those used in the polynomial expansion (\ding{109}), and those which are not (\ding{55}). Cyclic permutations of the Hermite tensors are omitted for clarity. Data is compiled from Ref~\cite{COREIXAS_PhD_2018}.}
\label{tab:lattices_hermite}
\end{table}

\subsection{Advanced models for accurate and robust simulations of low and high speed flows}

The equilibrium state is commonly approximated by a Hermite polynomial expansion of the Maxwell-Boltzmann distribution~\cite{GRADa_CPAM_2_1949,SHAN_PRL_80_1998}:
\begin{equation}
f_i^{eq} = \rho \exp\left[-(\bm{u}-\bm{\xi}_i)^2/T_0\right] \approx \displaystyle{\sum_{n=0}^N\dfrac{1}{n!T_0^{n}}\mathcal{H}^{(n)}_{i,\bm\alpha} : a^{(n)}_{eq,\bm\alpha}},
\end{equation}
where $\mathcal{H}^{(n)}_{i,\bm\alpha}$ and $a^{(n)}_{eq,\bm\alpha}$ are $n$th-order tensors, $\bm{\alpha}$ is a vector of coordinate indexes of size $n$, ``:'' is the Frobenius inner product, and the reference temperature $T_0$ can be replaced by $c_s^2$ for LBMs dedicated to low speed flow simulations. In practice, Hermite tensors $\mathcal{H}^{(n)}_{i,\bm\alpha}$ and their equilibrium moments $a^{(n)}_{eq,\bm\alpha}$ provide a convenient framework for deriving LBMs in a general way, regardless of the lattice used. The truncation order $N$, in conjunction with the orthogonality properties of the lattice, determines the macroscopic equations recovered by the LBM~\cite{SHAN_JFM_550_2006,PHILIPPI_PRE_73_2006}. It also determines whether additional degrees of freedom are required to account for, e.g., adjustable Prandtl number~\cite{SHAN_IJMPC_18_2007} or specific heat ratio~\cite{NIE_PRE_77_2008}. 

In terms of accuracy, BGK-LBMs benefits from impressive spectral properties whatever the lattice considered~\cite{MARIE_JCP_228_2009,COREIXAS_PhD_2018,WISSOCQ_JCP_380_2019,SUSS_JCP_485_2023}. However, the very small amount of numerical dissipation they introduce plays against them, as they are not able to overdamp spurious oscillations that appear in under-resolved conditions, particularly at finite Mach and high Reynolds numbers.
To remedy this issue, numerous collision models have been proposed over the years, and most of them aimed at increasing the amount of numerical dissipation in one way or another: multiple relaxation times, collision in a specific moment space, regularization of the non-equilibrium contribution of populations, dynamic relaxation times based on turbulence modeling or entropy principles, etc~\cite{COREIXAS_PRE_100_2019}. Systematic comparisons have identified several collision models that remain stable under severe conditions without significantly compromising LBM accuracy~\cite{COREIXAS_RSTA_378_2020,COREIXAS_ICMMES_Collision_2021}. Among these, the recursive regularization (RR) approach~\cite{MALASPINAS_ARXIV_2015,BROGI_JASA_142_2017} and its extensions~\cite{COREIXAS_PRE_96_2017,JACOB_JT_19_2018} have been shown to meet the needs of both academic and industrial groups for low-speed and, to some extent, high-speed flows.
In the latter case, coupling weakly compressible LBMs with finite difference or finite volume discretizations of the energy equation is almost mandatory to ensure sufficient robustness~\cite{RENARD_JCP_446_2021}, unless non-polynomial formulations of the collision model are employed~\cite{MIEUSSENS_MMMAS_10_2000,FRAPOLLI_PhD_2017,LATT_RSTA_378_2020,COREIXAS_PoF_32_2020,THYAGARAJAN_PoF_35_2023}.
All of this explains why RR-LBMs are now at the core of the most advanced commercial LB solvers: ProLB~\cite{JACOB_JT_19_2018,HOU_AIAA_2019_2555,GUO_JCP_418_2020,ASTOUL_PhD_2021,DEGRIGNY_PhD_2021,
NGUYEN_IJHMT_212_2023,TAHA_PhD_2023,WERNER_PoF_36_2024,MOZAFFARI_JCP_514_2024,DAVILLER_AIAA_3180_2024} and PowerFLOW~\cite{CHEN_PATENT_Collision_2015,CHEN_PS_95_2020,FARES_AIAA_0952_2014,CASALINO_AIAA_1834_2019,KHORRAMI_CEAS_AERO_2019,
GONZALEZMARTINO_AIAA_2585_2019,KOPRIVA_AIAA_3929_2019,JAMMALAMADAKA_AIAA_3055_2020,TERUNA_AIAA_2264_2021,KIRIS_AIAA_WMLES_LBM_2022}. 

As this work focuses solely on purely LBMs, we propose using RR collision models for low-speed flow simulations, while non-polynomial approaches will be employed for high-speed flow simulations.

For weakly compressible and low speed flow simulations, the collision step reads:
 \begin{equation}
 f_i^{*} = w_i \displaystyle{\sum_{n=0}^N\dfrac{1}{n!c_s^{2n}}\mathcal{H}^{(n)}_{i,\bm\alpha}:a^{(n)}_{*,\bm\alpha}},
 \label{eq:GenRegCollision}
 \end{equation}
where the post-collision Hermite moments read $a^{(n)}_{*,\bm\alpha} = a^{(n)}_{eq,\bm\alpha}+\left(1-\frac{1}{\tau_{n,\bm{\alpha}}}\right)a^{(n)}_{neq,\bm\alpha}$. The corresponding equilibrium state is
 \begin{equation}
 f_i^{eq} = w_i \displaystyle{\sum_{n=0}^N\dfrac{1}{n!c_s^{2n}}\mathcal{H}^{(n)}_{i,\bm\alpha}:a^{(n)}_{eq,\bm\alpha}}.
 \label{eq:GenEq}
 \end{equation}
In this work, we use full expansions up to third order for the D2Q13 and D2Q21 lattices, while a partial expansion up to fourth order is adopted for the D2Q9 lattice. This partial expansion is chosen due to the improved accuracy and robustness it offers for D2Q9-LBMs, regardless of the collision model used~\cite{COREIXAS_RSTA_378_2020}. In 2D, Hermite tensors read~\cite{GRADb_CPAM_2_1949}
\begin{equation}
\begin{array}{l @{\: = \:} l}
\mathcal{H}^{(0)}_i & 1, \vspace{0.1cm}\\ 
\mathcal{H}^{(1)}_{i,x} & \xi_{i,x}, \vspace{0.1cm}\\ 
\mathcal{H}^{(2)}_{i,xx} & \xi_{i,x}^2 - c_s^2, \quad\mathcal{H}^{(2)}_{i,xy} = \xi_{i,x}\xi_{i,y}, \vspace{0.1cm}\\ 
\mathcal{H}^{(3)}_{i,xxx} & \left(\xi_{i,x}^2 - 3c_s^2\right)\xi_{i,x}, \quad\mathcal{H}^{(3)}_{i,xxy} = \left(\xi_{i,x}^2 - c_s^2\right)\xi_{i,y}, \vspace{0.1cm}\\  
\mathcal{H}^{(4)}_{i,xxxx} & \xi_{i,x}^4  - 6 c_s^2 \xi_{i,x}^2 + 3 c_s^4, \quad\mathcal{H}^{(4)}_{i,xxxy} = \left(\xi_{i,x}^2  - 3c_s^2\right) \xi_{i,x}\xi_{i,y}, \quad\mathcal{H}^{(4)}_{i,xxyy} = (\xi_{i,x}^2-c_s^2)(\xi_{i,y}^2 -c_s^2).
\end{array}
\end{equation}
Corresponding equilibrium Hermite moments are computed through the recursive formula
\begin{equation}
	\label{eq:RecursivityEquilibrium}
	\forall n \ge 2,\quad a_{eq,\alpha_1..\alpha_n}^{(n)} = u_{\alpha_n} a_{eq,\alpha_1..\alpha_{n-1}}^{(n-1)},
\end{equation}
with $a_{eq}^{(0)}=\rho$.
Another recursive formula is used for the computation of non-equilibrium Hermite moments~\cite{MALASPINAS_ARXIV_2015}
\begin{equation}
	a_{neq,{\alpha_1..\alpha_n}}^{(n)}=u_{\alpha_{n}} a_{neq,{\alpha_1..\alpha_{n-1}}}^{(n-1)} + \frac{1}{\rho} \sum_{i=1}^{n-1} a_{eq, \beta_i} ^{(n-2)} a_{neq,\alpha_i \alpha_n}^{(2)}
\end{equation}
where $a_{neq}^{(0)}=a_{neq,\alpha}^{(1)}=0$ due to conservation laws. For $N=4$, we have
\begin{align}
a_{neq,xx}^{(2)} &= \Pi^{(1)}_{xx}, \quad a_{neq,xy}^{(2)} = \Pi^{(1)}_{xy}, \nonumber\\
a_{neq,xxx}^{(3)} &=  3 u_{x}a^{(2)}_{neq,xx},
\quad a_{neq,xxy}^{(3)} = 2 u_{x}a^{(2)}_{neq,xy} + u_{y}a^{(2)}_{neq,xx},\nonumber\\
a_{neq,xxxx}^{(4)} &= 4 u_{x}a_{neq,xxx}^{(3)} -6 u_{x}^2 a^{(2)}_{neq,xx}, \label{eq:a1Rec}\\
a_{neq,xxxy}^{(4)} &= \left(3 u_{x}a_{neq,xxy}^{(3)}+u_{y}a_{neq,xxx}^{(3)} \right) - 3u_{x}^2a^{(2)}_{neq,xy} - 3u_{x}u_{y}a^{(2)}_{neq,xx}, \nonumber\\
a_{neq,xxyy}^{(4)} &= 2\left(u_{x}a_{neq,yyx}^{(3)}+u_{y}a_{neq,xxy}^{(3)}\right) - u_{x}^2 a^{(2)}_{neq,yy} - u_{y}^2 a^{(2)}_{neq,xx} -4u_{x}u_{y} a^{(2)}_{neq,xy}. \nonumber
\end{align}
Here, $\Pi_{\alpha\beta} = -\rho c_s^4 \tau \left(\partial_{\alpha}u_{\beta} + \partial_{\beta}u_{\alpha} - \textstyle{\frac{2}{D}}\partial_{\gamma}u_{\gamma}\right) \approx \textstyle{\sum_i}\mathcal{H}^{(2)}_{\alpha\beta}\left(f_i-f_i^{eq}\right)$.
$\Pi_{\alpha\beta}$ can be computed using either LB data, finite differences, or a combination of both for enhanced stability~\cite{JACOB_JT_19_2018}. Here, \(\Pi_{\alpha\beta}\) is computed using only LB data. 
If additional robustness is required for purely aerodynamics simulations, equilibrating the trace of second-order (Hermite) moments is usually a good choice.
This approach is widely used in moment-based collision models by setting the bulk viscosity relaxation time to unity~\cite{GEIER_PRE_73_2006,ASINARI_PRE_78_2008,FEI_PRE_97_2018,FEI_PoF_31_2019,SAITO_PRE_98_2018,DEROSIS_PoF_31_2019,DEROSIS_PoF_32_2020}.
To ensure consistency and avoid case-by-case deliberations in Section~\ref{sec:validation_weakly_compressible}, we systematically apply this condition in aerodynamic simulations, \emph{while excluding it for aeroacoustic simulations}. \CC{The 3D simulation of the turbulent flow past a sphere is based on the same framework, with more details available in Ref.~\cite{MALASPINAS_ARXIV_2015}.}

Moving on to fully compressible and high speed flows, the polynomial expansion framework is abandoned in favor of more robust formulations of the collision term, particularly when temperature variations are non-negligible and velocities approach or exceed the speed of sound.
More precisely, equilibrium populations now have an exponential form
\begin{equation}\label{eq:exponential}
 f_i^{eq} = a \exp\left[-\left(1+{\sum_{n}}\lambda_{\bm\alpha}^{(n)}M_{i,\bm\alpha}^{(n)}\right)\right],
\end{equation}
that directly flows from the minimization of the $H$-functional~\cite{OTTINGER_RSTA_378_2020}
 \begin{equation}\label{eq:Hfunction}
 H=\sum_i f_i \big[\ln{(f_i/a)}\big]
 \end{equation}
under the constraints over the raw moments of the Maxwell-Boltzmann distribution, $M_{\bm\alpha}^{(n),\mathrm{MB}}$, 
\begin{equation}
G_{\bm\alpha}^{(n)}=M_{i,\bm\alpha}^{eq,(n)} - M_{\bm\alpha}^{(n),\mathrm{MB}}=0\label{eq:constraints}.
\end{equation}
This general formulation was proposed in a previous work~\cite{LATT_RSTA_378_2020}, building upon earlier research on lattice gas cellular automata~\cite{KORNREICH_PD_69_1993} and discrete velocity models~\cite{LEVERMORE_JSP_83_1996,LETALLEC_TechReport_1997,MIEUSSENS_MMMAS_10_2000,DUBROCA_ESAIM_10_2001} published during the 1990s and early 2000s.

In the 2D case, equilibrium raw moments are defined as $M_{i,\bm\alpha}^{eq,(n)}=\sum_i f_i^{eq} \xi_{i,x}^{n_x}\xi_{i,y}^{n_y}$, with $n_x+n_y=n$. Following previous studies~\cite{LATT_RSTA_378_2020,COREIXAS_PoF_32_2020}, we set $a=\rho$, and the Lagrange multipliers $\lambda_{\bm\alpha}^{(n)}$ are computed using a rapidly converging Newton-Raphson algorithm coupled with a line search method, as implemented in the GSL library~\cite{GSL_WEBSITE}. This approach ensures the exact definition of convective fluxes in the Navier-Stokes-Fourier equations --while reducing errors on diffusive fluxes to a minimum-- by imposing constraints on the first eight moments of the Maxwell-Boltzmann distribution~\cite{LATT_RSTA_378_2020,COREIXAS_PoF_32_2020}:

\begin{align}
    G^{(0)} &= \sum_i f_i^{eq} - \rho,\label{eq:G0_8mom}\\
    G^{(1)}_{x} &= \sum_i f_i^{eq} \xi_{ix}- \rho u_{x},\label{eq:G1x_8mom}\\
    G^{(1)}_{y} &= \sum_i f_i^{eq} \xi_{iy}- \rho u_{y},\label{eq:G1y_8mom}\\
    G^{(2)}_{xx} &= \sum_i f_i^{eq} \xi_{ix}^2 - \rho (u_x^2 + T)\label{eq:G2xx_8mom},\\
    G^{(2)}_{xy} &= \sum_i f_i^{eq} \xi_{ix}\xi_{iy} - \rho u_x u_y\label{eq:G2xy_8mom},\\
    G^{(2)}_{yy} &= \sum_i f_i^{eq} \xi_{iy}^2 - \rho (u_y^2 + T),\label{eq:G2yy_8mom}\\
    G^{(3)}_{x\alpha\alpha} &= \sum_i f_i^{eq} \xi_{ix}\xi_{i\alpha}^2 -  2 \rho u_x(E + T),\label{eq:G3xaa_8mom}\\
    G^{(3)}_{y\alpha\alpha} &= \sum_i f_i^{eq} \xi_{iy}\xi_{i\alpha}^2 -  2 \rho u_y(E + T),\label{eq:G3yaa_8mom}
\end{align}
where index repetition implies Einstein's summation rule. The latter methodology allows the user to compute $f_i^{eq}$ in a \textit{numerical} manner, i.e., even if the system (\ref{eq:G0_8mom})-(\ref{eq:G3yaa_8mom}) does not have an analytical solution.

However, the internal energy of the populations $f_i$ accounts only for the translational energy of molecules, leading to a total energy defined as $2E = u_x^2 + u_y^2 + DT$, where $D=2$. To model polyatomic gases, we use the double distribution function framework, which involves a second set of populations $g_i$ to impose the correct specific heat ratio $\gamma_r$~\cite{CHU_PoF_8_1965,RYKOV_FD_10_1975,DUBROCA_ESAIM_10_2001,TITAREV_ECCOMAS_2006,NIE_PRE_77_2008}. Interestingly, $g_i^{eq}$ is not computed via the root-finding solver; instead, it is derived from its monatomic counterpart,
\begin{equation}\label{eq:gEq}
    g_i^{eq} =(2C_v-D)T f_i^{eq},
\end{equation}
with the polyatomic heat capacity at constant volume given by $C_v=1/(\gamma_r - 1)$. This approach allows us to recover the correct polyatomic behavior with minimal computational overhead, though it requires doubled memory accesses and storage. While populations $f_i$ transfer monatomic information to $g_i$ via $g_i^{eq}$~(\ref{eq:gEq}), the feedback from $g_i$ to $f_i$ is provided through the computation of the temperature
\begin{equation}\label{eq:polyTemp}
    T = \dfrac{1}{2\rho C_v}\bigg[\sum_i (\xi_{i\alpha}^2 f_i + g_i) - \rho u_{\alpha}^2\bigg]
\end{equation}
with $C_v=1/(\gamma_r-1)$ representing the polyatomic heat capacity at constant volume. This ``polyatomic'' temperature is then incorporated into the constraints (\ref{eq:G0_8mom})-(\ref{eq:G3yaa_8mom}) used to compute $f_i^{eq}$~(\ref{eq:exponential}), thus, closing the loop.

As best practices, we recommend a value of $10^{-12}$ for the convergence threshold of the Newton-Raphson solver in case of double-precision arithmetic, so that constraint errors converge around $10^{-14}$, which is close to machine precision. For floating-point operations, this threshold should be set to $10^{-5}$ to maintain a safety margin relative to the machine precision, which is slightly less than $10^{-6}$.

\subsection{Boundary conditions and other features}
For both weakly and fully compressible LBMs, all populations are initialized at their equilibrium value (\ref{eq:GenEq}, \ref{eq:exponential} and \ref{eq:gEq}) based on initial macroscopic fields of each benchmark.
For supersonic inlet conditions, Dirichlet boundary conditions are imposed by setting populations to their equilibrium values where all macroscopic quantities are fixed to the desired values. For subsonic inlets, best practices recommend extrapolating one quantity from the fluid while imposing the remaining quantities (see Figure 1 and corresponding discussion in Ref.~\cite{POINSOT_JCP_101_1992}). In this study, $\rho$ is systematically extrapolated from the nearest fluid cell in the direction normal to the boundary condition. 
Equilibrium boundary conditions are chosen because they are very robust, and well-suited for external aerodynamics as the boundary conditions are very far from the geometry.
However, internal aerodynamics requires more accurate boundary conditions~\cite{KRUGER_Book_2017}. In such case, we rely on the velocity bounce-back approach by Ladd~\cite{LADD_JFM_271_1994a}. 
Our Neumann boundary conditions consists in copying all populations from the nearest fluid cells to the current one, effectively enforcing $\partial f_i/\partial n = \partial g_i/\partial n = 0$ where $n$ denotes the direction normal to the boundary. 
All these boundary conditions are applied across as many (ghost) layers of fluid as the lattice size (one for the D2Q9, two for the D2Q13, etc). This is consistent with our previous studies~\cite{LATT_RSTA_378_2020,COREIXAS_PoF_32_2020,THYAGARAJAN_PoF_35_2023} and works on high-order finite difference or finite volume schemes~\cite{TAM_JCP_107_1993,LEBRAS_AIAAJ_55_2017}.

Inviscid conditions are imposed by setting the relaxation time to $1/2$ (in LB units). For viscous conditions, the kinematic viscosity $\nu$ is adjusted according to the desired Reynolds number: $\nu = u_0 L / \mathrm{Re}$, where $u_0$ and $L$ are the characteristic velocity and length of the flow. Under inviscid conditions, LBMs often face severe stability issues, especially near discontinuities and strong gradients, since the collide-and-stream algorithm does not satisfy total-variation-diminishing conditions.
To improve robustness without compromising accuracy, we employ a shock-capturing technique based on a kinetic sensor from a previous study~\cite{LATT_RSTA_378_2020}. This sensor operates similarly to Jameson's shock sensor~\cite{GARNIER_JCP_153_1999}, which was employed in an earlier PhD work~\cite{COREIXAS_PhD_2018}. However, the present sensor requires only local computations, unlike Jameson's method which relies on non-local gradient calculations through finite differences.
The kinetic sensor reads
\begin{equation}
    \epsilon = \dfrac{1}{V}\sum_{i=0}^{V-1} \dfrac{\vert f_i-f_i^{eq}\vert}{f_i^{eq}},
\label{eq:DeltaAvg}
\end{equation}
where $V$ is the number of discrete velocities. $\epsilon$ is used to dynamically modify the relaxation time $\tau(\epsilon) = \tau \alpha(\epsilon)$, where $\alpha(\epsilon)$ takes the values:
\begin{equation}\label{eq:shock_function}
\alpha(\epsilon)=\left\{\begin{array}{rl}
1.05, &   0.01 \leqslant \epsilon < 0.10, \\
A, &  0.10 \leqslant \epsilon < 1.00, \\
1/\tau, &  1.00 \leqslant \epsilon.
\end{array}\right.
\end{equation} 
In this work, $A=1.50$ and $1.35$ for inviscid Riemann problems and the flow past an airfoil, regardless of the lattice. These values were selected based on simulations conducted with the D2Q13-LBM, which requires the highest level of artificial dissipation to maintain stability near discontinuities. Fine-tuning $A$ for each lattice, particularly for the D2Q21 and D2Q37 LBMs, would improve their accuracy. However, our primary goal is to validate the grid refinement strategy for compressible LBMs, not to optimize the values for the shock sensor.

Eventually, the Prandtl number is fixed to 1 due to the use of a single relaxation time. However, this limitation can be addressed by applying correction terms to the numerical equilibria (see Appendix E of Ref.~\cite{COREIXAS_PoF_32_2020}) or by relying on the numerical collision and choosing the second relaxation time accordingly (see Section III-B of Ref.~\cite{THYAGARAJAN_PoF_35_2023}). These options were not considered, as the Prandtl number has little impact on the physical phenomena encountered in the validation benchmarks.

\subsection{Known defects of grid refinement strategies for computational aeroacoustics \label{app:covo_schukmann}}

\begin{figure}[htb]
  \begin{minipage}{.2\linewidth}
    \centering
        \scriptsize
        \renewcommand{\arraystretch}{1.2} 
\begin{tabular}{@{}lcc@{}}
	\multicolumn{3}{c}{\text{OASPL (dB)}} \\ 
    \toprule 
               & \text{Mean} & \text{Max} \\
    \midrule
    \text{cm}       & 75.33       & 81.23       \\
    \text{vc (lag)} & 73.11       & 77.80       \\
    \text{vc (tou)} & 71.49       & 76.24       \\
    \text{cc (uni)} & 62.15       & 69.04       \\
    \text{cc (lin)} & 58.89       & 65.86       \\
    \text{vc (dc2)} & 59.89       & 65.61       \\
    \bottomrule
\end{tabular}
  \end{minipage}
  \begin{minipage}{.79\linewidth}
    \centering
    \includegraphics[width=\linewidth]{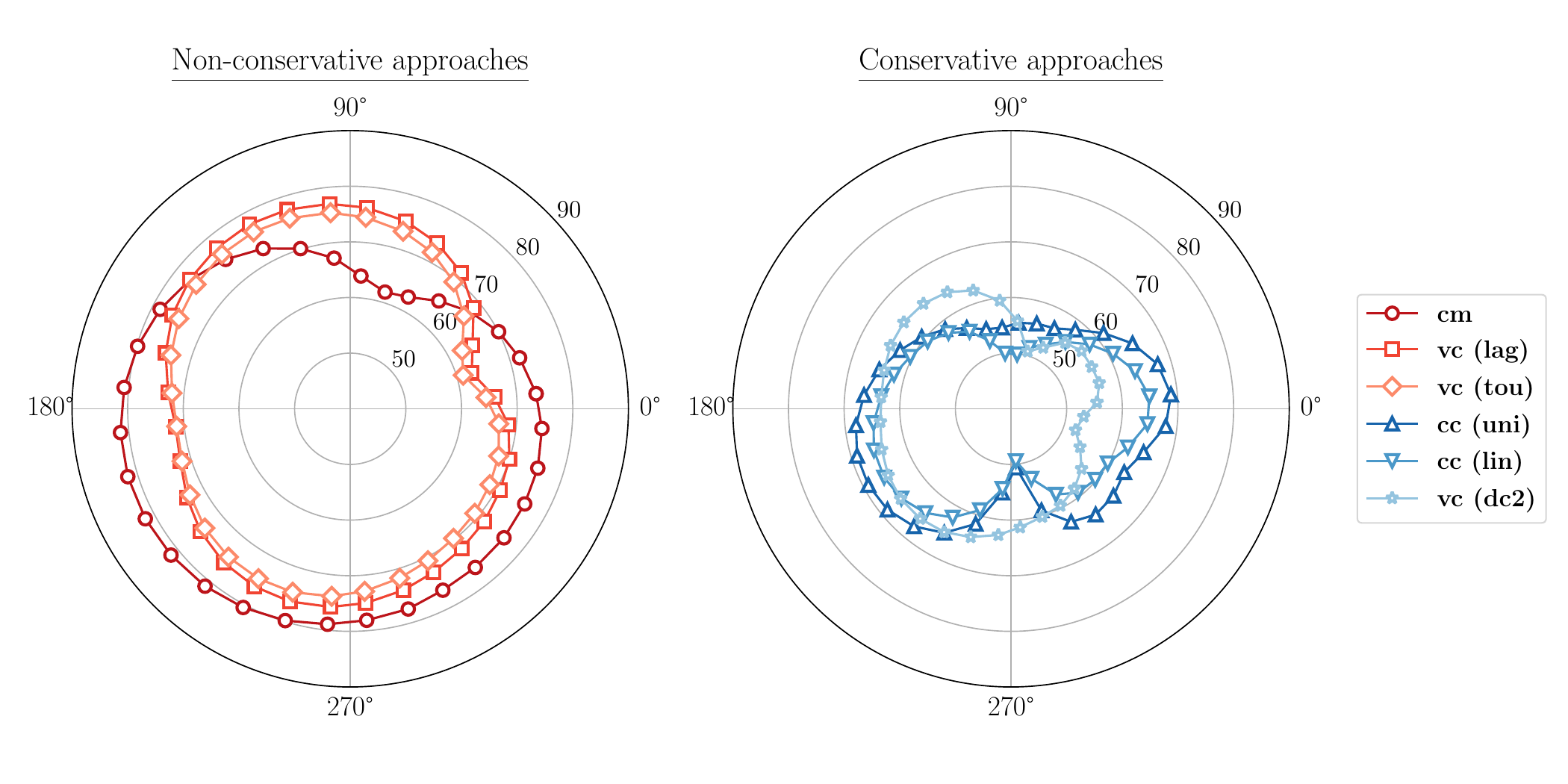}
  \end{minipage}
  \captionsetup{skip=0pt}
    \caption{Spurious noise generated when a vortex crosses a vertical grid refinement interface (lower is better). Left table: OASPL statistics for different methods in dB. Right figure: Directivity plots of OASPL in dB. Data is compiled from Schukmann et al.~\cite{SCHUKMANN_FLUIDS_10_2025}.}
    \label{fig:covo_oaspl_directivity_shuckmann}
\end{figure}

\begin{figure}[hbt]
    \centering
    \includegraphics[width=\textwidth,trim=0 1.3cm 0 0,clip]{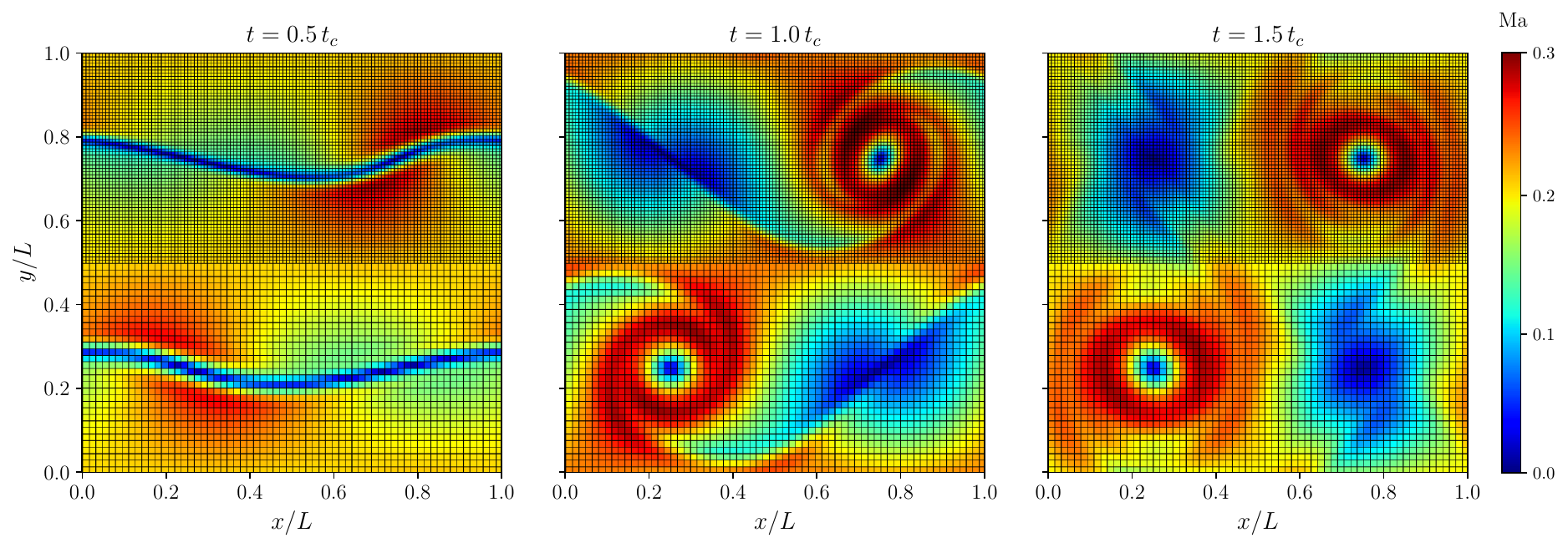}\\[0.2cm]
    \includegraphics[width=\textwidth,trim=0 0 0 0.95cm,clip]{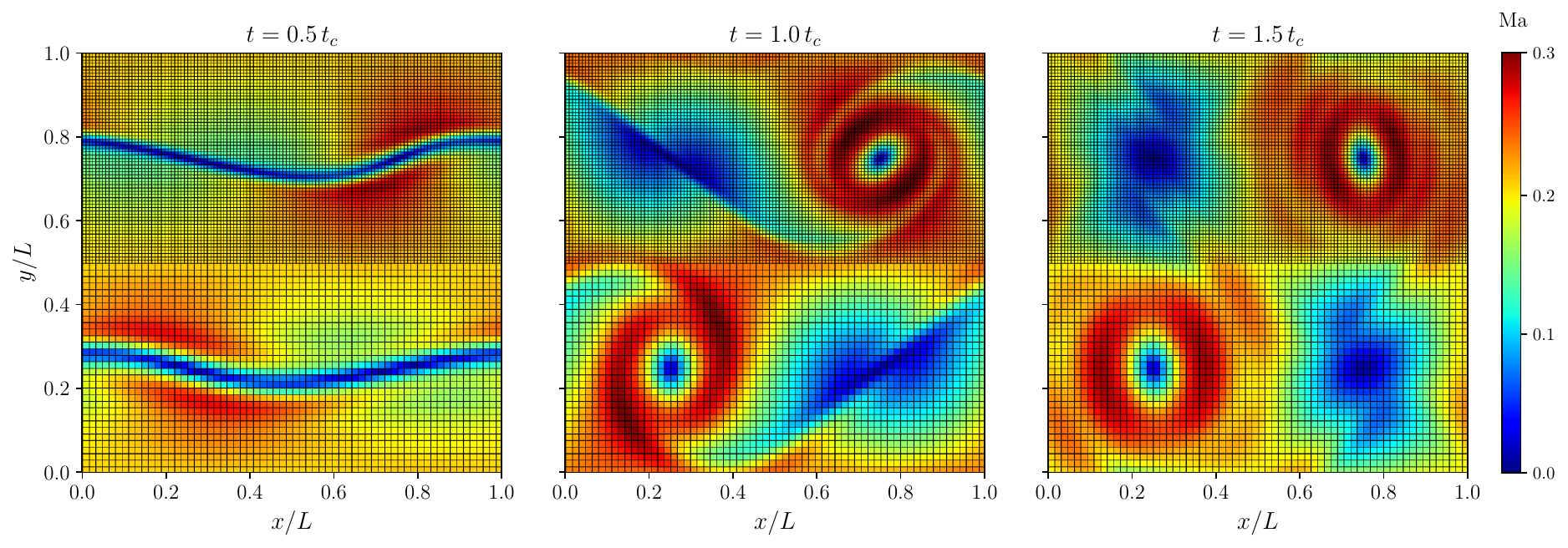}
    \captionsetup{skip=0pt}
    \caption{Time evolution of the double shear layer for $(\mathrm{Re},\mathrm{Ma},\Delta x_c)=(10^4, 0.2,L/64)$. Top panels: D2Q13 lattice. Bottom panels: D2Q21 lattice.}
    \label{fig:dsl_q13_q21_re1e4}
\end{figure}

Computational Aeroacoustics (CAA) is a specialized field within CFD dedicated to predicting noise generation mechanisms and simulating the propagation of sound waves far from their sources. Unlike conventional aerodynamics simulations, CAA requires numerical schemes with significantly higher accuracy. This is explains by the fact that acoustic waves have amplitudes several orders of magnitude smaller (approximately 0.1 to 1 Pascals) than those associated with aerodynamic phenomena such as vortices (typically 100 to 1000 Pascals). Because of that, spurious pressure fluctuations introduced by inadequate grid refinement strategies may go unnoticed in aerodynamics simulations but can completely contaminate the acoustic field in CAA.

In the specific context of CAA using LBMs, Schuckmann et al.~\cite{SCHUKMANN_FLUIDS_10_2025} conducted a comparative study of the most common grid refinement strategies found in the LBM literature. Below, we summarize some of the key methods evaluated in their work and the corresponding implementations:
\begin{itemize}
\setlength{\itemsep}{0pt} 
\setlength{\parskip}{0pt} 
\item \emph{VC approach with buffer layer}:
Proposed by Lagrava et al.~\cite{LAGRAVA_JCP_231_2012} and implemented in the open-source library Palabos~\cite{LATT_CMA_81_2021}, this approach employs buffer layers to manage data transfer between grid refinement levels. It is referred to here as `vc (lag)'.
\item \emph{VC with improved spatial filtering}:
Building on `vc (lag)', Touil et al.~\cite{TOUIL_JCP_256_2014} introduced enhanced spatial filtering techniques, implemented in a legacy version of the ProLB software~\cite{SENGISSEN_AIAA_2993_2015}. This method is referred to as `vc (tou)'.
\item \emph{VC with improved conservation and direct coupling}:
Further refining `vc (tou)', Astoul et al.~\cite{ASTOUL_JCP_447_2021} proposed two methods that enhances conservation and eliminates buffer layers through direct coupling. This strategy is implemented in the latest versions of ProLB~\cite{ASTOUL_PhD_2021}, and referred to as `vc (dc2)'.
\item \emph{CC approach with uniform explosion and a buffer layer}:
Rohde et al.~\cite{ROHDE_IJNMF_51_2006} proposed the CC approach with uniform explosion, which has been implemented in the open-source solver waLBerla~\cite{BAUER_CMA_81_2021}. This method is referred to as `cc (uni)'.
\item \emph{CC with linear explosion}:
Extending cc (uni)', Chen et al.~\cite{CHENb_PA_362_2006} introduced the linear explosion that accounts for local gradients of populations. This approach has been implemented in the commercial software PowerFLOW~\cite{KOPRIVA_AIAA_3929_2019} and is referred to as `cc (lin)'.
\item \emph{CM approach using local interpolation and a buffer layer}:
Geier et al.~\cite{GEIER_EPJST_171_2009} developed the CM approach, which employs local interpolations (bubble functions) to reconstruct missing information. This method has been implemented in the commercial software XFlow~\cite{CHAVEZ_Energies_13_2020} and is referred to as `cm'.
\end{itemize}

These approaches were evaluated across a series of benchmarks of increasing complexity. Here, we focus on the results obtained for the convected vortex benchmark. This test involves an under-resolved vortex, represented by six fine cells per radius, convected by a uniform velocity flow at $\mathrm{Ma}=0.1$ with the kinematic viscosity of air. The vortex interacts with a vertical refinement interface, generating a spurious sound wave that propagates into the far field. The amplitude of this wave depends on the refinement algorithm accuracy and conservativity, making it a valuable metric for assessing the performance of each method in handling CAA simulations on non-uniform grids. To avoid the interaction between spurious waves and the refinement interface, all algorithms were tested with the D3Q19-HRR-LBM which was designed to prevent such behavior.

Hereafter, the overall sound pressure level (OASPL) serves as a metric to quantify the sound intensity of the parasitic phenomena generated by the interaction between the vortex and the refinement interface. It is expressed in decibels (dB), and is calculated as:
\begin{equation*}
\mathrm{OASPL} = 10 \log_{10} \left( \frac{p_\text{rms}^2}{p_\text{ref}^2} \right) \, \mathrm{dB},
\end{equation*}
where $p_\text{rms}$ is the root mean square (rms) sound pressure, and $p_\text{ref}$ is the reference pressure, typically $20 \, \mu \mathrm{Pa}$ in air. In the context of this analysis, a difference of 3 dB in the results corresponds to a doubling or halving of the acoustic energy, representing a significant variation in the accuracy of the tested methods. 

Figure~\ref{fig:covo_oaspl_directivity_shuckmann} gathers statistics and directivity plots of the OASPL for the six grid refinement strategies discussed earlier. 
The contrast between conservative and non-conservative approaches is very significant, with differences ranging from 10 to 15 dB. Non-conservative methods cm, vc (lag), and vc (tou), generate considerably higher levels of spurious noise. In contrast, conservative strategies cc (uni), cc (lin), and vc (dc2), consistently yield better results, with OASPL values between 10 and 15 dB lower than their non-conservative counterparts.
More quantitative data, such as mean and maximal OASPL, reveals that cm is the worst strategy for CAA on non-uniform grids, followed closely by vc (lag) and vc (tou). The best approaches are vc (dc2) and cc (lin), respectively used in the commercial solvers ProLB and PowerFLOW. Surprisingly, the simpler and more general cc (uni) method delivers competitive results, despite a 3-4 dB gap compared to the state-of-the-art approaches.

In conclusion, this comparative study shows that \emph{enforcing conservation laws is the most critical feature a grid refinement strategy should prioritize}. This perspective might seem to go against the common approach in the LB community, which focuses more on rescaling $f_i^{neq}$ over ensuring mass, momentum, and energy conservation. However, errors in conservation directly affect macroscopic quantities and, in turn, $f_i^{eq}$. Since errors in $f_i^{eq}$ are much larger than those in $f_i^{neq}$, there is little point in fixing the latter if the former remain unaddressed.

\section{Complementary validation data\label{app:complementary_data}}

\subsection{Time evolution of the double shear layer using high-order lattices\label{app:dsl}}

To investigate the impact of the grid refinement interface on the time evolution of double shear layers, additional studies were conducted with the D2Q13 and D2Q21 lattices. Corresponding results are gathered in Figure~\ref{fig:dsl_q13_q21_re1e4} using the RR collision model. As for the D2Q9 lattice, no dispersion issues are observed which validates the accuracy of our general grid refinement strategy. The main discrepancies observed between the coarse and fine refinement patches can be attributed to an increased numerical dissipation. This was already pointed out in several previous works~\cite{COREIXAS_PhD_2018,WISSOCQ_PhD_2019,MASSET_JFM_897_2020,WISSOCQ_PRE_102_2020}.

\subsection{Lid-driven cavity simulations at higher Reynolds numbers\label{app:ldc}}

\begin{figure}[h]
    \centering
    \includegraphics[width=\textwidth]{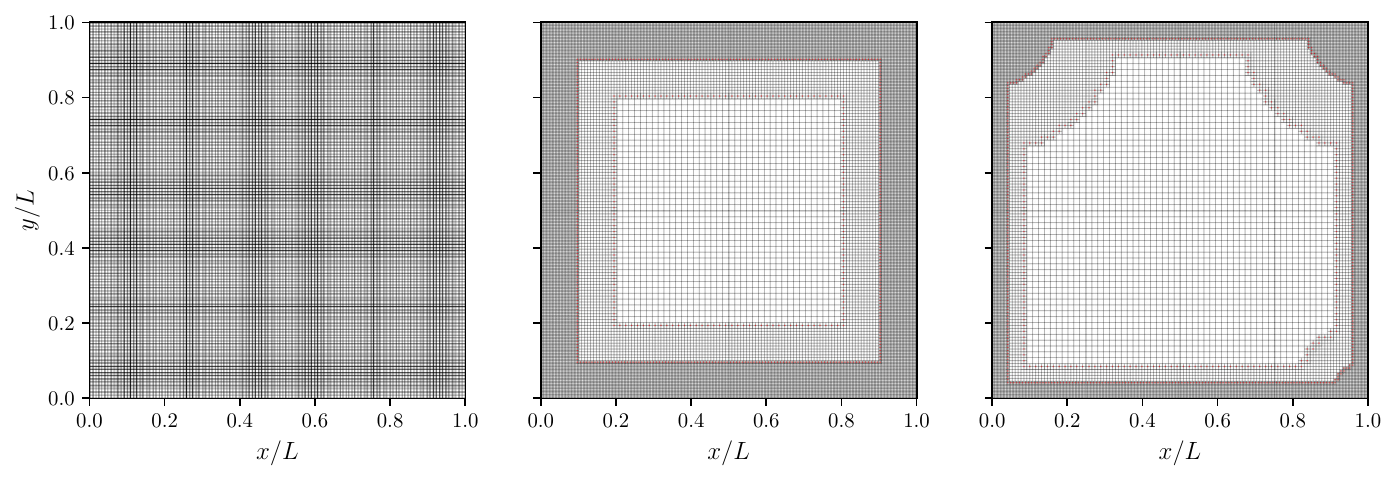}
	\captionsetup{skip=0pt}
    \caption{Mesh configurations used for the lid-driven cavity simulation at $(\mathrm{Re},\mathrm{Ma},\Delta x_f)=(7500, 0.1,1024)$. From left to right, \emph{unif}, \emph{3 lay}, and \emph{w{\&}c} configurations. For clarity, only one cell over four is shown. Red dots corresponds to coarse cell centers in the buffer layer. Only one layer is visible as meshes are plotted for the D2Q9-LBM.}
    \label{fig:cavity_grids_re7500}
\end{figure}

\begin{figure}[h]
    \centering
    \includegraphics[width=\textwidth]{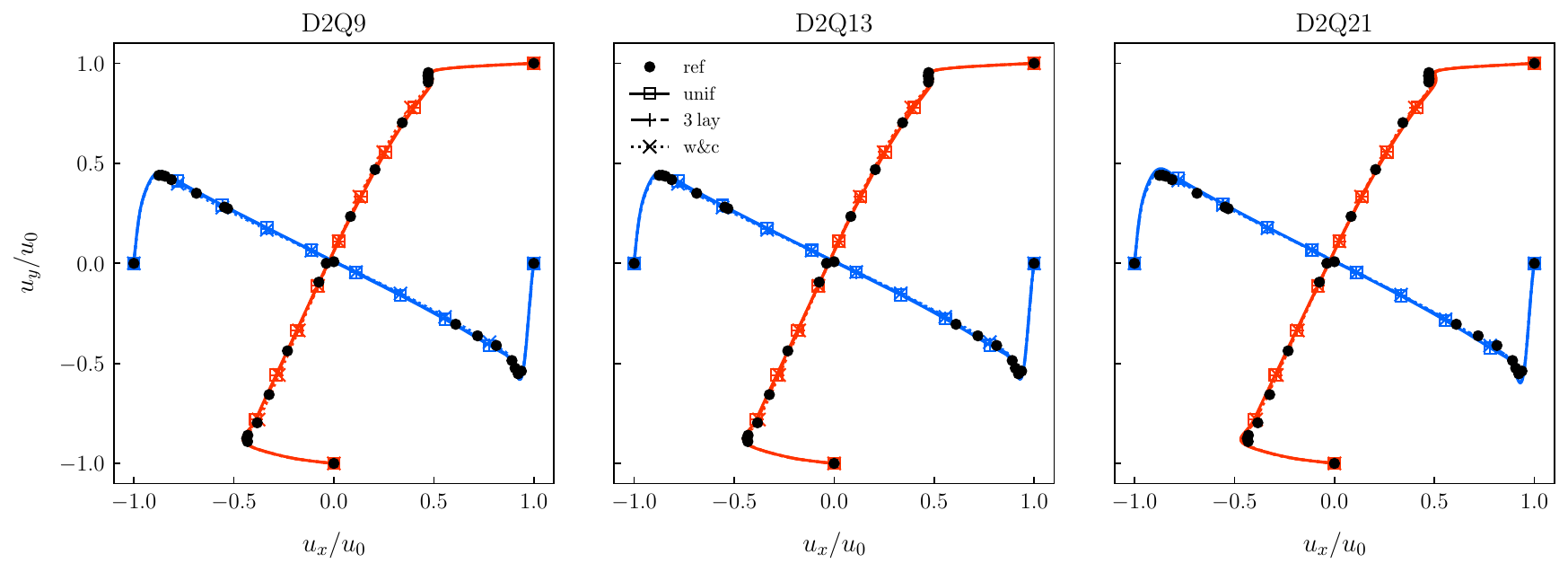}
	\captionsetup{skip=0pt}
    \caption{Velocity profiles along the centerlines of the lid-driven cavity benchmark for various lattices $(\mathrm{Re},\mathrm{Ma})=(7500, 0.1)$, and using the RR collision model. Results obtained with three different meshes (\emph{unif}, \emph{3 lay}, and \emph{w{\&}c}) are in excellent agreement with reference data from Ghia et al.~\cite{GHIA_JCP_48_1982}.}
    \label{fig:cavity_profiles_re7500}
\end{figure}

The fluid flow within the square lid-driven cavity becomes unsteady when the Reynolds number reaches the critical value of $8000$, a phenomenon known as Hopf bifurcation~\cite{FORTIN_IJNMF_24_1997,BRUNEAU_CF_35_2006}. Therefore, we simulate this benchmark at $\mathrm{Re}=7500$, the highest Reynolds number for which steady-state reference data is available~\cite{GHIA_JCP_48_1982}. At this higher Reynolds number, the boundary layer developing along the walls is much thinner than at $\mathrm{Re}=100$, hence, refinement patches should be added accordingly. Consequently, the \emph{vert} mesh configuration is no longer suitable, and another mesh composed of three nested refinement layers (\emph{3 lay}) is adopted instead. Figures~\ref{fig:cavity_grids_re7500} and~\ref{fig:cavity_profiles_re7500} display the meshes used for this configuration and the velocity profiles along the cavity centerlines, respectively. These profiles, plotted for the D2Q9, D2Q13, and D2Q21 lattices, show excellent agreement with the reference data.

\subsection{Aeolian noise: Impact of Reynolds and Mach numbers\label{app:aeolian_noise}}

\begin{figure}[hbt]
    \centering
    \includegraphics[width=\textwidth]{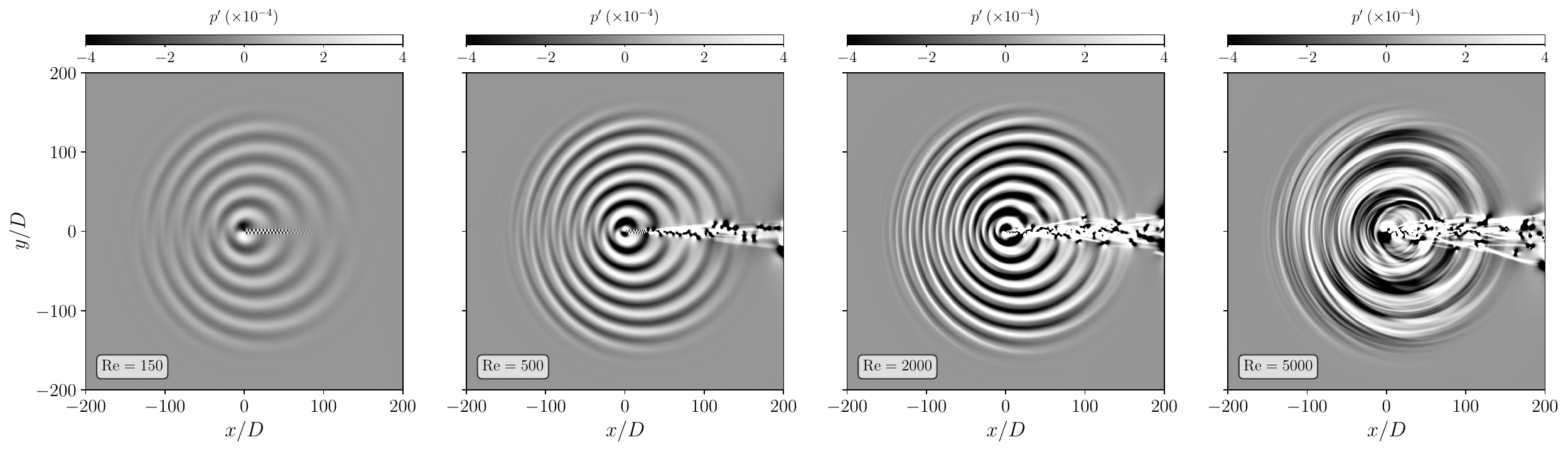}
	\captionsetup{skip=0pt}
    \caption{Effect of Reynolds number on far-field noise radiated by a circular cylinder at $(\mathrm{Ma}, \Delta x_f) = (0.2, D/160)$, simulated using the D2Q9-RR-LBM. At higher Reynolds numbers, the wake exhibits averaged-induced `streaks', reflecting more chaotic vortex shedding dynamics.}
    \label{fig:aeolian_noise_farfield_impact_re}
\end{figure}

\begin{figure}[hbt]
    \centering
    \includegraphics[width=\textwidth]{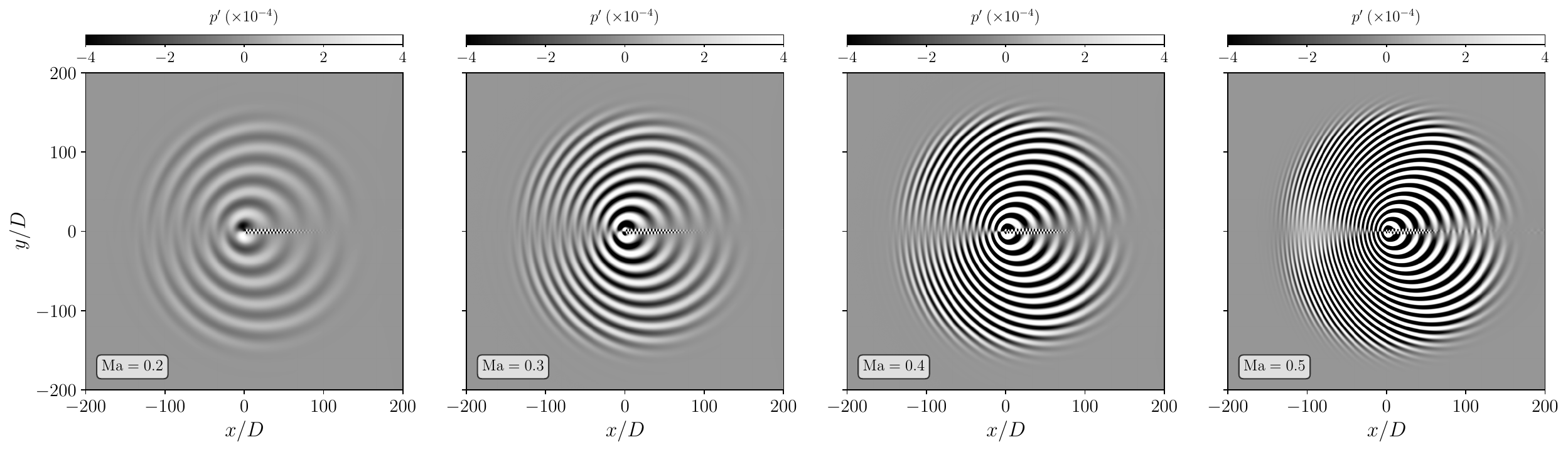}
	\captionsetup{skip=0pt}
    \caption{Effect of Mach number on far-field noise radiated by a circular cylinder at $(\mathrm{Re}, \Delta x_f) = (150, D/160)$, simulated using the D2Q9-RR-LBM. Higher Mach numbers amplify the Doppler effect which significantly influences noise directivity.}
    \label{fig:aeolian_noise_farfield_impact_ma}
\end{figure}

In this study, additional simulations were performed to investigate the influence of Reynolds and Mach numbers on the noise generated by the flow past a circular cylinder. Two parametric studies were conducted, and they are dedicated to: (1) examining the effect of varying the Reynolds number while keeping the Mach number constant at \( \mathrm{Ma} = 0.2 \), and (2) analyzing the effect of varying the Mach number while maintaining a fixed Reynolds number of \( \mathrm{Re} = 150 \). 
The second study also provides the opportunity to evaluate the accuracy and robustness of the refinement strategy near the stability limit of the D2Q9-RR-LBM. Notably, the case with \( \mathrm{Ma} = 0.5 \) results in a local Mach number of approximately 0.72, approaching the theoretical stability threshold for weakly compressible LBMs, \( 1 - \sqrt{3} \approx 0.73 \)~\cite{COREIXAS_PhD_2018,WISSOCQ_PhD_2019,MASSET_JFM_897_2020,COREIXAS_RSTA_378_2020,WISSOCQ_PRE_102_2020}.
Corresponding pressure fluctuation fields are shown in Figures~\ref{fig:aeolian_noise_farfield_impact_re} and~\ref{fig:aeolian_noise_farfield_impact_ma}.

Overall, Figure~\ref{fig:aeolian_noise_farfield_impact_re} demonstrates that increasing the Reynolds number from $\mathrm{Re} = 150$ to $2000$ at a fixed Mach number of $\mathrm{Ma} = 0.2$ has a more pronounced effect on the amplitude of Aeolian noise than on its frequency. Additionally, at higher Reynolds numbers, the wake transitions from a regular, periodic vortex shedding to a more chaotic behavior, causing interference with the initial tonal noise.
In contrast, Figure~\ref{fig:aeolian_noise_farfield_impact_ma} shows that increasing the Mach number from $\mathrm{Ma} = 0.2$ to $0.5$, while keeping the Reynolds number fixed at $\mathrm{Re} = 150$, results in increases in both the amplitude and frequency of the emitted noise. More precisely, the Doppler effect leads to asymmetric alterations in the perceived frequency and amplitude of the radiated sound waves for observers positioned at various angles relative to the cylinder. 

\begin{figure}[hbt]
    \centering
    \includegraphics[width=\textwidth]{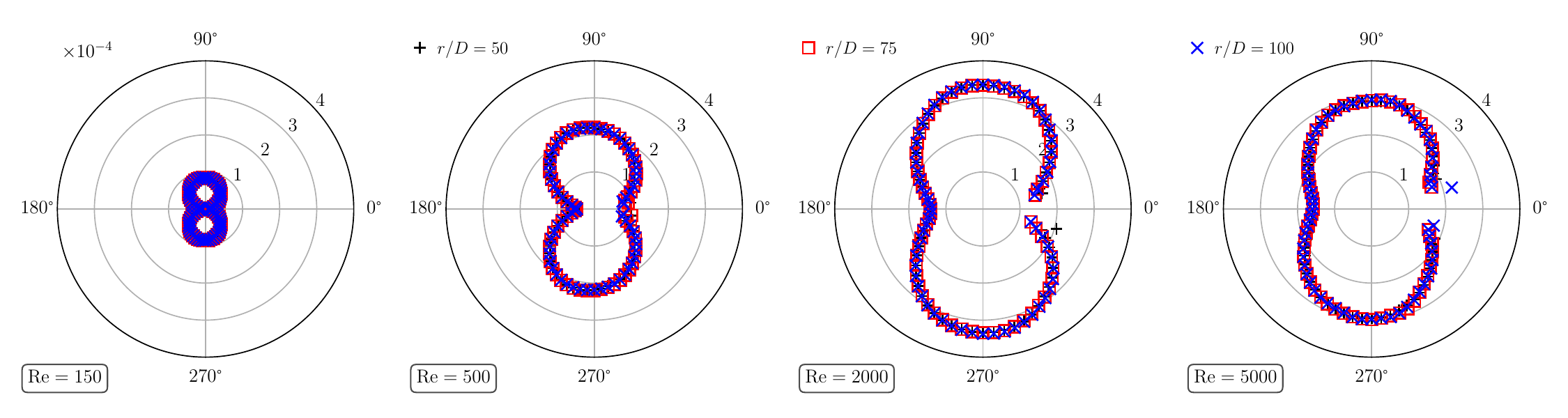}
	\captionsetup{skip=0pt}
    \caption{Impact of the Reynolds number on the Aeolian noise at $(\mathrm{Ma}, \Delta x_f) = (0.2, D/160)$ using the D2Q9-RR-LBM. Directivity of the normalized root-mean-square (rms) pressure, $p^{\mathrm{rms}}_{\mathrm{norm}} = p^{\mathrm{rms}} / p_{\infty}(r/r_{\mathrm{ref}})^{1/2}$, with the reference distance being $r_{\mathrm{ref}} / D = 75$.}
    \label{fig:aeolian_noise_rms_polar_impact_re}
\end{figure}

\begin{figure}[hbt]
    \centering
    \includegraphics[width=\textwidth]{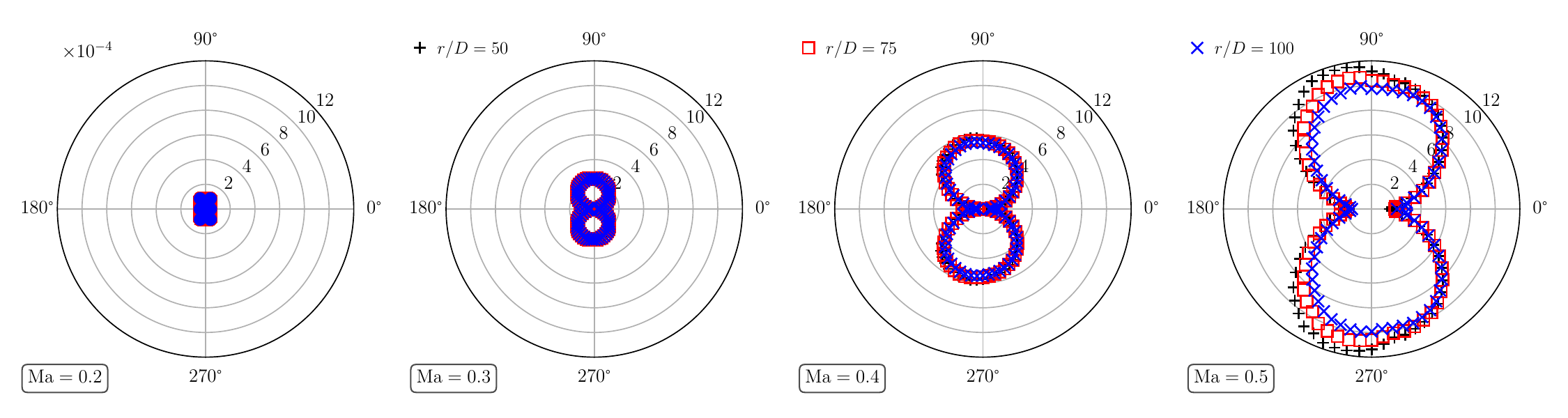}
	\captionsetup{skip=0pt}
    \caption{Impact of the Mach number on the Aeolian noise at $(\mathrm{Re}, \Delta x_f) = (150, D/160)$ using the D2Q9-RR-LBM. Directivity of the normalized root-mean-square (rms) pressure, $p^{\mathrm{rms}}_{\mathrm{norm}} = p^{\mathrm{rms}} / p_{\infty}(r/r_{\mathrm{ref}})^{1/2}$, with the reference distance being $r_{\mathrm{ref}} / D = 75$.}
    \label{fig:aeolian_noise_rms_polar_impact_ma}
\end{figure}

Quantitatively speaking, Figure~\ref{fig:aeolian_noise_rms_polar_impact_re} shows an increase in rms values of the pressure in all directions for $\mathrm{Re}\leq 2000$. At $\mathrm{Re} = 5000$, the mechanism behind the tonal noise is impacted by the transition of the wake to a more chaotic behavior, hence, reducing the amplitude of the sound waves in all directions compared to $\mathrm{Re} = 2000$ with the exception of the wake itself ($\vert\theta\vert \leq 10$). Despite this, the $1/r^2$ decrease of sound wave amplitude remains valid for all Reynolds numbers, and the lift dipole remains the primary noise emission mechanism. Figure~\ref{fig:aeolian_noise_rms_polar_impact_ma} confirms that the freestram Mach number impacts the magnitude of the Aeolian noise, its frequency, and its directivity because of a pronounced Doppler effect, especially for $\mathrm{Ma=0.5}$. As long as the Doppler effect remains negligible, the sound amplitude dependency with the Mach number follows a $\mathrm{Ma}^{2.5}$ scaling law. 
However, additional scaling and directivity adjustments are required for rms values to coincide for $\mathrm{Ma=0.5}$. 

In conclusion, both qualitative and quantitative results align very well with the DNS findings reported by Inoue et al., who employed a high-order finite-difference scheme to solve the 2D unsteady compressible Navier-Stokes equations for $\mathrm{Re} = 150$~\cite{INOUE_JFM_471_2002} and $\mathrm{Re} = 1000$~\cite{INOUE_AIAA_2132_2001} across various Mach numbers ($\mathrm{Ma}\leq 0.3$).

\subsection{Grid convergence analysis of the flow past the 30P30N airfoil\label{app:30p30n}}
\begin{figure}[hbt]
    \centering
    \includegraphics[width=\textwidth]{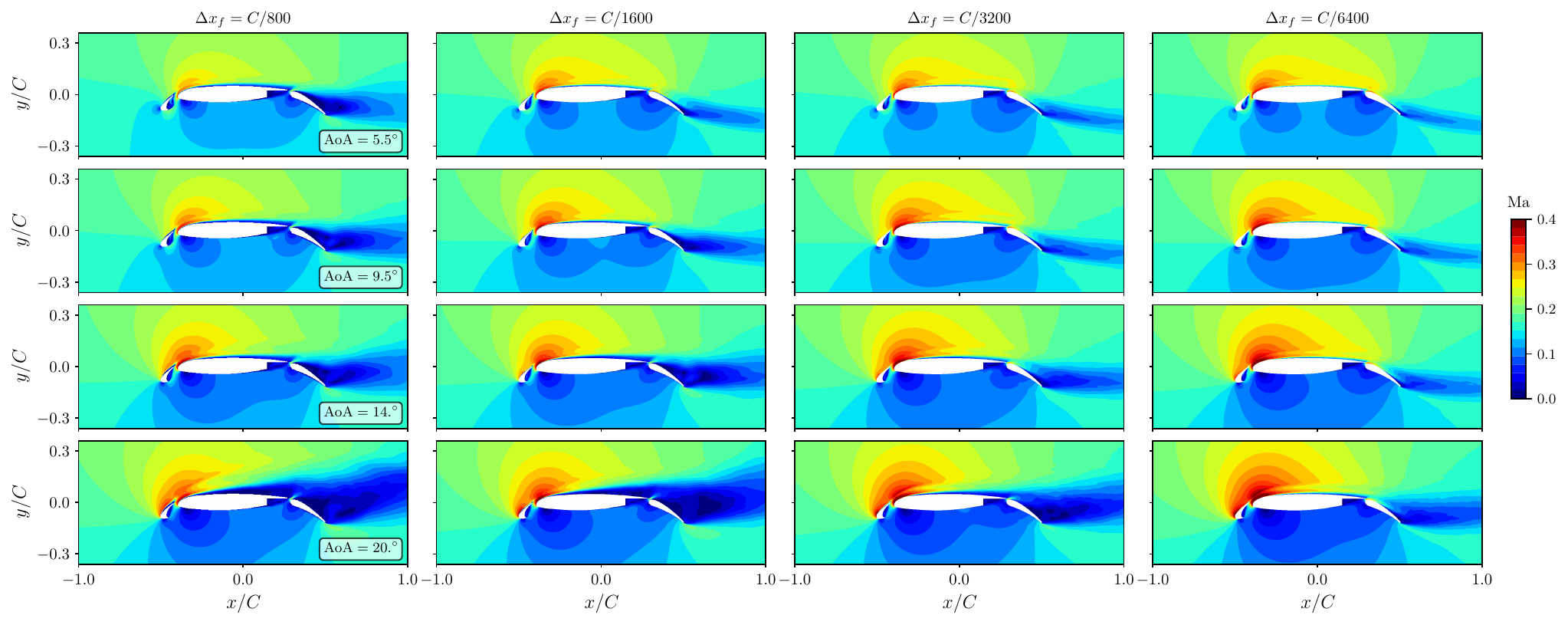}
	\captionsetup{skip=0pt}
    \caption{Mesh convergence study for the flow past the 30P30N three-element airfoil at $(\mathrm{Re},\mathrm{Ma})=(1.7\times 10^6 ,0.17)$ with the D2Q9-RR-LBM. Each column corresponds to a specific mesh resolution, and each row represents a given angle of attack (AoA). As the mesh is refined, early detachments progressively disappear, while the `high-speed bubble' increases in intensity and size, both indicating an enhanced accuracy.}
    \label{fig:30p30n_avg_mach_mesh_convergence}
\end{figure}

First simulations of this configuration were performed at a finest resolution of $\Delta x_f = C/800$. This led to early stall conditions induced by an early detachment of the boundary layer, even for an angle of attack (AoA) as low as $AoA=5.5^{\circ}$ (see left panels in Figure~\ref{fig:30p30n_avg_mach_mesh_convergence}). After doubling the resolution ($\Delta x_f = C/1600$), no severe early detachment is visible for the lowest AoA. However, the correct boundary layer behavior is only captured for $AoA \leq 9.5^{\circ}$ when $\Delta x_f = C/3200$. The final mesh considered in this work ($\Delta x_f = C/6400$) helps improving the global behavior of the boundary layer throughout the whole profile. 

It is worth noting that for $\Delta x_f = C/1600$, the simulations remained stable up to at least $150 t_c$ (we did not try going beyond this simulation time). As the mesh was further refined to $\Delta x_f = C/3200$, stability issues appeared for $t > 80 t_c$. The final grid led to stability issues occurring at $t > 45 t_c$. This is an excellent illustration of linear stability issues, induced by the high values of local Mach and Reynolds numbers, and that only appear after a finite time because of either the slow growth rate of the instability or the numerical dissipation of the scheme. Implementing an additional stabilization mechanism, such as a subgrid-scale model or spatial filtering, would enhance stability under these realistic conditions, but it is deferred to future work. 

\begin{figure}[hbt]
    \centering
    \includegraphics[width=\textwidth]{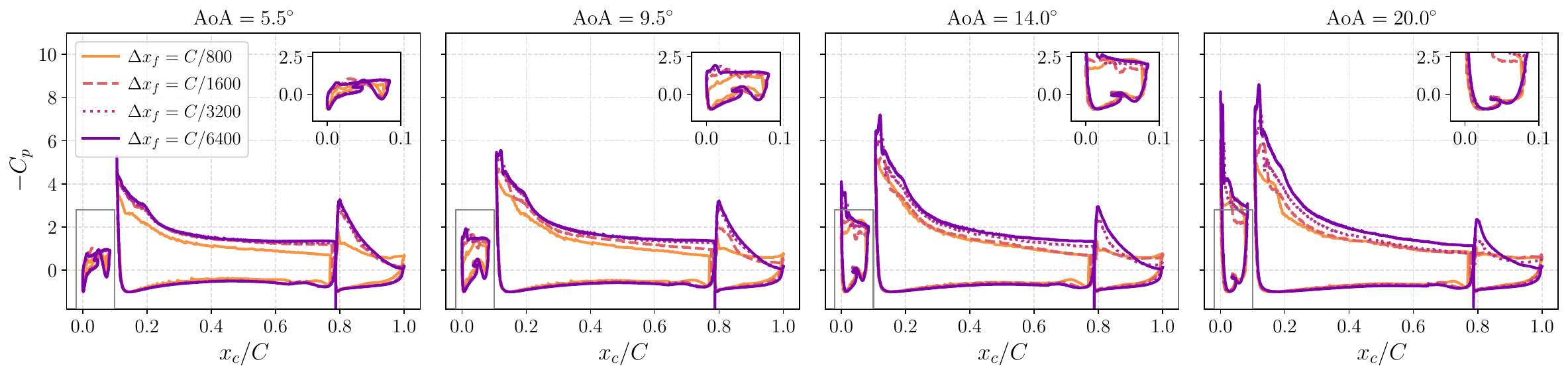}
	\captionsetup{skip=0pt}
    \caption{Mesh convergence study for the flow past the 30P30N three-element airfoil at $(\mathrm{Re},\mathrm{Ma})=(1.7\times 10^6 ,0.17)$ with the D2Q9-RR-LBM.}
    \label{fig:30p30n_nearfield_profiles_mesh_convergence}
\end{figure}

As a more quantitative analysis, Figure~\ref{fig:30p30n_nearfield_profiles_mesh_convergence} illustrates the impact of the mesh resolution on the wall pressure coefficient ($C_p$). For all AoAs, the $C_p$ profiles from the coarsest simulations ($\Delta x_f = C/800$) significantly differ from those with higher resolutions. For $\Delta x_f \geq C/1600$, the simulations yield similar profiles at $\mathrm{AoA} = 5.5^\circ$, indicating near mesh convergence for this configuration. As the AoA increases to $9.5^\circ$, a resolution of $\Delta x_f \geq C/3200$ becomes sufficient for most of the airfoil. For $AoA \geq 14^{\circ}$, $\Delta x_f \geq C/6400$ seems to be the minimal requirement to avoid early detachment. For near stall conditions ($\mathrm{AoA} = 20^\circ$), however, significant discrepancies remain between the two finest resolutions, suggesting that mesh convergence has not yet been achieved. 

As a last comment, the discrepancies observed on the slat, even when mesh convergence is reached for lower AoAs, highlight the limitations of the stair-cased no-slip boundary used in this study. To more accurately capture wall pressure on the slat, additional features, such as curved boundary conditions and wall modeling, are necessary.

\subsection{Inviscid Sod shock tube and 2D Riemann benchmarks with alternative lattices\label{app:riemann}}

\begin{figure}[hbt]
    \centering
    \includegraphics[width=\textwidth,trim=0 1.3cm 0 0,clip]{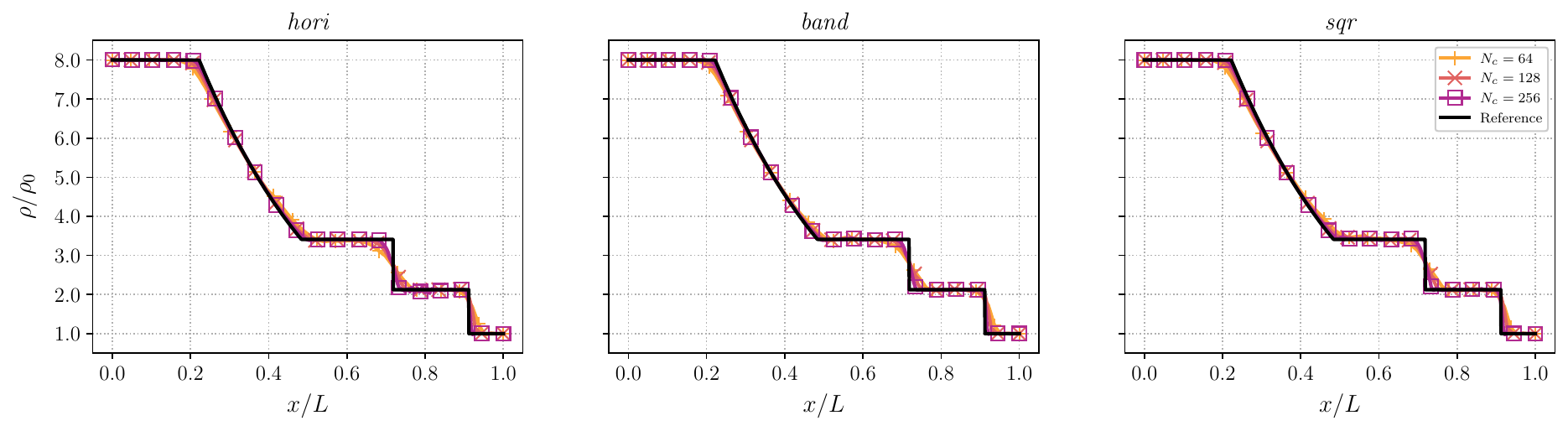}\vspace*{0.3cm}
    \includegraphics[width=\textwidth,trim=0 0 0 0.75cm,clip]{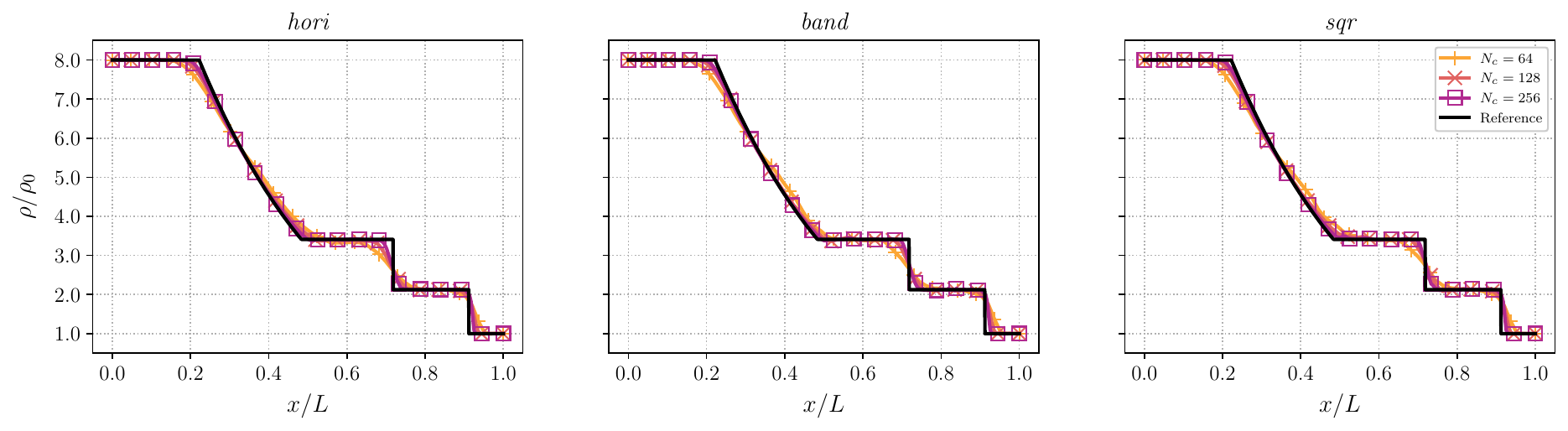}
	\captionsetup{skip=0pt}
    \caption{Normalized density profiles along the centerline ($y/L=0.5$) of the Sod shock tube benchmark for inviscid conditions ($\nu=\nu_T=0$), using the D2Q13-LBM (top) and the D2Q37-LBM (bottom) with a numerical equilibrium (8 moments).}
    \label{fig:riemann_profiles_q13_q37}
\end{figure}

In the main text, the Sod shock tube configuration has been simulated using the D2Q21 lattice (Section~\ref{subsec:sod}). Here, we extend the analysis to both a smaller (D2Q13) and a larger (D2Q37) lattice to demonstrate the universality of our framework while also highlighting the limitations of certain models used in this study. Normalized density profiles are gathered in 
Figure~\ref{fig:riemann_profiles_q13_q37}, respectively. Whatever the lattice, mesh configuration (\emph{hori}, \emph{band}, and \emph{sqr}) or mesh resolutions ($N_c=L/\Delta x_c$), numerical results are in very good agreement with the analytical solution. While this was expected for the D2Q37-LBM, as it is known to be more stable than its D2Q21 counterpart, the D2Q13-LBM based on a numerical equilibria~\cite{COREIXAS_PoF_32_2020} is way more robust than what is commonly reported in the literature, especially when the collide-and-stream algorithm and polynomial expansions of (equilibrium) distributions are considered~\cite{QIAN_JSC_8_1993,LALLEMAND_PRE_68_2003,LALLEMAND_CMA_65_2013,LALLEMAND_CCP_17_2015}. However, we do confirm that the present D2Q13-LBM is less stable than larger lattices, as already mentioned in Appendix D of Ref.~\cite{COREIXAS_PoF_32_2020}, and should be restricted to the simulation of low-amplitude shock waves.

\begin{figure}[bt!]
    \centering
	\includegraphics[height=0.35\textheight,trim=0 0 2.5cm 0,clip]{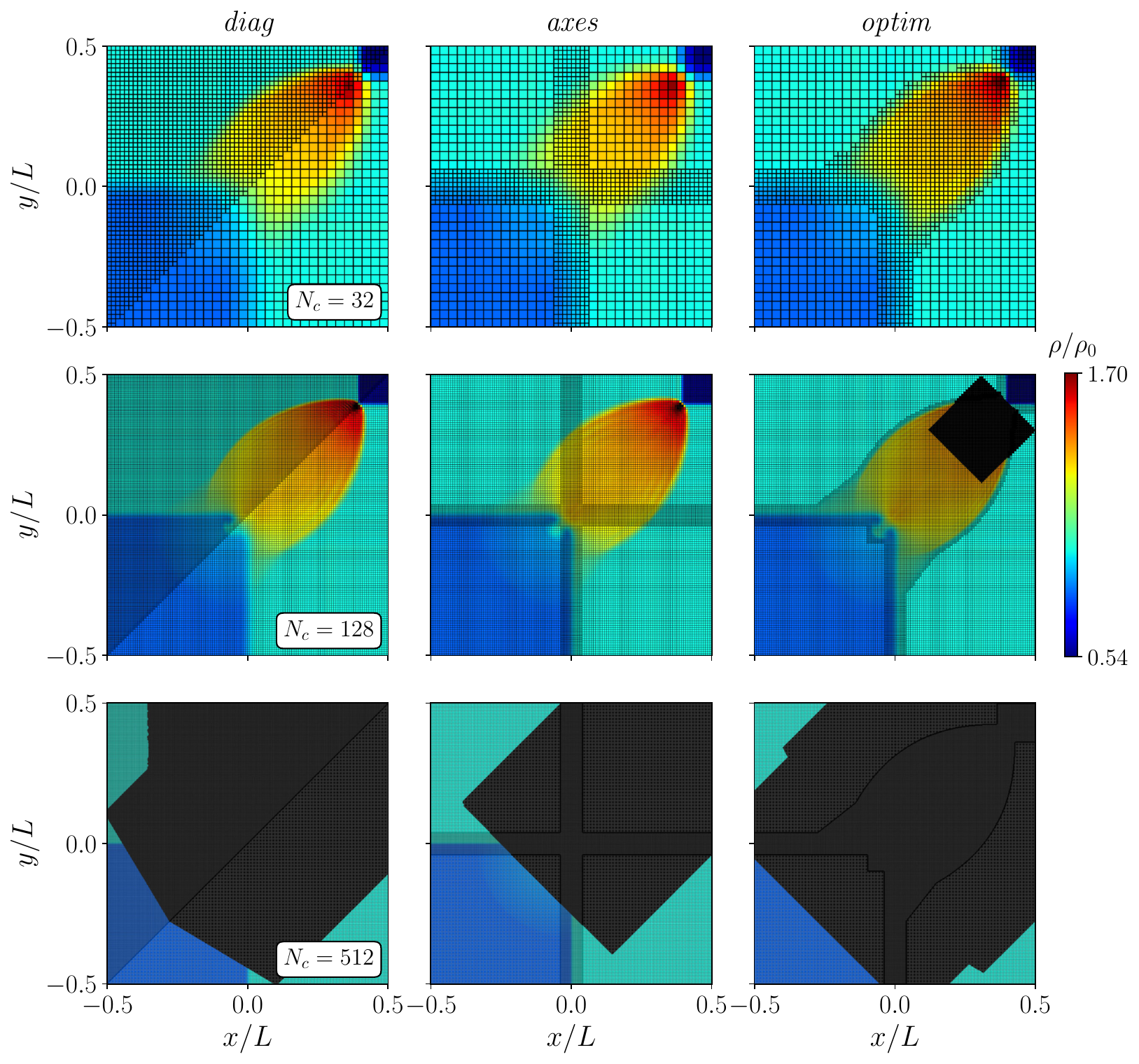}
	\hfill
    \includegraphics[height=0.35\textheight]{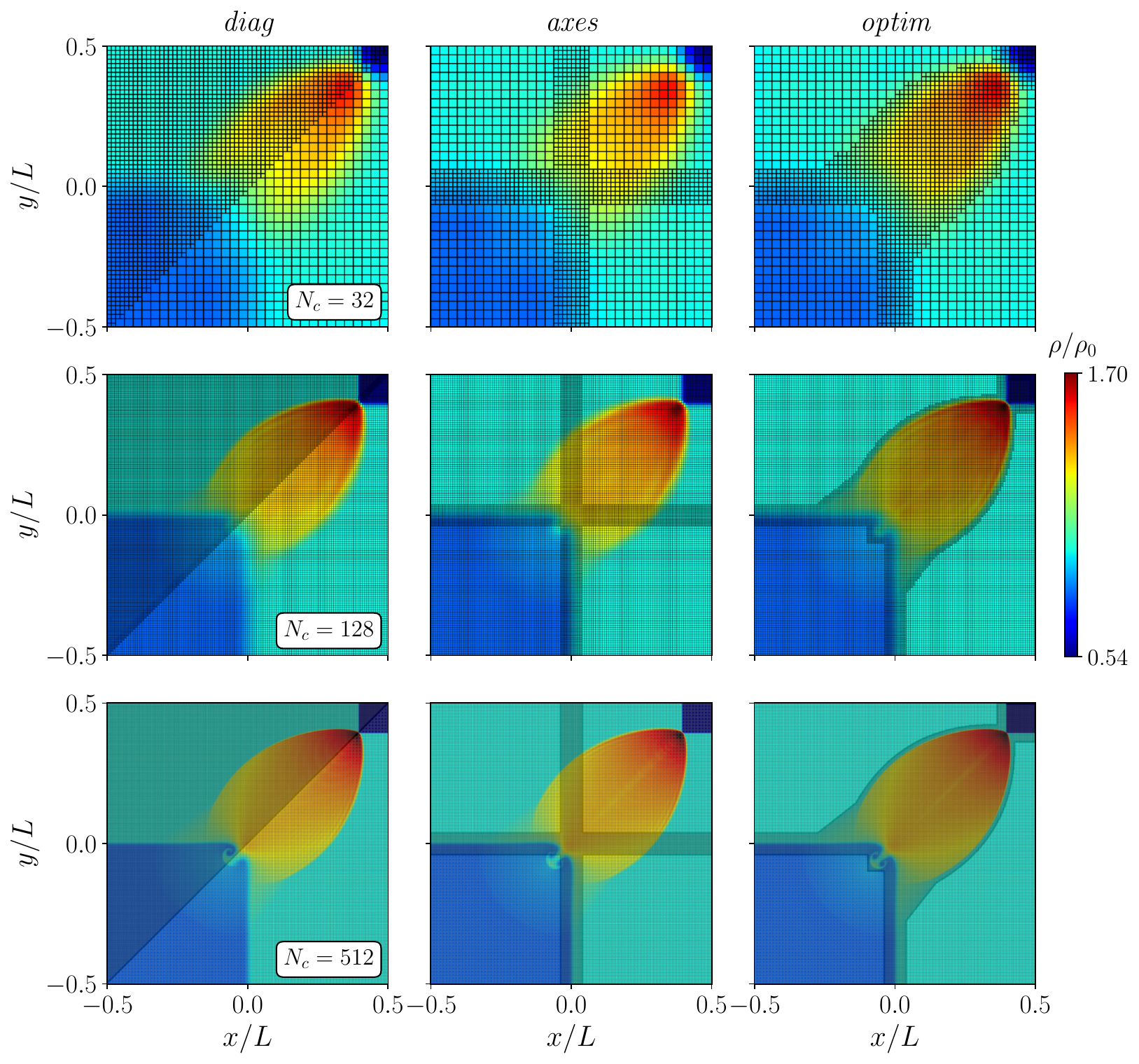}
	\captionsetup{skip=0pt}
    \caption{Normalized density fields of the 2D Riemann problem for inviscid conditions ($\nu=\nu_T=0$), using high-order LBMs with a numerical equilibrium (8 moments). Left panels: D2Q13. Right panels: D2Q37. Each column corresponds to a mesh configuration (\emph{diag}, \emph{axes}, and \emph{optim}), while each row stands for a given mesh resolution ($N_c=L/\Delta x_c$). Data is outputted at $t=0.30 t_c$ with the characteristic time being $t_c=L/\sqrt{\gamma_r T_0}$. The dark area corresponds to NaN values induced by stability issues encountered with the D2Q13 lattice.}
    \label{fig:riemann2d_grids_q13_q37}
\end{figure}

Additional results are also provided for the 2D Riemann problem discussed in Section~\ref{subsec:riemann2d}. Stability issues encountered with the D2Q13 lattice seem to originate from dispersion errors that take the form of high-frequency oscillations at the intersection between the two shockwaves (see left panels of Figure~\ref{fig:riemann2d_grids_q13_q37}). These issues are not seen with the D2Q21 (Figure~\ref{fig:riemann2d_grids_q21}) and D2Q37 lattices (right panels of Figure~\ref{fig:riemann2d_grids_q13_q37}), as they introduce more numerical dissipation in under-resolved conditions~\cite{COREIXAS_PhD_2018,WISSOCQ_PhD_2019,WISSOCQ_PRE_102_2020}. While adjusting the reference temperature usually helps reduce these oscillations, it was not possible to sufficiently attenuate them without slightly increasing viscosity coefficients, which conflicts with our will to enforce inviscid conditions. To improve the stability of the D2Q13-LBM, one would require further fine-tuning of the shock sensor, using a numerical collision model~\cite{THYAGARAJAN_PoF_35_2023}, or applying spatial filters~\cite{RICOT_JCP_228_2009,SAGAUT_CMA_59_2010}. 

\subsection{Transonic NACA flow: Preliminary simulations without local refinement at the leading edge\label{app:naca}}

\begin{figure}[bt!]
    \centering
    \includegraphics[width=\textwidth]{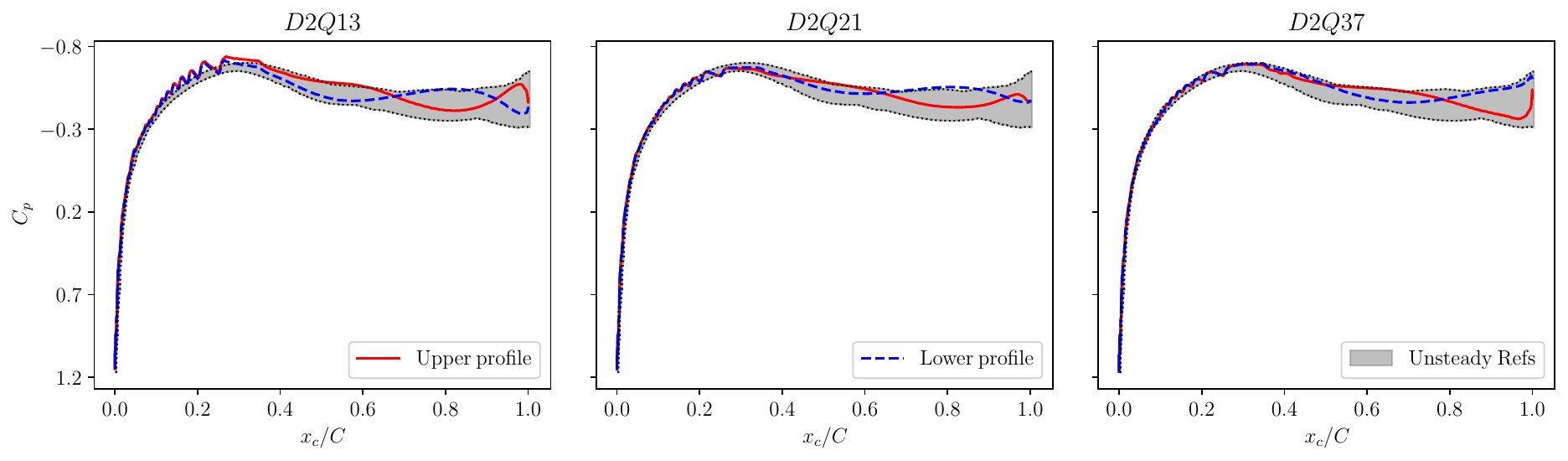}
	\captionsetup{skip=0pt}
    \caption{Wall $C_p$ profile comparisons for the viscous flow past a NACA0012 airfoil at transonic speeds with $(\mathrm{Re},\mathrm{Ma})=(10^4,0.85)$. The longitudinal coordinate $x_c$ is centered about the leading edge. The numerical equilibrium (8 moments) is used for all simulations. Unsteady reference data is gathered from the work by Satofuka et al.~\cite{SATOFUKA_GAMM_1987} and merged into a single grey area that highlights the unsteadiness of the flow.}
    \label{fig:naca_cp_profiles_ma085}
\end{figure}

In the early validation stage, simulations of the flow past a NACA0012 airfoil at transonic speeds were performed without any refinement patch close to the leading edge, and then, for a fine mesh resolution of $N_f=C/\Delta x_f\approx 550$ over the entire airfoil. Because of that, oscillations of higher amplitude could be found on the $C_p$ profiles, especially when simulations were performed with the D2Q13-LBM (see Figure~\ref{fig:naca_cp_profiles_ma085}). This is why, in the main text, additional simulations were performed with an extra refinement of the leading edge at a fine resolution of $N_f=C/\Delta x_f\approx 1100$.

\subsection{Performance of weakly compressible LBMs on matrix-like and tree-like structures with uniform grids: D2Q13 and D2Q21 lattices\label{app:perfo_weakly_tree_q13_q21}}

\begin{figure}[hbt]
    \centering
    \includegraphics[width=\textwidth]{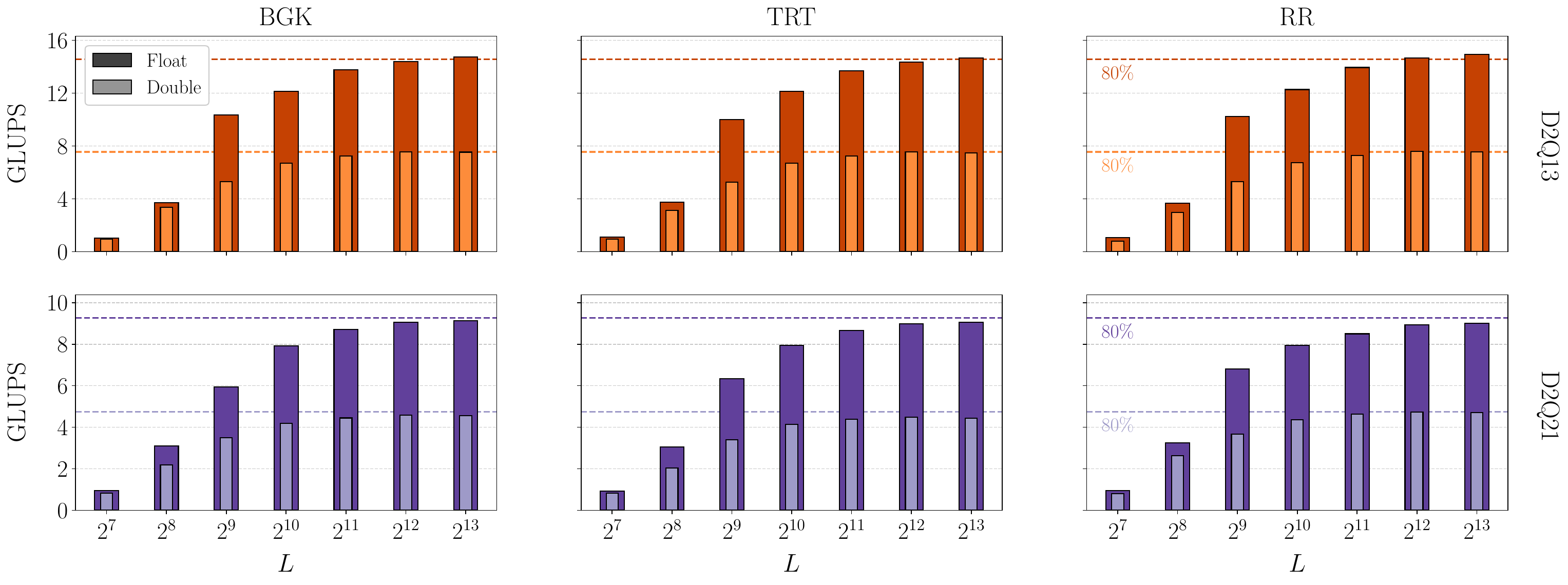}     
	\captionsetup{skip=0pt}
    \caption{Performance analysis of the lid-driven cavity simulation on a uniform grid, using an SoA and \emph{matrix-like} implementation of the high-order LBMs with various collision models, following the same methodology as for the D2Q9 study. Top: D2Q13-LBMs. Bottom: D2Q21-LBMs.}
    \label{fig:perfo_matrix_comp_lattices_coll_merged}
\end{figure}

On top of the performance analyses conducted with the D2Q9 lattice for several collision models (BGK, TRT and RR), we provide additional data corresponding to two high-order lattices, namely, the D2Q13 and D2Q21. First, we checked the implementation of these high-order LBMs on a matrix-like data structure, with SoA memory layout, and ``two-population'' strategy. This implementation is based on the STLBM framework~\cite{LATT_PLOSONE_16_2021}.
Figures~\ref{fig:perfo_matrix_comp_lattices_coll_merged} compares results obtained with the three collision models, as well as single and double precision arithmetics. As for the performance on D2Q9-LBMs (Section~\ref{subsec:perfo_matrix_uniform}), all configuration reach 80\% of the peak performance, $P_{\mathrm{peak}}^{\mathrm{matrix}}$, which is 18.21 (9.44) and 11.59 (5.93) for single (double) precision with the D2Q13 and D2Q21 LBMs, respectively.

\begin{figure}[hbt]
    \centering
    \includegraphics[width=\textwidth]{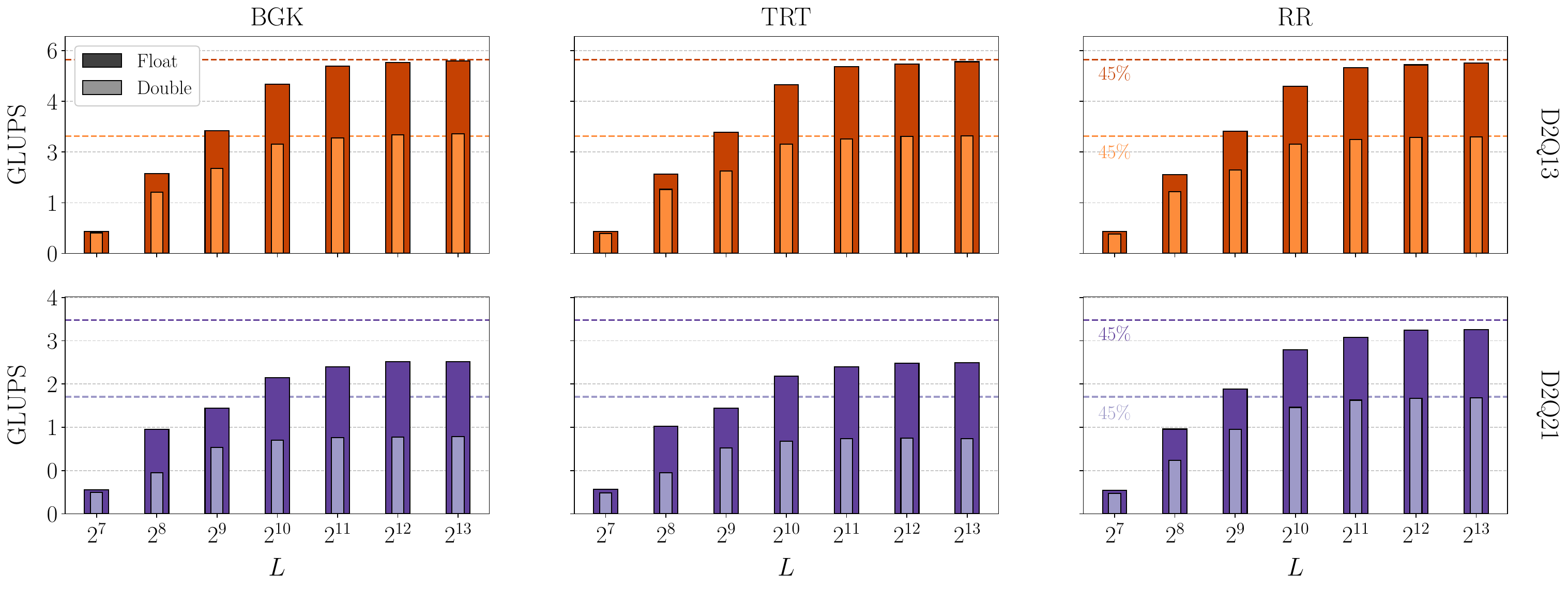}    
	\captionsetup{skip=0pt}
    \caption{Performance analysis of the lid-driven cavity simulation on a uniform grid, using an SoA and \emph{tree-like} implementation of high-order LBMs with various collision models, following the same methodology as for the D2Q9 study. Top: D2Q13-LBMs. Bottom: D2Q21-LBMs.}
    \label{fig:perfo_tree_comp_lattices_coll_merged}
\end{figure}

%

\CC{To isolate the impact of the neighbor-list on performance and to compare it with our theoretical predictions ($P_{\mathrm{peak}}^{\mathrm{tree}}$), we performed additional simulations using a tree-like structure while retaining the same uniform configuration. The corresponding results are shown in Figure~\ref{fig:perfo_tree_comp_lattices_coll_merged}.}
Due to partial code optimization, the high-order RR-LBMs reach 45\% of $P_{\mathrm{peak}}^{\mathrm{tree}}$. This corresponds to 12.74 (7.72) and 7.96 (4.81) GLUPS in single (double) precision for the D2Q13 and D2Q21 lattices, respectively. The variation in performance across collision models, despite the memory-bound nature of these schemes, further underlines the incomplete optimization of the collision step. 


\CC{Lastly, by comparing the measured performance between matrix-based and tree-like structures, we can assess the overhead introduced by the neighbor list. In the most optimized configuration, based on the RR collision model, the performance ratio ranges from 0.38 to 0.46 (see Table~\ref{tab:perfo_loss_neighbor_list_rr}). This indicates that employing neighbor lists reduces performance to a bit more than half of an equivalent solver operating on uniform grids. This observation, once again, aligns with results from highly optimized LBMs which rely on trees of blocks and GPU-specific languages~\cite{HOLZER_PhD_2025}.}

\begin{table}[h!]
\centering
\begin{tabular}{C{1.25cm} *{6}{C{1.25cm}}}
\toprule
& \multicolumn{3}{c}{Single Precision (GLUPS)} & \multicolumn{3}{c}{Double Precision (GLUPS)} \\
\cmidrule(lr){2-4}\cmidrule(lr){5-7}
Lattice & Matrix & Neighbor & $P_{\mathrm{ratio}}$ & Matrix & Neighbor & $P_{\mathrm{ratio}}$ \\
\midrule
Q9 & 20.584 & 10.541 & 0.51 & 10.795 & 7.041 & 0.65 \\
Q13 & 14.926 & 5.634 & 0.38 & 7.561 & 3.452 & 0.46 \\
Q21 & 9.003 & 3.405 & 0.38 & 4.716 & 2.140 & 0.45 \\
\bottomrule\\
\end{tabular}
\caption{Measured performance ratios $P_{\mathrm{ratio}}$ for RR-LBMs. Data is gathered from lid-driven cavity simulations using $L=2^{13}$.}
\label{tab:perfo_loss_neighbor_list_rr}
\end{table}

\subsection{Performance analysis of the Aeolian noise simulations: Additional explanations and results at higher resolutions\label{app:perfo_aeolian}}

\begin{table}[hbt]
\centering
\scriptsize
\begin{tabular}{crrrrrrrrrrr}
\specialrule{0.15em}{0em}{0.1em} 
\# & Depth & Level 0 & Level 1 & Level 2 & Level 3 & Level 4 & Level 5 & Level 6 & Tot Cells & Eq Fine Cells & \textpertenthousand \\
\toprule
13 & 14 & 1'320 & 1'104 & 1'320 & 25'352 & 7'056 & 3'984 & 63'696 & 103'832 & 6'931 & 4 \\
\midrule
14 &  & 1'320 & 1'104 & 1'320 & 25'352 & 7'056 & 258'768 & - & 294'920 & 13'898 & 11 \\
15 &  & 1'320 & 1'104 & 1'320 & 25'352 & 1'042'128 & - & - & 1'071'224 & 70'504 & 40 \\
16 &  & 1'320 & 1'104 & 1'320 & 4'193'864 & - & - & - & 4'197'608 & 526'435 & 156 \\
17 &  & 1'320 & 1'104 & 16'776'776 & - & - & - & - & 16'779'200 & 4'196'066 & 625 \\
18 &  & 1'320 & 67'108'208 & - & - & - & - & - & 67'109'528 & 33'555'424 & 2500 \\
\midrule
19 &  & 6'328 & 5'724 & 101'756 & 28'596 & 16'308 & 254'660 & - & 413'372 & 47'180 & 15 \\
20 &  & 30'328 & 423'716 & 118'918 & 67'718 & 1'017'470 & - & - & 1'658'150 & 343'972 & 62 \\
21 &  & 1'727'544 & 477'436 & 272'636 & 4'069'292 & - & - & - & 6'546'908 & 2'543'082 & 244 \\
22 &  & 3'704'112 & 1'107'894 & 16'268'654 & - & - & - & - & 21'080'660 & 8'325'222 & 785 \\
23 &  & 8'143'800 & 65'072'588 & - & - & - & - & - & 73'216'388 & 40'680'094 & 2728 \\
\specialrule{0.15em}{0.1em}{0em}
\\
\specialrule{0.15em}{0em}{0.1em} 
\# & Depth & Level 0 & Level 1 & Level 2 & Level 3 & Level 4 & Level 5 & Level 6 & Tot Cells & Eq Fine Cells & \textpertenthousand \\
\toprule
24 & 15 & 5'852 & 4'852 & 5'724 & 101'756 & 28'596 & 16'308 & 254'660 & 417'748 & 28'704 & 4 \\
\midrule
25 &  & 5'852 & 4'852 & 5'724 & 101'756 & 28'596 & 1'034'948 & - & 1'181'728 & 56'557 & 11 \\
26 &  & 5'852 & 4'852 & 5'724 & 101'756 & 4'168'388 & - & - & 4'286'572 & 282'952 & 40 \\
27 &  & 5'852 & 4'852 & 5'724 & 16'775'308 & - & - & - & 16'791'736 & 2'106'622 & 156 \\
28 &  & 5'852 & 4'852 & 67'106'956 & - & - & - & - & 67'117'660 & 16'785'017 & 625 \\
\midrule
29 &  & 26'364 & 23'724 & 423'716 & 118'918 & 67'718 & 1'017'470 & - & 1'677'910 & 195'048 & 16 \\
30 &  & 124'076 & 1'696'512 & 477'436 & 272'636 & 4'069'292 & - & - & 6'639'952 & 1'380'101 & 62 \\
31 &  & 7'042'484 & 1'943'478 & 1'107'894 & 16'268'654 & - & - & - & 26'362'510 & 10'324'778 & 246 \\
32 &  & 14'824'508 & 4'437'660 & 65'072'588 & - & - & - & - & 84'334'756 & 33'311'485 & 785 \\
\specialrule{0.15em}{0.1em}{0em}\\
\end{tabular}
\caption{Overview of the mesh configurations for Aeolian noise simulations with the D2Q9-RR-LBM. Tree depths range from $14$ to $15$, with corresponding finest grid resolutions of $\Delta x_f \in \{D/40,D/80\}$, and total number of time iterations of $n_{\mathrm{ite}}\in\{$283'840, 567'616$\}$.}
\label{tab:perfo_configs_details_14_15}
\end{table}

\begin{figure}[hbt!]
    \centering
    \includegraphics[width=\textwidth]{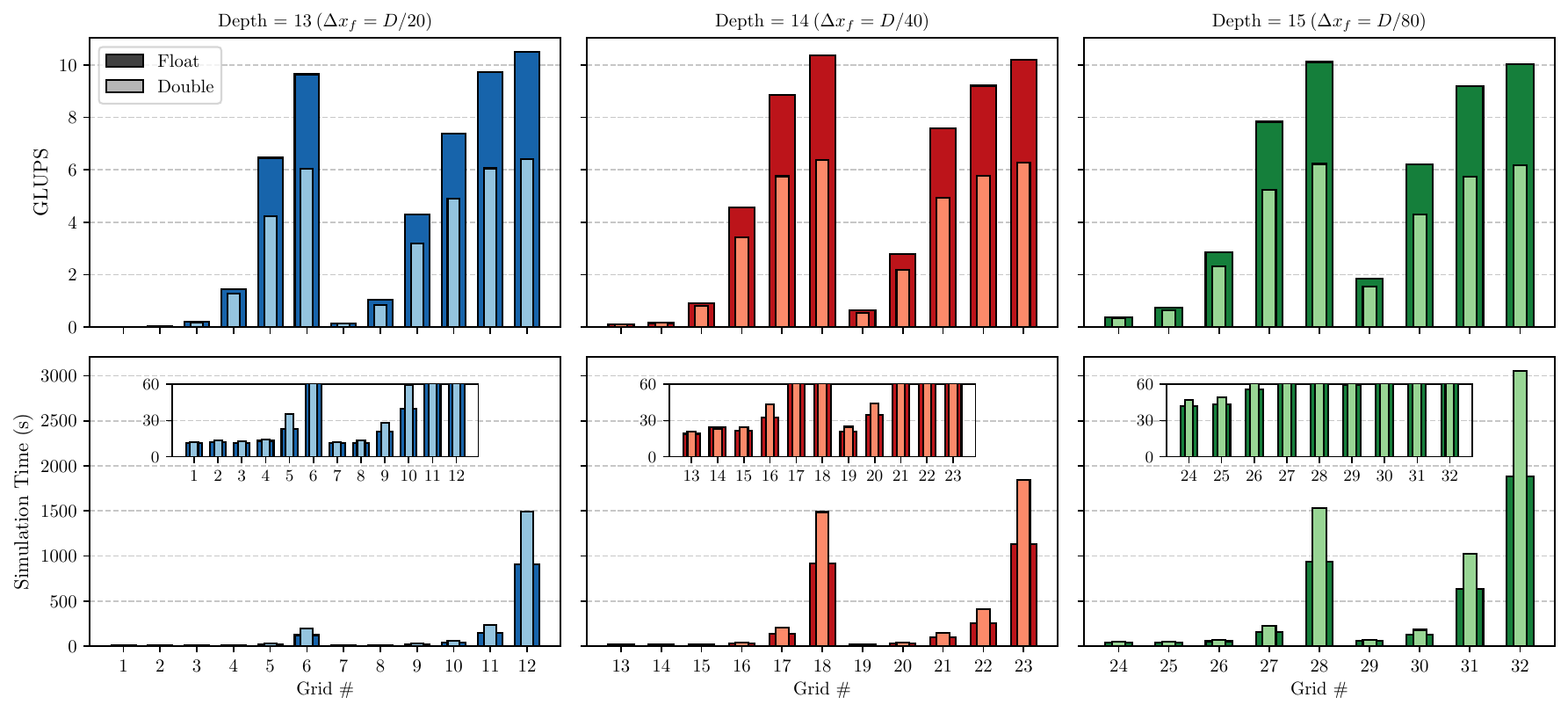}    
	\captionsetup{skip=0pt}
    \caption{Performance data (top) and simulation times (bottom) measured for Aeolian noise simulations based on the D2Q9-RR and with several finest resolution of $\Delta x_f \in \{D/20,D/40,D/80\}$.}
    \label{fig:perfo_tree_aeolian_all_resolutions}
\end{figure}

Let us explain in more details how the two parametric studies use grid configurations that are build from the one used in Section~\ref{subsec:aeolian_noise}.
The first study converts each of the coarsest cells (level $i+1$) to four finer cells (level $i$), hence adding four times more cells to the count of cells at level $i$: $n_{\mathrm{cells},i}^{\mathrm{new}}=n_{\mathrm{cells},i}^{\mathrm{old}}+4n_{\mathrm{cells},i+1}^{\mathrm{old}}$. As such, only the cell count of level $i$ increases as level $i+1$ disappears. Instead, the second study merges cells from the finest level and its coarse counterpart, while keeping the finest resolution. This leads to a cascade of level conversions that ensures the ratio between consecutive refinement levels remains equal to two. As such, the new cell count per level is computed as $n_{\mathrm{cells},i}^{\mathrm{new}}=4n_{\mathrm{cells},i+1}^{\mathrm{old}}$ except for the finest level for which $n_{\mathrm{cells},0}^{\mathrm{new}}=n_{\mathrm{cells},0}^{\mathrm{old}}+4n_{\mathrm{cells},1}^{\mathrm{old}}$. 
In practice, as many grids as the number of configurations were generated prior to the performance analysis. As each refinement level is constructed via mesh density function computed via analytical formulas (e.g., circles for circular patches), the count of cells per level \emph{does not exactly} follow the above rules when moving from one configuration to the other (notably because of round of errors), but they remain close enough to reality so that they can be used to illustrate the philosophy behind each parametric study. 

These parametric studies were performed for two additional finest mesh resolutions $\Delta x_f \in \{D/40,D/80\}$ corresponding to tree depths of 14 and 15. The final configuration for each depth was chosen in such a way that it would have a similar footprint whatever the tree depth. Key features of these configurations are gathered in Table~\ref{tab:perfo_configs_details_14_15}, and (unsorted) performance results are shown in Figure~\ref{fig:perfo_tree_aeolian_all_resolutions} for all configurations investigated in this work.  
All trends identified for the depth-13 case ($\Delta x_f = D/20$) remain unchanged with the increased cylinder discretization:
\begin{enumerate}
\item As grid size increases, performance again converges to approximately 60\% of the peak performance.
\item For smaller grids, performance continues to be independent of arithmetic precision, with single and double precision yielding comparable results far from the theoretical peak.
\item Near-peak performance is again associated with the largest grids, which require longer simulation times.
\end{enumerate}

The consistent findings across all finest grid resolutions ($\Delta x_f \in \{D/20,D/40,D/80\}$) underline the importance of tailoring the grid to the specific simulation objectives. Whether the focus is on \CC{memory footprint,} near-field aerodynamics, or far-field acoustics, adopting best practices to balance accuracy, performance, and simulation time remains crucial. The performance model developed here and the insights from the case studies offer a robust methodology for optimizing grid generation in a semi-automated manner, even for more complex scenarios requiring large domains or localized refinement patches.

\bibliographystyle{elsarticle-num}

\begin{thebibliography}{100}
\expandafter\ifx\csname url\endcsname\relax
  \def\url#1{\texttt{#1}}\fi
\expandafter\ifx\csname urlprefix\endcsname\relax\def\urlprefix{URL }\fi
\expandafter\ifx\csname href\endcsname\relax
  \def\href#1#2{#2} \def\path#1{#1}\fi

\bibitem{MANOHA_AIAA_2846_2015}
E.~Manoha, B.~Caruelle, \href{10.2514/6.2015-2846}{Summary of the {LAGOON}
  solutions from the benchmark problems for airframe noise computations-{III}
  workshop}, in: 21st AIAA/CEAS Aeroacoustics Conference, 2015, p. 2846.

\bibitem{ASTOUL_PhD_2021}
T.~Astoul, \href{http://www.theses.fr/2021AIXM0270}{Towards improved lattice
  {B}oltzmann aeroacoustic simulations with non-uniform grids: application to
  landing gears noise prediction}, Ph.D. thesis, Aix-Marseille Universit\'{e}
  (2021).

\bibitem{DEGRIGNY_PhD_2021}
J.~Degrigny, \href{http://www.theses.fr/2021AIXM0537}{{Towards the
  computational prediction of low-speed buffet : Improved wall modeling for the
  lattice Boltzmann method}}, Ph.D. thesis, Aix-Marseille University (2021).

\bibitem{ANIELLO_CF_241_2022}
A.~Aniello, D.~Schuster, P.~Werner, J.~Boussuge, M.~Gatti, C.~Mirat, L.~Selle,
  T.~Schuller, T.~Poinsot, U.~Rüde,
  \href{https://www.sciencedirect.com/science/article/pii/S0045793022001128}{Comparison
  of a finite volume and two lattice {B}oltzmann solvers for swirled confined
  flows}, Comput. Fluids 241 (2022) 105463.

\bibitem{KIRIS_AIAA_WMLES_LBM_2022}
C.~C. Kiris, A.~S. Ghate, O.~M. Browne, J.~P. Slotnick, J.~Larsson,
  \href{https://arc.aiaa.org/doi/abs/10.2514/6.2022-3294}{{HLPW-4/GMGW-3:
  Wall-Modeled LES and Lattice-Boltzmann Technology Focus Group Workshop
  Summary}}, 2022.

\bibitem{MARIE_JCP_228_2009}
S.~Mari{\'e}, D.~Ricot, P.~Sagaut,
  \href{http://www.sciencedirect.com/science/article/pii/S002199910800538X}{Comparison
  between lattice {B}oltzmann method and {N}avier–{S}tokes high order schemes
  for computational aeroacoustics}, J. Comput. Phys. 228~(4) (2009) 1056 --
  1070.

\bibitem{WISSOCQ_PRE_102_2020}
G.~Wissocq, C.~Coreixas, J.-F. Boussuge,
  \href{https://link.aps.org/doi/10.1103/PhysRevE.102.053305}{Linear stability
  and isotropy properties of athermal regularized lattice {B}oltzmann methods},
  Phys. Rev. E 102 (2020) 053305.

\bibitem{SUSS_JCP_485_2023}
A.~Suss, I.~Mary, T.~{Le Garrec}, S.~Mari\'{e},
  \href{https://www.sciencedirect.com/science/article/pii/S0021999123001936}{{A
  hybrid lattice Boltzmann--Navier-Stokes method for unsteady aerodynamic and
  aeroacoustic computations}}, J. Comput. Phys. 485 (2023) 112098.

\bibitem{HOLZER_IJHPCA_2021}
M.~Holzer, M.~Bauer, H.~K\"{o}stler, U.~R\"{u}de,
  \href{https://doi.org/10.1177/10943420211016525}{Highly efficient lattice
  {B}oltzmann multiphase simulations of immiscible fluids at high-density
  ratios on {CPUs and GPUs} through code generation}, Int. J. High Perform.
  Comput. Appl. (2021).

\bibitem{HOLZER_PhD_2025}
M.~Holzer, \href{https://open.fau.de/handle/openfau/33506}{{Code generation in
  a lattice Boltzmann framework for exascale computing}}, Ph.D. thesis,
  Friedrich-Alexander-Universit\"{a}t Erlangen-N\"{u}rnberg (2025).

\bibitem{LATT_PLOSONE_16_2021}
J.~Latt, C.~Coreixas, J.~Beny,
  \href{https://doi.org/10.1371/journal.pone.0250306}{Cross-platform
  programming model for many-core lattice {B}oltzmann simulations}, PLoS One
  16~(4) (2021) 1--29.

\bibitem{LATT_ARXIV_09242_2025}
J.~Latt, C.~Coreixas, \href{https://arxiv.org/abs/2506.09242}{{Multi-GPU acceleration of PALABOS fluid solver using C++ standard parallelism}},
  arXiv:2506.09242 (2025).

\bibitem{SCHUKMANN_FLUIDS_8_2023}
A.~Schukmann, A.~Schneider, V.~Haas, M.~B\"{o}hle,
  \href{https://www.mdpi.com/2311-5521/8/3/103}{{Analysis of hierarchical grid
  refinement techniques for the lattice Boltzmann method by numerical
  experiments}}, Fluids 8~(3) (2023).

\bibitem{SCHUKMANN_FLUIDS_10_2025}
A.~Schukmann, V.~Haas, A.~Schneider,
  \href{https://www.mdpi.com/2311-5521/10/2/31}{{Spurious aeroacoustic
  emissions in lattice Boltzmann simulations on non-uniform grids}}, Fluids
  10~(2) (2025).

\bibitem{FILIPPOVA_JCP_147_1998}
O.~Filippova, D.~H\"{a}nel,
  \href{http://www.sciencedirect.com/science/article/pii/S0021999198960892}{Grid
  refinement for lattice{-BGK} models}, J. Comput. Phys. 147~(1) (1998) 219 --
  228.

\bibitem{YU_IJNMF_39_2002}
D.~Yu, R.~Mei, W.~Shyy, \href{10.1002/fld.280}{A multi-block lattice
  {B}oltzmann method for viscous fluid flows}, Int. J. Numer. Meth. Fluids
  39~(2) (2002) 99--120.
\bibitem{DUPUIS_PRE_67_2003}
A.~Dupuis, B.~Chopard,
  \href{https://link.aps.org/doi/10.1103/PhysRevE.67.066707}{Theory and
  applications of an alternative lattice {B}oltzmann grid refinement
  algorithm}, Phys. Rev. E 67 (2003) 066707.

\bibitem{LAGRAVA_JCP_231_2012}
D.~Lagrava, O.~Malaspinas, J.~Latt, B.~Chopard,
  \href{http://www.sciencedirect.com/science/article/pii/S002199911200157X}{Advances
  in multi-domain lattice {B}oltzmann grid refinement}, J. Comput. Phys.
  231~(14) (2012) 4808 -- 4822.

\bibitem{GENDRE_PRE_96_2017}
F.~Gendre, D.~Ricot, G.~Fritz, P.~Sagaut,
  \href{https://link.aps.org/doi/10.1103/PhysRevE.96.023311}{Grid refinement
  for aeroacoustics in the lattice {B}oltzmann method: A directional splitting
  approach}, Phys. Rev. E 96 (2017) 023311.

\bibitem{ASTOUL_JCP_418_2020}
T.~Astoul, G.~Wissocq, J.-F. Boussuge, A.~Sengissen, P.~Sagaut,
  \href{http://www.sciencedirect.com/science/article/pii/S0021999120304198}{Analysis
  and reduction of spurious noise generated at grid refinement interfaces with
  the lattice {B}oltzmann method}, J. Comput. Phys. 418 (2020) 109645.

\bibitem{ASTOUL_JCP_447_2021}
T.~Astoul, G.~Wissocq, J.-F. Boussuge, A.~Sengissen, P.~Sagaut,
  \href{https://www.sciencedirect.com/science/article/pii/S0021999121005623}{Lattice
  {B}oltzmann method for computational aeroacoustics on non-uniform meshes: {A}
  direct grid coupling approach}, J. Comput. Phys. 447 (2021) 110667.

\bibitem{FENG_PRE_101_2020}
Y.~Feng, S.~Guo, J.~Jacob, P.~Sagaut,
  \href{https://link.aps.org/doi/10.1103/PhysRevE.101.063302}{Grid refinement
  in the three-dimensional hybrid recursive regularized lattice {B}oltzmann
  method for compressible aerodynamics}, Phys. Rev. E 101 (2020) 063302.

\bibitem{LYU_PoF_35_2023}
C.~Lyu, P.~Liu, T.~Hu, X.~Geng, Q.~Qu, T.~Sun, R.~A.~D. Akkermans,
  \href{https://doi.org/10.1063/5.0130467}{Hybrid method for wall local
  refinement in lattice boltzmann method simulation}, Phys. Fluids 35~(1)
  (2023) 017103.

\bibitem{ROHDE_IJNMF_51_2006}
M.~Rohde, D.~Kandhai, J.~J. Derksen, H.~E.~A. van~den Akker,
  \href{https://onlinelibrary.wiley.com/doi/abs/10.1002/fld.1140}{A generic,
  mass conservative local grid refinement technique for lattice-{B}oltzmann
  schemes}, Int. J. Numer. Meth. Fluids 51~(4) (2006) 439--468.

\bibitem{CHENb_PA_362_2006}
H.~Chen, O.~Filippova, J.~Hoch, K.~Molvig, R.~Shock, C.~Teixeira, R.~Zhang,
  \href{http://www.sciencedirect.com/science/article/pii/S0378437105009696}{Grid
  refinement in lattice {B}oltzmann methods based on volumetric formulation},
  Physica A 362~(1) (2006) 158 -- 167, discrete Simulation of Fluid
  DynamicsProceedings of the 13th International Conference on Discrete
  Simulation of Fluid Dynamics13th International Conference on Discrete
  Simulation of Fluid Dynamics.

\bibitem{LI_EPJST_171_2009}
Y.~Li, R.~Zhang, R.~Shock, H.~Chen,
  \href{https://doi.org/10.1140/epjst/e2009-01015-9}{Prediction of vortex
  shedding from a circular cylinder using a volumetric lattice-boltzmann
  boundary approach}, Eur. Phys. J. Spec. Top. 171~(1) (2009) 91--97.

\bibitem{DEITERDING_NRMEFMX_2016}
R.~Deiterding, S.~L. Wood,
  \href{https://doi.org/10.1007/978-3-319-27279-5_74}{An adaptive lattice
  {B}oltzmann method for predicting wake fields behind wind turbines}, in:
  A.~Dillmann, G.~Heller, E.~Kr{\"a}mer, C.~Wagner, C.~Breitsamter (Eds.), New
  Results in Numerical and Experimental Fluid Mechanics X, Springer
  International Publishing, Cham, 2016, pp. 845--857.

\bibitem{SCHORNBAUM_SIAM_40_2018}
F.~Schornbaum, U.~R\"{u}de,
  \href{https://doi.org/10.1137/17M1128411}{Extreme-scale block-structured
  adaptive mesh refinement}, SIAM J. Sci. Comput. 40~(3) (2018) C358--C387.

\bibitem{LI_PhD_2011}
Y.~Li, \href{https://doi.org/10.31274/etd-180810-2467}{{An improved volumetric
  {LBM} boundary approach and its extension for sliding mesh simulation}},
  Ph.D. thesis, {Iowa State University} (2011).

\bibitem{FARES_AIAA_0952_2014}
E.~Fares, M.~Wessels, R.~Zhang, C.~Sun, N.~Gopalaswamy, P.~Roberts, J.~Hoch,
  H.~Chen, \href{10.2514/6.2014-0952}{Validation of a lattice-{B}oltzmann
  approach for transonic and supersonic flow simulations}, 52nd AIAA Aerospace
  Sciences Meeting 0952 (2014).

\bibitem{RIBEIRO_AIAA_1438_2017}
A.~F. Ribeiro, B.~Konig, D.~Singh, E.~Fares, R.~Zhang, P.~Gopalakrishnan,
  A.~Jammalamadaka, Y.~Li, H.~Chen, \href{10.2514/6.2017-1438}{Buffet
  simulations with a lattice-{B}oltzmann based transonic solver}, 55th AIAA
  Aerosp. Sci. Meet. (2017) 1438.

\bibitem{GONZALEZMARTINO_AIAA_3919_2018}
I.~Gonzalez-Martino, D.~Casalino,
  \href{https://arc.aiaa.org/doi/abs/10.2514/6.2018-3919}{Fan tonal and
  broadband noise simulations at transonic operating conditions using
  lattice-{B}oltzmann methods}, in: {2018 AIAA/CEAS Aeroacoustics Conference},
  2018, p. 3919.

\bibitem{KOPRIVA_AIAA_3929_2019}
J.~Kopriva, F.~Polidoro, C.~Nardari, L.~Gregory,
  \href{https://arc.aiaa.org/doi/abs/10.2514/6.2019-3929}{Lattice-{B}oltzmann
  very large eddy simulations of an underexpanded jet from a rectangular nozzle
  with and without aft-deck}, in: AIAA Propulsion and Energy 2019 Forum, 2019.

\bibitem{LI_AIAAJ_J063199_2023}
W.~Li, J.~Wang, P.~Gopalakrishnan, Y.~Li, R.~Zhang, H.~Chen,
  \href{10.2514/1.J063199}{Hybrid lattice boltzmann approach for simulation of
  high-speed flows}, AIAA J. 0~(0) (2023) 1--10.

\bibitem{YU_JCP_228_2009}
Z.~Yu, L.-S. Fan,
  \href{https://www.sciencedirect.com/science/article/pii/S0021999109002903}{An
  interaction potential based lattice {B}oltzmann method with adaptive mesh
  refinement ({AMR}) for two-phase flow simulation}, J. Comput. Phys. 228~(17)
  (2009) 6456--6478.

\bibitem{WERNER_IJNMF_93_2021}
L.~Werner, C.~Rettinger, U.~Rüde,
  \href{https://onlinelibrary.wiley.com/doi/abs/10.1002/fld.5034}{Coupling
  fully resolved light particles with the lattice boltzmann method on
  adaptively refined grids}, Int. J. Numer. Meth. Fl. 93~(11) (2021)
  3280--3303.

\bibitem{WATANABE_CPC_264_2021}
S.~Watanabe, T.~Aoki,
  \href{https://www.sciencedirect.com/science/article/pii/S0010465521000291}{{Large-scale
  flow simulations using lattice Boltzmann method with AMR following
  free-surface on multiple GPUs}}, Comput. Phys. Commun. 264 (2021) 107871.

\bibitem{GEIER_EPJST_171_2009b}
M.~Geier, A.~Greiner, J.~G. Korvink,
  \href{https://doi.org/10.1140/epjst/e2009-01026-6}{Bubble functions for the
  lattice {B}oltzmann method and their application to grid refinement}, Eur.
  Phys. J. Spec. Top. 171~(1) (2009) 173--179.

\bibitem{FAR_CMWA_79_2020}
E.~K. Far, M.~Geier, M.~Krafczyk,
  \href{http://www.sciencedirect.com/science/article/pii/S0898122118304814}{Simulation
  of rotating objects in fluids with the cumulant lattice {B}oltzmann model on
  sliding meshes}, Comput. Math. Appl. 79~(1) (2020) 3 -- 16, mesoscopic
  Methods in Engineering and Science.

\bibitem{FEUCHTER_CF_224_2021}
C.~Feuchter,
  \href{https://www.sciencedirect.com/science/article/pii/S0045793021001377}{Direct
  aeroacoustic simulation with a cumulant lattice-{B}oltzmann model}, Comput.
  Fluids 224 (2021) 104970.

\bibitem{NICKOLLS_QUEUE_2008}
J.~Nickolls, I.~Buck, M.~Garland, K.~Skadron,
  \href{https://doi.org/10.1145/1365490.1365500}{{Scalable parallel programming
  with CUDA: Is CUDA the parallel programming model that application developers
  have been waiting for?}}, Queue 6~(2) (2008) 40–53.

\bibitem{HIP_WEBSITE}
{HIP (C++ Heterogeneous-Compute Interface for Portability):
  https://rocm.docs.amd.com/projects/HIP/en/latest/}.
\newblock \href{https://rocm.docs.amd.com/projects/HIP/en/latest/}{[link]}.

\bibitem{WIENKE_EUROPAR_2012}
S.~Wienke, P.~Springer, C.~Terboven, D.~an~Mey, {OpenACC -- First Experiences
  with Real-World Applications}, in: C.~Kaklamanis, T.~Papatheodorou, P.~G.
  Spirakis (Eds.), {Euro-Par 2012 Parallel Processing}, Springer Berlin
  Heidelberg, Berlin, Heidelberg, 2012, pp. 859--870.

\bibitem{CHANDRA_Book_2001}
R.~Chandra, R.~Menon, L.~Dagum, D.~Kohr, D.~Maydan, J.~McDonald, {Parallel
  Programming in OpenMP}, 1st Edition, Elsevier, 2001.

\bibitem{MUNSHI_IEEE_2009}
A.~Munshi, {The OpenCL specification}, in: {IEEE Hot Chips 21 Symposium (HCS)},
  2009, pp. 1--314.

\bibitem{ALPAY_PIWOCL_2020}
A.~Alpay, V.~Heuveline, \href{https://doi.org/10.1145/3388333.3388658}{{SYCL
  beyond OpenCL: The architecture, current state and future direction of
  hipSYCL}}, in: {Proceedings of the International Workshop on OpenCL}, IWOCL
  '20, Association for Computing Machinery, New York, NY, USA, 2020.

\bibitem{HWU_Chapter_2012}
N.~Bell, J.~Hoberock,
  \href{https://www.sciencedirect.com/science/article/pii/B9780123859631000265}{{Thrust:
  A Productivity-Oriented Library for CUDA}}, in: W.~mei W.~Hwu (Ed.), GPU
  Computing Gems Jade Edition, Applications of GPU Computing Series, Morgan
  Kaufmann, Boston, 2012, pp. 359--371.

\bibitem{CARTEREDWRADS_JPDC_74_2014}
H.~{Carter Edwards}, C.~R. Trott, D.~Sunderland,
  \href{https://www.sciencedirect.com/science/article/pii/S0743731514001257}{{Kokkos:
  Enabling manycore performance portability through polymorphic memory access
  patterns}}, J. Parallel Distr. Com. 74~(12) (2014) 3202--3216,
  domain-Specific Languages and High-Level Frameworks for High-Performance
  Computing.

\bibitem{LARKIN_GTC_2022}
J.~Larkin,
  \href{https://www.nvidia.com/en-us/on-demand/session/gtcspring22-s41496/}{{No
  More Porting: Coding for GPUs with Standard C++, Fortran, and Python}}, in:
  {NVIDIA GTC (spring)}, 2022.

\bibitem{LARKIN_GTC_2024}
J.~Larkin, A.~Stulova,
  \href{https://www.nvidia.com/en-us/on-demand/session/gtc24-s61204/}{{No more
  porting: Accelerated computing with Standard C++, Fortran, and Python}}, in:
  {NVIDIA GTC (spring)}, 2024.

\bibitem{THYAGARAJAN_PoF_35_2023}
K.~Thyagarajan, C.~Coreixas, J.~Latt,
  \href{https://doi.org/10.1063/5.0175908}{{Exponential distribution functions
  for positivity-preserving lattice Boltzmann schemes: Application to 2D
  compressible flow simulations}}, Phys. Fluids 35~(12) (2023) 126115.

\bibitem{COREIXAS_HIFILED_LBFDGPU_2022}
C.~Coreixas, J.~Latt, X.~Shan, {No more porting: GPU acceleration of lattice
  Boltzmann and finite difference solvers via C++ parallel algorithms}, in:
  {Symposium on high fidelity LES/DNS (HiFiLeD) for industrial relevant flow
  configurations}, 2022.

\bibitem{CAPLAN_SC_2025}
R.~M. Caplan, M.~M. Stulajter, J.~A. Linker, J.~Larkin, H.~A. Gabb, S.~Su,
  I.~Rodriguez, Z.~Tschirhart, N.~Malaya, {Portability of Fortran’s ‘do
  concurrent’ on GPUs}, in: {SC24-W: Workshops of the International
  Conference for High Performance Computing, Networking, Storage and Analysis},
  2024, pp. 1904--1913.

\bibitem{CAPLAN_ARXIV_2501_2025}
R.~M. Caplan, M.~M. Stulajter, J.~A. Linker, C.~Downs, L.~A. Upton, R.~Attie,
  C.~N. Arge, C.~J. Henney,
  \href{https://arxiv.org/abs/2501.06377}{{Open-source Flux Transport (OFT). I.
  HipFT -- High-performance Flux Transport}} (2025).

\bibitem{MAGGIOAPRILE_Master_2023}
R.~{Maggio-Aprile},
  \href{https://archive-ouverte.unige.ch/unige:174443/}{{GPU-based simulation
  of dense suspensions: Coupling the DEM and the LBM in standard C++}}, Master
  thesis (2023).

\bibitem{LIN_IEEE_2022}
W.-C. Lin, T.~Deakin, S.~McIntosh-Smith, {Evaluating ISO C++ parallel
  algorithms on heterogeneous HPC systems}, in: {IEEE/ACM International
  Workshop on Performance Modeling, Benchmarking and Simulation of High
  Performance Computer Systems (PMBS)}, 2022, pp. 36--47.

\bibitem{LIN_ARXIV_02680_2024}
W.-C. Lin, S.~McIntosh-Smith, T.~Deakin,
  \href{https://doi.org/10.48550/arXiv.2401.02680}{{Preliminary report: Initial
  evaluation of StdPar implementations on AMD GPUs for HPC}}, arXiv preprint
  arXiv:2401.02680 (2024).

\bibitem{DELLAR_CMA_65_2013}
P.~J. Dellar,
  \href{http://www.sciencedirect.com/science/article/pii/S0898122111007206}{An
  interpretation and derivation of the lattice {B}oltzmann method using
  {S}trang splitting}, Comput. Math. Appl. 65~(2) (2013) 129 -- 141.

\bibitem{BHATNAGAR_PR_94_1954}
P.~Bhatnagar, E.~Gross, M.~Krook,
  \href{http://link.aps.org/doi/10.1103/PhysRev.94.511}{A model for collision
  processes in gases. {I}. {S}mall amplitude processes in charged and neutral
  one-component systems}, Phys. Rev. 94 (1954) 511--525.

\bibitem{SHAN_PRL_80_1998}
X.~Shan, X.~He,
  \href{http://link.aps.org/doi/10.1103/PhysRevLett.80.65}{Discretization of
  the velocity space in the solution of the {B}oltzmann equation}, Phys. Rev.
  Lett. 80 (1998) 65--68.

\bibitem{SHAN_JFM_550_2006}
X.~Shan, X.-F. Yuan, H.~Chen,
  \href{http://journals.cambridge.org/article_S0022112005008153}{Kinetic theory
  representation of hydrodynamics: {A} way beyond the {N}avier–{S}tokes
  equation}, J. Fluid Mech. 550 (2006) 413--441.

\bibitem{GUO_Book_2013}
Z.~Guo, C.~Shu, Lattice {B}oltzmann Method and Its Applications in Engineering,
  World Scientific, 2013.

\bibitem{PHILIPPI_PRE_73_2006}
P.~C. Philippi, L.~A. Hegele, L.~O.~E. dos Santos, R.~Surmas,
  \href{http://link.aps.org/doi/10.1103/PhysRevE.73.056702}{From the continuous
  to the lattice {B}oltzmann equation: The discretization problem and thermal
  models}, Phys. Rev. E 73 (2006) 056702.

\bibitem{COREIXAS_PRE_100_2019}
C.~Coreixas, B.~Chopard, J.~Latt,
  \href{https://link.aps.org/doi/10.1103/PhysRevE.100.033305}{Comprehensive
  comparison of collision models in the lattice {B}oltzmann framework:
  Theoretical investigations}, Phys. Rev. E 100 (2019) 033305.

\bibitem{LIa_PRE_76_2007}
Q.~Li, Y.~L. He, Y.~Wang, W.~Q. Tao,
  \href{http://link.aps.org/doi/10.1103/PhysRevE.76.056705}{Coupled
  double-distribution-function lattice {B}oltzmann method for the compressible
  {N}avier-{S}tokes equations}, Phys. Rev. E 76 (2007) 056705.

\bibitem{JAMMALAMADAKA_AIAA_3055_2020}
A.~Jammalamadaka, G.~M. Laskowski, Y.~Li, J.~Kopriva, P.~Gopalakrishnan,
  R.~Zhang, H.~Chen,
  \href{https://arc.aiaa.org/doi/abs/10.2514/6.2020-3055}{Lattice-{B}oltzmann
  very large eddy simulations of fluidic thrust vectoring in a
  converging/diverging nozzle} (2020).

\bibitem{LATT_RSTA_378_2020}
J.~Latt, C.~Coreixas, J.~Beny, A.~Parmigiani,
  \href{https://royalsocietypublishing.org/doi/abs/10.1098/rsta.2019.0559}{Efficient
  supersonic flow simulations using lattice {B}oltzmann methods based on
  numerical equilibria}, Phil. Trans. R. Soc. A 378~(2175) (2020) 20190559.

\bibitem{COREIXAS_PoF_32_2020}
C.~Coreixas, J.~Latt, \href{https://doi.org/10.1063/5.0027986}{Compressible
  lattice {B}oltzmann methods with adaptive velocity stencils: {A}n
  interpolation-free formulation}, Phys. Fluids 32~(11) (2020) 116102.

\bibitem{NIE_AIAA_139_2009}
X.~Nie, X.~Shan, H.~Chen, \href{10.2514/6.2009-139}{A lattice-{B}oltzmann /
  finite-difference hybrid simulation of transonic flow}, in: Aerospace
  Sciences Meetings, Vol. 139, American Institute of Aeronautics and
  Astronautics, 2009.

\bibitem{RENARD_PhD_2021}
F.~Renard,
  \href{https://www.researchgate.net/publication/351348385_Hybrid_Lattice_Boltzmann_Method_for_Compressible_Flows}{Hybrid
  lattice boltzmann method for compressible flows}, Ph.D. thesis, Aix-Marseille
  Universit\'{e} (2021).

\bibitem{GUO_AA_3_2021}
Z.~Guo, K.~Xu, \href{https://doi.org/10.1186/s42774-020-00058-3}{Progress of
  discrete unified gas-kinetic scheme for multiscale flows}, Advances in
  Aerodynamics 3 (2021) 1--42.

\bibitem{MALASPINAS_ARXIV_2015}
O.~{Malaspinas}, \href{http://arxiv.org/pdf/1505.06900v1.pdf}{{Increasing
  stability and accuracy of the lattice {B}oltzmann scheme: {R}ecursivity and
  regularization}}, arXiv preprint arXiv:1505.06900 (2015).

\bibitem{BROGI_JASA_142_2017}
F.~Brogi, O.~Malaspinas, B.~Chopard, C.~Bonadonna, \href{https://doi.org/10.1121/1.5006900}{{H}ermite regularization of
  the lattice {B}oltzmann method for open source computational aeroacoustics},
  J. Acoust. Soc. Am 142~(4) (2017) 2332--2345.

\bibitem{COREIXAS_PRE_96_2017}
C.~Coreixas, G.~Wissocq, G.~Puigt, J.-F. Boussuge, P.~Sagaut,
  \href{https://link.aps.org/doi/10.1103/PhysRevE.96.033306}{Recursive
  regularization step for high-order lattice {B}oltzmann methods}, Phys. Rev. E
  96 (2017) 033306.

\bibitem{COREIXAS_RSTA_378_2020}
C.~Coreixas, G.~Wissocq, B.~Chopard, J.~Latt,
  \href{https://royalsocietypublishing.org/doi/abs/10.1098/rsta.2019.0397}{Impact
  of collision models on the physical properties and the stability of lattice
  {B}oltzmann methods}, Phil. Trans. R. Soc. A 378~(2175) (2020) 20190397.

\bibitem{JACOB_JT_19_2018}
J.~Jacob, O.~Malaspinas, P.~Sagaut,
  \href{https://doi.org/10.1080/14685248.2018.1540879}{A new hybrid recursive
  regularised {B}hatnagar-{G}ross-{K}rook collision model for lattice
  {B}oltzmann method-based {L}arge {E}ddy {S}imulation}, J. Turb. 19~(11-12)
  (2018) 1051--1076.

\bibitem{CHEN_PS_95_2020}
H.~Chen, R.~Zhang, P.~Gopalakrishnan,
  \href{https://doi.org/10.1088%2F1402-4896%2Fab4b4d}{Filtered lattice
  {B}oltzmann collision formulation enforcing isotropy and {G}alilean
  invariance}, Phys. Scr. 95~(3) (2020) 034003.

\bibitem{NGUYEN_PhD_2023}
M.~Nguyen,
  \href{https://cerfacs.fr/wp-content/uploads/2023/07/Phd_-Minh_Nguyen.pdf}{{Investigation
  of the lattice Boltzmann method for the simulation of turbine active
  clearance control systems}}, Phd thesis, Aix-Marseille Universit'{e} (2023).

\bibitem{GIANOLI_PhD_2024}
T.~Gianoli,
  \href{https://cerfacs.fr/wp-content/uploads/2024/05/Phd_Gianoli_TH_CFD_24_138.pdf}{{Development
  of a lattice-Boltzmann method for turbomachinery: Towards S-duct
  simulations}}, Phd thesis, Aix-Marseille Universit\'{e} (2024).

\bibitem{WERNER_PoF_36_2024}
P.~Werner, J.~F. Boussuge, C.~Scholtes, P.~Sagaut,
  \href{https://doi.org/10.1063/5.0182741}{{Lattice-Boltzmann modeling of
  centrifugal buoyancy-induced flows in rotating compressor cavities}}, Phys.
  Fluids 36~(1) (2024) 015147.

\bibitem{CHEN_PATENT_Collision_2015}
H.~Chen, R.~Zhang, P.~Gopalakrishnan,
  \href{http://www.google.com/patents/CA2919062A1?cl=en}{Lattice {B}oltzmann
  collision operators enforcing isotropy and {G}alilean invariance}, {CA}
  {P}atent {A}pp. {CA} 2,919,062 (Jan.~29 2015).

\bibitem{CASALINO_AIAA_1834_2019}
D.~Casalino, W.~C. van~der Velden, G.~Romani, Community noise of urban air
  transportation vehicles, in: AIAA Scitech 2019 Forum, 2019, p. 1834.

\bibitem{KHORRAMI_CEAS_AERO_2019}
M.~R. Khorrami, E.~Fares,
  \href{https://doi.org/10.1007/s13272-019-00378-1}{Toward noise certification
  during design: airframe noise simulations for full-scale, complete aircraft},
  CEAS Aeronaut. J. (Mar 2019).

\bibitem{ROMANI_PhD_2022}
G.~Romani,
  \href{https://doi.org/10.4233/uuid:5d36b4de-8593-4f7e-bc92-a7ae175a0900}{Computational
  aeroacoustics of rotor noise in novel aircraft configurations: {A}
  lattice-{B}oltzmann method-based study}, Ph.D. thesis, Delft University of
  Technology (2022).

\bibitem{KRUGER_Book_2017}
T.~Kr\"{u}ger, H.~Kusumaatmaja, A.~Kuzmin, O.~Shardt, G.~Silva, E.~M. Viggen,
  \href{http://www.springer.com/la/book/9783319446479}{The Lattice {B}oltzmann
  Method: Principles and Practice}, Springer International Publishing, 2017.

\bibitem{WEIMAR_PA_224_1996}
J.~R. Weimar, J.~P. Boon,
  \href{http://www.sciencedirect.com/science/article/pii/037843719500355X}{Nonlinear
  reactions advected by a flow}, Physica A 224~(1–2) (1996) 207 -- 215,
  dynamics of Complex Systems.

\bibitem{ZHANG_PRE_74_2006}
R.~Zhang, X.~Shan, H.~Chen,
  \href{http://link.aps.org/doi/10.1103/PhysRevE.74.046703}{Efficient kinetic
  method for fluid simulation beyond the {N}avier-{S}tokes equation}, Phys.
  Rev. E 74 (2006) 046703.

\bibitem{EITELAMOR_CF_75_2013}
G.~Eitel-Amor, M.~Meinke, W.~Schr\"{o}der,
  \href{https://www.sciencedirect.com/science/article/pii/S0045793013000315}{A
  lattice-{B}oltzmann method with hierarchically refined meshes}, Comput.
  Fluids 75 (2013) 127--139.

\bibitem{TOUIL_JCP_256_2014}
H.~Touil, D.~Ricot, E.~L\'ev\^eque,
  \href{http://www.sciencedirect.com/science/article/pii/S0021999113005299}{Direct
  and large-eddy simulation of turbulent flows on composite multi-resolution
  grids by the lattice {B}oltzmann method}, J. Comput. Phys. 256~(0) (2014) 220
  -- 233.

\bibitem{MACNEICE_CPC_126_2000}
P.~MacNeice, K.~M. Olson, C.~Mobarry, R.~{de Fainchtein}, C.~Packer,
  \href{https://www.sciencedirect.com/science/article/pii/S0010465599005019}{{PARAMESH:
  A parallel adaptive mesh refinement community toolkit}}, Comput. Phys.
  Commun. 126~(3) (2000) 330--354.

\bibitem{TOMIDA_AJSS_266_2023}
K.~Tomida, J.~M. Stone, \href{https://dx.doi.org/10.3847/1538-4365/acc2c0}{{The
  Athena++ adaptive mesh refinement framework: Multigrid solvers for
  self-gravity}}, Astrophys. J. Suppl. Ser. 266~(1) (2023) 7.

\bibitem{LATT_NVIDIA_BLOG_1_2022}
J.~Latt, C.~Coreixas, G.~Brito, J.~Larkin,
  \href{https://developer.nvidia.com/blog/multi-gpu-programming-with-standard-parallel-c-part-1/}{{Multi-GPU
  Programming with Standard Parallel C++ (Part 1)}}, Tech. rep. (2022).

\bibitem{LATT_NVIDIA_BLOG_2_2022}
J.~Latt, C.~Coreixas, G.~Brito, J.~Larkin,
  \href{https://developer.nvidia.com/blog/multi-gpu-programming-with-standard-parallel-c-part-2/}{{Multi-GPU
  Programming with Standard Parallel C++ (Part 2)}}, Tech. rep. (2022).

\bibitem{ZIER_MNRAS_533_2024}
O.~Zier, R.~Kannan, A.~Smith, M.~Vogelsberger, E.~Verbeek,
  \href{https://doi.org/10.1093/mnras/stae1837}{{Adapting arepo-rt for exascale
  computing: GPU acceleration and efficient communication}}, MNRAS 533~(1)
  (2024) 268--286.

\bibitem{CAMBIER_MI_14_2013}
L.~Cambier, S.~Heib, S.~Plot, \href{https://doi.org/10.1051/MECA/2013056}{{The
  Onera elsA CFD software: Input from research and feedback from industry}},
  Mech. Ind. 14 (2013) 159--174.

\bibitem{ZHANG_IJHPCA_35_2021}
W.~Zhang, A.~Myers, K.~Gott, A.~Almgren, J.~Bell,
  \href{https://doi.org/10.1177/10943420211022811}{{AMReX: Block-structured
  adaptive mesh refinement for multiphysics applications}}, The International
  Journal of High Performance Computing Applications 35~(6) (2021) 508--526.

\bibitem{LATT_CMA_81_2021}
J.~Latt, O.~Malaspinas, D.~Kontaxakis, A.~Parmigiani, D.~Lagrava, F.~Brogi,
  M.~B. Belgacem, Y.~Thorimbert, S.~Leclaire, S.~Li, F.~Marson, J.~Lemus,
  C.~Kotsalos, R.~Conradin, C.~Coreixas, R.~Petkantchin, F.~Raynaud, J.~Beny,
  B.~Chopard,
  \href{http://www.sciencedirect.com/science/article/pii/S0898122120301267}{Palabos:
  {P}arallel lattice {B}oltzmann solver}, Comput. Math. Appl. 81 (2021) 334 --
  350.

\bibitem{ZHANG_AJSS_164_2006}
W.~Zhang, A.~I. MacFadyen, \href{https://dx.doi.org/10.1086/500792}{{RAM: A
  relativistic adaptive mesh refinement hydrodynamics code}}, Astrophys. J.
  Suppl. Ser. 164~(1) (2006) 255.

\bibitem{JUDE_JSC_2022}
D.~Jude, J.~Sitaraman, A.~Wissink,
  \href{https://doi.org/10.1007/s11227-022-04324-7}{An octree-based, cartesian
  navier-stokes solver for modern cluster architectures}, J. Supercomput.
  (2022) 1--32.

\bibitem{BAUER_CMA_81_2021}
M.~Bauer, S.~Eibl, C.~Godenschwager, N.~Kohl, M.~Kuron, C.~Rettinger,
  F.~Schornbaum, C.~Schwarzmeier, D.~Th\"{o}nnes, H.~K\"{o}stler, U.~R\"{u}de,
  \href{https://www.sciencedirect.com/science/article/pii/S0898122120300146}{{waLBerla:
  A block-structured high-performance framework for multiphysics simulations}},
  Comput. Math. Appl. 81 (2021) 478--501.

\bibitem{ASHTON_AIAA_0888_2025}
N.~Ashton, A.~Ghate, G.~Kenway, J.~Angel, M.~L.~Wong, C.~C.~Kiris,  
  \href{https://arc.aiaa.org/doi/abs/10.2514/6.2025-0888}{{Immersed boundary wall-modeled large eddy simulations for the 5th High-Lift Prediction Workshop}},  
AIAA SCITECH Forum, (2025).

\bibitem{JABER_CPC_311_2025}
K.~Jaber, E.~E.~Essel, P.~E.~Sullivan,  
\href{https://www.sciencedirect.com/science/article/pii/S0010465525000463}{{GPU-native adaptive mesh refinement with application to lattice Boltzmann simulations}},  
Comput. Phys. Commun. 311 (2025) 109543.

\bibitem{HOPPE_CMAME_391_2022}
N.~Hoppe, S.~Adami, N.~A. Adams,
  \href{https://www.sciencedirect.com/science/article/pii/S004578252100699X}{A
  parallel modular computing environment for three-dimensional multiresolution
  simulations of compressible flows}, Comput. Methods in Appl. Mech. Eng. 391
  (2022) 114486.

\bibitem{BELL_JCP_85_1989}
J.~B. Bell, P.~Colella, H.~M. Glaz,
  \href{https://www.sciencedirect.com/science/article/pii/0021999189901514}{{A
  second-order projection method for the incompressible Navier-Stokes
  equations}}, J. Comput. Phys. 85~(2) (1989) 257--283.

\bibitem{BROWN_JCP_122_1995}
D.~L. Brown, M.~L. Minion,
  \href{http://www.sciencedirect.com/science/article/pii/S0021999185712053}{Performance
  of under-resolved two-dimensional incompressible flow simulations}, J.
  Comput. Phys. 122~(1) (1995) 165 -- 183.

\bibitem{MINION_JCP_138_1997}
M.~L. Minion, D.~L. Brown,
  \href{http://www.sciencedirect.com/science/article/pii/S0021999197958435}{Performance
  of under-resolved two-dimensional incompressible flow simulations, {II}}, J.
  Comput. Phys. 138~(2) (1997) 734 -- 765.

\bibitem{SHU_Chapter_1998}
C.-W. Shu, \href{https://doi.org/10.1007/BFb0096355}{Essentially
  non-oscillatory and weighted essentially non-oscillatory schemes for
  hyperbolic conservation laws}, Springer Berlin Heidelberg, Berlin,
  Heidelberg, 1998, pp. 325--432.

\bibitem{DELLAR_PRE_64_2001}
P.~J. Dellar, \href{http://link.aps.org/doi/10.1103/PhysRevE.64.031203}{Bulk
  and shear viscosities in lattice {B}oltzmann equations}, Phys. Rev. E 64
  (2001) 031203.

\bibitem{MATTILA_PRE_91_2015}
K.~K. Mattila, L.~A. Hegele, P.~C. Philippi,
  \href{http://link.aps.org/doi/10.1103/PhysRevE.91.063010}{Investigation of an
  entropic stabilizer for the lattice-{B}oltzmann method}, Phys. Rev. E 91
  (2015) 063010.

\bibitem{EZZATNESHAN_MCS_156_2019}
E.~Ezzatneshan,
  \href{http://www.sciencedirect.com/science/article/pii/S0378475418301976}{Comparative
  study of the lattice {B}oltzmann collision models for simulation of
  incompressible fluid flows}, Math. Comput. Simul. 156 (2019) 158 -- 177.

\bibitem{SHAN_PRE_100_2019}
X.~Shan,
  \href{https://link.aps.org/doi/10.1103/PhysRevE.100.043308}{Central-moment-based
  galilean-invariant multiple-relaxation-time collision model}, Phys. Rev. E
  100 (2019) 043308.

\bibitem{LI_PRE_100_2019}
X.~Li, Y.~Shi, X.~Shan,
  \href{https://link.aps.org/doi/10.1103/PhysRevE.100.013301}{Temperature-scaled
  collision process for the high-order lattice {B}oltzmann model}, Phys. Rev. E
  100 (2019) 013301.

\bibitem{FORTIN_IJNMF_24_1997}
A.~Fortin, M.~Jardak, J.~J. Gervais, R.~Pierre,
  \href{https://onlinelibrary.wiley.com/doi/abs/10.1002/%28SICI%291097-0363%2819970615%2924%3A11%3C1185%3A%3AAID-FLD535%3E3.0.CO%3B2-X}{{Localization
  of Hopf bifurcations in fluid flow problems}}, Int. J. Numer. Meth. Fl.
  24~(11) (1997) 1185--1210.

\bibitem{BRUNEAU_CF_35_2006}
C.-H. Bruneau, M.~Saad,
  \href{https://www.sciencedirect.com/science/article/pii/S0045793005000368}{{The
  2D lid-driven cavity problem revisited}}, Comput. Fluids 35~(3) (2006)
  326--348.

\bibitem{LADD_JFM_271_1994a}
A.~J.~C. Ladd, Numerical simulations of particulate suspensions via a
  discretized {B}oltzmann equation. {P}art 1. {T}heoretical foundation, J.
  Fluid Mech. 271 (1994) 285–309.

\bibitem{GHIA_JCP_48_1982}
U.~Ghia, K.~Ghia, C.~Shin,
  \href{http://www.sciencedirect.com/science/article/pii/0021999182900584}{{High-Re}
  solutions for incompressible flow using the {N}avier-{S}tokes equations and a
  multigrid method}, J. Comput. Phys. 48~(3) (1982) 387 -- 411.

\bibitem{STROUHAL_AP_241_1878}
V.~Strouhal,
  \href{https://onlinelibrary.wiley.com/doi/abs/10.1002/andp.18782411005}{{Ueber
  eine besondere Art der Tonerregung}}, Ann. Phys. (Berl.) 241~(10) (1878)
  216--251.

\bibitem{GERRARD_PPSSB_68_1955}
J.~H. Gerrard,
  \href{https://dx.doi.org/10.1088/0370-1301/68/7/307}{Measurements of the
  sound from circular cylinders in an air stream}, Proc. Phys. Soc. London,
  Sect. B 68~(7) (1955) 453.

\bibitem{INOUE_JFM_471_2002}
O.~Inoue, N.~Hatakeyama, Sound generation by a two-dimensional circular
  cylinder in a uniform flow, J. Fluid Mech. 471 (2002) 285–314.

\bibitem{INOUE_AIAA_2132_2001}
O.~Inoue, N.~Hatakeyama, H.~Hosoya, H.~Shoji,
  \href{https://arc.aiaa.org/doi/abs/10.2514/6.2001-2132}{Direct numerical
  simulation of {A}eolian tones}, {7th AIAA/CEAS Aeroacoustics Conference and
  Exhibit} (2001).

\bibitem{POINSOT_JCP_101_1992}
T.~Poinsot, S.~Lelef,
  \href{http://www.sciencedirect.com/science/article/pii/0021999192900462}{Boundary
  conditions for direct simulations of compressible viscous flows}, J. Comput.
  Phys. 101~(1) (1992) 104 -- 129.

\bibitem{WISSOCQ_JCP_331_2017}
G.~Wissocq, N.~Gourdain, O.~Malaspinas, A.~Eyssartier,
  \href{http://www.sciencedirect.com/science/article/pii/S0021999116306295}{Regularized
  characteristic boundary conditions for the lattice-{B}oltzmann methods at
  high {R}eynolds number flows}, J. Comput. Phys. 331 (2017) 1 -- 18.

\bibitem{FENG_PF_31_2019}
Y.~Feng, S.~Guo, J.~Jacob, P.~Sagaut,
  \href{https://doi.org/10.1063/1.5129138}{Solid wall and open boundary
  conditions in hybrid recursive regularized lattice {B}oltzmann method for
  compressible flows}, Phys. Fluids 31~(12) (2019) 126103.

\bibitem{CHEN_JCP_490_2023}
X.~Chen, K.~Yang, X.~Shan,
  \href{https://www.sciencedirect.com/science/article/pii/S0021999123003972}{{Characteristic
  boundary condition for multispeed lattice Boltzmann model in acoustic
  problems}}, J. Comput. Phys. 490 (2023).

\bibitem{ISHIDA_AIAA_0259_2016}
T.~Ishida, \href{https://arc.aiaa.org/doi/abs/10.2514/6.2016-0259}{Lattice
  {B}oltzmann method for aeroacoustic simulations with block-structured
  {C}artesian grid}, in: {54th AIAA Aerospace Sciences Meeting}, 2016.

\bibitem{LANDAU_Book_2nd_1987}
L.~Landau, E.~M. Lifshitz, Fluid Mechanics. Landau and Lifshitz: Course of
  Theoretical Physics, Volume 6, Butterworth-Heinemann Ltd, 1987.

\bibitem{PRESS_Book_3rd_2007}
W.~H. Press, S.~A. Teukolsky, W.~T. Vetterling, B.~P. Flannery, Numerical
  recipes: {T}he art of scientific computing, 3rd Edition, Cambridge University
  Press, 2007.

\bibitem{WILLIAMSON_JFM_206_1989}
C.~H.~K. Williamson, Oblique and parallel modes of vortex shedding in the wake
  of a circular cylinder at low reynolds numbers, J. Fluid Mech. 206 (1989)
  579–627.

\bibitem{VALAREZO_JA_30_1993}
W.~O. Valarezo, C.~J. Dominik, R.~J. McGhee,
  \href{https://doi.org/10.2514/3.46399}{{Multi-element airfoil performance due
  to Reynolds and Mach number variations}}, J. Aircraft 30~(5) (1993) 689--694.
\bibitem{PASCIONI_AIAA_3062_2014}
K.~Pascioni, L.~N. Cattafesta, M.~M. Choudhari,
  \href{https://arc.aiaa.org/doi/abs/10.2514/6.2014-3062}{An experimental
  investigation of the 30p30n multi-element high-lift airfoil}.

\bibitem{MURAYAMA_AIAA_2080_2014}
M.~Murayama, K.~Nakakita, K.~Yamamoto, H.~Ura, Y.~Ito, M.~M. Choudhari,
  \href{https://arc.aiaa.org/doi/abs/10.2514/6.2014-2080}{{Experimental study
  on slat noise from 30P30N three-element high-lift airfoil at JAXA hard-wall
  lowspeed wind tunnel}}.

\bibitem{MURAYAMA_AIAA_3460_2018}
M.~Murayama, Y.~Yokokawa, H.~Ura, K.~Nakakita, K.~Yamamoto, Y.~Ito,
  T.~Takaishi, R.~Sakai, K.~Shimoda, T.~Kato, T.~Homma,
  \href{https://arc.aiaa.org/doi/abs/10.2514/6.2018-3460}{Experimental study of
  slat noise from 30p30n three-element high-lift airfoil in jaxa kevlar-wall
  low-speed wind tunnel}.

\bibitem{VALAREZO_JA_32_1995}
W.~O. Valarezo, D.~J. Mavriplis,
  \href{https://doi.org/10.2514/3.46764}{Navier-stokes applications to
  high-lift airfoil analysis}, J. Aircraft 32~(3) (1995) 618--624.
\bibitem{RUMSEY_PAS_38_2002}
C.~L. Rumsey, S.~X. Ying,
  \href{https://www.sciencedirect.com/science/article/pii/S0376042102000039}{{Prediction
  of high lift: Review of present CFD capability}}, Prog. Aerosp. Sci. 38~(2)
  (2002) 145--180.

\bibitem{KHORRAMI_AIAA_2802_2004}
M.~Khorrami, M.~Choudhari, L.~Jenkins,
  \href{https://arc.aiaa.org/doi/abs/10.2514/6.2004-2802}{{Characterization of
  unsteady flow structures near leading-edge slat: Part II: 2D Computations}},
  {10th AIAA/CEAS Aeroacoustics Conference}.

\bibitem{CHOUDHARI_AIAA_2844_2015}
M.~M. Choudhari, D.~P. Lockard,
  \href{https://arc.aiaa.org/doi/abs/10.2514/6.2015-2844}{{Assessment of slat
  noise predictions for 30P30N high-lift configuration from BANC-III
  workshop}}, {21st AIAA/CEAS Aeroacoustics Conference} (2015).

\bibitem{HOUSMAN_AIAA_2963_2016}
J.~A. Housman, C.~C. Kiris,
  \href{https://arc.aiaa.org/doi/abs/10.2514/6.2016-2963}{Slat noise
  predictions using higher-order finite-difference methods on overset grids},
  2016.

\bibitem{MONTALA_PoF_36_2024}
R.~Montal\`{a}, O.~Lehmkuhl, I.~Rodriguez,
  \href{https://doi.org/10.1063/5.0182215}{{On the dynamics of the turbulent
  flow past a three-element wing}}, Phys. Fluids 36~(2) (2024) 025125.

\bibitem{ISHIDA_AIAA_2306_2019}
T.~Ishida, Aerodynamic simulations of a high-lift configuration by lattice
  {B}oltzmann method with block-structured {C}artesian grid, in: {AIAA Scitech
  2019 Forum}, 2019, p. 2306.

\bibitem{RUMSEY_JA_56_2018}
C.~L. Rumsey, J.~P. Slotnick, A.~J. Sclafani, Overview and summary of the third
  aiaa high lift prediction workshop, J. Aircraft 56~(2) (2018) 621--644.

\bibitem{MAEYAMA_CF_233_2022}
H.~Maeyama, T.~Imamura, J.~Osaka, N.~Kurimoto,
  \href{https://www.sciencedirect.com/science/article/pii/S0045793021003522}{{Unsteady
  aerodynamic simulations by the lattice Boltzmann method with near-wall
  modeling on hierarchical Cartesian grids}}, Comput. Fluids 233 (2022) 105249.

\bibitem{COREIXAS_Master_2014}
C.~Coreixas,
  \href{http://www.cerfacs.fr/~cfdbib/repository/WN_CFD_14_76.pdf}{Simulations
  a\'{e}roacoustiques aux grandes \'{e}chelles par les m\'{e}thodes lattice
  {B}oltzmann}, Master's thesis, ISAE-ENSICA Toulouse (2014).

\bibitem{GEIER_PRE_73_2006}
M.~Geier, A.~Greiner, J.~G. Korvink,
  \href{http://link.aps.org/doi/10.1103/PhysRevE.73.066705}{Cascaded digital
  lattice {B}oltzmann automata for high {R}eynolds number flow}, Phys. Rev. E
  73 (2006) 066705.

\bibitem{ASINARI_PRE_78_2008}
P.~Asinari,
  \href{https://link.aps.org/doi/10.1103/PhysRevE.78.016701}{Generalized local
  equilibrium in the cascaded lattice {B}oltzmann method}, Phys. Rev. E 78
  (2008) 016701.

\bibitem{FEI_PRE_97_2018}
L.~Fei, K.~H. Luo, Q.~Li,
  \href{https://link.aps.org/doi/10.1103/PhysRevE.97.053309}{Three-dimensional
  cascaded lattice {B}oltzmann method: Improved implementation and consistent
  forcing scheme}, Phys. Rev. E 97 (2018) 053309.

\bibitem{FEI_PoF_31_2019}
L.~Fei, J.~Du, K.~H. Luo, S.~Succi, M.~Lauricella, A.~Montessori, Q.~W~ang,
  \href{https://doi.org/10.1063/1.5087266}{Modeling realistic multiphase flows
  using a non-orthogonal multiple-relaxation-time lattice {B}oltzmann method},
  Phys. Fluids 31~(4) (2019) 042105.

\bibitem{SAITO_PRE_98_2018}
S.~Saito, A.~De~Rosis, A.~Festuccia, A.~Kaneko, Y.~Abe, K.~Koyama,
  \href{https://link.aps.org/doi/10.1103/PhysRevE.98.013305}{Color-gradient
  lattice {B}oltzmann model with nonorthogonal central moments: Hydrodynamic
  melt-jet breakup simulations}, Phys. Rev. E 98 (2018) 013305.

\bibitem{DEROSIS_PoF_31_2019}
A.~De~Rosis, R.~Huang, C.~Coreixas,
  \href{https://doi.org/10.1063/1.5124719}{Universal formulation of
  central-moments-based lattice {B}oltzmann method with external forcing for
  the simulation of multiphysics phenomena}, Phys. Fluids 31~(11) (2019)
  117102.

\bibitem{DEROSIS_PoF_32_2020}
A.~De~Rosis, C.~Coreixas,
  \href{https://aip.scitation.org/doi/10.1063/5.0026316}{Multiphysics flow
  simulations using {D3Q19} lattice {B}oltzmann methods based on central
  moments}, Phys. Fluids 32~(11) (2020) 117101.

\bibitem{WISSOCQ_JCP_450_2022}
G.~Wissocq, P.~Sagaut,
  \href{https://www.sciencedirect.com/science/article/pii/S0021999121007531}{{Hydrodynamic
  limits and numerical errors of isothermal lattice Boltzmann schemes}}, J.
  Comput. Phys. 450 (2022) 110858.

\bibitem{SUCCI_Book_2018}
S.~Succi,
  The Lattice {B}oltzmann Equation: For Complex States of Flowing Matter, Oxford
  University Press, 2018.

\bibitem{SAGAUT_CMA_59_2010}
P.~Sagaut,
  \href{http://www.sciencedirect.com/science/article/pii/S0898122109006385}{Toward
  advanced subgrid models for lattice-{B}oltzmann-based large-eddy simulation:
  Theoretical formulations}, Comput. Math. Appl. 59~(7) (2010) 2194 -- 2199.

\bibitem{DELLAR_JCP_259_2014}
P.~J. Dellar,
  \href{http://www.sciencedirect.com/science/article/pii/S0021999113007833}{Lattice
  {B}oltzmann algorithms without cubic defects in galilean invariance on
  standard lattices}, J. Comput. Phys. 259~(0) (2014) 270 -- 283.

\bibitem{SOD_JCP_27_1978}
G.~A. Sod,
  \href{http://www.sciencedirect.com/science/article/pii/0021999178900232}{A
  survey of several finite difference methods for systems of nonlinear
  hyperbolic conservation laws}, J. Comput. Phys. 27~(1) (1978) 1 -- 31.

\bibitem{JIANG_JCP_126_1996}
G.-S. Jiang, C.-W. Shu,
  \href{https://www.sciencedirect.com/science/article/pii/S0021999196901308}{Efficient
  implementation of weighted {ENO} schemes}, J. Comput. Phys. 126~(1) (1996)
  202--228.

\bibitem{BORGES_JCP_227_2008}
R.~Borges, M.~Carmona, B.~Costa, W.~S. Don,
  \href{https://www.sciencedirect.com/science/article/pii/S0021999107005232}{An
  improved weighted essentially non-oscillatory scheme for hyperbolic
  conservation laws}, J. Comput. Phys. 227~(6) (2008) 3191--3211.

\bibitem{GEROLYMOS_JCP_228_2009}
G.~Gerolymos, D.~S\'{e}n\'{e}chal, I.~Vallet,
  \href{https://www.sciencedirect.com/science/article/pii/S0021999109003908}{{Very-high-order
  WENO schemes}}, J. Comput. Phys. 228~(23) (2009) 8481--8524.

\bibitem{FU_CiCP_26_2019}
L.~Fu, \href{http://global-sci.org/intro/article_detail/cicp/13226.html}{{A
  hybrid method with TENO based discontinuity indicator for hyperbolic
  conservation laws}}, Commun. Comput. Phys. 26~(4) (2019) 973--1007.

\bibitem{DIRENZO_CPC_255_2020}
M.~{Di Renzo}, L.~Fu, J.~Urzay,
  \href{https://www.sciencedirect.com/science/article/pii/S0010465520300837}{{HTR
  solver: An open-source exascale-oriented task-based multi-GPU high-order code
  for hypersonic aerothermodynamics}}, Comput. Phys. Commun 255 (2020) 107262.

\bibitem{LUSHER_CPC_267_2021}
D.~J. Lusher, S.~P. Jammy, N.~D. Sandham,
  \href{https://www.sciencedirect.com/science/article/pii/S0010465521001752}{{OpenSBLI:
  Automated code-generation for heterogeneous computing architectures applied
  to compressible fluid dynamics on structured grids}}, Comput. Phys. Commun
  267 (2021) 108063.

\bibitem{NADIGA_JSP_81_1995}
B.~Nadiga, \href{https://link.springer.com/article/10.1007/BF02179972}{An euler
  solver based on locally adaptive discrete velocities}, J. Stat. Phys.
  81~(1-2) (1995) 129--146.

\bibitem{CHEN_CTP_52_2009}
F.~Chen, A.-G. Xu, G.-C. Zhang, Y.-B. Gan, T.~Cheng, Y.-J. Li,
  \href{http://stacks.iop.org/0253-6102/52/i=4/a=25}{Highly efficient lattice
  {B}oltzmann model for compressible fluids: Two-dimensional case}, Commun.
  Theor. Phys. 52~(4) (2009) 681.

\bibitem{WANG_IJMPC_21_2010}
Y.~Wang, Y.~L. He, Q.~Li, G.~H. Tang, W.~Q. Tao,
  \href{https://doi.org/10.1142/S0129183110015178}{Lattice boltzmann model for
  simulating viscous compressible flows}, Int. J. Mod. Phys. C 21~(03) (2010)
  383--407.

\bibitem{YANG_CF_79_2013}
L.~Yang, C.~Shu, J.~Wu,
  \href{http://www.sciencedirect.com/science/article/pii/S0045793013001187}{A
  moment conservation-based non-free parameter compressible lattice {B}oltzmann
  model and its application for flux evaluation at cell interface}, Comput.
  Fluids 79~(0) (2013) 190 -- 199.

\bibitem{GUO_PRE_91_2015}
Z.~Guo, R.~Wang, K.~Xu,
  \href{https://link.aps.org/doi/10.1103/PhysRevE.91.033313}{Discrete unified
  gas kinetic scheme for all {K}nudsen number flows. {II}. {T}hermal
  compressible case}, Phys. Rev. E 91 (2015) 033313.

\bibitem{LIU_PRE_98_2018}
H.~Liu, M.~Kong, Q.~Chen, L.~Zheng, Y.~Cao,
  \href{https://link.aps.org/doi/10.1103/PhysRevE.98.053310}{Coupled discrete
  unified gas kinetic scheme for the thermal compressible flows in all knudsen
  number regimes}, Phys. Rev. E 98 (2018) 053310.

\bibitem{DZANIC_JCP_486_2023}
T.~Dzanic, F.~Witherden, L.~Martinelli,
  \href{https://www.sciencedirect.com/science/article/pii/S0021999123002413}{{A
  positivity-preserving and conservative high-order flux reconstruction method
  for the polyatomic Boltzmann-BGK equation}}, J. Comput. Phys. 486 (2023)
  112146.

\bibitem{MATTILA_PF_29_2017}
K.~K. Mattila, P.~C. Philippi, L.~A. Hegele~Jr.,
  \href{10.1063/1.4981227}{High-order regularization in lattice-{B}oltzmann
  equations}, Phys. Fluids 29~(4) (2017) 046103.

\bibitem{SUN_PRE_58_1998}
C.~Sun,
  \href{http://link.aps.org/doi/10.1103/PhysRevE.58.7283}{Lattice-{B}oltzmann
  models for high speed flows}, Phys. Rev. E 58 (1998) 7283--7287.

\bibitem{SUN_TST_5_2000}
C.~Sun, B.~Wang, M.~Shen,
  \href{https://ieeexplore.ieee.org/abstract/document/6083247}{Adaptive lattice
  {B}oltzmann model for compressible flows}, Tsinghua Sci. Technol. 5~(1)
  (2000) 43--46.

\bibitem{LI_AIAA_4128_2015}
Y.~Li, A.~Jammalamadaka, P.~Gopalakrishnan, N.~Gopalaswamy, C.~Sun, R.~Zhang,
  H.~Chen, \href{10.2514/6.2015-4128}{Exploring an {LBM}-{VLES} based {CFD}
  approach for predictions of aero-thermal flows in generic turbo-machinery
  devices}, 51st AIAA/SAE/ASEE Jt. Propuls. Conf. (July 27-29 2015).

\bibitem{WISSOCQ_PoF_34_2022}
G.~Wissocq, T.~Coratger, G.~Farag, S.~Zhao, P.~Boivin, P.~Sagaut,
  \href{https://doi.org/10.1063/5.0083377}{Restoring the conservativity of
  characteristic-based segregated models: {A}pplication to the hybrid lattice
  {B}oltzmann method}, Phys. Fluids 34~(4) (2022) 046102.

\bibitem{THORIMBERT_JoCS_64_2022}
Y.~Thorimbert, D.~Lagrava, O.~Malaspinas, B.~Chopard, C.~Coreixas, J.~{de
  Santana Neto}, R.~Deiterding, J.~Latt,
  \href{https://www.sciencedirect.com/science/article/pii/S187775032200223X}{Local
  mesh refinement sensor for the lattice {B}oltzmann method}, J. Comput. Sci.
  64 (2022) 101864.

\bibitem{SHU_SIAM_51_2009}
C.-W. Shu, \href{https://doi.org/10.1137/070679065}{High order weighted
  essentially nonoscillatory schemes for convection dominated problems}, SIAM
  Rev. 51~(1) (2009) 82--126.

\bibitem{SCHULZRINNE_SIAM_14_1993}
C.~W. Schulz-Rinne, J.~P. Collins, H.~M. Glaz,
  \href{https://epubs.siam.org/doi/10.1137/0914082}{Numerical solution of the
  {R}iemann problem for two-dimensional gas dynamics}, SIAM J. Sci. Comput.
  14~(6) (1993) 1394--1414.
\bibitem{LAX_SIAM_19_1998}
P.~D. Lax, X.-D. Liu, \href{10.1137/S1064827595291819}{Solution of
  two-dimensional {R}iemann problems of gas dynamics by positive schemes}, SIAM
  J. Sci. Comput 19~(2) (1998) 319--340.

\bibitem{KURGANOV_NMPDE_18_2002}
A.~Kurganov, E.~Tadmor, \href{https://doi.org/10.1002/num.10025}{Solution of
  two-dimensional {R}iemann problems for gas dynamics without {R}iemann problem
  solvers}, Numer. Methods Partial Differ. Equ. 18~(5) (2002) 584--608.

\bibitem{SATOFUKA_GAMM_1987}
N.~Satofuka, K.~Morinishi, Y.~Nishida,
  \href{https://doi.org/10.1007/978-3-322-87873-1_12}{{Numerical solution of
  two-dimensional compressible Navier-Stokes equations using rational
  Runge-Kutta method}}, Vieweg+Teubner Verlag, Wiesbaden, 1987, pp. 201--218.

\bibitem{FRAPOLLI_PhD_2017}
N.~Frapolli,
  \href{https://www.research-collection.ethz.ch/handle/20.500.11850/130664}{Entropic
  lattice {B}oltzmann models for thermal and compressible flows}, Ph.D. thesis,
  ETH-Z\"{u}rich (2017).

\bibitem{KIM_IEEE_2011}
W.~Kim, M.~Voss, {Multicore desktop programming with Intel Threading Building
  Blocks}, {IEEE Software} 28~(1) (2011) 23--31.

\bibitem{BAUER_JCS_49_2021}
M.~Bauer, H.~K\"{o}stler, U.~R\"{u}de,
  \href{https://www.sciencedirect.com/science/article/pii/S1877750320305652}{lbmpy:
  {A}utomatic code generation for efficient parallel lattice {B}oltzmann
  methods}, J. Comput. Sci. 49 (2021) 101269.

\bibitem{BERGER_PhD_1982}
M.~J. Berger, Adaptive mesh refinement for hyperbolic partial differential
  equations, Ph.D. thesis (1982).

\bibitem{BERGER_JCP_53_1984}
M.~J. Berger, J.~Oliger,
  \href{https://www.sciencedirect.com/science/article/pii/0021999184900731}{Adaptive
  mesh refinement for hyperbolic partial differential equations}, J. Comput.
  Phys. 53~(3) (1984) 484--512.

\bibitem{BERGER_JCP_82_1989}
M.~Berger, P.~Colella,
  \href{https://www.sciencedirect.com/science/article/pii/0021999189900351}{Local
  adaptive mesh refinement for shock hydrodynamics}, J. Comput. Phys. 82~(1)
  (1989) 64--84.

\bibitem{BERGER_SIAM_35_1998}
M.~J. Berger, R.~J. LeVeque,
  \href{https://doi.org/10.1137/S0036142997315974}{Adaptive mesh refinement
  using wave-propagation algorithms for hyperbolic systems}, SIAM J. Numer.
  Anal. 35~(6) (1998) 2298--2316.

\bibitem{BALSARA_JCP_174_2001}
D.~S. Balsara,
  \href{https://www.sciencedirect.com/science/article/pii/S0021999101969177}{{Divergence-free
  adaptive mesh refinement for magnetohydrodynamics}}, Journal of Computational
  Physics 174~(2) (2001) 614--648.

\bibitem{DEITERDING_JPCS_753_2016}
R.~Deiterding, S.~L. Wood,
  \href{https://doi.org/10.1088/1742-6596/753/8/082005}{Predictive wind turbine
  simulation with an adaptive lattice {B}oltzmann method for moving
  boundaries}, J. Phys.: Conf. Ser. 753 (2016) 082005.

\bibitem{ARSLANBEKOV_PRE_88_2013}
R.~R. Arslanbekov, V.~I. Kolobov, A.~A. Frolova,
  \href{https://link.aps.org/doi/10.1103/PhysRevE.88.063301}{Kinetic solvers
  with adaptive mesh in phase space}, Phys. Rev. E 88 (2013) 063301.

\bibitem{BARANGER_JCP_257_2014}
C.~Baranger, J.~Claudel, N.~Hérouard, L.~Mieussens,
  \href{http://www.sciencedirect.com/science/article/pii/S0021999113006827}{Locally
  refined discrete velocity grids for stationary rarefied flow simulations}, J.
  Comput. Phys. 257 (2014) 572 -- 593.

\bibitem{BRULL_JCP_266_2014}
S.~Brull, L.~Mieussens,
  \href{https://www.sciencedirect.com/science/article/pii/S0021999114001016}{Local
  discrete velocity grids for deterministic rarefied flow simulations}, J.
  Comput. Phys. 266 (2014) 22--46.

\bibitem{SHARMA_AST_141_2023}
P.~K. Sharma, R.~Deiterding, A.~Cerminara, N.~Sandham,
  \href{https://www.sciencedirect.com/science/article/pii/S1270963823004789}{Numerical
  simulation of transpiration cooling for a high-speed boundary layer
  undergoing transition to turbulence}, Aerospace Science and Technology 141
  (2023) 108581.

\bibitem{ZHAO_AST_136_2023}
W.~Zhao, R.~Deiterding, J.~Liang, X.~Wang, X.~Cai, J.~Duell,
  \href{https://www.sciencedirect.com/science/article/pii/S1270963823001025}{Adaptive
  simulations of flame acceleration and detonation transition in subsonic and
  supersonic mixtures}, Aerospace Science and Technology 136 (2023) 108205.

\bibitem{LOHNER_CMAME_61_1987}
R.~L\"{o}hner,
  \href{https://www.sciencedirect.com/science/article/pii/0045782587900983}{{An
  adaptive finite element scheme for transient problems in CFD}}, Computer
  Methods in Applied Mechanics and Engineering 61~(3) (1987) 323--338.

\bibitem{HARTMANN_SIAM_24_2003}
R.~Hartmann, P.~Houston,
  \href{https://doi.org/10.1137/S1064827501389084}{Adaptive discontinuous
  galerkin finite element methods for nonlinear hyperbolic conservation laws},
  SIAM Journal on Scientific Computing 24~(3) (2003) 979--1004.

\bibitem{MOSSIER_JoCS_97_2023}
P.~Mossier, D.~Appel, A.~D. Beck, C.-D. Munz,
  \href{https://doi.org/10.1007/s10915-023-02363-7}{An efficient hp-adaptive
  strategy for a level-set ghost-fluid method}, Journal of Scientific Computing
  97~(2) (2023).

\bibitem{MOSSIER_JCP_520_2025}
P.~Mossier, S.~J\"{o}ns, S.~Chiocchetti, A.~D. Beck, C.-D. Munz,
  \href{https://www.sciencedirect.com/science/article/pii/S0021999124007629}{{Numerical
  simulation of phase transition with the hyperbolic Godunov-Peshkov-Romenski
  model}}, Journal of Computational Physics 520 (2025) 113514.

\bibitem{SCHORNBAUM_PhD_2018}
F.~Schornbaum,
  \href{https://www10.cs.fau.de/publications/dissertations/Diss{\_}2018-Schornbaum.pdf}{Block-structured
  adaptive mesh refinement for simulations on extreme-scale supercomputers},
  Ph.D. thesis, Friedrich-Alexander-Universit\"{a}t Erlangen-N\"{u}rnberg
  (2018).

\bibitem{CHEN_JCP_231_2012b}
S.~Chen, K.~Xu, C.~Lee, Q.~Cai,
  \href{https://www.sciencedirect.com/science/article/pii/S0021999112002616}{A
  unified gas kinetic scheme with moving mesh and velocity space adaptation},
  J. Comput. Phys. 231~(20) (2012) 6643--6664.

\bibitem{XIAO_JCP_415_2020}
T.~Xiao, C.~Liu, K.~Xu, Q.~Cai,
  \href{https://www.sciencedirect.com/science/article/pii/S0021999120303090}{A
  velocity-space adaptive unified gas kinetic scheme for continuum and rarefied
  flows}, J. Comput. Phys. 415 (2020) 109535.

\bibitem{DEITERDING_HPCC_2005}
L.~T. Yang, O.~F. Rana, B.~Di~Martino, J.~Dongarra (Eds.), {Detonation
  structure simulation with AMROC}, Springer Berlin Heidelberg, Berlin,
  Heidelberg, 2005.

\bibitem{WISSINK_SC_2001}
A.~M. Wissink, R.~D. Hornung, S.~R. Kohn, S.~S. Smith, N.~Elliott,
  \href{https://doi.org/10.1145/582034.582040}{{Large scale parallel structured
  AMR calculations using the SAMRAI framework}}, in: Proceedings of the 2001
  ACM/IEEE Conference on Supercomputing, SC '01, Association for Computing
  Machinery, New York, NY, USA, 2001, p.~6.

\bibitem{VINOD_AIAA_0028_2018}
V.~K. Lakshminarayan, J.~Sitaraman, B.~Roget, A.~M. Wissink,
  \href{https://arc.aiaa.org/doi/abs/10.2514/6.2018-0028}{Simulation of complex
  geometries using automatically generated strand meshes}, 2018.

\bibitem{CONVERGE_WEBSITE}
{CONVERGE CFD: https://convergecfd.com/}.
\newblock \href{https://convergecfd.com/}{[link]}.

\bibitem{XFLOW_WEBSITE}
{XFlow:
  https://www.3ds.com/products-services/simulia/products/xflow/latest-release/}.
\newblock
  \href{https://www.3ds.com/products-services/simulia/products/xflow/latest-release/}{[link]}.

\bibitem{ASMUTH_WES_5_2020}
H.~Asmuth, H.~Olivares-Espinosa, S.~Ivanell,
  \href{https://wes.copernicus.org/articles/5/623/2020/}{Actuator line simulations 
  of wind turbine wakes using the lattice {B}oltzmann method}, Wind Energ. Sci. 5 
  (2020), 623--645.
  
\bibitem{WISSOCQ_PRE_101_2020}
G.~Wissocq, J.-F. Boussuge, P.~Sagaut,
  \href{https://link.aps.org/doi/10.1103/PhysRevE.101.043306}{Consistent vortex
  initialization for the athermal lattice {B}oltzmann method}, Phys. Rev. E 101
  (2020) 043306.

\bibitem{BAILEY_ICPP_2009}
P.~Bailey, J.~Myre, S.~D. Walsh, D.~J. Lilja, M.~O. Saar,
  \href{10.1109/ICPP.2009.38}{Accelerating lattice {B}oltzmann fluid flow
  simulations using graphics processors}, in: International Conference on
  Parallel Processing, 2009, pp. 550--557.

\bibitem{LATT_ARXIV_09242_2025}
J.~Latt and C.~Coreixas,
  \href{https://arxiv.org/abs/2506.09242}{{Multi-GPU acceleration of PALABOS fluid solver using C++ standard parallelism}},
  arXiv preprint (2025) 2506.09242.

\bibitem{BOUZIDI_PoF_13_2001}
M.~Bouzidi, M.~Firdaouss, P.~Lallemand,
  \href{https://doi.org/10.1063/1.1399290}{{Momentum transfer of a
  Boltzmann-lattice fluid with boundaries}}, Phys. Fluids 13~(11) (2001)
  3452--3459.

\bibitem{MALASPINAS_JCP_275_2014}
O.~Malaspinas, P.~Sagaut,
  \href{http://www.sciencedirect.com/science/article/pii/S0021999114004276}{Wall
  model for large-eddy simulation based on the lattice {B}oltzmann method}, J.
  Comput. Phys. 275~(0) (2014) 25 -- 40.

\bibitem{SENGISSEN_AIAA_2993_2015}
A.~Sengissen, J.-C. Giret, C.~Coreixas, J.-F. Boussuge,
  \href{10.2514/6.2015-2993}{Simulations of {LAGOON} landing-gear noise using
  lattice {B}oltzmann solver}, in: 21st AIAA/CEAS Aeroacoustics Conference,
  2015, p. 2993.

\bibitem{POPE_Book_2000}
S.~B. Pope, \href{https://doi.org/10.1017/CBO9780511840531}{Turbulent Flows},
  Cambridge University Press, 2000.

\bibitem{HE_JCP_146_1998}
X.~He, S.~Chen, G.~D. Doolen,
  \href{http://www.sciencedirect.com/science/article/pii/S0021999198960570}{A
  novel thermal model for the lattice {B}oltzmann method in incompressible
  limit}, J. Comput. Phys. 146~(1) (1998) 282 -- 300.

\bibitem{COREIXAS_PhD_2018}
C.~Coreixas, \href{http://oatao.univ-toulouse.fr/19861/}{High-order extension
  of the recursive regularized lattice {B}oltzmann method}, Ph.D. thesis, {INP
  Toulouse} (2018).

\bibitem{GRADa_CPAM_2_1949}
H.~Grad, \href{10.1002/cpa.3160020402}{Note on {N}-dimensional {H}ermite
  polynomials}, Commun. Pure Appl. Math. 2~(4) (1949) 325--330.

\bibitem{SHAN_IJMPC_18_2007}
X.~Shan, H.~Chen, A general multiple-relaxation-time {B}oltzmann collision
  model, Int. J. Mod. Phys. C 18~(04) (2007) 635--643.

\bibitem{NIE_PRE_77_2008}
X.~Nie, X.~Shan, H.~Chen,
  \href{http://link.aps.org/doi/10.1103/PhysRevE.77.035701}{Thermal lattice
  {B}oltzmann model for gases with internal degrees of freedom}, Phys. Rev. E
  77 (2008) 035701.

\bibitem{WISSOCQ_JCP_380_2019}
G.~Wissocq, P.~Sagaut, J.-F. Boussuge,
  \href{http://www.sciencedirect.com/science/article/pii/S0021999118308118}{An
  extended spectral analysis of the lattice {B}oltzmann method: modal
  interactions and stability issues}, J. Comput. Phys. 380 (2019) 311 -- 333.

\bibitem{COREIXAS_ICMMES_Collision_2021}
C.~Coreixas, B.~Chopard, J.~Latt,
  \href{https://doi.org/10.13140/RG.2.2.13457.35684/1}{{Collision models in the
  lattice Boltzmann framework: Accuracy, stability, and performance comparisons
  on standard lattices}}, in: {17th International Conference for Mesoscopic
  Methods in Engineering and Science (ICMMES)}, 2021.

\bibitem{RENARD_JCP_446_2021}
F.~Renard, G.~Wissocq, J.-F. Boussuge, P.~Sagaut,
  \href{https://www.sciencedirect.com/science/article/pii/S0021999121005441}{A
  linear stability analysis of compressible hybrid lattice boltzmann methods},
  J. Comput. Phys. 446 (2021) 110649.

\bibitem{MIEUSSENS_MMMAS_10_2000}
L.~Mieussens, \href{https://doi.org/10.1142/S0218202500000562}{Discrete
  velocity model and implicit scheme for the {BGK} equation of rarefied gas
  dynamics}, Math. Models Methods Appl. Sci. 10~(08) (2000) 1121--1149.

\bibitem{HOU_AIAA_2019_2555}
Y.~Hou, D.~Angland, A.~Sengissen, A.~Scotto, Lattice-{B}oltzmann and
  navier-stokes simulations of the partially dressed, cavity-closed nose
  landing gear benchmark case, in: 25th AIAA/CEAS Aeroacoustics Conference,
  2019, p. 2555.

\bibitem{GUO_JCP_418_2020}
S.~Guo, Y.~Feng, J.~Jacob, F.~Renard, P.~Sagaut,
  \href{http://www.sciencedirect.com/science/article/pii/S0021999120303442}{An
  efficient lattice {B}oltzmann method for compressible aerodynamics on {D3Q19}
  lattice}, J. Comput. Phys. 418 (2020) 109570.

\bibitem{NGUYEN_IJHMT_212_2023}
M.~Nguyen, J.-F. Boussuge, P.~Sagaut, J.-C. Larroya-Huguet,
  \href{https://www.sciencedirect.com/science/article/pii/S0017931023004088}{{Large
  eddy simulation of a row of impinging jets with upstream crossflow using the
  lattice Boltzmann method}}, Int. J. Heat Mass Tran. 212 (2023) 124256.

\bibitem{TAHA_PhD_2023}
M.~Taha, \href{http://www.theses.fr/2023AIXM0057}{{Modelling fire-induced flows
  using Lattice Boltzmann methods}}, Ph.D. thesis, Aix-Marseille Universit\'{e}
  (2023).

\bibitem{MOZAFFARI_JCP_514_2024}
S.~Mozaffari, S.-G. Cai, J.~Jacob, P.~Sagaut,
  \href{https://www.sciencedirect.com/science/article/pii/S0021999124005175}{{Lattice
  Boltzmann k-w SST based hybrid RANS/LES simulations of turbulent flows}}, J.
  Comput. Phys. 514 (2024) 113269.

\bibitem{DAVILLER_AIAA_3180_2024}
G.~Daviller, E.~Charles, J.~F. Boussuge, F.~Renard, J.~Huber,
  \href{https://arc.aiaa.org/doi/abs/10.2514/6.2024-3180}{{Investigation of
  jet-pylon interaction noise using LBM}}, {30th AIAA/CEAS Aeroacoustics
  Conference} (2024).

\bibitem{GONZALEZMARTINO_AIAA_2585_2019}
I.~Gonzalez-Martino, D.~Casalino,
  \href{https://arc.aiaa.org/doi/abs/10.2514/6.2019-2585}{Noise from a rotor
  ingesting a turbulent boundary layer using very-large eddy simulations}, in:
  {25th AIAA/CEAS Aeroacoustics Conference}, 2019, p. 2585.

\bibitem{TERUNA_AIAA_2264_2021}
C.~Teruna, F.~Avallone, D.~Ragni, D.~Casalino, A.~R. Carpio,
  \href{https://arc.aiaa.org/doi/abs/10.2514/6.2021-2264}{Numerical study on
  trailing-edge noise attenuation using {3D}-printed porous insert} (2021).

\bibitem{GRADb_CPAM_2_1949}
H.~Grad, \href{10.1002/cpa.3160020403}{On the kinetic theory of rarefied
  gases}, Commun. Pure Appl. Math. 2~(4) (1949) 331--407.

\bibitem{OTTINGER_RSTA_378_2020}
H.~C. \"{O}ttinger, H.~Struchtrup, M.~Torrilhon,
  \href{https://royalsocietypublishing.org/doi/abs/10.1098/rsta.2019.0174}{Formulation
  of moment equations for rarefied gases within two frameworks of
  non-equilibrium thermodynamics: {RET} and {GENERIC}}, Philos. Trans. R. Soc.
  A 378~(2170) (2020) 20190174.

\bibitem{KORNREICH_PD_69_1993}
P.~J. Kornreich, J.~Scalo,
  \href{http://www.sciencedirect.com/science/article/pii/016727899390097K}{Supersonic
  lattice gases: {R}estoration of {G}alilean invariance by nonlinear resonance
  effects}, Physica D 69~(3) (1993) 333 -- 344.

\bibitem{LEVERMORE_JSP_83_1996}
C.~D. Levermore, \href{https://doi.org/10.1007/BF02179552}{Moment closure
  hierarchies for kinetic theories}, J. Stat. Phys. 83~(5) (1996) 1021--1065.

\bibitem{LETALLEC_TechReport_1997}
P.~Le~Tallec, J.-P. Perlat,
  \href{https://hal.inria.fr/inria-00073565}{Numerical analysis of
  {L}evermore's moment system}, Research Report {RR-3124}, {INRIA}, projet
  {M3N} (1997).

\bibitem{DUBROCA_ESAIM_10_2001}
B.~Dubroca, L.~Mieussens,
  \href{https://hal.archives-ouvertes.fr/hal-00385476}{A conservative and
  entropic discrete-velocity model for rarefied polyatomic gases}, {ESAIM}:
  {P}roceedings 10 (2001) 127--139.

\bibitem{GSL_WEBSITE}
{GSL library multi-root solver:
  https://www.gnu.org/software/gsl/doc/html/multiroots.html}.
\newblock
  \href{https://www.gnu.org/software/gsl/doc/html/multiroots.html}{[link]}.

\bibitem{CHU_PoF_8_1965}
C.~K. Chu,
  \href{https://aip.scitation.org/doi/abs/10.1063/1.1761077}{Kinetic-theoretic
  description of the formation of a shock wave}, The Physics of Fluids 8~(1)
  (1965) 12--22.

\bibitem{RYKOV_FD_10_1975}
V.~A. Rykov, \href{https://doi.org/10.1007/BF01023275}{A model kinetic equation
  for a gas with rotational degrees of freedom}, Fluid Dyn. 10~(6) (1975)
  959--966.

\bibitem{TITAREV_ECCOMAS_2006}
V.~Titarev,
  \href{http://resolver.tudelft.nl/uuid:a4f5da7c-4104-4ade-bd9c-77a0e3526221}{Conservative
  numerical methods for advanced model kinetic equations}, in: ECCOMAS CFD
  2006: Proceedings of the European Conference on Computational Fluid Dynamics,
  Egmond aan Zee, The Netherlands, September 5-8, 2006, 2006.

\bibitem{TAM_JCP_107_1993}
C.~K. Tam, J.~C. Webb,
  \href{http://www.sciencedirect.com/science/article/pii/S0021999183711423}{Dispersion-relation-preserving
  finite difference schemes for computational acoustics}, J. Comput. Phys.
  107~(2) (1993) 262 -- 281.

\bibitem{LEBRAS_AIAAJ_55_2017}
S.~Le~Bras, H.~Deniau, C.~Bogey, G.~Daviller,
  \href{https://doi.org/10.2514/1.J055107}{Development of compressible
  large-eddy simulations combining high-order schemes and wall modeling}, AIAA
  J. 55~(4) (2017) 1152--1163.

\bibitem{GARNIER_JCP_153_1999}
E.~Garnier, M.~Mossi, P.~Sagaut, P.~Comte, M.~Deville,
  \href{http://www.sciencedirect.com/science/article/pii/S002199919996268X}{On
  the use of shock-capturing schemes for large-eddy simulation}, J. Comput.
  Phys. 153~(2) (1999) 273 -- 311.

\bibitem{GEIER_EPJST_171_2009}
M.~Geier, A.~Greiner, J.~G. Korvink, \href{https://link.springer.com/article/10.1140/epjst/e2009-01011-1}{A
  factorized central moment lattice {B}oltzmann method}, Eur. Phys. J. Spec.
  Top. 171~(1) (2009) 55--61.

\bibitem{CHAVEZ_Energies_13_2020}
M.~Ch\'{a}vez-Modena, J.~L. Mart\'{i}nez, J.~A. Cabello, E.~Ferrer,
  \href{https://www.mdpi.com/1996-1073/13/19/5146}{Simulations of aerodynamic
  separated flows using the lattice {B}oltzmann solver {XF}low}, Energies
  13~(19) (2020).

\bibitem{WISSOCQ_PhD_2019}
G.~Wissocq, \href{https://www.theses.fr/2019AIXM0635}{Investigating lattice
  {B}oltzmann methods for turbomachinery secondary air system simulations},
  Ph.D. thesis, Aix-Marseille Universit\'{e} (2019).

\bibitem{MASSET_JFM_897_2020}
P.-A. Masset, G.~Wissocq, \href{https://doi.org/10.1017/jfm.2020.374}{Linear
  hydrodynamics and stability of the discrete velocity boltzmann equations}, J.
  Fluid Mech. 897 (2020) A29.

\bibitem{QIAN_JSC_8_1993}
Y.~Qian, \href{https://doi.org/10.1007/BF01060932}{Simulating
  thermohydrodynamics with lattice {BGK} models}, J. Sci. Comput. 8~(3) (1993)
  231--242.

\bibitem{LALLEMAND_PRE_68_2003}
P.~Lallemand, L.-S. Luo,
  \href{http://link.aps.org/doi/10.1103/PhysRevE.68.036706}{Theory of the
  lattice {B}oltzmann method: Acoustic and thermal properties in two and three
  dimensions}, Phys. Rev. E 68 (2003) 036706.

\bibitem{LALLEMAND_CMA_65_2013}
P.~Lallemand, F.~Dubois,
  \href{http://www.sciencedirect.com/science/article/pii/S0898122112006530}{Some
  results on energy-conserving lattice {B}oltzmann models}, Comput. Math. Appl.
  65~(6) (2013) 831 -- 844, mesoscopic Methods in Engineering and Science.

\bibitem{LALLEMAND_CCP_17_2015}
P.~Lallemand, F.~Dubois,
  \href{http://journals.cambridge.org/article_S1815240615000389}{Comparison of
  simulations of convective flows}, Commun. Comput. Phys. 17 (2015) 1169--1184.

\bibitem{RICOT_JCP_228_2009}
D.~Ricot, S.~Mari\'{e}, P.~Sagaut, C.~Bailly,
  \href{http://www.sciencedirect.com/science/article/pii/S0021999109001399}{Lattice
  {B}oltzmann method with selective viscosity filter}, J. Comput. Phys.
  228~(12) (2009) 4478 -- 4490.

\end{thebibliography}

\end{document}